\documentclass[aoas,11pt]{imsart}
\usepackage{amsmath}
\usepackage{amsfonts}
\usepackage{amssymb}
\usepackage{amsfonts}
\usepackage{amssymb}
\usepackage{amsthm}
\usepackage{mathrsfs}
\usepackage[pdftex]{graphicx}
\usepackage{enumerate}
\usepackage[numbers]{natbib}
\usepackage[colorlinks,citecolor=blue,urlcolor=blue]{hyperref}
\usepackage{caption}
\usepackage{subcaption}
\RequirePackage[OT1]{fontenc}
\allowdisplaybreaks
\usepackage[top=1.2in, bottom=1.2in, left=1.2in, right=1.2in]{geometry}

\startlocaldefs
\newcommand{\argmin}{\operatornamewithlimits{arg \,min}}

\DeclareMathOperator{\tr}{tr}
\numberwithin{equation}{section}
\theoremstyle{plain}

\begin{document}

\begin{frontmatter}

\title{A random effects stochastic block model for joint community detection in multiple networks with applications to neuroimaging \thanksref{T1}}

\runtitle{Random effects stochastic block model}
\thankstext{T1}{
This work was supported in part by National Science Foundation grants DMS-1406455 and DMS-1830412. This work made use of the Illinois Campus Cluster, a computing resource that is operated by the Illinois Campus Cluster Program (ICCP) in conjunction with the National Center for Supercomputing Applications (NCSA), which is supported by funds from the University of Illinois at Urbana-Champaign.}

\author{\fnms{Subhadeep} \snm{Paul} \thanksref{m1}\ead[label=e1]{paul.963@osu.edu}}

\and
\author{\fnms{Yuguo} \snm{Chen} \thanksref{m2} \ead[label=e2]{yuguo@illinois.edu}}
\affiliation{The Ohio State University \thanksmark{m1} and University of Illinois at Urbana-Champaign \thanksmark{m2}}

\runauthor{S. Paul and Y. Chen}

\address{Subhadeep Paul\\Department of Statistics\\ The Ohio State University\\
Columbus, OH 43210, USA \\
\printead{e1}\\
}

\address{Yuguo Chen \\ Department of Statistics\\ University of Illinois at Urbana-Champaign\\
Champaign, IL 61820, USA \\
\printead{e2}\\
}

\begin{abstract}

To analyze data from multi-subject experiments in neuroimaging studies, we develop a modeling framework for joint community detection in a group of related networks that can be considered as a sample from a population of networks. The proposed random effects stochastic block model facilitates the study of group differences and subject-specific variations in the community structure. The model proposes a putative mean community structure which is representative of the group or the population under consideration, but is not the community structure of any individual component network. Instead, the community memberships of nodes vary in each component network with a transition matrix, thus modeling the variation in community structure across a group of subjects. To estimate the quantities of interest, we propose two methods: a variational EM algorithm, and a model-free ``two-step" method called Co-OSNTF which is based on non-negative matrix factorization. We also develop a resampling-based hypothesis test for differences between community structure in two populations both at the whole network level and node level. The methodology is applied to the COBRE dataset, a publicly available fMRI dataset from multi-subject experiments involving schizophrenia patients. Our methods reveal an overall putative community structure representative of the group as well as subject-specific variations within each of the two groups, healthy controls and schizophrenia patients. The model has good predictive ability for predicting community structure in subjects from the same population but outside the training sample. Using our network level hypothesis tests we are able to ascertain statistically significant difference in community structure between the two groups, while our node level tests help determine the nodes that are driving the difference.

\end{abstract}



\end{frontmatter}

\section{Introduction: Networks and neuroimaging data analysis}

Network analysis has received a plethora of multi-disciplinary interest in the last few decades due to its various scientific and industrial applications in a variety of fields including genetics, neuroscience, ecology, economics and social sciences. A rapidly growing application area of network science is in the analysis of neuroimaging data, where it is used to analyze anatomical and functional connectivity among the brain regions (See \citet{rubinov10,bullmore09,hutchison13,sporns2014contributions,fornito2013graph,stam2014modern,fornito2015connectomics} for reviews). In network neuroscience, a typical approach is to construct functional brain networks based on measures of inter-regional associations obtained from various sources of measurements, including the functional magnetic resonance imaging (fMRI) with blood oxygen level dependent (BOLD) signals \citep{bassett11,simpson13}. Modern neuroimaging experiments typically involve multiple subjects and/or multiple trials across the subjects. Various properties of the functional network (e.g., modularity, connectivity, degree, rich club organization, clustering coefficient, average path length, small-world property) are then investigated and contrasted among the subjects or groups of subjects \citep{bullmore09, van10}. The inter-subject and inter-group variations in many such network metrics have been related to cognitive ability and diseases in the literature \citep{bassett11,braun15, stevens12,hutchison13,jones12,wang13,braun15,yu2012brain,lynall2010functional,alexander2012discovery}.

A community or module of vertices in a network is defined as a group of vertices which are more connected among themselves than they are to the rest of the network. Many real world networks, including the functional brain networks, are known to exhibit community structure, whereby the vertices in the network can be roughly divided into such communities or modules.  Several authors have uncovered intrinsic module structures in the functional organization of brain regions through network based analysis of spontaneous neuronal activity in resting state fMRI \citep{he2009uncovering,meunier2010modular,power2011functional,yu2012modular,moussa12}. The identified modules are consistent with several functionally connected subsystems generating spontaneous activities, e.g., motor functions, auditory, visual, attention and default mode.

 However, due to physiological differences among the subjects, responses to changing environmental conditions, and variations in imaging instruments, the measured networks and consequently the network community structures vary from subject to subject within a group of subjects or even from trial to trial within a subject \citep{stevens12,simpson13perm,moussa12,weber13}. In neuroimaging studies, often researchers are interested in studying brain functional connectivity patterns in two groups of subjects, one is a group of patients diagnosed with a certain condition and the other group is healthy controls. In such studies jointly analyzing these multi-subject networks and comparing between the two groups of networks is of interest \cite{zalesky2010network,ginestet2013statistical,chen2015parsimonious,stam2014modern,fornito2015connectomics}. In a typical experimental setup, the subjects in the two groups serve as random samples from these two populations. Hence to facilitate comparison of populations beyond those in terms of single-value network summary measures (e.g., modularity), it is important to build statistical models for a random sample of networks from a population of networks.

Unfortunately most of the literature on networks deal with a single instance of a network. This is primarily due to the fact that network data collection usually involves observing a network at one time point or tracking its evolution over time. However modern application of networks in neuroscience brings a unique challenge and opportunity in terms of multiple instances of interactions among the same set of nodes through measurements on multiple subjects. The central question then is how to quantify the uncertainty in community structure due to subject specific variations within a group of subjects, so that two groups (populations) of subjects can be statistically compared.

Simultaneously the problem of statistically testing for differences in community structure has received considerable attention recently in the neuroimaging literature \citep{alexander2012discovery,gadelkarim2012framework,fujita2014non,glerean2016reorganization,kujala2016graph}. These papers argue that differences between populations might not be well captured using a single network property measure like modularity, but it might be more meaningful to look at some measure of how different the module structures in the populations are. A permutation test using the average of Normalized Mutual Information (NMI) between pairs of network community assignments was proposed in \citet{alexander2012discovery}. \citet{gadelkarim2012framework} used a node-wise community consistency measure between two community partitions, called ``Scaled Inclusivity (SI)" defined in \citet{steen2011assessing}, to assess how consistent a node's module is in a subject network with a mean module assignment for the whole group (obtained from community detection in the mean connectivity matrix). \citet{gadelkarim2012framework} then proposed statistical hypothesis tests based on the SI vectors for the two groups. \citet{glerean2016reorganization} first obtained a consensus partition across the group using a ``clustering of clusters" technique similar in function to module allegiance method in \citet{braun15}, and then determined SI of nodes across subjects with this consensus. The more general problem of hypothesis testing involving networks in functional neuroimaging has also been addressed in the literature \citep{ginestet2013statistical,ginestet2014hypothesis,narayan2016mixed,chen2015parsimonious}.

We approach these problems by developing methodology in the spirit of random effects models, which will help us separate systematic variations between two populations of networks  from variations due to subject specific ``noise". In particular we propose a random effects stochastic block model (RESBM), parameterized by a putative mean community assignment matrix, a transition probability matrix, and appropriate block parameters.
We develop two estimation strategies to estimate the parameters and other quantities of interest, a variational EM algorithm and a model-free two-step approach based on non-negative matrix factorization (NMF). Finally, we develop resampling based two-sample hypothesis tests to compare two populations of networks in terms of their network level and node level community structures.

 The rest of the article is organized as follows. In Section 2 we describe the application to schizophrenia data. In Section 3 we describe the random effects stochastic block model, the estimation strategies, and the two-sample hypothesis tests. Section 4 studies the estimation and inference methods in simulated networks under several scenarios and several metrics. Finally in Section 5 we present our results on the schizophrenia data.

\section{Application to a resting state fMRI study on schizophrenia: COBRE dataset}

We apply the methods developed in this article to a publicly available dataset from resting state fMRI experiment performed on subjects diagnosed with schizophrenia along with healthy controls. Network analysis is a key tool employed in analysis of functional connectivity in fMRI based experiments on schizophrenia \citep{lynall2010functional,liu2008disrupted,bassett2012altered,yu2012brain,van2010aberrant,van2013abnormal,alexander2010disrupted,alexander2012discovery}. Analysis of modular organization (community structure) of brain networks is of particular importance in understanding schizophrenia, since it has been hypothesized that schizophrenia is associated with neurodevelopment and evolution of brain, which in turn is influenced by modular organization (see \citet{yu2012brain} and references therein).

The dataset we analyze is the COBRE dataset publicly available to download from International Neuroimaging Data-sharing Initiative (INDI, 1000 Functional Connectomes project,  \url{http://fcon_1000.projects.nitrc.org/indi/retro/cobre.html}), that consists of anatomical magnetic resonance and resting state functional magnetic resonance scans from 72 patients diagnosed with schizophrenia and 75 healthy controls with ages ranging from 18 to 65 years. A detailed description of experimental conditions and equipment is available in the aforementioned webpage.

We used an automated preprocessing pipeline (Statistical Parametric Mapping's (SPM) default preprocessing pipeline for volume based analyses \citep{penny2011statistical}) implemented in Matlab toolbox CONN \citep{whitfield2012conn}  with parameters similar to earlier studies in \citet{lynall2010functional} and \citet{bassett2012altered}. In particular the steps included, deleting first three volumes, correcting for head motion by functional realignment and unwarping, functional slice timing correction, functional and structural image co-registration, structural segmentation and normalization, functional normalization to the standard Montreal Neurological Institute (MNI) space, functional outlier detection, and spatial smoothing with a Gaussian kernel with 6mm full width at half maximum (FWHM). Temporal filtering was performed with a high pass filter with cutoff at 0.008 Hz. The regions of interest (ROIs) were determined from a whole brain image percellation into anatomically defined regions described in the Automated Anatomical Labeling (AAL) atlas \citep{tzourio2002automated}. We exclude the cerebellum and vermis, and concentrate on the remaining 90 cortical and subcortical ROIs, similar to previous studies \citep{he2009uncovering,lynall2010functional,bassett2012altered}. Mean ROI time series was obtained for each of the ROIs by averaging the time series in the voxels within the ROI.

\subsection{Exploratory analysis}

To compute functional connectivity among the ROIs, we first decompose the mean time series in each ROI to a 4 scale maximal overlap discrete wavelet transform \citep{percival2006wavelet}, and then take the second scale which roughly corresponds to the frequency range 0.060-0.125 Hz. A ROI to ROI correlation matrix is subsequently constructed from the pairwise correlations among this scale 2 wavelet transformed time series. The second scale of the wavelet transformation is chosen to strike a balance between minimizing the impact of physiological noise that might confound the higher frequencies and not having enough samples to compute correlation matrix in lower frequency ranges \citep{lynall2010functional,bassett2012altered,alexander2012discovery}. Binary connection matrices were created for each subject through thresholding at 12 levels between 0.05 and 0.60 with increments of 0.05, resulting in graphs of different network connection densities.

\begin{figure}[h]
\centering{}
\begin{subfigure}{0.25 \textwidth}
\includegraphics[width=\linewidth]{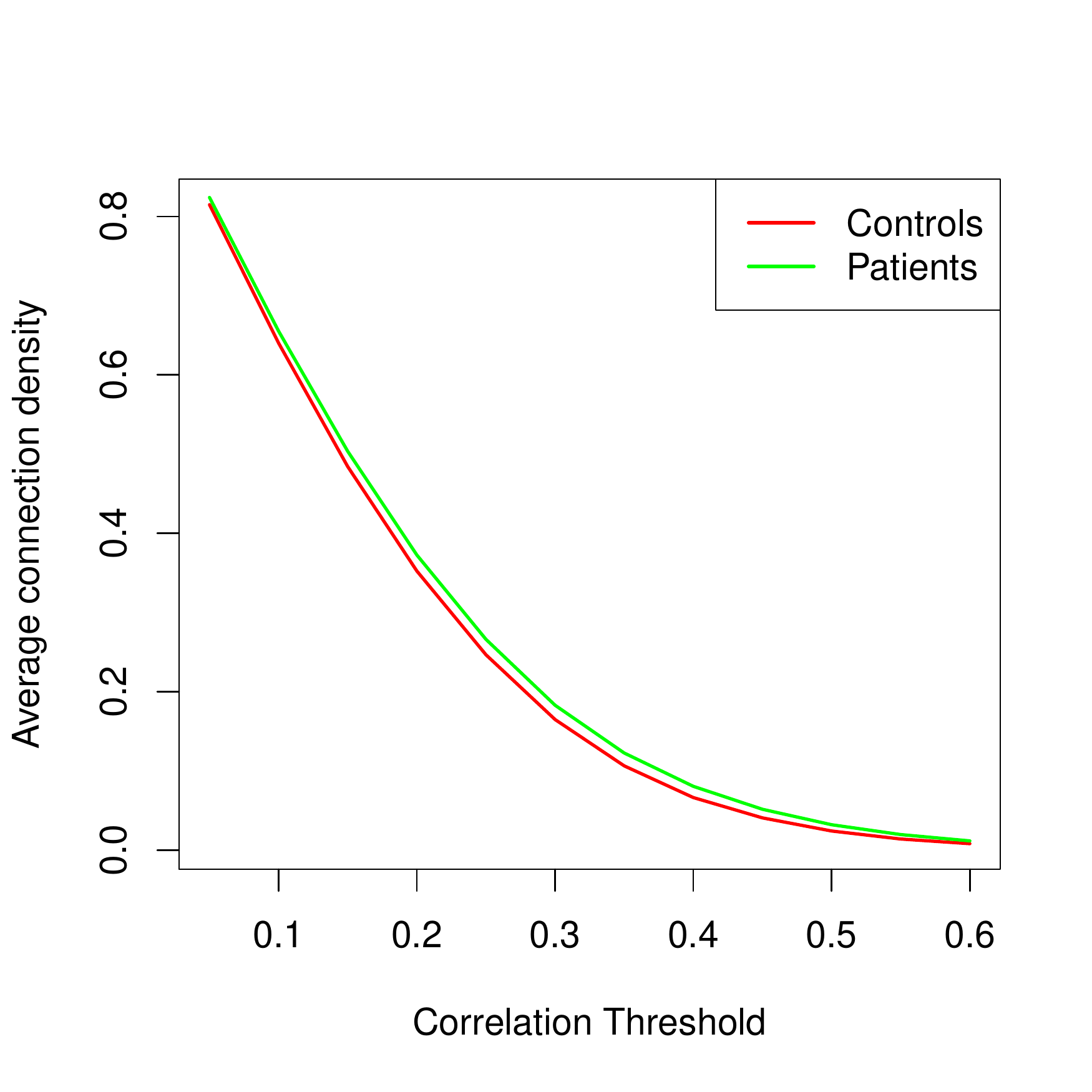}
\end{subfigure}%
\begin{subfigure}{0.25 \textwidth}
\includegraphics[width=\linewidth]{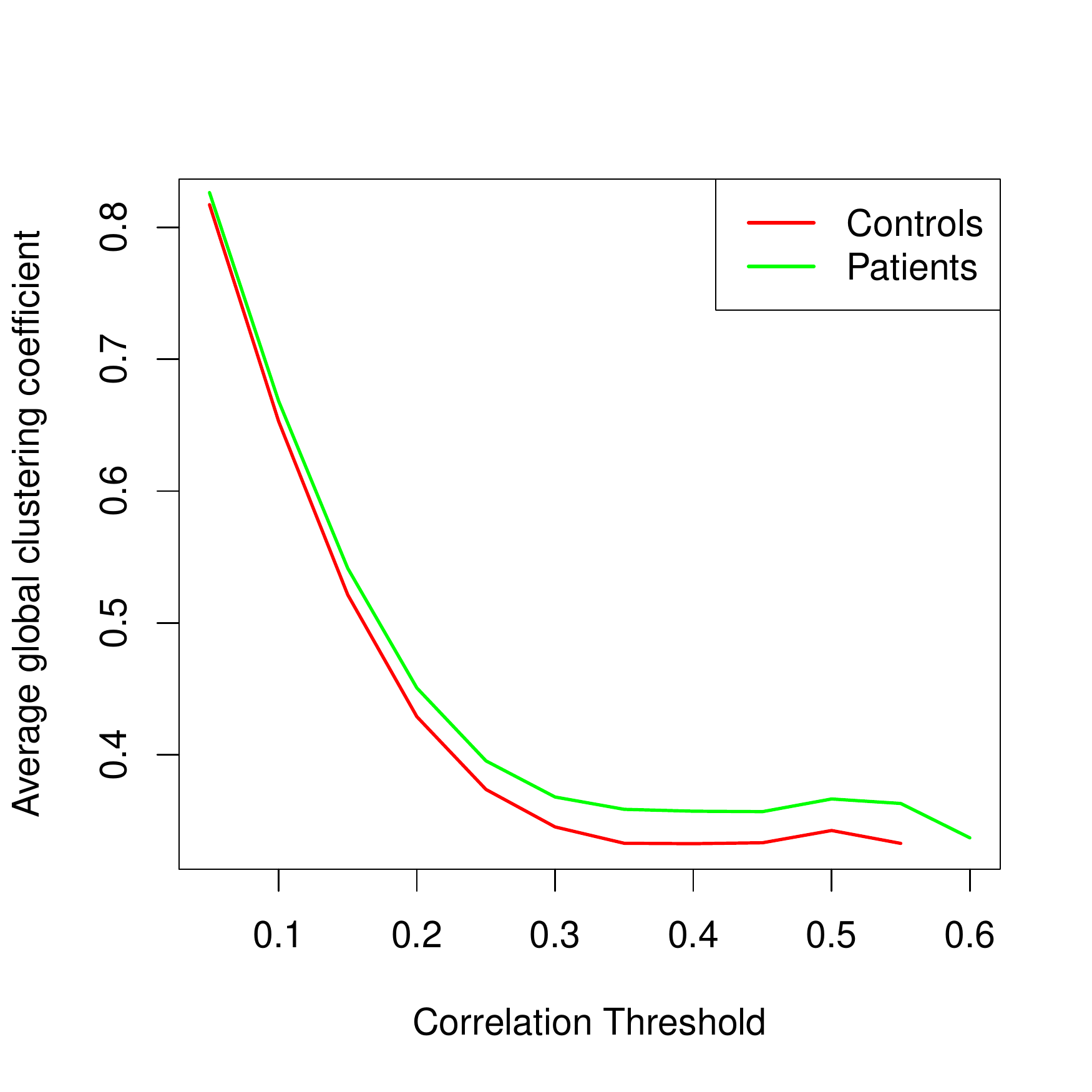}
\end{subfigure}%
\begin{subfigure}{0.25 \textwidth}
\includegraphics[width=\linewidth]{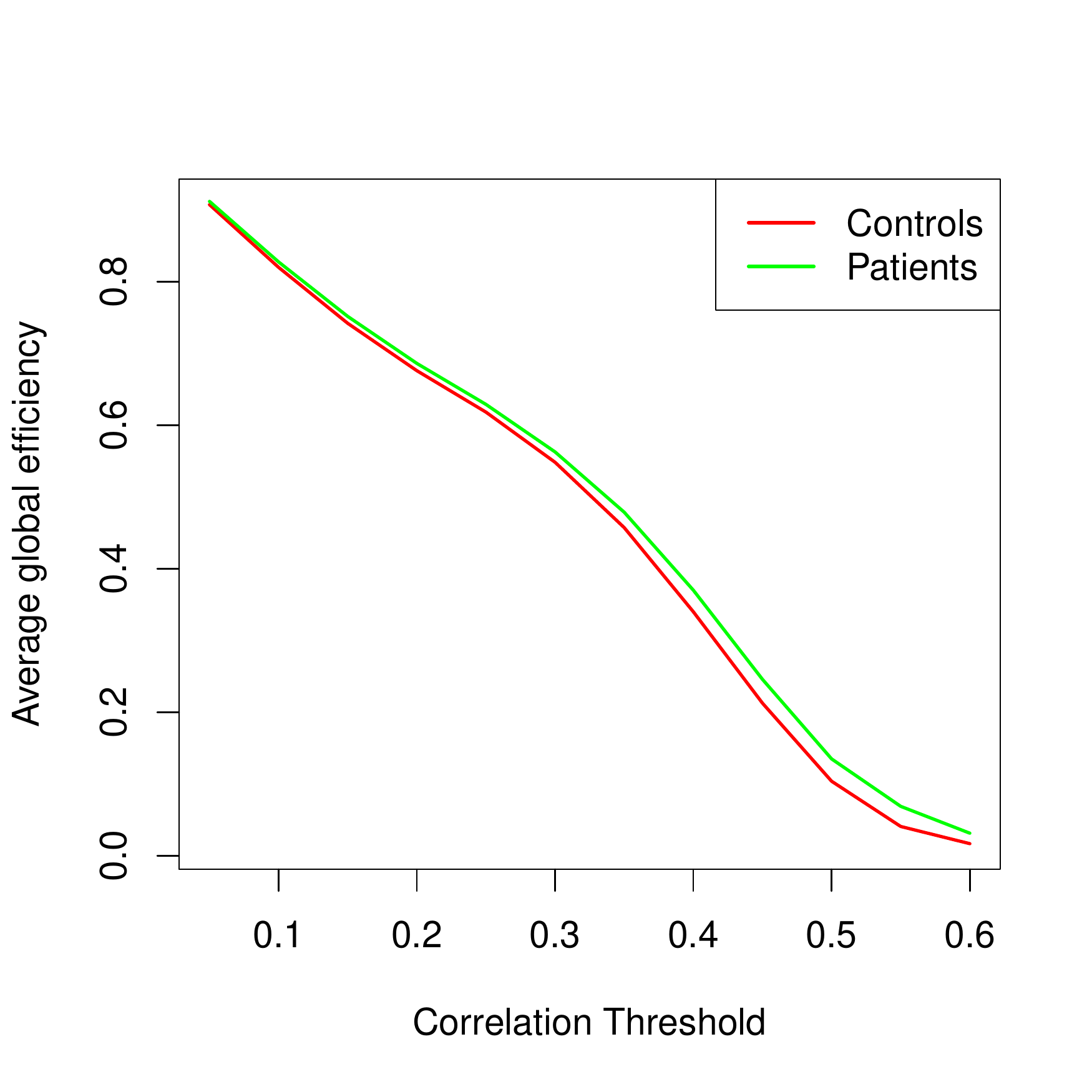}
\end{subfigure}
\begin{center}
(A) \hspace{100pt} (B) \hspace{100pt} (C)
\end{center}
\vspace{-10pt}
\begin{subfigure}{0.25 \textwidth}
\includegraphics[width=\linewidth]{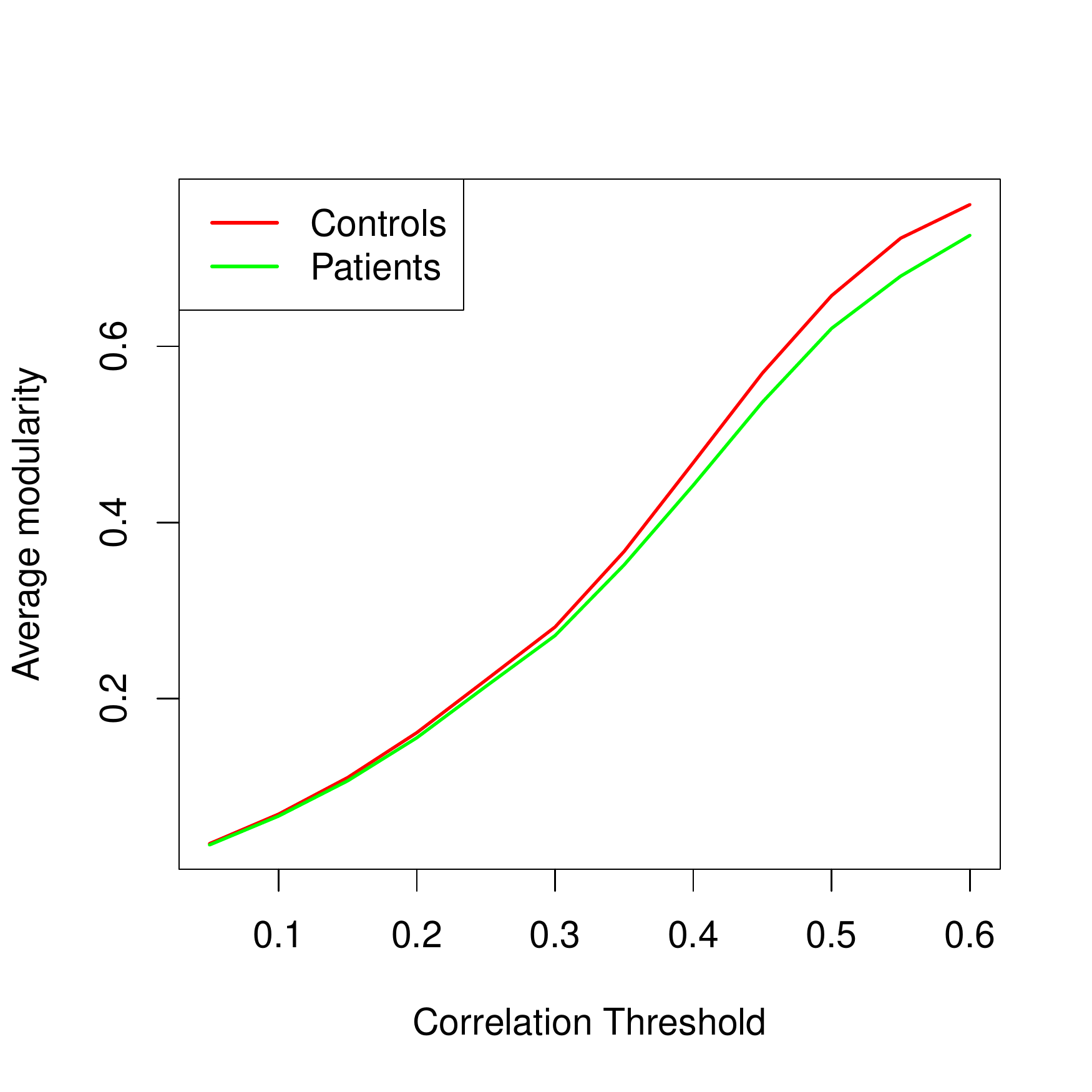}
\end{subfigure}%
\begin{subfigure}{0.25 \textwidth}
\includegraphics[width=\linewidth]{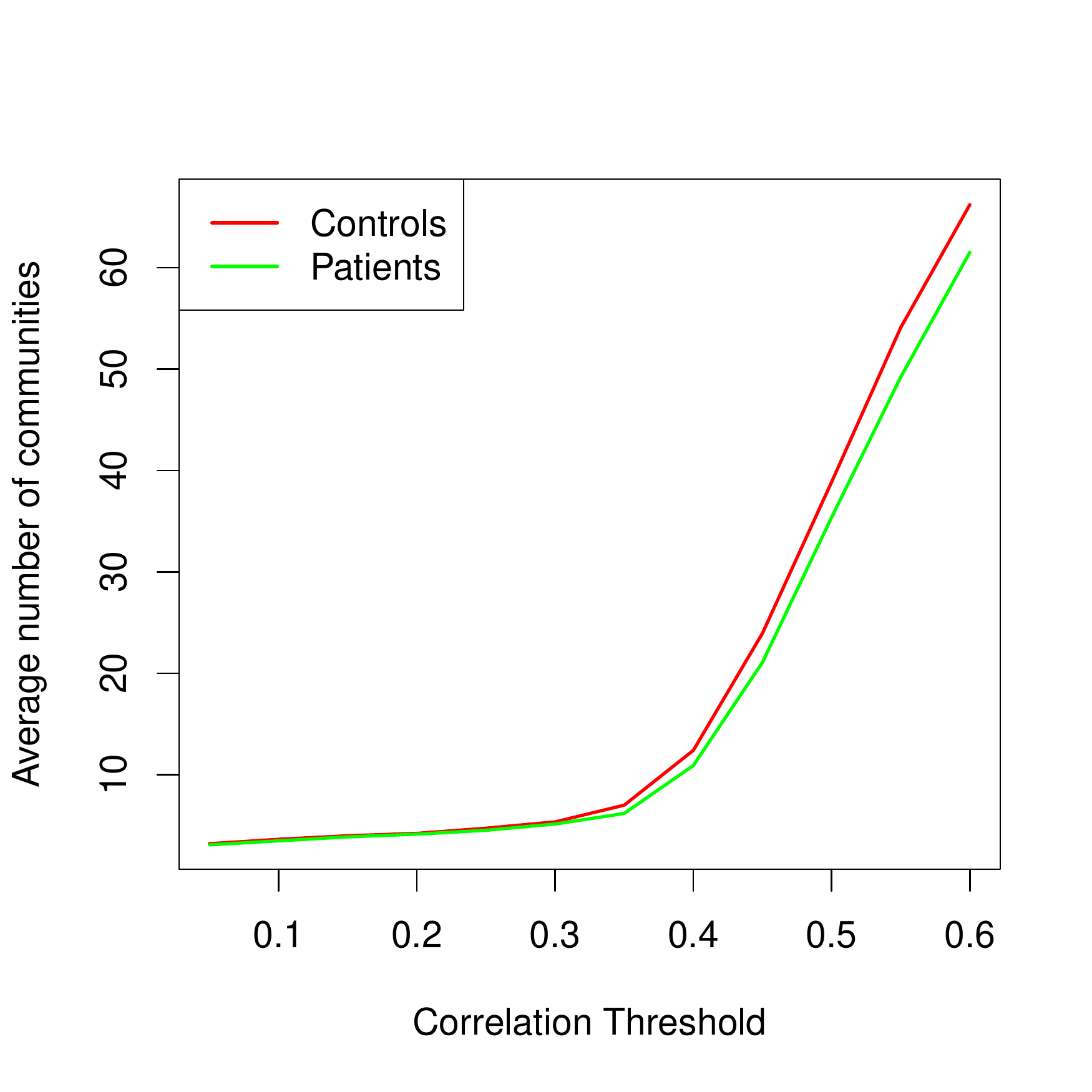}
\end{subfigure}%
\begin{subfigure}{0.25 \textwidth}
\includegraphics[width=\linewidth]{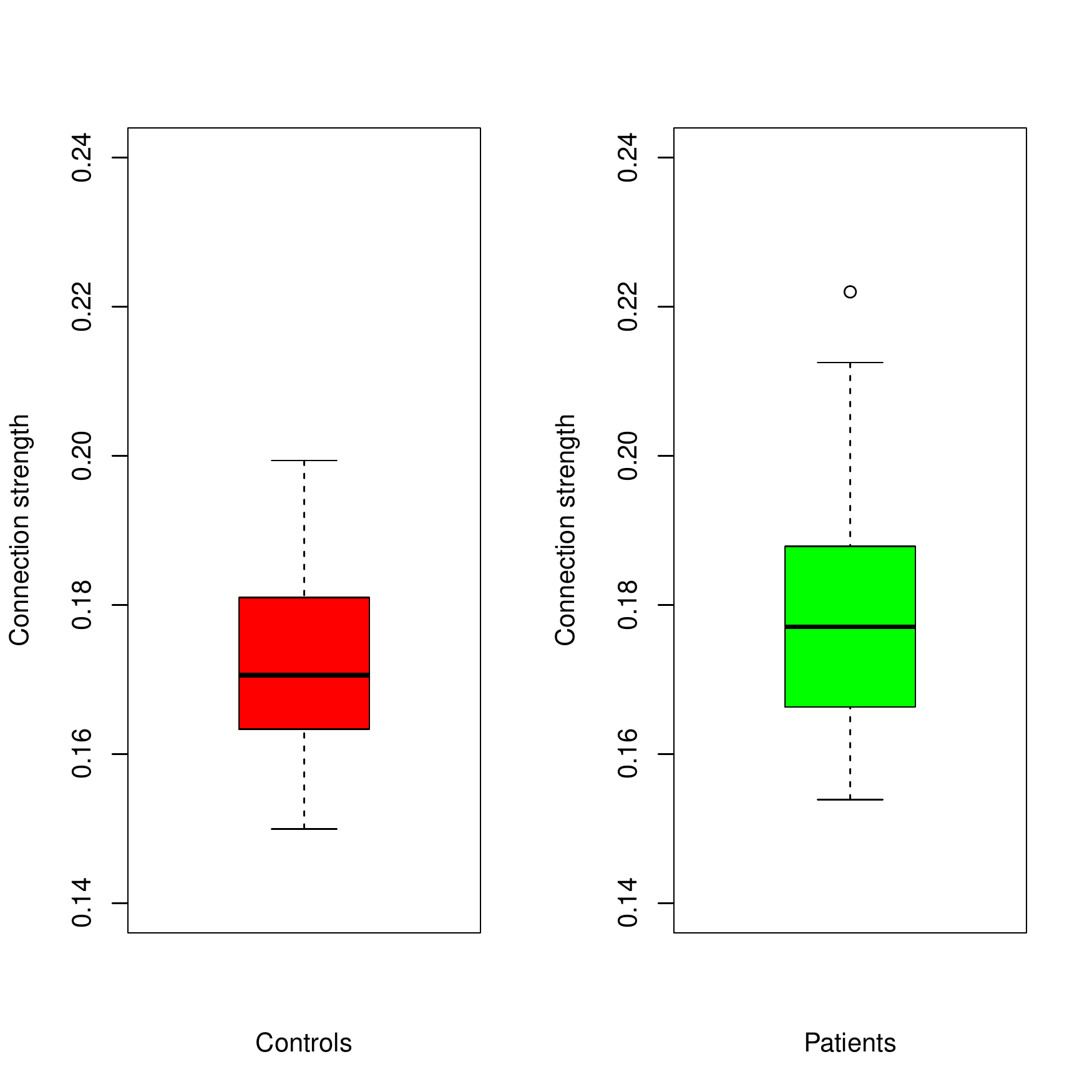}
\end{subfigure}
\begin{center}
(D) \hspace{100pt} (E) \hspace{100pt} (F)
\end{center}
\vspace{-10pt}
\caption{Exploratory analysis 1: Average values of global network summary measures with varying thresholds (A) network density, (B) global clustering coefficient, (C) global efficiency, (D) modularity and (E) number of communities. A box plot of average strength of absolute correlation (without thresholding) is displayed in (F).}
\label{explore1}
\end{figure}

\begin{table}[h]
\caption{Average modularity and average number of communities detected by modularity maximization (Spin-glass and Louvain algorithms) in the two groups of subjects for different thresholds. The columns of p-value indicate p-values obtained from Welch two sample t-test for difference in means. A * and ** indicate statistically significant at 5\% and 1\% FDR multiple comparison corrected levels.}
\begin{center}
\begin{tabular}{c|ccl|ccl}
\multicolumn{1}{c}{Threshold} &  \multicolumn{3}{|c}{Modularity} & \multicolumn{3}{|c}{Communities}\tabularnewline
\hline
 & Controls & Patients & p-value& Controls & Patients & p-value \tabularnewline
\hline
0.05  & 0.0354 & 0.0339 & 0.0014** & 3.20 & 3.08 & 0.0668 \tabularnewline
0.10  & 0.0689 & 0.0667 & 0.0013** & 3.63 & 3.47 & 0.0875 \tabularnewline
0.15  & 0.1105 & 0.1067 & 0.0019** & 3.98 & 3.87 & 0.2032 \tabularnewline
0.20  & 0.1615 & 0.1556 & 0.0028** & 4.21 & 4.14 &  0.4581 \tabularnewline
0.25  & 0.2212 & 0.2141 & 0.0197* & 4.72 & 4.52 & 0.0710 \tabularnewline
0.30  & 0.2814 & 0.2717 & 0.0483* & 5.35 & 5.14 & 0.1681 \tabularnewline
0.35  & 0.3673  & 0.3520 & 0.0385* & 7.00 & 6.18 & 0.0007* \tabularnewline
0.40  & 0.4681 & 0.4425 & 0.0262* & 12.40 & 10.91 & 0.0291  \tabularnewline
0.45 & 0.5697 & 0.5368 & 0.0329* & 23.95 & 21.08 & 0.0414 \tabularnewline
0.50  & 0.6576 & 0.6206 & 0.0282* & 38.86 & 35.38 & 0.0903 \tabularnewline
0.55  & 0.7229 & 0.6798 & 0.0116* & 54.09 & 49.21 & 0.0291 \tabularnewline
0.60  & 0.7609 & 0.7261 & 0.0373* & 66.21 & 61.52 & 0.0250 \tabularnewline
\hline
\end{tabular}
\end{center}
\label{tab:mod}
\end{table}

\subsubsection{Varying the thresholds}

Figures \ref{explore1}(A)-(E) compare the average values of a number of network global summary measures, namely, network density, global clustering coefficient, global efficiency, modularity, and number of communities in the two populations, the controls and the patients over a range of threshold values. The modularity values and number of communities were detected from modularity maximization using spin-glass \citep{reichardt2006statistical} and Louvain algorithms \citep{blondel08} applied to the individual subject networks. Generally we observe, the patient group on average had a higher network density, higher global clustering coefficient, higher global efficiency, and lower modularity as compared to the control group across thresholds. Figure \ref{explore1}(F) presents a box plot of the distribution of strength of absolute correlation across the subjects in control and patient groups respectively. This measure is sometimes reported in the literature as connection strength.
The higher efficiency and lower modularity imply the networks  in patients are more integrated as opposed to controls where the networks are more modular and hence segregated \cite{rubinov10}. While this observation is in agreement with some previous studies on schizophrenia, it is also in contrast to some findings \citep{yu2012brain}. We also see higher clustering coefficient in the patients, but this observation could also be because of higher network density. Since we thresholded all correlation matrices at the same value, the resulting binary networks may have different densities. The optimal number of communities (as detected by modularity optimization methods) appears to be the same for both groups until the threshold of 0.35 after which the number of communities increases exponentially for both groups and there is some difference in the number between the two groups (Figure \ref{explore1}(E)). Our observation on reduced modularity yet similar number of communities is in agreement with most previously reported studies in schizophrenia \citep{yu2012brain}.

\begin{figure}[h]
\centering{}
\begin{subfigure}{0.3 \textwidth}
\includegraphics[width=\linewidth]{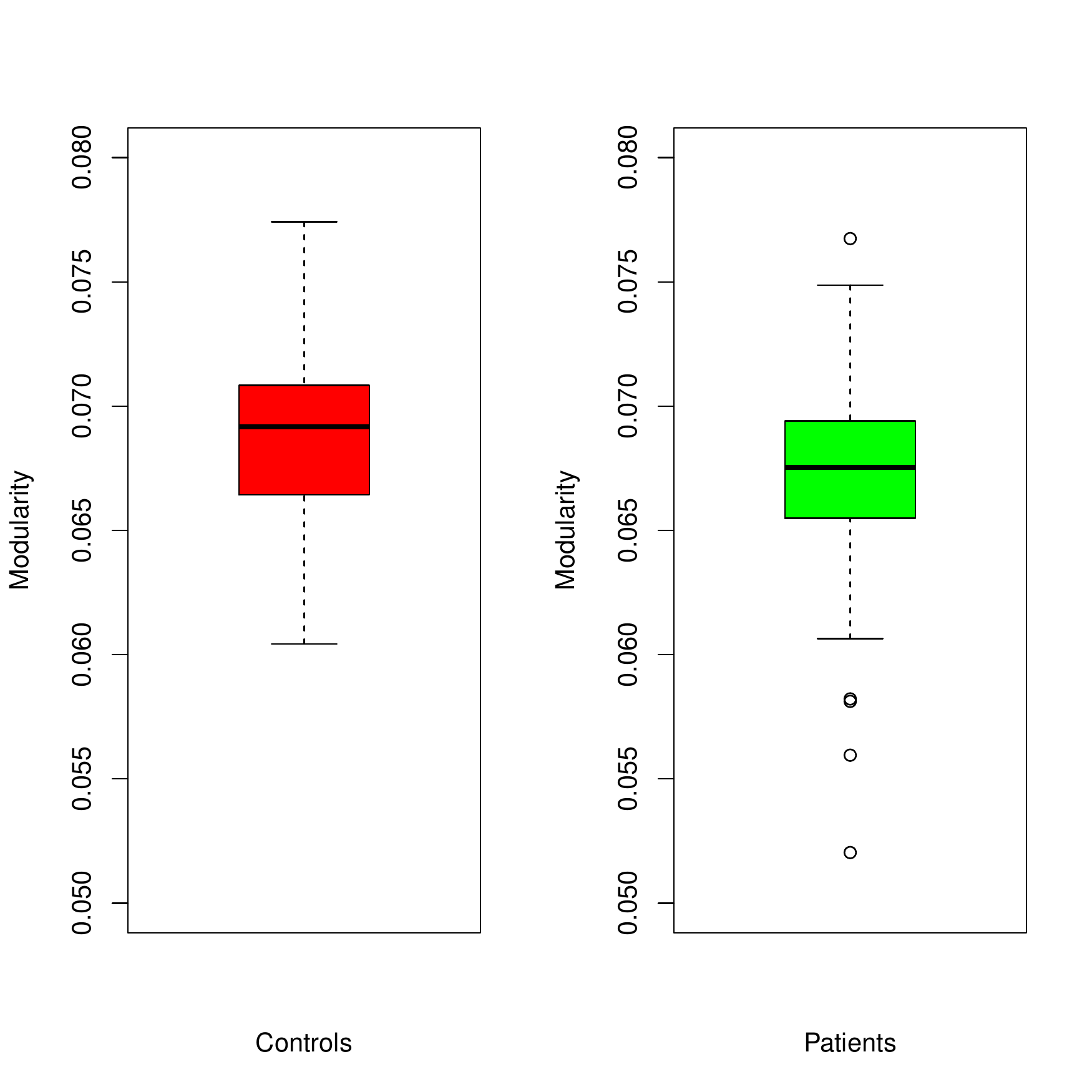}
\end{subfigure}%
\hspace{10pt}
\begin{subfigure}{0.3 \textwidth}
\includegraphics[width=\linewidth]{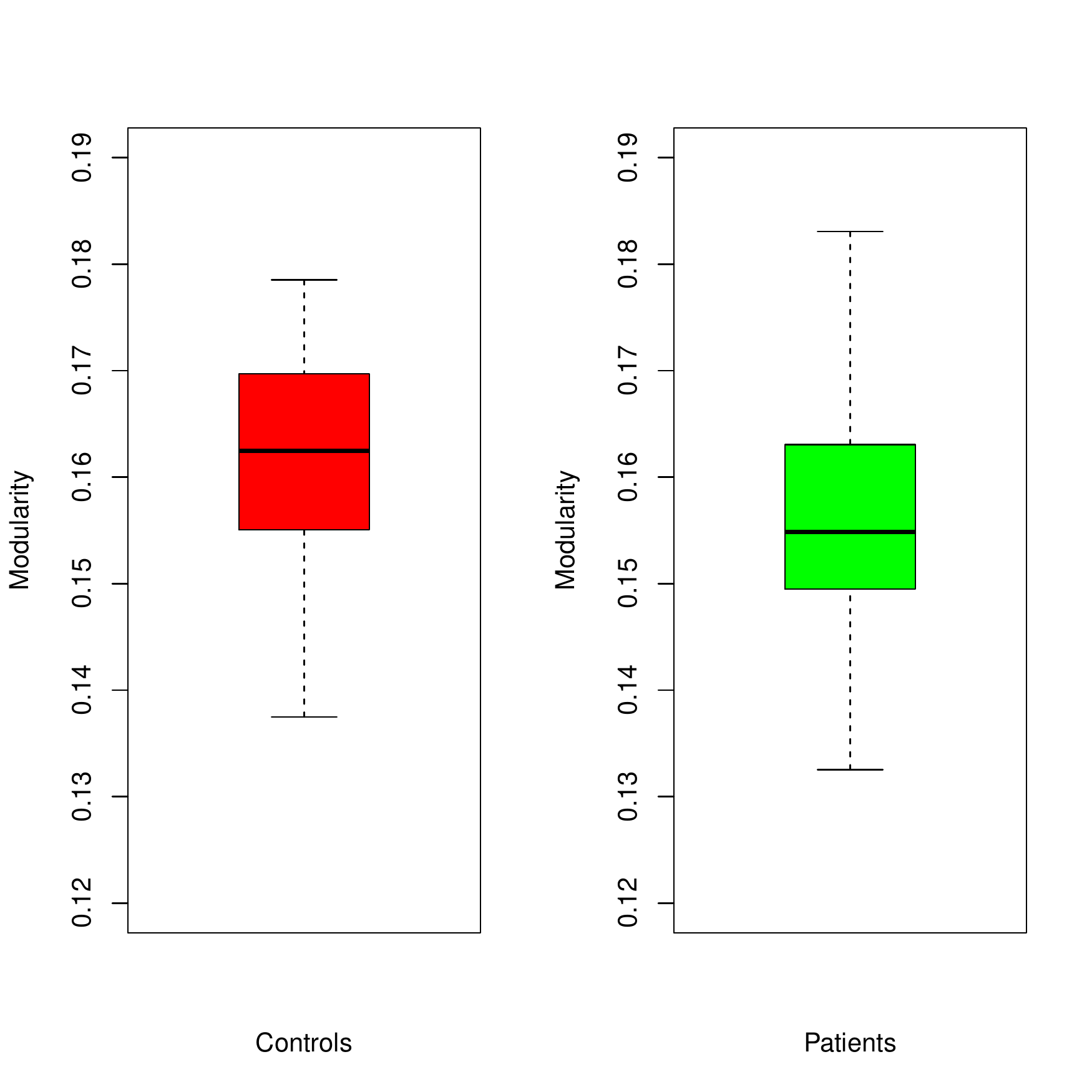}
\end{subfigure}%
\hspace{10pt}
\begin{subfigure}{0.3 \textwidth}
\includegraphics[width=\linewidth]{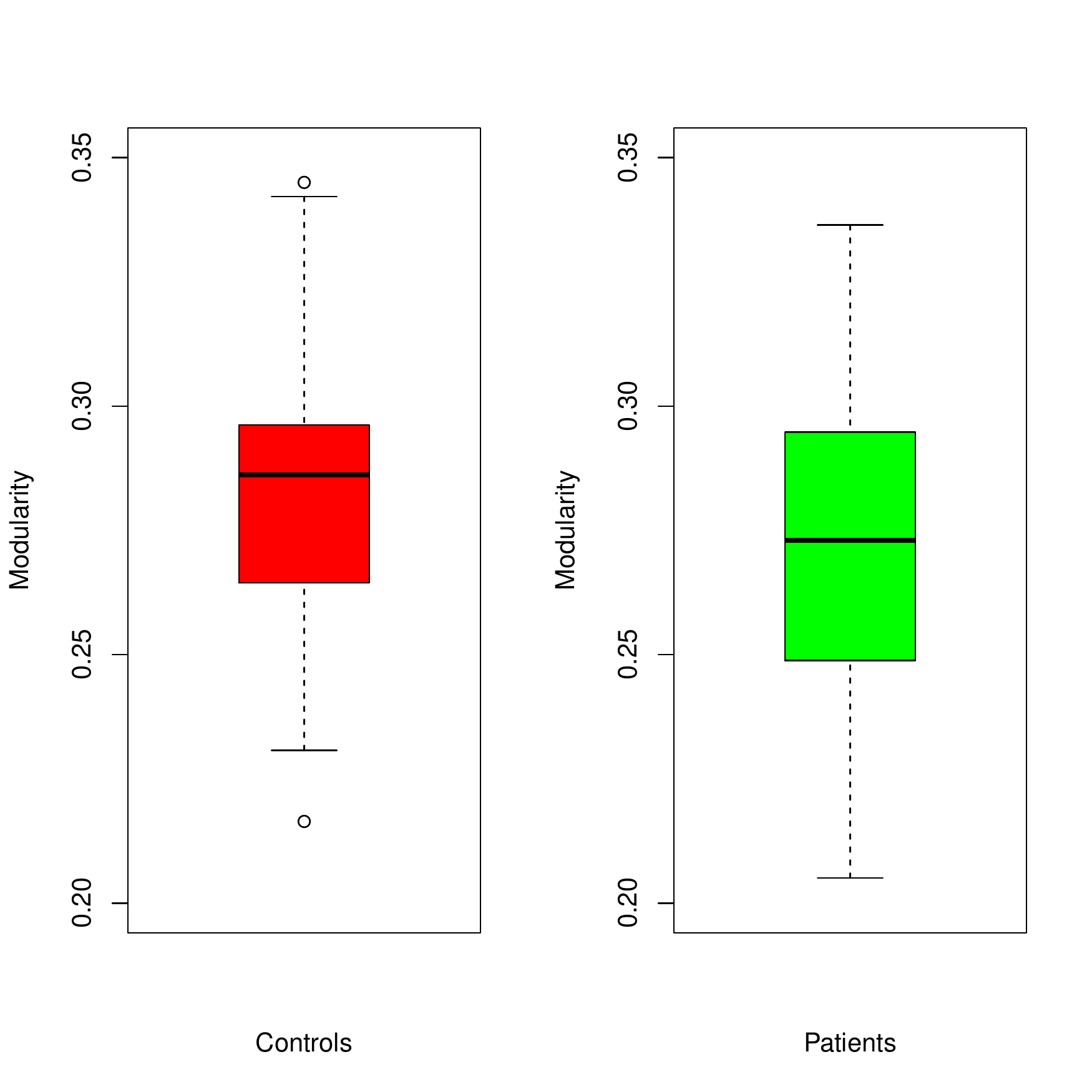}
\end{subfigure}
\begin{center}
(A) \hspace{100pt} (B) \hspace{100pt} (C)
\end{center}
\begin{subfigure}{0.3 \textwidth}
\includegraphics[width=\linewidth]{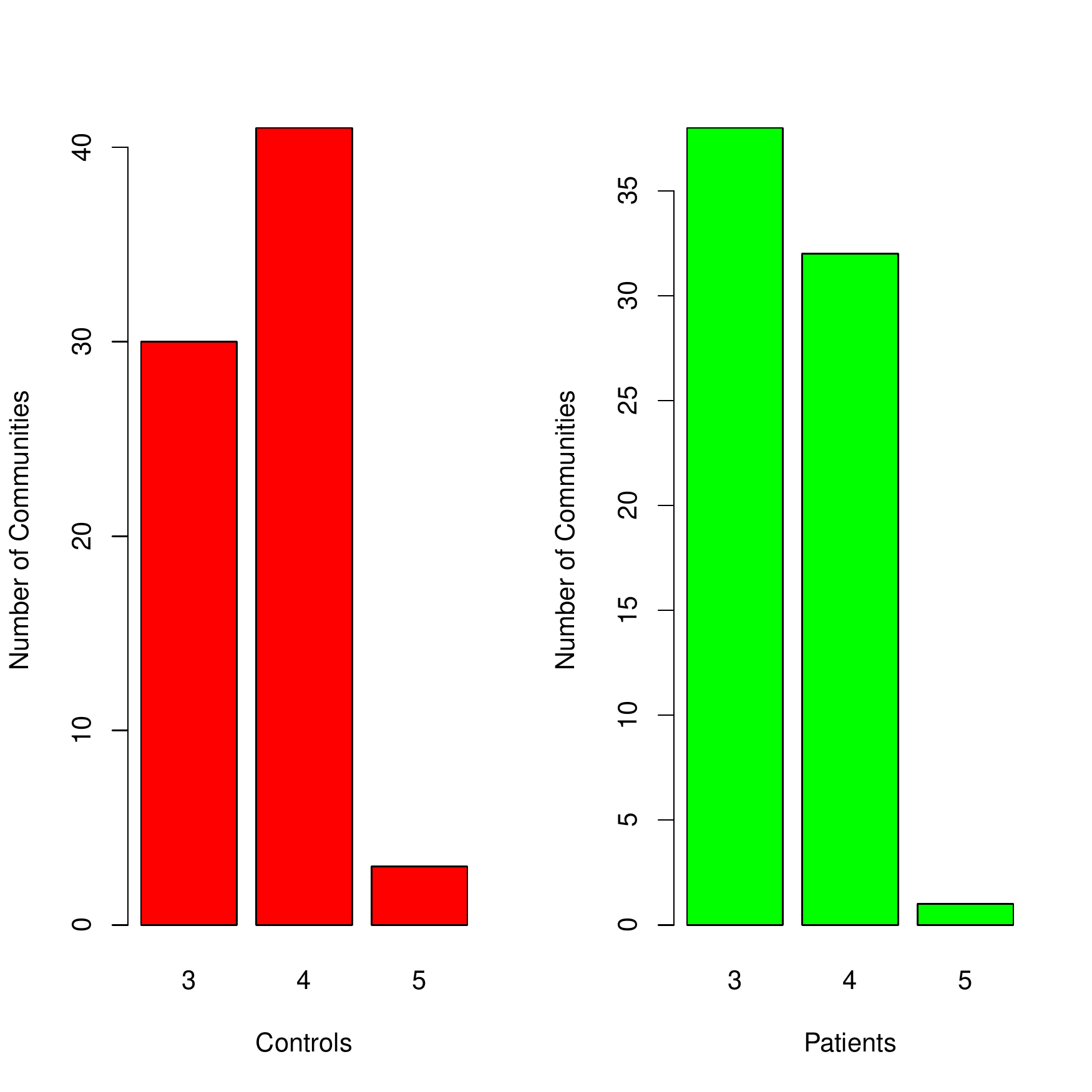}
\end{subfigure}%
\hspace{10pt}
\begin{subfigure}{0.3 \textwidth}
\includegraphics[width=\linewidth]{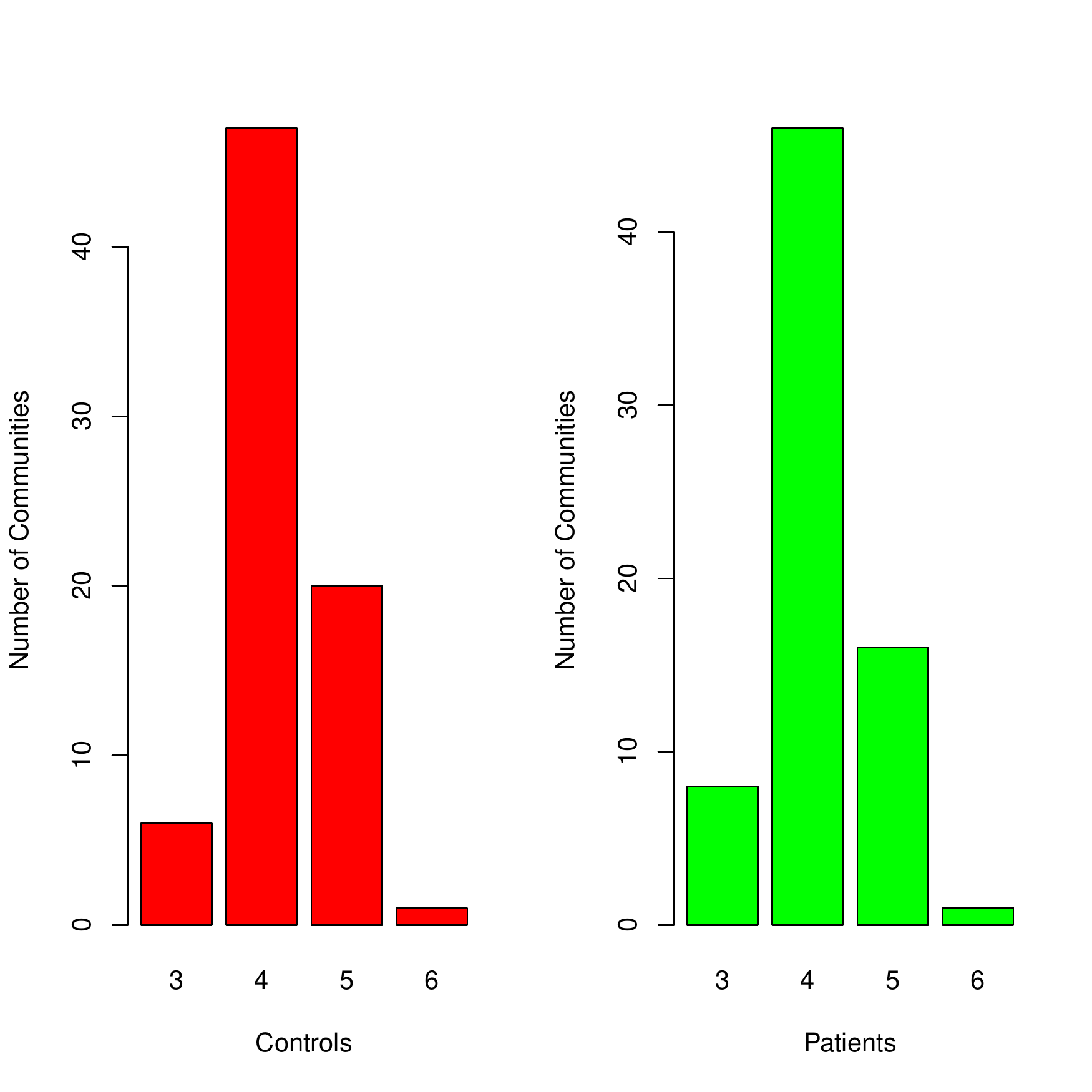}
\end{subfigure}%
\hspace{10pt}
\begin{subfigure}{0.3 \textwidth}
\includegraphics[width=\linewidth]{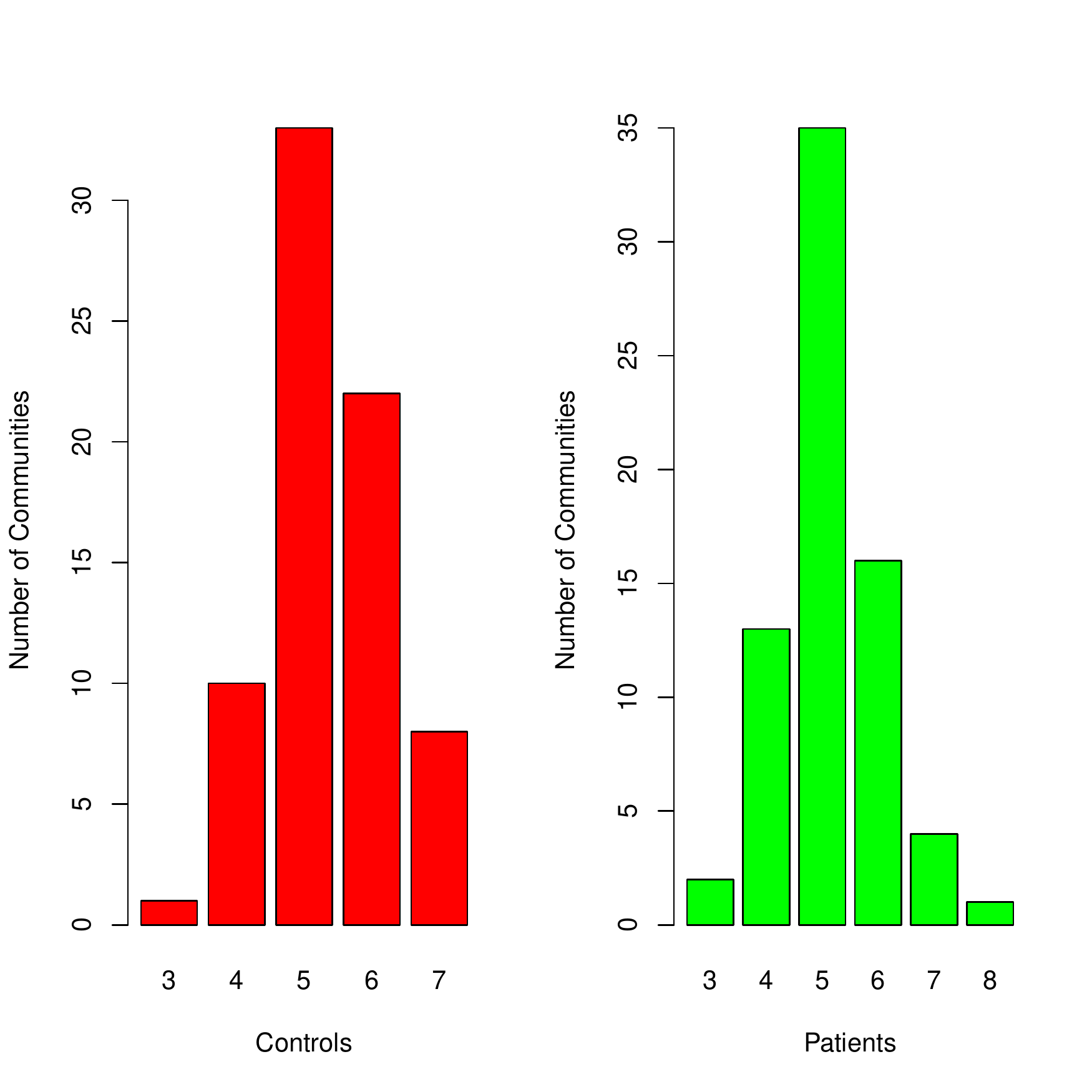}
\end{subfigure}
\begin{center}
(D) \hspace{100pt} (E) \hspace{100pt} (F)
\end{center}
\vspace{-10pt}
\caption{Exploratory analysis 2: (A)-(C) boxplots of distribution of maximum modularity across subjects for three different thresholds 0.1, 0.2 and 0.3. (D)-(F) histogram plot of distribution of optimal number of communities  across subjects for three different thresholds 0.1, 0.2 and 0.3. }
\label{explore2}
\end{figure}

For each of these thresholds, Table \ref{tab:mod} presents the average of modularity values, the number of communities detected and p-values from Welch two sample t-test for difference in means. We note that the modularity in patients is consistently lower than in controls across all thresholds. The p-values indicate that modularity in patients is significantly lower at a 5\% False Discovery Rate (FDR, \cite{benjamini1995controlling}) corrected significance level for all thresholds, while it is lower at 1\% statistical significance level at the first 4 thresholds. However, the number of communities detected are not significantly different between the two groups at the 5\% FDR corrected significance level for any  threshold value except 0.35.

In Figure \ref{explore2} we assess the differences between the two populations in terms of distribution of modularity and number of communities through box plots and histogram plots respectively at thresholds 0.1, 0.2 and 0.3. We note that at the threshold of 0.2, which we will primary focus on later in the main analysis, the boxplot of modularity for controls lies substantially above those of patients, giving an indication that the community structure might be quite different at this threshold. On the other hand, at this threshold, the distribution of the number of communities appear almost identical with the most common number of communities being 4 in both cases.

\subsubsection{Removing the effects of covariates}

As a robustness check in our main analysis we will also perform the analysis on a ``residualized" correlation matrix obtained by removing the effect of some of the known covariates from the estimated correlations. For this purpose,  we run a linear regression of the correlations pulled together across subjects with a number of subject level covariates: age (continuous), gender (categorical with 2 levels, male and female), and handedness (categorical with 3 levels, left, right and both). The overall effect of the covariates were statistically significant (test for regression model: $F$ statistic 68.38, p-value $<0.0001$), however the very low adjusted $R$ squared (0.00022) indicates the effect of the covariates is rather low.  Individually all the covariates were found to be statistically significant. To asses the effect of the covariates in our analysis, we take the residuals from this regression and add the intercept term with it to create new residualized correlations. We repeat the above exploratory analysis with these residualized correlations. We do not find any significant differences in the global summary measures and hence omit those results. However, in Section 5.4 we run our methods on this residualized correlation matrix and compare the results with those obtained from unresidualized correlation matrix.

\section{Models and methods}

Suppose we observe a sample of $M$ graphs or networks $\mathcal{G}=\{G^{(1)},\ldots, G^{(M)}\}$ with a common set of nodes $V$ of size $n$ from a population of networks. We assume the component graphs to be unweighted and undirected. Hence to each component graph in the sample $G^{(m)}$, we can associate an $n \times n$ square and symmetric adjacency matrix $A^{(m)}$, such that $ A^{(m)}_{ij}$ is $1$ if there is an edge between nodes $i$ and $j$ in  $G^{(m)}$ and $0$ otherwise. For each component, we can also define a normalized Laplacian matrix as $L^{(m)}=D^{(m)-1/2}A^{(m)}D^{(m)-1/2}$, where $D^{(m)}$ is a diagonal matrix whose elements are the degrees of the nodes in the $m$th component defined as $D^{(m)}_{ii}=\sum_{j}A^{(m)}_{ij}$. Our primary goal in this paper is to model the community structure of the population of networks from which the sample is generated.

A problem closely related to our problem is that of community detection in multi-layer networks, that has received considerable attention in the literature in the last decade \citep{kivela14,nicosia2015measuring,boccaletti14,pc15}. A group of related interactions on the same set of nodes can also be represented as a multi-layer network, where each network layer or type of edge represents a component network of the group.  The multi-layer networks observed in the nature are also known to exhibit community structure \citep{mucha10,bazzi2016community,kivela14,nicosia2015measuring,boccaletti14,peixoto15}. The multi-layer stochastic block model (MLSBM) is a statistical model for such multi-layer networks with community structure \citep{hxa14,valles14, pc15,stanley15,peixoto15,barbillon2017stochastic,paul2016null,paul2017spectral}. Recently \citet{de2017community} proposed a more flexible multi-layer mixed membership stochastic block model that allows overlapping clusters. Most of the models described in the literature, with the exception of the strata-MLSBM of \citet{stanley15}, are constrained by the fact that they assume the community structure to be the same across all layers. The estimation task is usually then to estimate this consensus community structure by fusing information from all layers.

However, in many situations, e.g., in the multi-subject neuroimaging studies, it may be desirable to model the variation in community structure in different subjects along with finding a consensus clustering. The existing models are not flexible enough to model such data. In a partial remedy of the situation, \citet{stanley15} introduced the strata-MLSBM where the community assignments vary across stratas but stay the same within the same strata. Within a strata, they further constraint the block model probability matrix to be identical across layers. A Bayesian nonparametric mixture model for jointly estimating community structure and identifying groups of networks with similar community structure in a collection of exchangeable networks was proposed in \citet{reyes2016stochastic}. A model similar in spirit to our proposed model is the Bayesian hierarchical mixed membership stochastic block model for an ensemble of networks proposed in \citet{sweet2014hierarchical}, however the MCMC estimation method is more computationally expensive and difficult to apply for larger networks and no hypothesis testing procedure was provided.

In this paper we propose a random effects stochastic block model (RESBM) to model the community structure of a population of networks. The RESBM is a general and flexible modeling framework which contains the MLSBM as a special case.  The model assumes the existence of a putative mean community structure which is a group level parameter representative of the group or the population, but is not necessarily the actual community structure in any of the subjects in the sample. The true community assignment matrices for each of the component networks are random variables generated from this putative mean community assignment through the transition probability matrix. We formally define the model in the next section.

\subsection{Random effects stochastic block model}

 We define the $n$ node $k$ block RESBM as follows. To each node $i$ of a network we associate a $k$ dimensional \textit{community assignment vector} $x_i$, which takes value $1$ in exactly one place and $0$'s everywhere else. The location of $1$ in the vector indicates the community the node belongs to. We call a matrix $X \in [0,1]^{n \times k}$ a \textit{community assignment matrix} of the nodes of a network, if each row of the matrix is a community assignment vector for one of the $n$ nodes in the network. Let $\bar{Z} \in [0,1]^{n \times k}$ be a \textit{putative mean community assignment matrix} for a group or population of $M$ networks. This is a fixed group or population level parameter. For each member network $G^{(m)}$ in the sample, each node $i$ can randomly switch its community label from this putative mean community label independent of other networks and other nodes.

Formally we use the multinomial unit vector and a transition probability matrix to describe the random deviation from the putative mean structure. A unit vector, similar to the community assignment vector, is a vector whose components are all zeros except for one component that is one. A $k$-dimensional random unit vector $Y$ follows a multinomial distribution (some authors also call it multinoulli distribution to draw parallel with Bernoulli distribution) with parameters $N=1$ and $ \mathbf{p}=(p_1,\ldots,p_k)$, $\sum_{i=1}^{k}p_i=1$, if for unit vector $\mathbf{u}=(u_1,\ldots,u_k)$,
\[ P(Y={\bf u}) = p_i, \ \ \ \ \mbox{if } u_i=1 \mbox{ and } u_j=0 \mbox{ for } j\not=i.\]

For each member network $G^{(m)}$, the community assignment matrix of the network $Z^{(m)}$ is drawn in the following way. For each node $i$, the vector of community assignments is
\begin{equation}
    Z_{i}^{(m)} \sim \mbox{Multinomial}(1,\bar{Z}_{i}T), \quad i=1,\ldots,n, \ \ m=1,\ldots,M,
    \label{VCMLSBM}
\end{equation}
where $Z_{i}^{(m)}$ and $\bar{Z}_{i}$ are the $i$th row of $Z^{(m)}$ and $\bar{Z}$, respectively. Here $T$ denotes the $k \times k$ non-negative transition probability matrix among the communities, with its diagonal elements being $\{\eta_1,\ldots,\eta_k\}$ and the off-diagonal elements in each row $q$ summing to $1-\eta_q$. The vectors $Z^{(m)}_{i}$ are independent for all $i$ and $m$. However, clearly the elements of the vector, i.e., $Z^{(m)}_{iq}, \, q=1,\ldots, k$, are dependent since $Z^{(m)}_{iq}=1$ mandates that $Z^{(m)}_{iq'}=0$ for any $q'\neq q$.

We can also write the  expectation and variance of the random vectors $Z^{(m)}_{i}$s as follows,
\begin{equation}
E[Z_i^{(m)}]=\bar{Z}_i T=T_q, \ \ Var[Z_i^{(m)}] = diag(\bar{Z}_i T)-  T^T\bar{Z}_i^{T} \bar{Z}_i T= diag(T_q)- T_q^{T}T_q,
\label{Zexpectation}
\end{equation}
where $q$ is the putative mean community label of node $i$, $T_q$ is the $q$th row of $T$ and $diag(X)$ for any vector $X$ represents the diagonal matrix whose diagonal is the vector $X$. Here, and throughout the paper $T$ in superscript denotes the matrix transpose. Consequently $Z^{(m)}_{i}$ is a multinomial random unit vector with $T_q$ as the vector of parameters (probabilities). This implies that for each member network, a node $i$ that belongs to the mean community $q$, gets assigned to the same community as its mean community assignment with probability $\eta_q$ and a different community with probability $1-\eta_q$. While we do not put any restriction on the transition probability matrix, we note that the model is most interesting when $\eta_q$ is large as compared to the remaining elements in the $q$th row. The model then can be interpreted in the context of multi-subject networks as follows. While most nodes retain their putative group mean community memberships for the individual subject networks, a few randomly selected nodes change their memberships to another community according to probabilities from the transition probability matrix.

\begin{figure}
\centering
\includegraphics[height=2.5in]{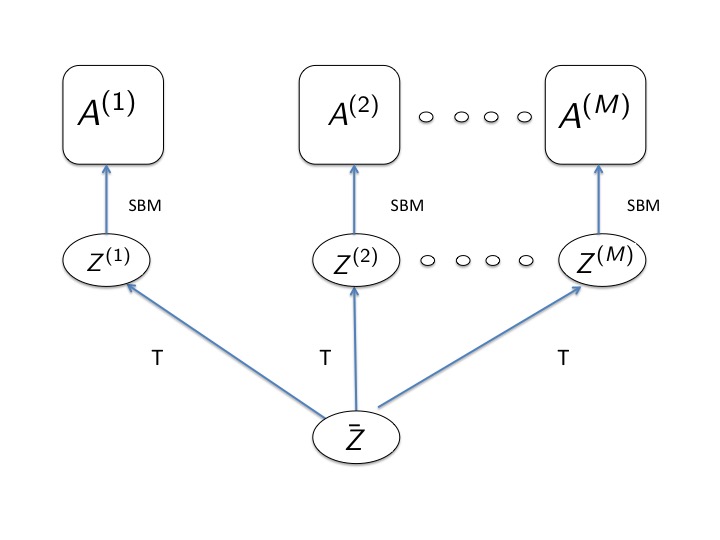}
\vspace{-20pt}
 \caption{Schematic diagram of the RESBM}
 \label{resbmpic}
\end{figure}

Given the community assignment matrices $\{Z^{(1)}, \ldots, Z^{(M)}\}$, the edges in the $M$ member networks are independently generated following a Bernoulli distribution,
\begin{equation}
  A_{ij}^{(m)} | Z^{(m)} \sim Bernoulli (P_{ij}^{(m)}), \quad i,j=1,\ldots,n, \quad m=1,\ldots,M.  \label{generative}
\end{equation}
The Bernoulli probabilities can be modeled as a $k$ class stochastic block model (SBM) or a $k$ class degree-corrected stochastic block model with appropriate identifiability constraints. We focus only on the SBM in this paper. We have
\[
A^{(m)}_{ij}|(Z_{iq}^{(m)}=1,Z_{jl}^{(m)}=1) \sim Bernoulli(\pi^{(m)}_{ql}), \ \  q,l=1,\ldots, k, \ \  m=1,\ldots,M.
\]
The model is schematically represented in Figure \ref{resbmpic}. However the model is not identifiable without further constraints. Similar to the discussion in \citet{matias15} in the context of dynamic networks, the community labels might get switched between two member networks and still give the same model, leading to incorrect inference. Hence we need certain constraints on the matrices $\{\pi^{(1)},\ldots,\pi^{(M)}\}$ such that the communities are identifiable at all member networks. We use the constraint that the diagonal elements of the matrices are identical in each member network, i.e., the vector $\{\pi^{(m)}_{11},\dots,\pi^{(m)}_{kk}\}$ is the same for all $m=1,\ldots,M$ \citep{matias15}.

\subsubsection{Interpretation in fMRI studies}

In fMRI neuroimaging studies, there is often interest in detecting a group community structure that is representative of the whole population, either a group of healthy control subjects or a group of patients with a certain condition, and then contrast such group community assignments between groups of interest. However, it is also widely recognized that all subjects in a group will not have the same community structure due to individual differences \cite{alexander2012discovery, betzel2019community}.  The parameter $\bar{Z}$ denotes such an overall group community assignment. This is a ``hard" community assignment where we designate a single community structure to be representative of the group. However, the consistency of this structure across subjects is estimated through the transition probability matrix $T$.  Then $\bar{Z}T$ is the expected community assignment for a randomly selected subject from the group. This can be thought of as ``mixed" or ``overlapping" membership or ``soft communities", with each row $(\bar{Z}T)_i$ providing probabilities with which the ROI $i$ belongs to different communities. This is also our prediction for a new subject who is not part of the sample. Hence the parameter $T$ controls how much variation is inherent in the group, i.e., how much do we expect a randomly chosen subject to match the putative group mean. In addition, the  estimate for $Var[Z_i^{(m)}]$ in Equation (\ref{Zexpectation}) lets us  derive confidence intervals for each ROI in the network. Thus from this model we can infer an overall community assignment for the group of networks and obtain a quantification of the variability in this group putative community structure through the transition probability matrix.

\subsection{Estimation}

Our estimation goals from this model include estimation of community assignments in each member network $Z^{(m)}$, the putative mean community assignment matrix $\bar{Z}$, and the transition probability matrix $T$. In what follows we describe two methods to perform these estimation goals.

\subsubsection{A variational expectation-maximization estimator}

The first method we describe is approximate maximum likelihood estimation (MLE) through variational expectation-maximization (EM) algorithm, first introduced in the context of standard SBM in \citet{dpr08} and later extended to multi-layer SBMs in \citet{hxa14}, \citet{pc15} and \citet{barbillon2017stochastic}, and to dynamic SBM in \citet{matias15}. The asymptotic consistency and limiting distribution of the parameter estimates for the case of standard SBM are investigated in \citet{cdp11} and \citet{bccz13} respectively.

Below we derive the update rules for variational EM algorithm approximation to the MLE of the RESBM parameters. For this purpose we view the RESBM from a mixture model perspective and pose the parameters $\bar{Z}_i$s as random variables generated from a multinomial distribution with parameters $\alpha=\{\alpha_1,\ldots,\alpha_k\}$. We denote the unobserved variables $\bar{Z}$ and $Z^{(m)}$s together as $X$ and the model parameters $T$, $\pi$ and $\alpha$ together as $\theta$. Further we denote the observed log-likelihood of the model as $l(A,\theta)$. The complete data log-likelihood, which is the joint likelihood of the observed data and the unobserved model community assignment variables, is given by
\begin{align*}
 \log P(A,X,\theta)  =& \sum_{i=1}^{n} \sum_{q=1}^{k} \bar{Z}_{iq} \log \alpha_{q} +\sum_{m=1}^{M}\sum_{i=1}^{n}\sum_{1\leq q,l \leq k}\bar{Z}_{iq}Z^{(m)}_{il}\log T_{ql}\\
& \quad +\sum_{m=1}^{M}\sum_{1\leq i<j\leq n}\sum_{1\leq q,l \leq k}Z^{(m)}_{iq}Z^{(m)}_{jl}\{A_{ij}^{(m)}\log (\pi_{ql}^{(m)})-(1-A_{ij}^{(m)}) \log (1-\pi_{ql}^{(m)})\}.
\end{align*}

We define the variational distribution $R(X)$ to have
the following form of the product of multinomial densities:
\[
R(X)=\prod_{i=1}^{n}\prod_{q=1}^{k} \bar{\tau}_{iq}^{\bar{Z}_{iq}} \times \prod_{i=1}^{n} \prod_{m=1}^{M} \prod_{1 \leq q,l \leq k} (\epsilon_{iql}^{(m)})^{Z_{il}^{(m)}\bar{Z}_{iq}},
\]
where $\bar{\tau}$ and $\epsilon$ are the variational parameters. We delegate the remaining details of the derivation to the supplementary materials. The following fixed point equations are used to update the variational parameters iteratively in the Variational Expectation (VE) step:
\begin{align}
\hat{\epsilon}_{iql}^{(m)}  & \propto \exp [\log (T_{ql}) + \sum_{j \neq i} \sum_{p=1}^{k} \tau_{jp}^{(m)} \log (b_{ijlp}^{(m)})], \\
\bar{\tau}_{iq} & \propto \exp [\log \alpha_q + \sum_{m} \sum_{l} \epsilon_{iql}^{(m)} \log \left(\frac{T_{ql}}{\epsilon_{iql}^{(m)}}\right) + \sum_{m} \sum_{l} \sum_{j \neq i} \epsilon_{iql}^{(m)} \tau_{jl}^{(m)} \log (b_{ijql}^{(m)})], \\
    \tau_{il}^{(m)} & \propto \sum_{q=1}^{k} \bar{\tau}_{iq}\epsilon_{iql}^{(m)}.
\end{align}
For the M step we have the following closed form update steps:
\begin{equation}
T_{ql} \propto \sum_{m} \sum_{i} \bar{\tau}_{iq}\epsilon_{iql}^{(m)}, \quad \quad   \alpha_q = \frac{1}{n}\sum_{i}\bar{\tau}_{iq},
\end{equation}
\begin{equation}
\pi_{ql}^{(m)} = \frac{\sum_{1 \leq i<j \leq n} \tau_{iq}^{(m)}\tau_{jl}^{(m)}A_{ij}^{(m)}}{\sum_{1 \leq i<j \leq n} \tau_{iq}^{(m)}\tau_{jl}^{(m)}}, \quad q \neq l, \quad \quad \pi_{qq}^{(m)} = \frac{\sum_{m}\sum_{1 \leq i<j \leq n} \tau_{iq}^{(m)}\tau_{jq}^{(m)}A_{ij}^{(m)}}{\sum_{m}\sum_{1 \leq i<j \leq n} \tau_{iq}^{(m)}\tau_{jq}^{(m)}}.
\end{equation}

The proportionalities in the above algorithm are turned into equalities through normalization using the constraints on the parameters.

\subsection{Two-step matrix factorization and maximum likelihood method}

While the variational EM proposed in the previous section attempts to estimate the unknown quantities in RESBM by approximately maximizing the model likelihood, there are two concerns with the approach. It can be computationally expensive and being a  likelihood based method, it can be susceptible to misspecification of the model. Therefore, next we describe a two-step matrix factorization based approach to nonparametrically estimate some of the model quantities, namely $\bar{Z}$ and $Z^{(m)}$s. We intuitively argue in the next section why this method is appropriate for those tasks under the RESBM despite not being based on the model.

The two steps are as follows:
\begin{enumerate}
\item Solve an optimization problem involving a joint objective function similar to Equation (\ref{conmf}) to obtain the community assignment matrices in the subjects $Z^{(1)},\ldots,Z^{(M)}$, and the putative mean community assignment matrix $\bar{Z}$ simultaneously. This is a fully non-parametric step not dependent on any model and hence is expected to be less influenced by model misspecification.
\item The transition probability matrix $T$ is obtained through conditional maximum likelihood (ML) conditioned on $\bar{Z}$ and $Z^{(m)}$s following Equation (\ref{condML}).
\end{enumerate}

\subsubsection{Co-regularized orthogonal symmetric non-negative matrix tri-factorization algorithm}

We propose an algorithm for the first step of the two-step method based on orthogonal symmetric non-negative matrix tri-factorization (OSNTF) method, described for single networks in \citet{pc16}. The proposed algorithm combines the ideas in linked matrix factorization \citep{tang09} and co-regularized spectral clustering \citep{kumar2011co}. Similar to the co-regularized spectral clustering, the proposed method involves minimizing an  objective function with two terms. The first term is the sum of Frobenius norm objective functions for single network OSNTF \citep{pc16}, and the second term is a smoothness penalty on the factor matrices obtained at each network to make the subspaces spanned by the factor matrices closer to each other.

The global optimizer of the OSNTF objective function was shown in \cite{pc16} to consistently recover the community structure from networks generated from stochastic block model and degree-corrected block model. Hence we expect the first part of the objective function attempts to recover the community structures of the subjects $Z^{(1)},\ldots, Z^{(M)}$ from the individual subject network adjacency matrices. However the second part, which is the smoothness penalty, attempts to find $\bar{Z}$, the group mean community structure such that the individual community structures are ``shrunk" to this community structure. This penalty allows for both information sharing across subjects and a regularization to a group mean. Intuitively then this method has similar goals as the RESBM and therefore is a reasonable non-parametric approach to estimate the group and subject specific community structures. Finally, we note that this method is also in the same spirit as other non-negative matrix factorization based multi-view learning methods proposed in the literature \citep{liu13,mankad2013structural}. However, the orthogonality constraints on the factor matrices, despite being restrictive, makes the factors sparse which is more appropriate for the task of community detection.

We denote a matrix $U\geq 0$ and call it a non-negative matrix if all its elements are non-negative and $U>0$ if all its elements are strictly positive. The method, which we call co-regularized orthogonal symmetric non-negative matrix tri-factorization (Co-OSNTF), solves the following optimization problem on the collection of Laplacian matrices:
\begin{eqnarray}
\nonumber
  [\hat{U}^{(1)},\ldots, \hat{U}^{(m)},\hat{U}^{*}] = \argmin_{\substack{U^{(m)},S^{(m)} \geq 0,\\ U^{(m)T}U^{(m)}=I, \, \forall m, \\ U^{*} \geq 0, \, U^{*T}U^{*}=I}} \sum_{m=1}^{M} & \Big \{\|L^{(m)}-U^{(m)}S^{(m)}U^{(m)T}\|_F^2\\
  & + \lambda_{m}(k-\|U^{(m)T}U^{*}\|_F^2)\Big \},
  \label{conmf}
\end{eqnarray}
where $U^{(1)},\ldots, U^{(m)},U^{*}$ are $n\times k$ non-negative matrices with orthonormal columns and $\lambda_{m} \geq 0$ are user-chosen tuning parameters (scalars). Note that this is a constraint optimization problem on Stiefel manifold with additional non-negativity constraints. To solve this optimization problem we employ the method of Lagrange multipliers for the orthogonality constraints and then derive multiplicative update rules. The Lagrangian objective function with the orthogonality constraints incorporated is
\begin{align}
  \xi & \triangleq  \tr \bigg ( \sum_{m=1}^{M} \{-2U^{(m)T}L^{(m)}U^{(m)}S^{(m)} + U^{(m)}S^{(m)}U^{(m)T}U^{(m)}S^{(m)}U^{(m)T} \nonumber \\
  & \quad  \quad  -2 \lambda_{m} U^{(m)T}U^{*}U^{*T}U^{(m)}
   + \Lambda_{U^{(m)}}U^{(m)T}U^{(m)} \}+   \Lambda_{U^{*}}U^{*T}U^{*}\bigg ),
 \label{conmf2}
\end{align}
where $\Lambda_{U^{*}} \geq 0$ and $\Lambda_{U^{(m)}} \geq 0$ are $k \times k$ non-negative symmetric matrices of Lagrange multiplier parameters. Here and throughout the paper $\tr(\cdot)$ denotes the matrix trace. The goal is now to minimize this new objective function under the constraints that $U^{(m)} \geq 0, U^{*} \geq 0,S^{(m)} \geq 0$.

To derive an algorithm for the optimization problem, we follow the derivation techniques for multiplicative updates described in \citet{lee01}, \citet{ding06} and \citet{mirzal2014convergent}. Briefly, the technique is as follows. The KKT conditions for the objective in (\ref{conmf2}) are
\begin{align*}
   & U^{(m)} \geq 0, \quad U^{*} \geq 0, \quad S^{(m)} \geq 0, \\
   & \nabla \xi|_{U^{(m)}} \geq 0, \quad \nabla \xi|_{U^{*}} \geq 0, \quad \nabla \xi|_{S^{(m)}} \geq 0, \\
   & U^{(m)} \odot \nabla \xi|_{U^{(m)}} = 0, \quad U^{*} \odot \nabla \xi|_{U^{*}} = 0, \quad S^{(m)} \odot \nabla \xi|_{S^{(m)}} = 0,
\end{align*}
where $\nabla \xi|_{U^{(m)}}, \nabla \xi|_{U^{*}}, \nabla \xi|_{S^{(m)}}$ are gradients of the objective function (\ref{conmf2}) with respect to $U^{(m)},U^{*}$ and $S^{(m)}$ and $\odot$ represents the Hadamard (element-wise) product. The equations in the last line of the KKT conditions are known as the complimentary slackness conditions which can be used to derive the multiplicative update rules (MUR). If the gradient $\nabla \xi$ with respect to a parameter can be written in the form $\nabla \xi= [\nabla \xi]^{+} - [\nabla \xi]^{-}$, where $[\nabla \xi]^{+}, [\nabla \xi]^{-} \geq 0 $, then the multiplicative update rule would be
\[
 X \leftarrow X \odot \left(\frac{([\nabla \xi]^{-})_{ik}}{([\nabla \xi]^{+})_{ik}}\right)^\eta,
\]
 where $(\cdot)^{\eta}$ represents raising each element of the matrix to the power $\eta$, and $0 <\eta \leq 1 $ is the learning rate. The inner division in the rightmost term is also element-wise.  We take $\eta=1/2$ as learning rate and prove the convergence of the resulting update rules to a stationary point of the objective function.

The details of the derivation are once again delegated to the supplementary materials. The  algorithm consists of iteratively updating $U^{(m)}$s, $S^{(m)}$s and  $U^{*}$ according to the following update rules:
\begin{equation}
 S^{(m)} \leftarrow S^{(m)} \odot \left(\frac{ (U^{(m)T} L^{(m)}U^{(m)})_{ik}}{(U^{(m)T}U^{(m)}S^{(m)}U^{(m)T}U^{(m)})_{ik}}\right)^{1/2},
 \label{MUR3}
\end{equation}
\begin{equation}
 U^{(m)} \leftarrow U^{(m)} \odot \left(\frac{ (L^{(m)}U^{(m)}S^{(m)} +  \lambda_{m} U^{*}U^{*T}U^{(m)})_{ik}}{(U^{(m)} U^{(m)T} L^{(m)}U^{(m)}S^{(m)}+\lambda_{m} U^{(m)} U^{(m)T}U^{*}U^{*T}U^{(m)})_{ik}}\right)^{1/2},
 \label{algo1}
\end{equation}
\begin{equation}
 U^{*} \leftarrow U^{*} \odot \left(\frac{ (\sum_{m} \lambda_{m} U^{(m)}U^{(m)T}U^{*})_{ik}}{(\sum_{m} \lambda_{m} U^{*} U^{*T} U^{(m)}U^{(m)T}U^{*})_{ik}}\right)^{1/2}.
 \label{algo2}
\end{equation}
Note that $\lambda_{m}$s are user specified parameters that should be chosen to address the user's preference in the trade-off between optimizing within a member network of the sample and increasing subspace cohesion as well as to reflect the relative importance of different members. The value of $\lambda$ can also be chosen using cross validation method to minimize some notion of prediction error. For our numerical experiments and real data analysis in this paper we choose $\lambda_{m}$s to be 0.01 for all $m$. This choice of $\lambda_{m}$s works well in our synthetic experiments. In real data analysis in Section 5.7, we display a plot of out of sample prediction error as a function of $\lambda$ and we find $0.01$ to be very close to the optimum.
A similar observation was made for joint non-negative factorization in \citet{liu13} for a number of different datasets.

In the supplemental materials we prove convergence results  and comment on some implementation issues. In particular, we prove two results that together imply that the update rules can reach a stationary point of the objective function $\xi$ provided the solution lies on the positive orthant of the feasible region (all elements of all matrices in the solution contain strictly positive entries) and we start with an initial solution which is in the positive orthant.

 An alternative method to estimate the parameters in first step is the (centroid) co-regularized spectral clustering method. The method was proposed in \citet{kumar2011co} and its theoretical properties under the MLSBM were studied in \citet{paul2017spectral}. We propose to use this alternative method in conjunction with the conditional MLE described below and call the resulting method Co-Spectral. In our experiments we use the regularization parameters for Co-Spectral as 0.05 for all subject networks.

\subsubsection{The conditional maximum likelihood step}
The first step of the two-step method obtains $\hat{U}^{(1)},\ldots, \hat{U}^{(m)},\hat{U}^{*}$, which are matrices in the Grassmann manifold. In the second step, we first obtain community assignments from each of the $\hat{U}^{(m)}$s and $\hat{U}^{*}$, and then the transition probability matrix is obtained from the estimated matrices through a conditional maximum likelihood (ML) estimator. For Co-OSNTF community assignment is performed by assigning each row to the community corresponding to the largest element in a row, while for co-regularized spectral clustering this is accomplished through the k-means clustering of the rows of the matrices. Let $\hat{Z}^{(1)},\ldots,\hat{Z}^{(M)}$ be the member-wise community assignment matrices and $\hat{Z}^{*}$ be the mean community assignment matrix obtained in this way. Then the conditional ML estimate of $T$ is given by
\begin{equation}
  \hat{T}_{ql}=\frac{n_{ql}}{\sum_{l}n_{{ql}}},
  \label{condML}
\end{equation}
where $n_{ql}=\frac{1}{M}\sum_{m}\sum_{i}I\{\hat{Z}^{*}_{iq}=1,\hat{Z}^{(m)}_{il}=1\}$, with $I(\cdot)$ being the indicator function.
We note that the same estimator for $T$ can be obtained by equating the first moments as well.

\subsection{Two-sample hypothesis testing: whole network-level and node-level tests}

We next develop procedures for performing a two-sample hypothesis test between network community structures of two populations of subjects. Let the two populations to be compared be denoted as $A$ and $B$ (in neuroimaging experiments these correspond to a group of healthy controls and a group of patients), and the sample size from each group be $M_1$ and $M_2$ respectively. Further let the estimated model parameters for the two groups be  $\bar{Z}_{(A)},T_{(A)},\{Z^{(1)}_{(A)},\ldots,Z^{(M_1)}_{(A)}\}$ and $\bar{Z}_{(B)},T_{(B)},\{Z^{(1)}_{(B)},\ldots,Z^{(M_2)}_{(B)}\}$. There are two notions of ``group mean" of community structure giving rise to two natural ways of testing the difference between the two populations in modular organization. The first notion of ``group mean" is $\bar{Z}$, the putative mean community assignment matrix of the group of networks. Accordingly the notion of group mean at the node level is $\bar{Z}_i$ for node $i$. However, note that $\bar{Z}$ is not the expectation of the community assignment matrices $Z^{(m)}$s in the group. Instead the expectation is $\bar{Z}T$. This implies that for any node, its expected community assignment vector in any network within the group is $\bar{Z}_iT$. Hence the second notion of ``group mean" is represented by $\bar{Z}T$ at the network level and $\bar{Z}_iT$s at the node level.

The first network level test statistic we propose is the distance between $\mathcal{R}(\bar{Z}_{(A)})$ and $\mathcal{R}(\bar{Z}_{(B)})$, the subspaces spanned by the columns of $\bar{Z}_{(A)}$ and $\bar{Z}_{(B)}$ respectively. Note that the columns of both matrices $\bar{Z}_{(A)}$ and $\bar{Z}_{(B)}$ span subspaces in the Grassmann manifold $\mathcal{G}(k,n)$ \citep{edelman1998geometry}. A common notion of distance between  subspaces in the Grassmann manifold  is in terms of norms of the matrix of sines of canonical angles between the subspaces \citep{stewart,edelman1998geometry,dong2014clustering}. Formally, we define the ``Sine test" statistic as the Grassmann subspace distance between the putative mean community assignment subspaces,
\begin{align*}
S=\|\sin(\Theta(\mathcal{R}(\bar{Z}_{(A)}),\mathcal{R}(\bar{Z}_{(B)})))\|_F^2  =\frac{1}{2}\|\bar{Z}_{(A)}Q_A\bar{Z}_{(A)}^T-\bar{Z}_{(B)}Q_B\bar{Z}_{(B)}^T\|_F^2
\end{align*}
where $\Theta(\mathcal{R}(\bar{Z}_{(A)}),\mathcal{R}(\bar{Z}_{(B)}))$ is the matrix canonical angles between subspaces spanned by the columns of $\bar{Z}_{(A)}$ and $\bar{Z}_{(B)}$,  $Q_A=(\bar{Z}_{(A)}^T\bar{Z}_{(A)})^{-1}$ and $Q_B=(\bar{Z}_{(B)}^T\bar{Z}_{(B)})^{-1}$ are diagonal ``scaling" matrices whose elements are the sizes of the communities.

The second test statistic corresponds to the second notion of ``group mean" as described earlier. Following similar intuition as the Sine test, the network level ``multinomial unit vector (MUV)" statistic is:
\[
MUV=\|\bar{Z}_{(A)}T_{(A)}T_{(A)}^{T}\bar{Z}^{T}_{(A)} - \bar{Z}_{(B)}T_{(B)}T_{(B)}^{T}\bar{Z}^{T}_{(B)}\|_F^2.
\]
Note in this case we have dropped the diagonal scaling matrices that we had for Sine test. Large values of the test statistic will then indicate that the mean module structures in the two groups are markedly different. For each node $i$, define the node level MUV test statistic
\[
MUV(i)=(\bar{Z}_{(A)i}T_{A} - \bar{Z}_{(B)i}T_{B})^T (\bar{Z}_{(A)i}T_{A} - \bar{Z}_{(B)i}T_{B}).
\]

Intuitively, $\bar{Z}T$ is a better measure of the group mean since it is the expectation of the subject's community assignment matrices. Consequently, we expect MUV test to be a better test than Sine test, since it takes into account the variation in community structure present in the population. Indeed we observe the same in our simulations (omitted in the interest of space) and henceforth only use the MUV test.

In all the above cases it is difficult to derive asymptotic distribution of the test statistics.  Hence we construct p-value for the test through a permutation test based on re-sampling from the observed networks. We combine the network samples together and sample without replacement from the combined sample of $M_1+M_2$ networks to create two samples of sizes $M_1$ and $M_2$ respectively. We fit the RESBM to both samples using variational EM and two-step methods and compute the Sine test and MUV test statistics in each case. We repeat the procedure many times to construct the empirical distribution of the test statistic under the null hypothesis. Comparing the observed value of the test statistics with the constructed empirical distribution yields the p-values. For the node level tests, we can perform the same procedures. However, when we make inference using the p-values we need to account for multiple comparisons through a Family Wise Error Rate (FWER) or False Discovery Rate (FDR) correction \citep{benjamini1995controlling}.

\section{Performance on simulated networks}
\label{sec:simulation}

In this section we numerically compare the performance of the proposed methods along with some already available or baseline methods in samples of networks simulated from the RESBM. We compare the performance under three metrics: the average performance in community detection across the members of the network sample $Z^{(1)},\ldots Z^{(m)}$, the performance in estimating the putative mean community structure $\bar{Z}$, and the  accuracy in estimating the transition probability matrix $T$. The normalized mutual information (NMI), an information theoretic measure of similarity between two community structures, is used to assess the accuracy of the estimated community assignments against the true assignments. The NMI between two community assignment vectors measures the degree to which one community assignment vector can be obtained from the knowledge of the other vector. It takes values between $0$ and $1$, where $0$ indicates a random assignment (no overlap of information) and $1$ indicates a perfect match, and the higher the value, the better the match. The accuracy of the estimated transition probability matrix is measured in terms of difference in Frobenius norm.

We generate the sample of networks from the RESBM in the following way. We first generate the group putative community labels $\bar{Z}_i$s from a multinomial distribution with $k$ classes and equal probability for each class. To generate the community assignments of the member networks, the community labels of a fraction $\kappa$ of nodes are then randomly changed to one of the communities other than its original community. Hence the transition probability matrix $T$ will have $1-\kappa$ as the diagonal elements, while the $k-1$ off-diagonal elements in each row sum to $\kappa$. We call the fraction $\kappa$ ``variation factor" since it is an indicator of how much variation there is in the community structure among the members of the sample.

For each of the members the edges between the nodes are then drawn from a stochastic block model in the following fashion. We first generate the vector of $k$ diagonal elements which is common for all members (required for the model to be identifiable) as $\lambda \sim U(a,b)$, where $U(a,b)$ denotes the continuous uniform distribution with parameters $a$ and $b$. Next in each member the lower half of the $k^2-k$ off-diagonal elements are generated from $U(a/\rho,b/\rho)$ while the upper half are identical to the lower half. The parameter $\rho$ controls the signal to noise ratio (SNR), and in all our experiments we keep the value close to 2 in all members. The average density of the networks is controlled by another parameter called degree multiplier. Roughly speaking, increasing the degree multiplier by 1 corresponds to an increase of 2\% of maximum degree in the degree density per network. As an example if we have 500 nodes in a network, then a degree multiplier of 3 corresponds to average degree density per network being $500 \times 0.06=30$.

\subsection{Methods compared}
In our simulations, we compare the performance of three algorithms from the two proposed methods along with a number of baseline and other available methods. In particular, the following methods are compared: (a)
\textbf{VarEM:} The variational EM algorithm for computing approximate MLE in RESBM, (b) \textbf{Co-Spectral:} The two-step co-regularized spectral clustering with conditional MLE, (c) \textbf{Co-OSNTF:} The two-step co-regularized orthogonal non-negative matrix tri-factorization with conditional MLE, (d) \textbf{Ind. Spectral:} Spectral clustering in each member network performed independently \citep{rcy11}. This method is used only for the comparison on performance in individual networks, (e) \textbf{Mean Spectral:} Spectral clustering of the mean adjacency matrix \citep{hxa14,paul2017spectral}, (f) \textbf{SpectralK:} The spectral kernel method for mean community detection \citep{paul2017spectral}, which is similar to the module allegiance matrix technique in \citet{braun15}, and (g) \textbf{MLSBM:} The variational EM algorithm in MLSBM \citep{hxa14,pc15,barbillon2017stochastic}. This method is used only for comparison in terms of putative mean community assignments.

In our comparisons we also include an estimate for the $T$ matrix obtained by using Ind. Spectral in the individual networks and SpectralK for the mean community assignments. Some comments on the initialization and practical implementation of the methods are made in the supplementary materials.

\begin{figure}[htbp]
\centering{}
\includegraphics[width=.95\linewidth]{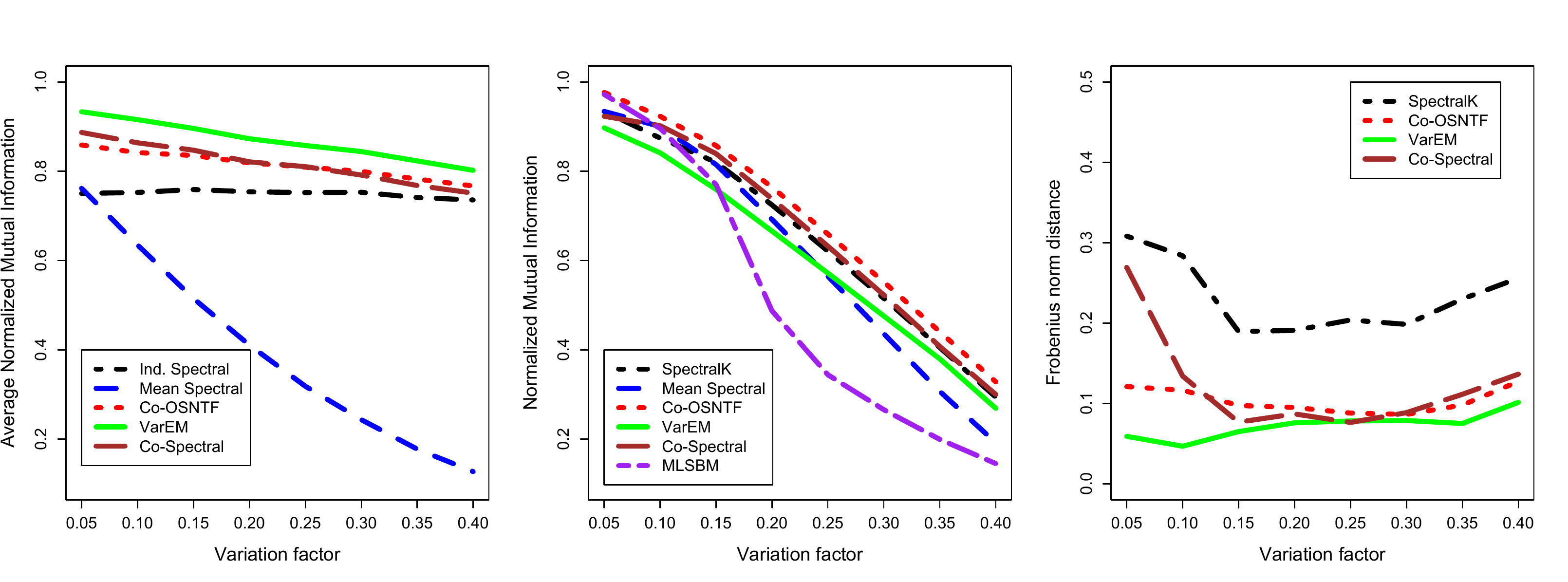}
\vspace{-10pt}
\begin{center}  (a) \hspace{120pt} (b) \hspace{120pt} (c)  \end{center}
\vspace{-10pt}
\caption{Performance of various methods across three metrics: (a) Average clustering performance across all member networks (in NMI), (b) performance in detecting the mean community structure (in NMI) and (c) accuracy in estimating the transition probability matrix (in Frobenius norm) with increasing variation factor from 0.05 to 0.40. The number of nodes is 500, the number of communities is 3, the number of member networks is 5 and the average degree per networks is 40.}
\label{varfac}
\end{figure}

\subsection{Increasing variation across members of the sample}

Our first simulation setup fixes $n$ at 500, $M$ at 5, $k$ at 3, and the average degree in each network at 40 (which is about 8\% degree density), while it varies the variation factor across the networks from 0.05 to 0.40, in steps of 0.05. Figure \ref{varfac} shows the results of this simulation across the three aforementioned metrics of comparison.

We note that in terms of community detection in member networks, the
performance of all methods, except Ind. Spectral, steadily
falls as the variation factor increases (see Figure \ref{varfac}(a)). This is because with increasing variation factor, the networks are increasingly dissimilar and hence information sharing across networks does not improve performance as much as it does for low variation factor. The VarEM consistently outperforms all other methods in this metric, while the performance of Co-Spectral and Co-OSNTF trails closely. The Ind. Spectral does not share information across networks and hence is agnostic to variation factor. Consequently, although it gives inferior performance initially, its performance almost catches up with Co-Spectral and Co-OSNTF with increasing variation factor. Assuming the same community structure in all networks and using the community assignments obtained from Mean Spectral for all member networks gives considerably worse performance, especially when the variation factor is large, since the networks become very dissimilar.

For the putative mean community assignments all methods, except MLSBM, behave similarly (see Figure \ref{varfac}(b)). The two-step methods (Co-Spectral and Co-OSNTF) and SpectralK slightly outperform VarEM in this case. While the performance of all methods falls with increasing variation factor, the fall is slightly steeper for Mean Spectral as compared to the two-step methods. Finally the estimate of $T$ is most accurate with VarEM and stays almost flat with increasing variation factor. The performance of Co-Spectral is poor initially at low variation factor but quickly improves as the variation factor increases. The performance of Co-OSNTF is slightly worse than that of VarEM and also stays flat with increasing variation factor. The estimate of $T$ from the combination of Ind. Spectral and SpectralK (labeled as SpectralK in Figure \ref{varfac}(c)) performs poorly compared to the other three methods throughout.

\begin{figure}[h]
\centering{}
\includegraphics[width=.95\linewidth]{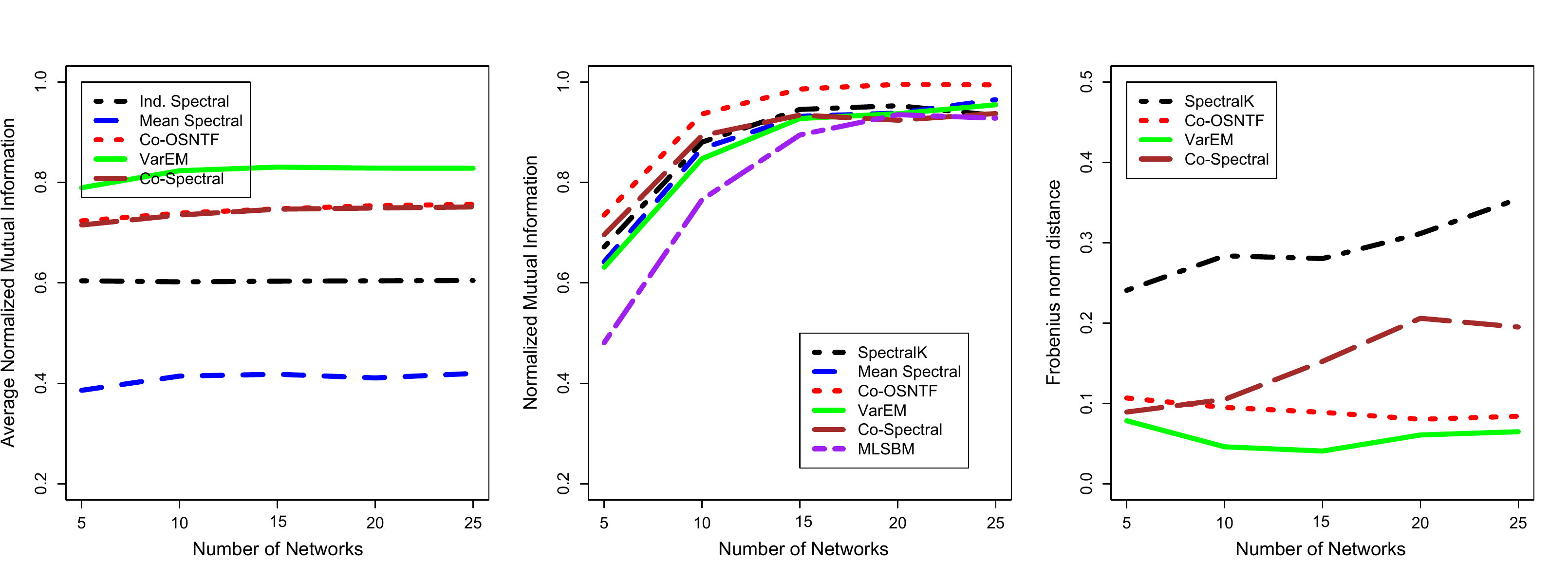}
\vspace{-10pt}
\begin{center}  (a) \hspace{120pt} (b) \hspace{120pt} (c)  \end{center}
\vspace{-10pt}
\caption{Performance of  various methods across three metrics: (a) Average clustering performance across all member networks (in NMI), (b) performance in detecting the mean community structure (in NMI) and (c) accuracy in estimating the transition probability matrix (in Frobenius norm) with increasing number of member networks from 5 to 25. The number of nodes is 300, the number of communities is 3, the average degree per  networks is 25, and the variation factor is 0.20.}
\label{incm}
\end{figure}

\subsection{Increasing size of sample}
This simulation is to assess the effect of increasing the sample size, i.e., the number of member networks in the sample on the performance of the methods. We fix $n$ at 300, $k$ at 3, the average degree per network at 25 (about 8\% degree density), and the variation factor at 0.20, while we vary the number of networks from 5 to 25. The three plots in Figure \ref{incm} display the results on the three metrics of comparison. The VarEM outperforms all competing methods in terms of average performance in the individual networks (Figure \ref{incm}(a)) and the accuracy of estimating the $T$ matrix (Figure \ref{incm}(c)). However, it underperforms in estimating the mean community assignments as compared to Mean Spectral, SpectralK, Co-Spectral and Co-OSNTF (Figure \ref{incm}(b)). Among the two-step algorithms the newly proposed Co-OSNTF performs better than Co-Spectral algorithm in estimating both the mean community assignment and $T$ matrix, while remaining competitive with Co-Spectral in member-wise performance. The SpectralK algorithm is competitive to Co-Spectral in detecting the mean community structure. However its member-wise counterpart, the Ind. Spectral, performs poorly in detecting the member-wise community structures. Consequently the estimates of $T$, derived from a combination of Ind. Spectral and SpectralK, are also less accurate compared to other methods.

In the supplementary materials we present another simulation where we vary the density of the member networks while fixing the variation factor, number of nodes, number of communities and number of member networks.

\subsection{Performance of hypothesis testing procedures on synthetic networks}

We next numerically compare the  MUV tests using the estimates from the various methods developed in this article in a hypothesis testing problem on synthetic networks generated from the RESBM. We generate two samples of sizes 20 and 25 from a RESBM with 100 nodes, 3 communities and a $T$ matrix whose diagonal elements are $0.8$ and the off diagonal elements are $0.2$. Hence there is a $20 \%$ chance that a node will not retain its $\bar{Z}$ community in $Z^{(m)}$. The putative mean community assignment matrices for the two populations $\bar{Z}_{A}$ and $\bar{Z}_{B}$ are changed from being identical to being different in approximately $30 \%$ of nodes at intervals of $5\%$. A good test should not detect any difference either at network level or node level when the samples come from the same population RESBM, but detect differences as the populations from which the samples are drawn differ. The distribution of all the test statistics under the null hypothesis was computed by resampling from the combined sample 10000 times. The computation time was about 300 CPU core-hours (using HP X5650 2.66 GHz cores in a computing cluster) for each of the seven cases. Due to large number of function calls to the VarEM method, this test is computationally expensive, however evaluation in the resampled datasets can be performed in parallel. The results from the network level tests are presented in Table \ref{tab:network_test} and those from the node level tests are presented in Table \ref{tab:node_test}.

\begin{table}[h]
\caption{Network level tests: p-values of various test statistics on two synthetic network samples of sizes 20 and 25 drawn from a 100-node, 3-community RESBM. The columns represent results for different fraction of nodes that were changed to obtain the second $\bar{Z}$ from the first. 10000 resamples was used to compute the p-values, a * indicates significant at 1\% level}
\begin{center}
\begin{tabular}{lccccccc}
\hline
 Test & 0 & 0.05 & 0.10 & 0.15 & 0.20 & 0.25 & 0.30  \tabularnewline
\hline
MUV test (VarEM) & 0.6797 & 0.0013* & 0.0000* & 0.0000* & 0.0000* & 0.0000* & 0.0000* \tabularnewline
MUV test (Co-Spectral) & 0.3437 & 0.0087* & 0.0021* & 0.0000* & 0.0000* & 0.0000* & 0.0000* \tabularnewline
MUV test(Co-OSNTF) & 0.1316 & 0.0051* & 0.0012* & 0.0000* & 0.0000* & 0.0000* & 0.0000* \tabularnewline
MUV test (SpectralK)  & 0.2341 & 0.0037* & 0.0061* & 0.0000* & 0.0000* & 0.0000* & 0.0000* \tabularnewline
\hline
\end{tabular}
\end{center}
\label{tab:network_test}
\end{table}

\begin{table}[!h]
\caption{Performance of node level tests (total errors and false positives) on testing between two samples of synthetic networks of sizes 20 and 25 drawn from a 100-node, 3-community RESBM at 0.05 FDR threshold. ``Number of nodes changed" is the actual number of nodes whose putative mean community assignment varied between the two groups. The best performance in terms of least total errors in each column is indicated in bold.}
\vspace{-10pt}

\begin{center}
\resizebox{\textwidth}{!}{
\begin{tabular}{lccccccc|ccccccc}
\hline
& \multicolumn{7}{c}{False positives} & \multicolumn{7}{|c}{Total errors}\tabularnewline
\hline
Number of nodes changed & 0 & 6 & 7 & 12 & 23 & 24 & 32 & 0 & 6 & 7 & 12 & 23 & 24 & 32\tabularnewline
\hline
MUV test (VarEM) & 0 & 0 & 2 & 1 & 1 & 5 & 2 & 0 & \textbf{4} & \textbf{5} & \textbf{4} & 3 & 5 & 2 \tabularnewline
MUV test (Co-Spectral) & 0 & 0 & 0 & 1 & 0 & 1 & 1 & 0 & 6 & 7 & 10 & \textbf{0} & 1 & 1 \tabularnewline
MUV test(Co-OSNTF) & 0 & 0 & 0 & 1 & 0 & 0 & 0 & 0 & 5 & 7 & 7 & \textbf{0} & \textbf{0} & \textbf{0} \tabularnewline
MUV test (SpectralK)  & 0 & 0 & 0 & 1 & 0 & 2 & 1 & 0 & 5 & 6 & 11 & 1 & 2 & 1 \tabularnewline
\hline
\end{tabular}
}
\end{center}
\label{tab:node_test}
\end{table}

First we note from Table \ref{tab:network_test} that all the MUV tests can detect statistically significant difference at the 1\% level when the $\bar{Z}$ matrices differ by 5\% nodes. When the $\bar{Z}$ matrices differ by 10\% or more nodes, all MUV tests give extremely small p-values. Given that each component network community structure is expected to differ from its group mean in about 20\% of the nodes, the tests seem to be quite powerful. We also note that all tests correctly give large p-values when there is no difference between the $\bar{Z}$ matrices.

Table \ref{tab:node_test} presents the performance of the tests to detect node level differences. We correct for multiple comparison using a threshold of  0.05 FDR, and present results for thresholds of 0.05 FWER and 0.10 FDR in the supplementary materials. All MUV tests correctly fail to detect any node-level differences when no nodes were changed. However, the procedures start and continue to make errors as the number of nodes changed increases from 0 to 12, but then quickly reduce to almost no error as the number of nodes changed increases from 23 to 32. Among the MUV tests, both the VarEM based test and the Co-OSNTF based test perform particularly well. Moreover, the Co-OSNTF test detects all the node-level differences correctly when the number of nodes is 23 or more. However, the VarEM test suffers from increased false positives as the number of nodes changed becomes high, despite FWER and FDR controls.  We note that switching from 0.05 FDR to 0.10 FDR does not increase the number of false positives much but improves performance, especially when the number of nodes truly changed is low (see supplementary materials). Based on the observations from the performance of the competing methods and test statistics on the synthetic networks, we recommend the MUV test with VarEM and Co-OSNTF as the two best performing tests.

\section{Application of the methods to COBRE data}

From the exploratory analysis in Section 2, it is clear that the control group has higher modularity values as compared to the patient group, irrespective of our choice of threshold (and statistically significant at most of the thresholds). The average number of communities detected in each of the groups is not significantly different at any of the thresholds (Table \ref{tab:mod}). Both observations in terms of modularity and number of communities are consistent with previously reported results in the context of childhood-onset schizophrenia \citep{alexander2012discovery}.

\begin{figure}[h]
\centering{}
\begin{subfigure}{0.33\textwidth}
\centering{}
\includegraphics[width=\linewidth]{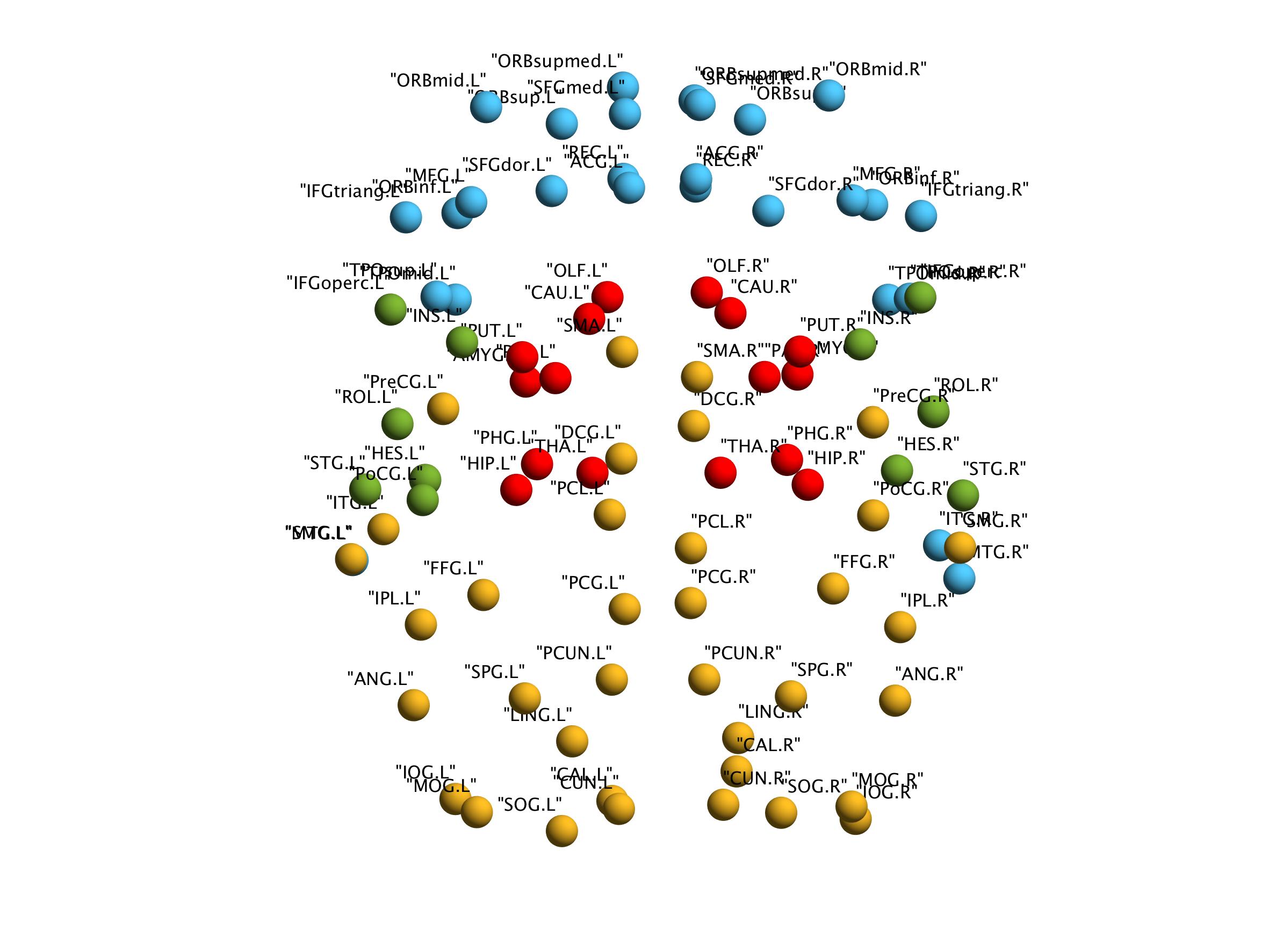}
\end{subfigure}%
\begin{subfigure}{0.33\textwidth}
\centering{}
\includegraphics[width=\linewidth]{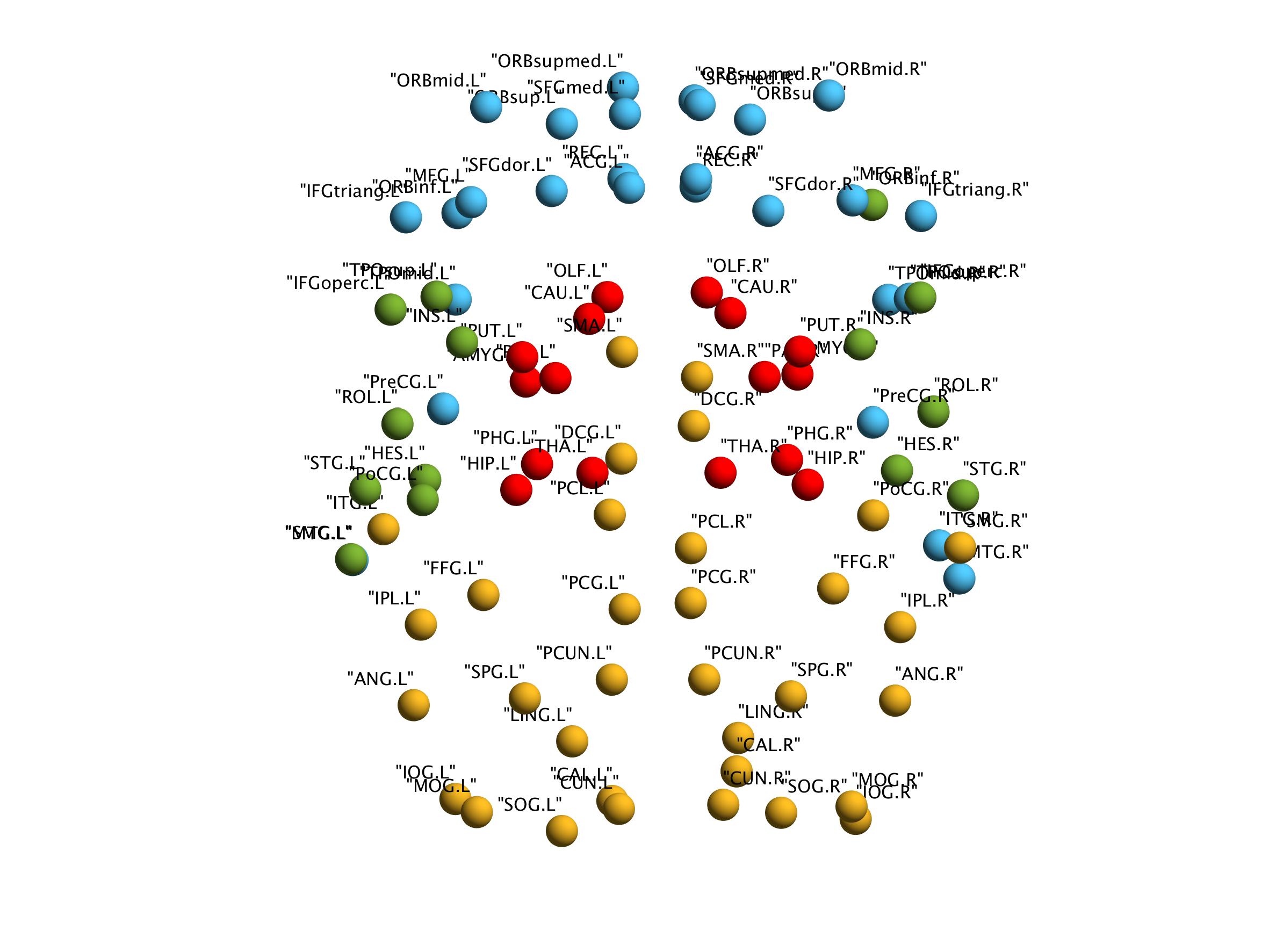}
\end{subfigure}%
\begin{subfigure}{0.33\textwidth}
\centering{}
\includegraphics[width=\linewidth]{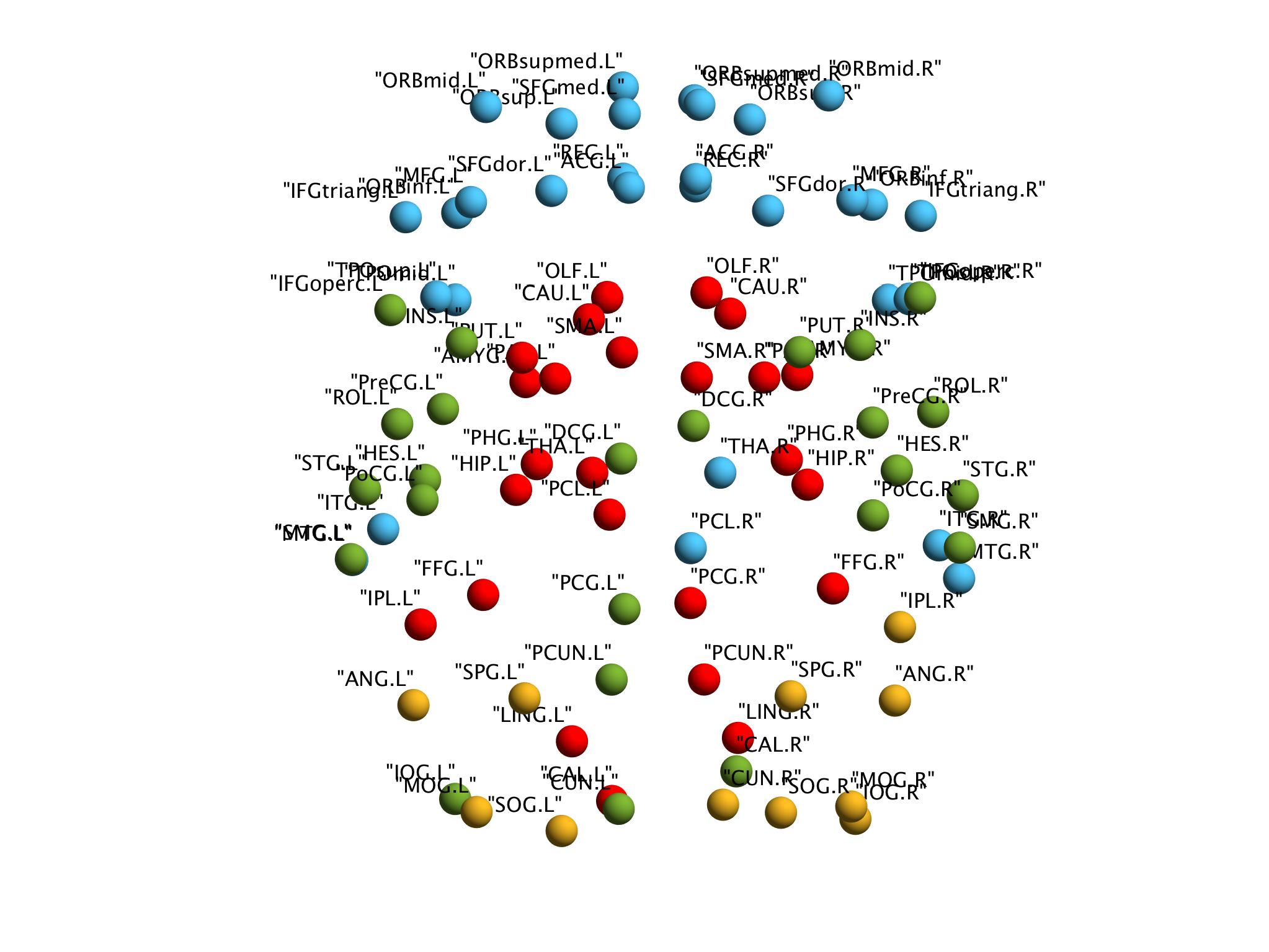}
\end{subfigure}
\begin{center} (a) Co-Spectral: controls \hspace{60pt} (b) Co-OSNTF: controls \hspace{60pt} (c) VarEM: controls \end{center}
\begin{subfigure}{0.33\textwidth}
\centering{}
\includegraphics[width=\linewidth]{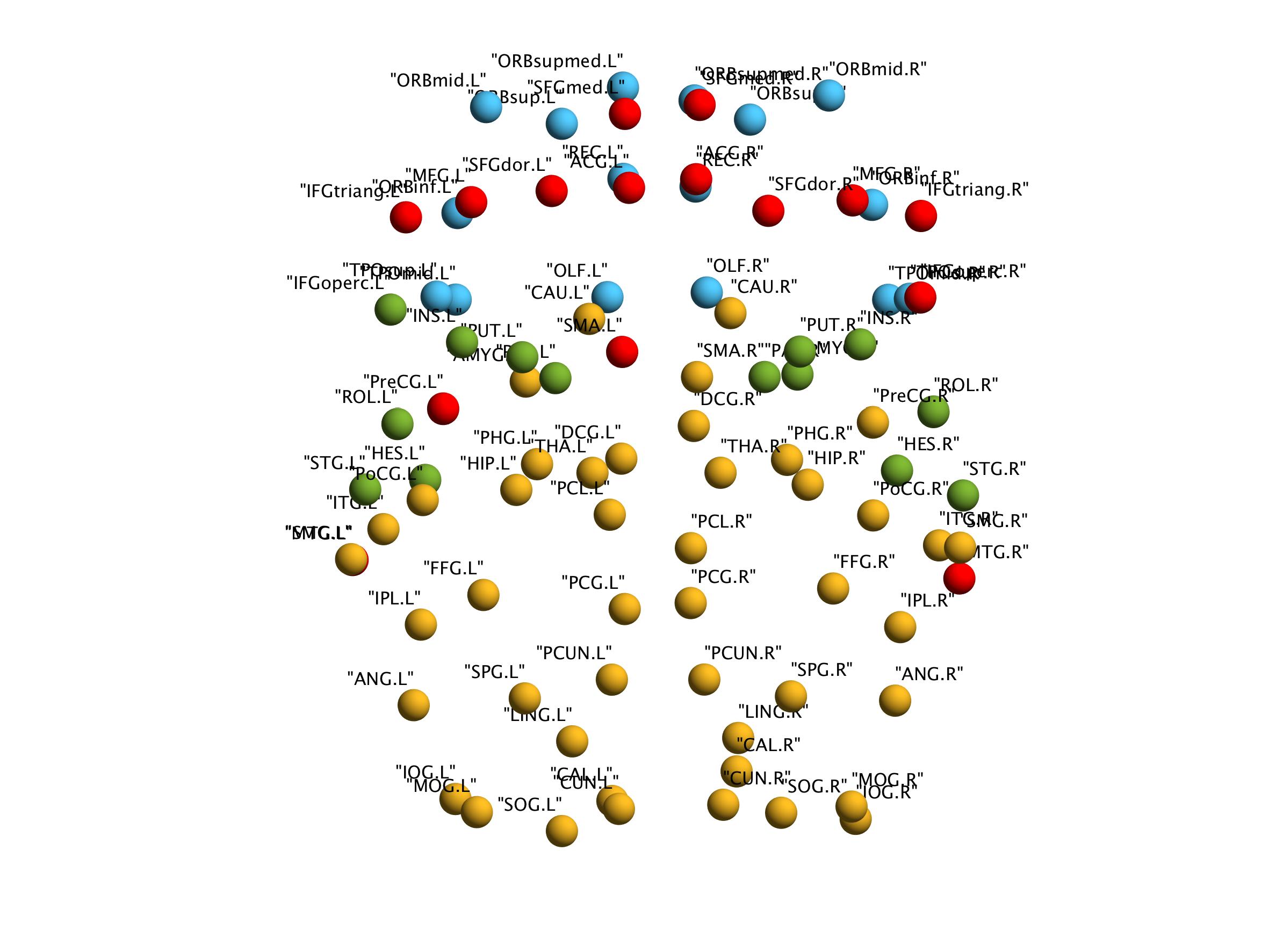}
\end{subfigure}%
\begin{subfigure}{0.33\textwidth}
\centering{}
\includegraphics[width=\linewidth]{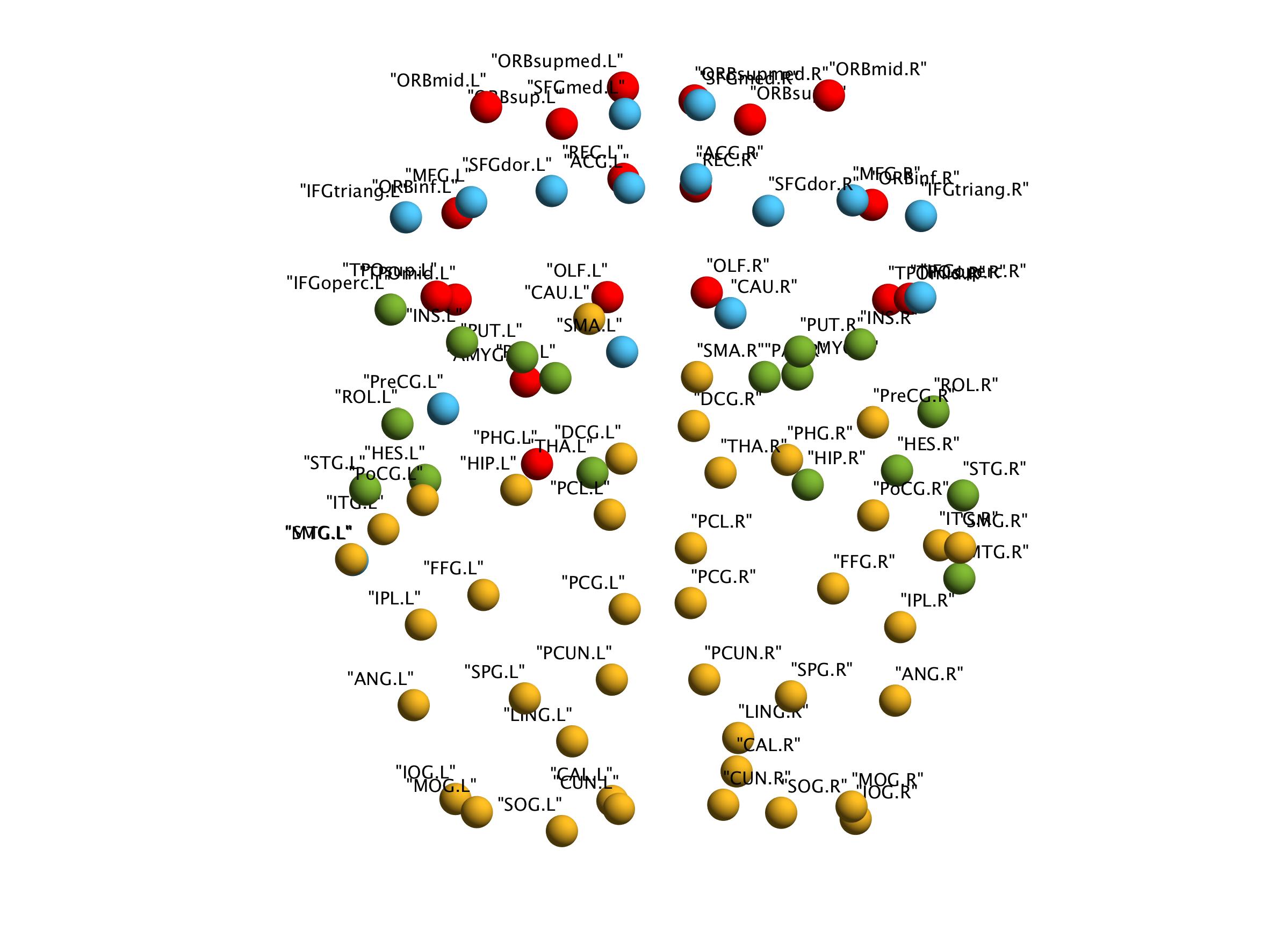}
\end{subfigure}%
\begin{subfigure}{0.33\textwidth}
\centering{}
\includegraphics[width=\linewidth]{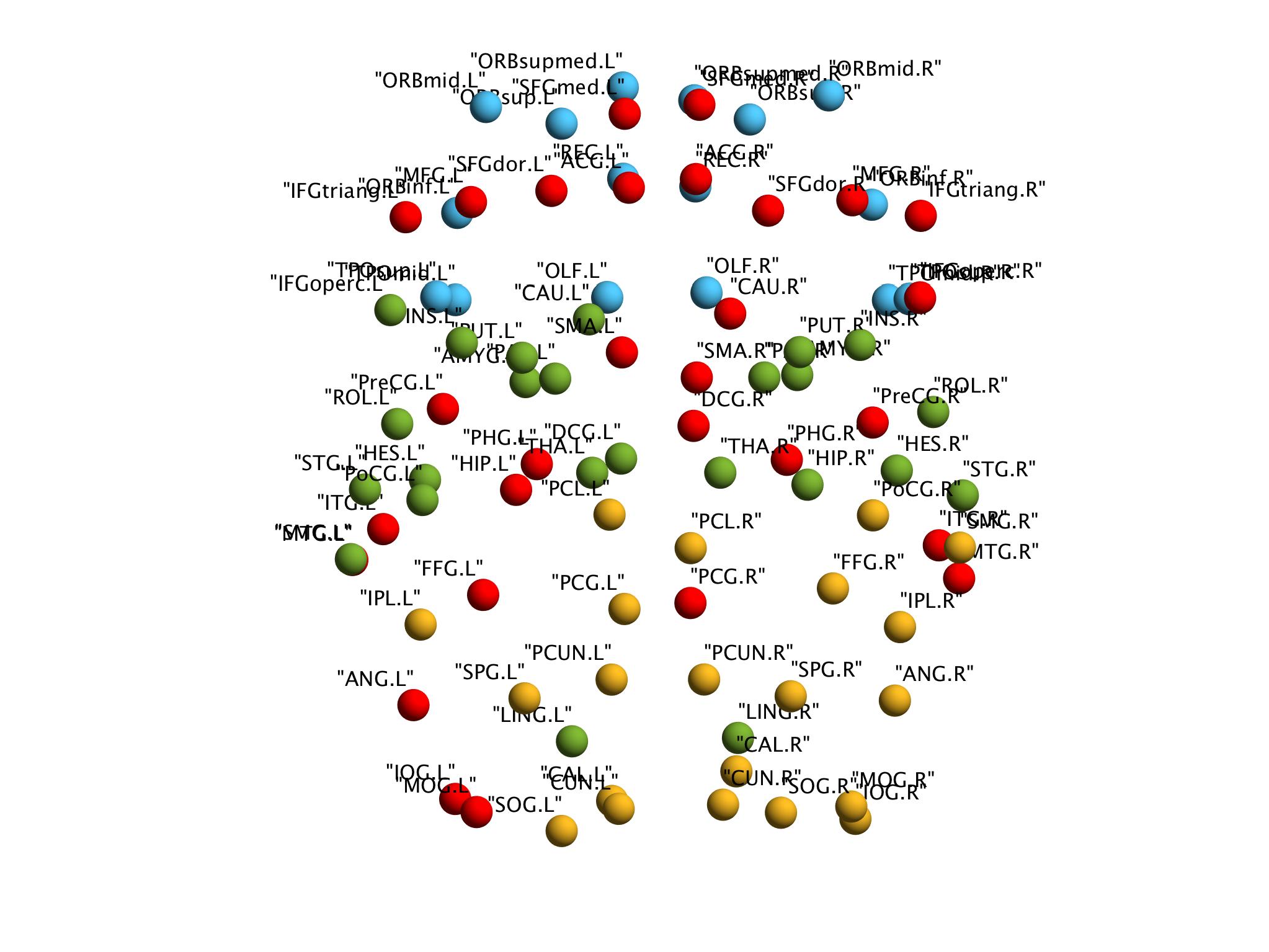}
\end{subfigure}
\begin{center} (d) Co-Spectral: patients \hspace{60pt} (e) Co-OSNTF: patients \hspace{60pt} (f) VarEM: patients \end{center}
\caption{Group putative community structure of resting state network based on AAL ROIs in (a, b, c) healthy controls, and (d, e, f) patients with schizophrenia. Nodes are colored according to their group putative community obtained from the following methods: Co-Spectral for the first column (a, d), Co-OSNTF for the second column (b, e) and VarEM for the third column (c, f).}
\label{putative}
\end{figure}

 \begin{table}[h]
\caption{Nodes of Interest: Nodes which are significant at 0.1 FDR correction.}

\begin{center}
\begin{tabular}{cccc}
Node & Uncorrected & FDR corrected & FWER
\tabularnewline
\hline
Pallidum.L (PAL.L) & 0.0004 & 0.0360 & 0.0360  \tabularnewline
Amygdala.R (AMYG.R) & 0.0020 & 0.0774 & 0.1780 \tabularnewline
Pallidum.R (PAL.R) & 0.0035 & 0.0774 & 0.3080 \tabularnewline
Putamen.L (PUT.L) & 0.0039 & 0.0774 & 0.3393 \tabularnewline
Thalamus.L (THA.L) & 0.0043 & 0.0774 & 0.3698\tabularnewline
\hline
\end{tabular}
\end{center}
\begin{center}
(A) Co-Spectral: Network level p-value: 0.0105
\end{center}
\begin{center}
\begin{tabular}{cccc}
Node & Uncorrected & FDR corrected & FWER
\tabularnewline
\hline
Caudate.R (CAU.R) & 0.0013 & 0.0742 & 0.1170 \tabularnewline
Temporal.Pole.Sup.L  (TPOsup.L)& 0.0017 & 0.0742 & 0.1513\tabularnewline
Hippocampus.R (HIP.R) & 0.0031 & 0.0742 & 0.2728  \tabularnewline
Pallidum.L (PAL.L) & 0.0033 & 0.0742 & 0.2871 \tabularnewline
\hline
\end{tabular}
\end{center}
\begin{center}
(B) Co-OSNTF: Network level p-value: 0.0042
\end{center}
\label{tab:scztest}
\end{table}

 We primarily focus our analysis on the threshold of 0.2 and later in Section 5.8 we repeat some of our analyses for two other thresholds 0.3 and 0.4 as robustness checks. Note that the methods developed in this paper require the number of communities to be supplied as input. We obtain the number of communities in each case from the average number of communities detected in Table \ref{tab:mod}. For the threshold of 0.2, we observe that the average number of communities in both groups is about 4. The histogram of number of communities in Figure \ref{explore2} also shows 4 as the most common number of communities for subjects in both control and patient groups. Hence we fit RESBMs with 4 communities to the networks from the control group and patient group subjects using both the varEM and the two-step methods.

 Although \citet{alexander2012discovery} have shown that group differences in both modularity value and modular organization become more pronounced at higher values of the threshold, we use a relatively smaller value of threshold since at higher values the optimum number of communities is quite high which are difficult to interpret and visualize. For example, for thresholds between 0.5-0.6, the network  divides into 30-60 communities (Table \ref{tab:mod}). In a network of 90 nodes, they are really high number of communities, and the network is essentially disintegrated in very local sub modules. Consequently, large differences are
expected between the community structures of
any two subjects, no matter which populations they are from, and therefore a consistent picture of the modular disruption in schizophrenia cannot be obtained.

\subsection{Group differences in community structure}
\begin{figure}[h]
\centering{}
\begin{subfigure}{0.48\textwidth}
\centering{}
\includegraphics[width=.95\linewidth]{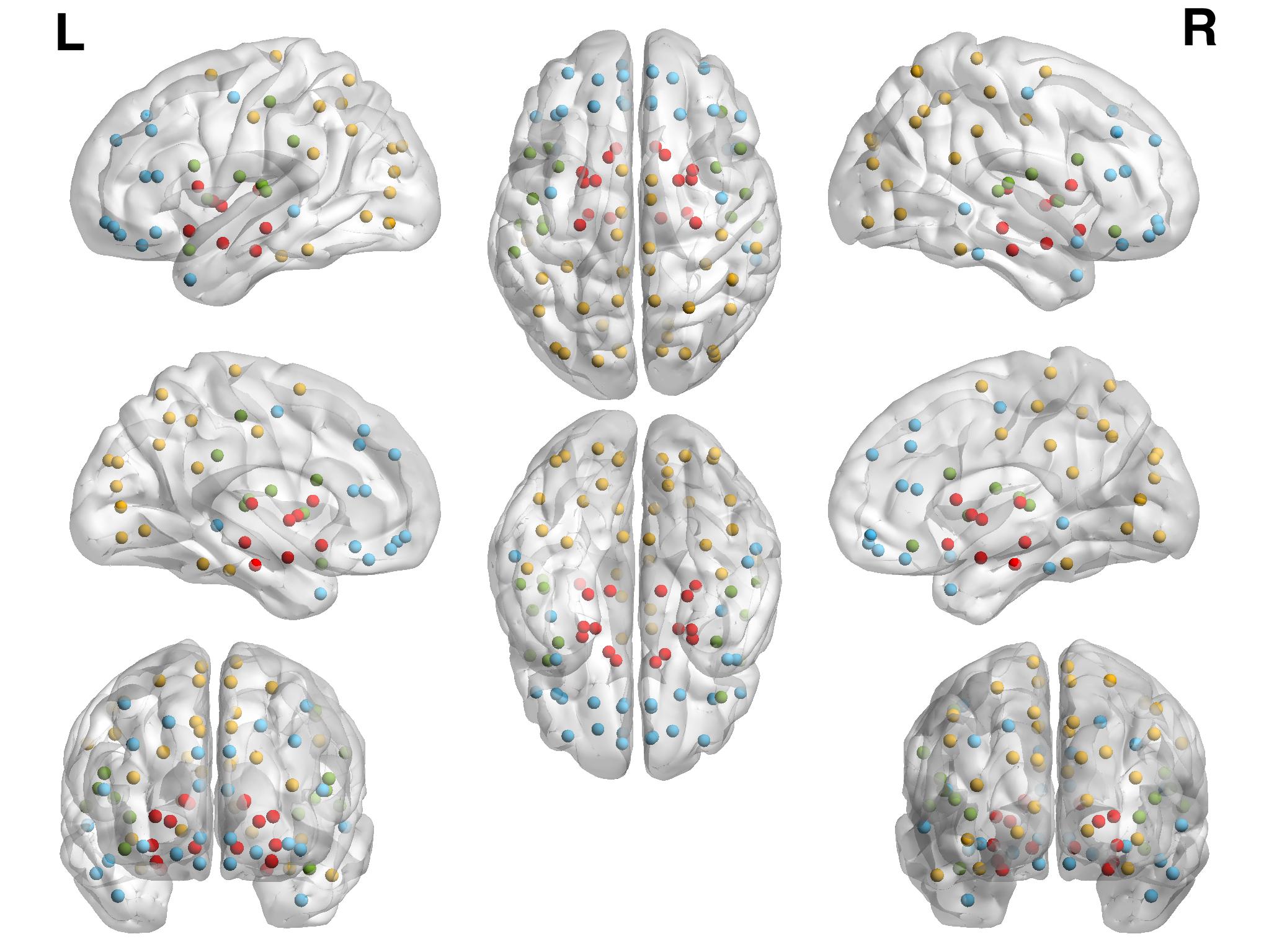}
\end{subfigure}%
\begin{subfigure}{0.48\textwidth}
\centering{}
\includegraphics[width=.95\linewidth]{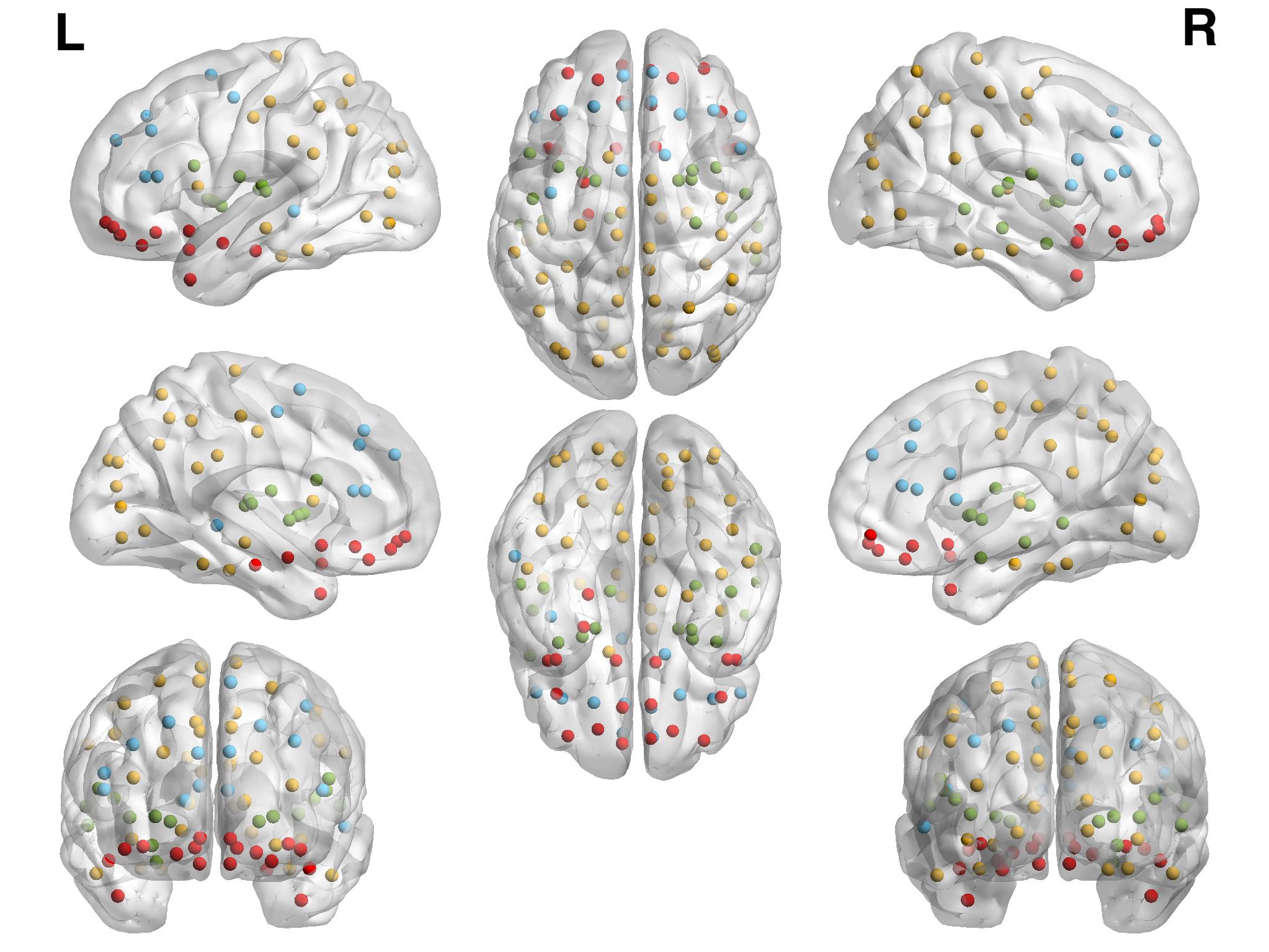}
\end{subfigure}
\begin{center} (a) Controls \hspace{140pt} (b) Patients \end{center}
\caption{A visualization of the group putative community structure in healthy controls and patients. The ROIs are colored according to the community memberships detected using Co-OSNTF method.}
\label{3D}
\end{figure}

The group putative community structure obtained for each group from each of the three methods is illustrated on a brain surface template in Figure \ref{putative}. In addition, Figure \ref{3D} displays a visualization of the locations of the ROI groups, in a three dimensional 8-view layout of the brain. All visualizations have been produced using the BrainNet Viewer software (\url{http://www.nitrc.org/projects/bnv/}) \citep{xia2013brainnet}. In Figures \ref{putative} and \ref{3D}, the ROIs are colored according to their community membership detected in the respective group putative community structures $\bar{Z}$. In each of the three methods, the community labels between controls and patients are matched by maximizing overlap through an algorithm that solves the Linear Sum Association Problem (LSAP) \cite{papadimitriou2003computational,kuhn1955hungarian}. This is necessary because community labels are recovered by any method only up to the ambiguity of label permutations, and hence two community partitions need to be aligned in terms of labels before they can be compared.  We first note that community structure for the control group from Co-Spectral and Co-OSNTF almost agree with each other. The four modules obtained roughly correspond to the modules detected consistently across different subjects in resting state in \citet{moussa12} using a voxel level network approach on a very large scale dataset.
Specifically, in the control group our modules blue (module 1), red (module 2), yellow (module 3) and green (module 4) have large overlaps with \citet{moussa12}'s default mode module, basal ganglia module, visual module and sensory/motor module respectively. We refer the reader to  \citet{moussa12} for a comprehensive list of modules detected in other works in the literature using both a network approach and an Independent Component Analysis (ICA) approach in resting state, to which these modules are equivalent or related to. The modules detected by the VarEM algorithm differs from the ones detected by Co-Spectral and Co-OSNTF, however there is a large overlap. Later in Sections 5.4 and 5.5 on prediction and validation, we note the performance of the VarEM algorithm is poor compared to Co-OSNTF and Co-Spectral for predicting community structure of held out subjects in this data. Therefore we primarily focus on the Co-OSNTF method in interpreting the results.

 For both Co-Spectral and Co-OSNTF the red group in controls gets disrupted in patients, and nodes belonging to the module get distributed to the yellow, green and blue modules in patients indicating disruption of the functional cohesion of these nodes. The blue module in controls also gets divided into two modules in patients, while the yellow and green groups remain relatively intact. Disruption of module structures has been reported in the literature on schizophrenia before, however, our methods provide visual identification of the disruption. In the subsequent analysis we investigate the disruption at the ROI level.

 To statistically test the disruption in community structure, we compute the p-values for the MUV test with 10000 permutation resamples. In both Co-Spectral and Co-OSNTF, the MUV test rejects the null hypothesis of no difference in the community structure between the controls and patients with p-values of 0.0105 and 0.0042 respectively (Table \ref{tab:scztest}). The node level MUV test with Co-Spectral and Co-OSNTF found a number of ROIs to be statistically significantly altered at 0.1 FDR corrected p-values, which we call nodes of interest in Table \ref{tab:scztest}. All the ROIs in cases of Co-Spectral and Co-OSNTF, except Temporal.Pole.Sup.L from Co-OSNTF, belong to the red module, as can be seen from Figure \ref{putative}.

 \begin{figure}[h]
\centering{}
\begin{subfigure}{0.25\textwidth}
\centering{}
\includegraphics[width=.95\linewidth]{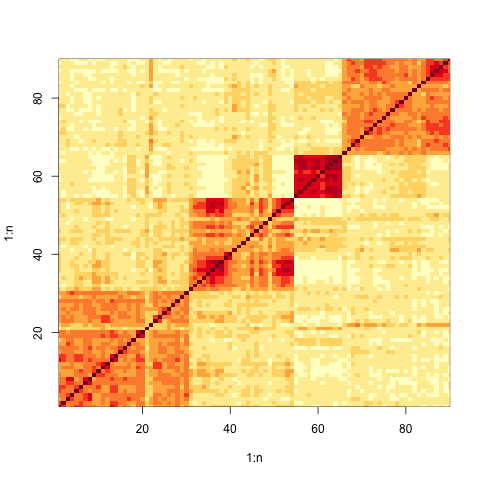}
\end{subfigure}%
\begin{subfigure}{0.25\textwidth}
\centering{}
\includegraphics[width=.95\linewidth]{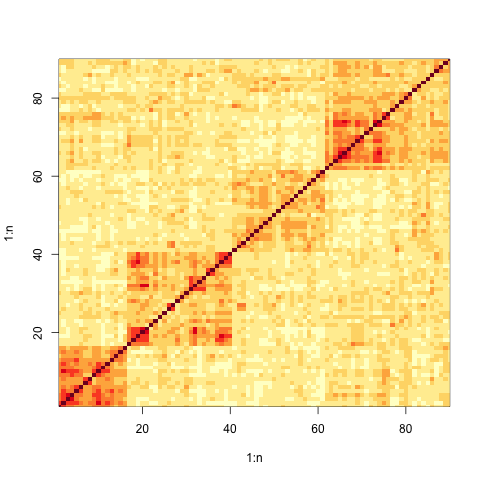}
\end{subfigure}%
\begin{subfigure}{0.25\textwidth}
\centering{}
\includegraphics[width=.95\linewidth]{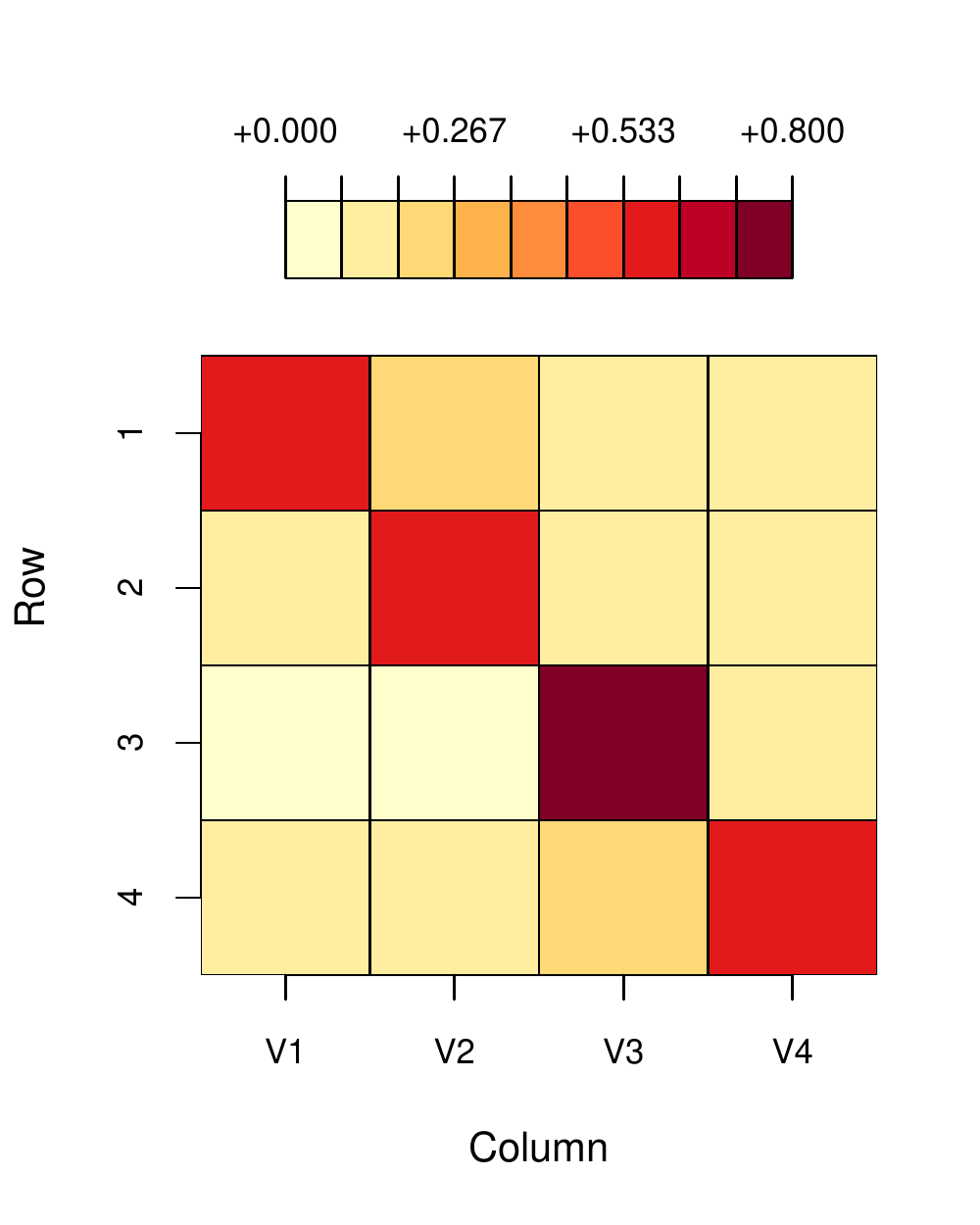}
\end{subfigure}%
\begin{subfigure}{0.25\textwidth}
\centering{}
\includegraphics[width=.95\linewidth]{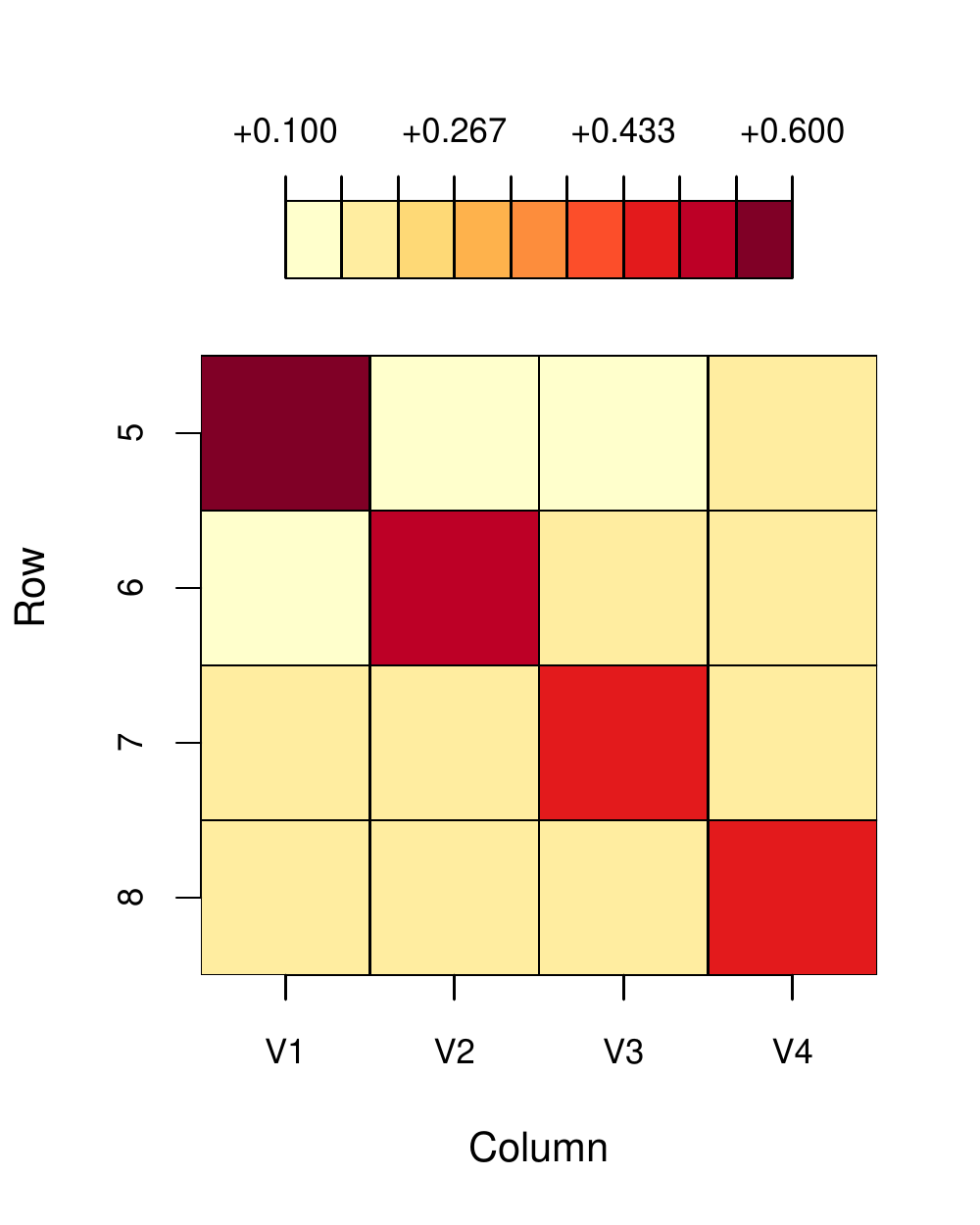}
\end{subfigure}
\vspace{-10pt}
\begin{center} (a) Controls \hspace{38pt} (b) Patients \hspace{38pt} (c) Controls \hspace{38pt} (d) Patients \end{center}
\vspace{-5pt}
\begin{subfigure}{0.25\textwidth}
\centering{}
\includegraphics[width=.95\linewidth]{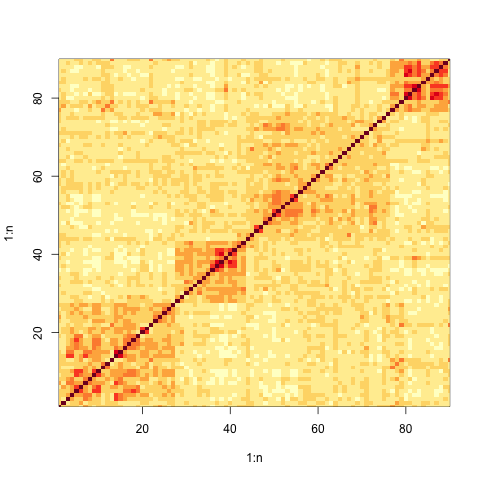}
\end{subfigure}%
\begin{subfigure}{0.25\textwidth}
\centering{}
\includegraphics[width=.95\linewidth]{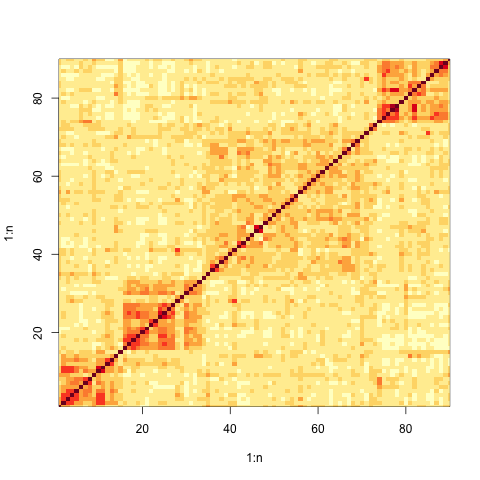}
\end{subfigure}%
\begin{subfigure}{0.25\textwidth}
\centering{}
\includegraphics[width=.95\linewidth]{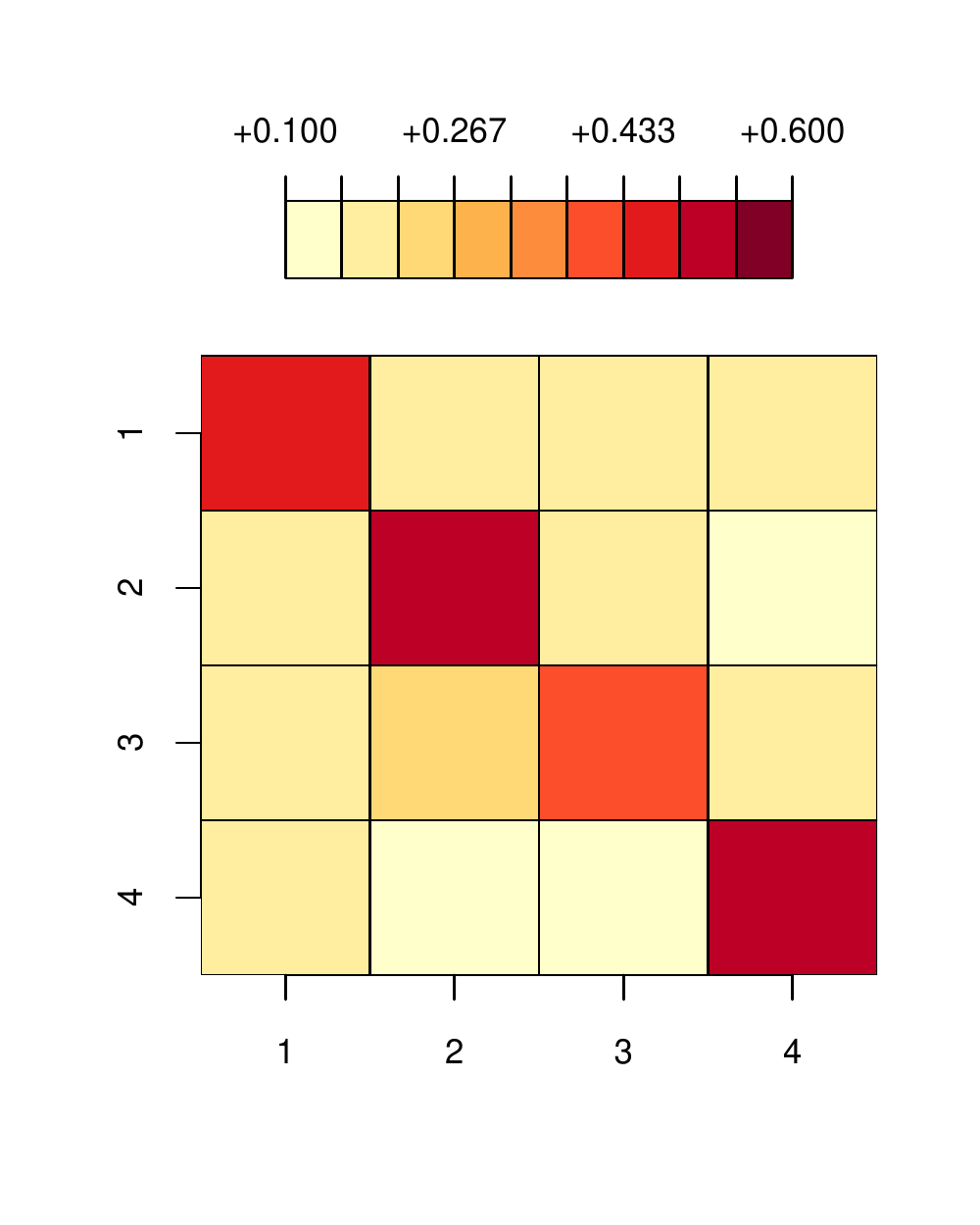}
\end{subfigure}%
\begin{subfigure}{0.25\textwidth}
\centering{}
\includegraphics[width=.95\linewidth]{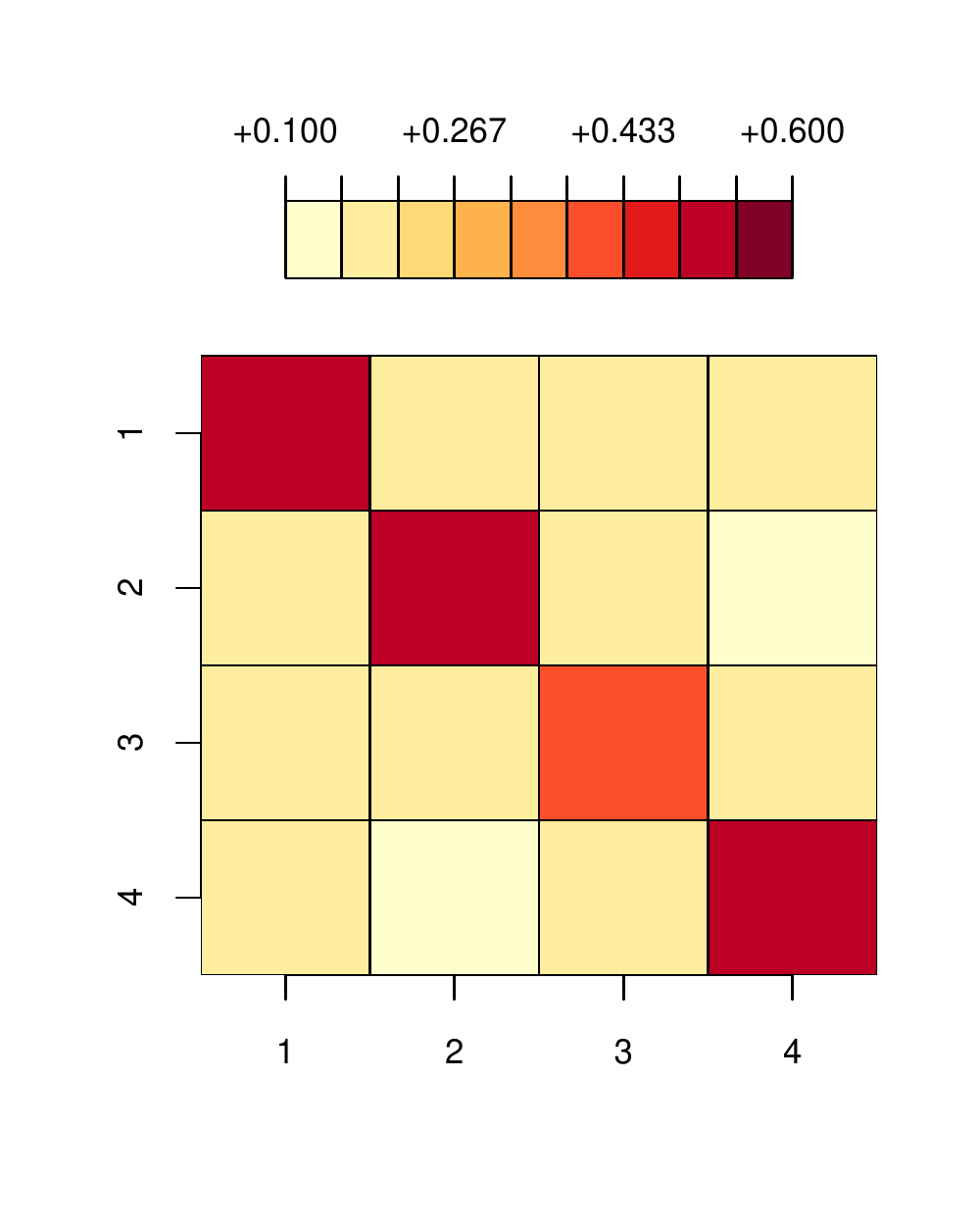}
\end{subfigure}
\vspace{-10pt}
\begin{center} (e) Controls \hspace{38pt} (f) Patients \hspace{38pt} (g) Controls \hspace{38pt} (h) Patients \end{center}
\vspace{-5pt}
\caption{ Stability of the group putative community assignments for VarEM (upper row (a)-(d)) and Co-OSTNF (lower row (e)-(h)): (a), (b), (e), and (f) are the matrices with elements as fraction of subjects for which two nodes (ROIs) are in the same community sorted according to the putative community structure for healthy controls and patients respectively; (c), (d), (g), and (h) are the estimated transition matrices among modules for healthy controls and patients respectively.}
\label{flex}
\end{figure}

Our findings add to the growing evidence that module structure in brain functional networks gets disrupted in schizophrenia. Not only are we able to highlight the nature of the disruption at the global scale, but also more locally infer the ROIs where the differences are the most.

\subsection{Consistency of community structure across subjects}

It is also important to understand the variability of the different modules in the group putative community structure across the different subjects within the group. In our model, it can be assessed through the estimated transition probability matrices, as well as a module consistency matrix that measures the fraction of subjects for which two nodes or ROIs are in the same community. The module consistency matrix is similar to the module allegiance matrix in \citet{braun15} and Scaled Inclusivity measure in \citet{steen2011assessing} and \citet{moussa12}. Figure \ref{flex} displays the two measures for the results from VarEM (Figure \ref{flex} (a)-(d)) and Co-OSNTF (Figure \ref{flex} (e)-(h)) methods. In Figures \ref{flex}(a), (b), (e), and (f) the ROIs are sorted according to their community label in increasing order (i.e., module 1 (blue module) in the bottom left corner and module 4 (green module) in the top right corner). In Figures \ref{flex}(c), (d), (g), and (h), the $4 \times 4$ module transition matrix is plotted with the putative modules arranged from 1 to 4 along the row and column (i.e., module 1 (blue module) in the top left corner and module 4 (green module) in the bottom right corner).  From Figures \ref{flex} (a)-(d) the module structure appears to be more consistent in controls than in patients for varEM. The patients show greater variability in the functional connectivity between any two regions, leading them to be classified into different modules more often than controls. This is possibly because of the variation in severity of the underlying conditions in the patient group. In particular, it can be seen from both the metrics in Figure \ref{flex} that the yellow module (module 3) is highly consistent in controls as has been observed in many previous resting state studies (see \citet{moussa12} and references therein), however, in patients it is much less consistent. We do not see much difference in consistency of module structure between controls and patients using Co-OSNTF method (Figure \ref{flex} (e)-(f)).

\begin{figure}[!h]
\centering{}
\begin{subfigure}{0.24 \textwidth}
\includegraphics[width=\linewidth]{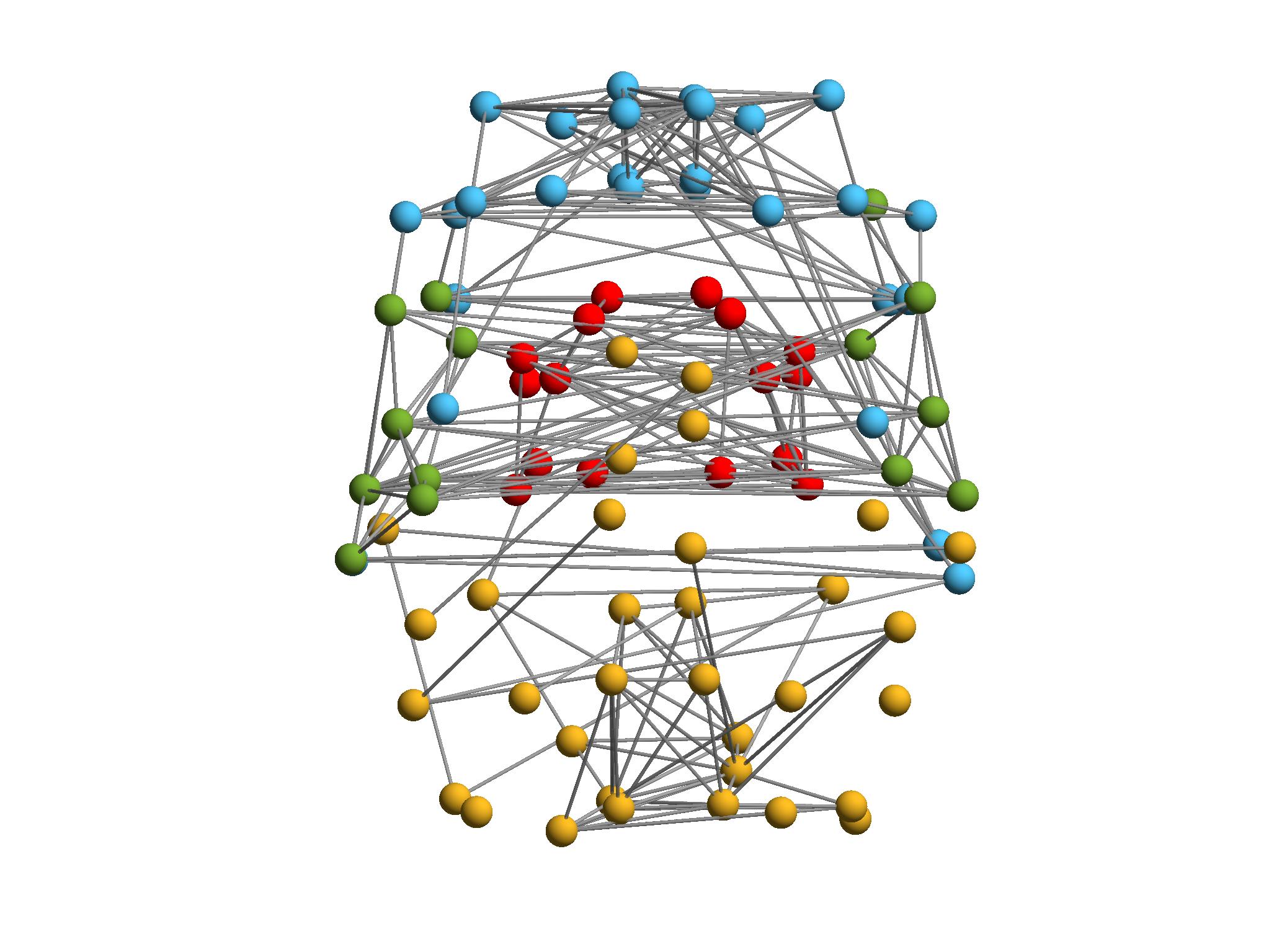}
\end{subfigure}%
\begin{subfigure}{0.24 \textwidth}
\includegraphics[width=\linewidth]{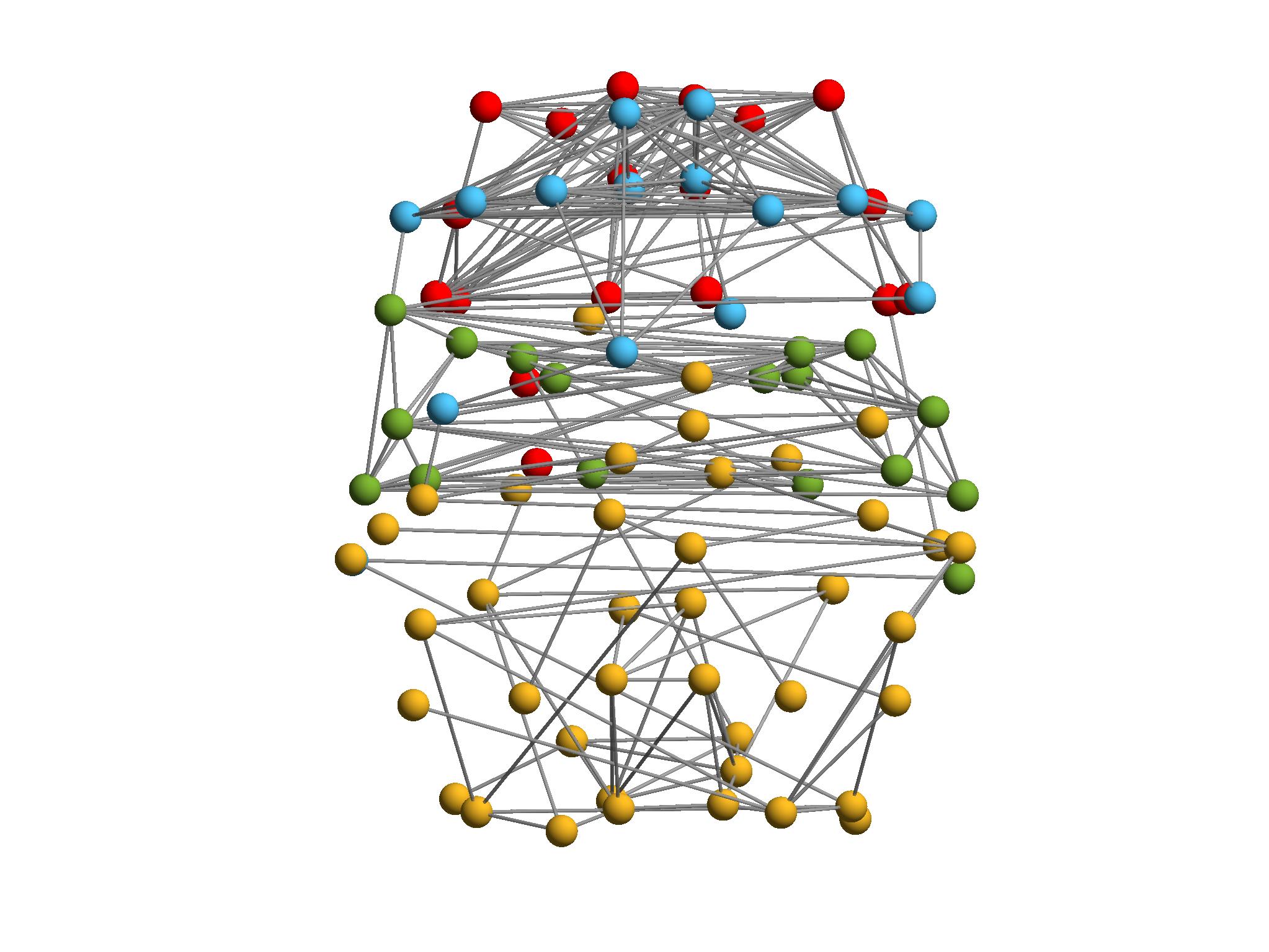}
\end{subfigure}\hfill%
\begin{subfigure}{0.24 \textwidth}
\includegraphics[width=\linewidth]{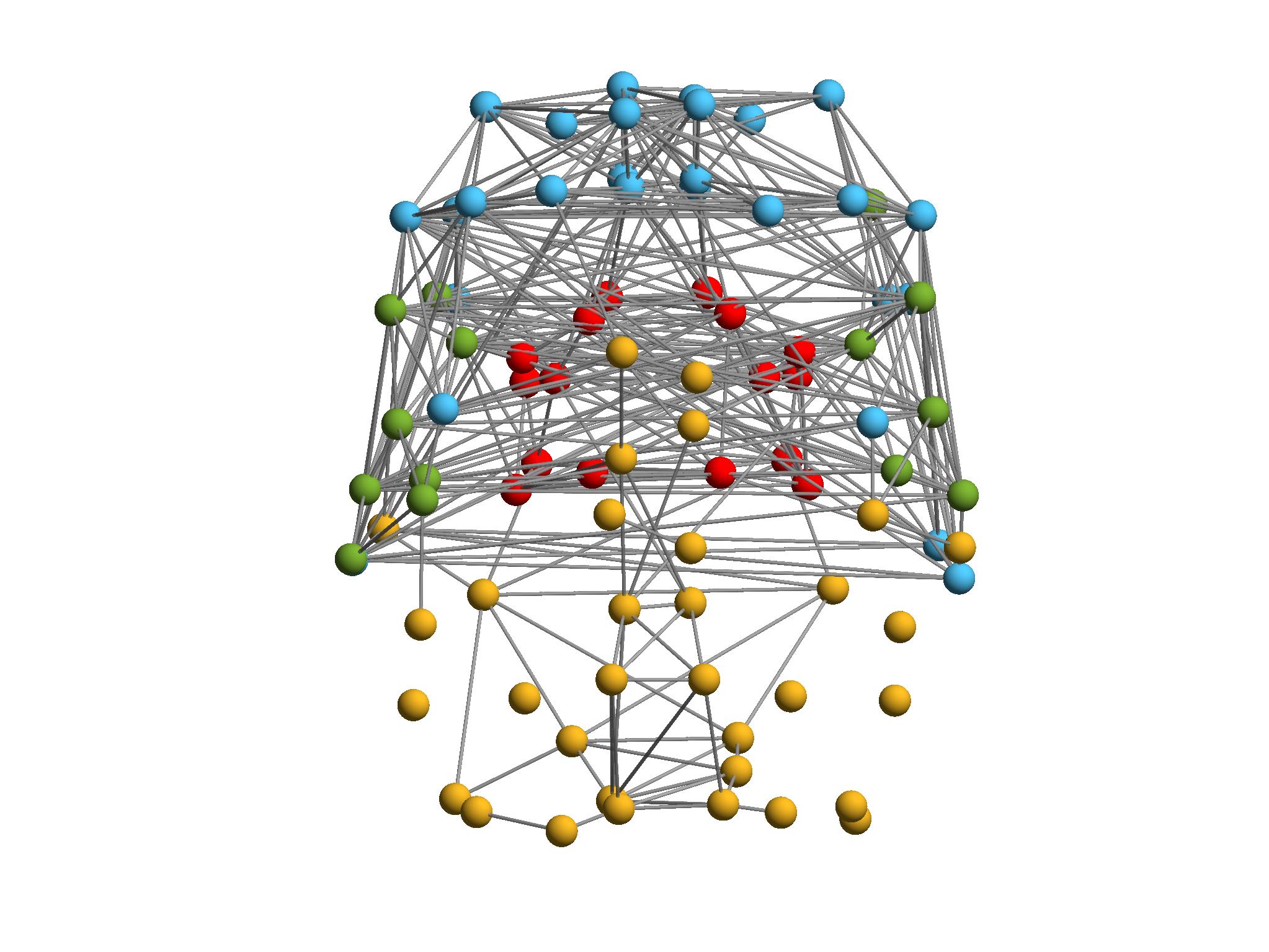}
\end{subfigure}%
\begin{subfigure}{0.24 \textwidth}
\includegraphics[width=\linewidth]{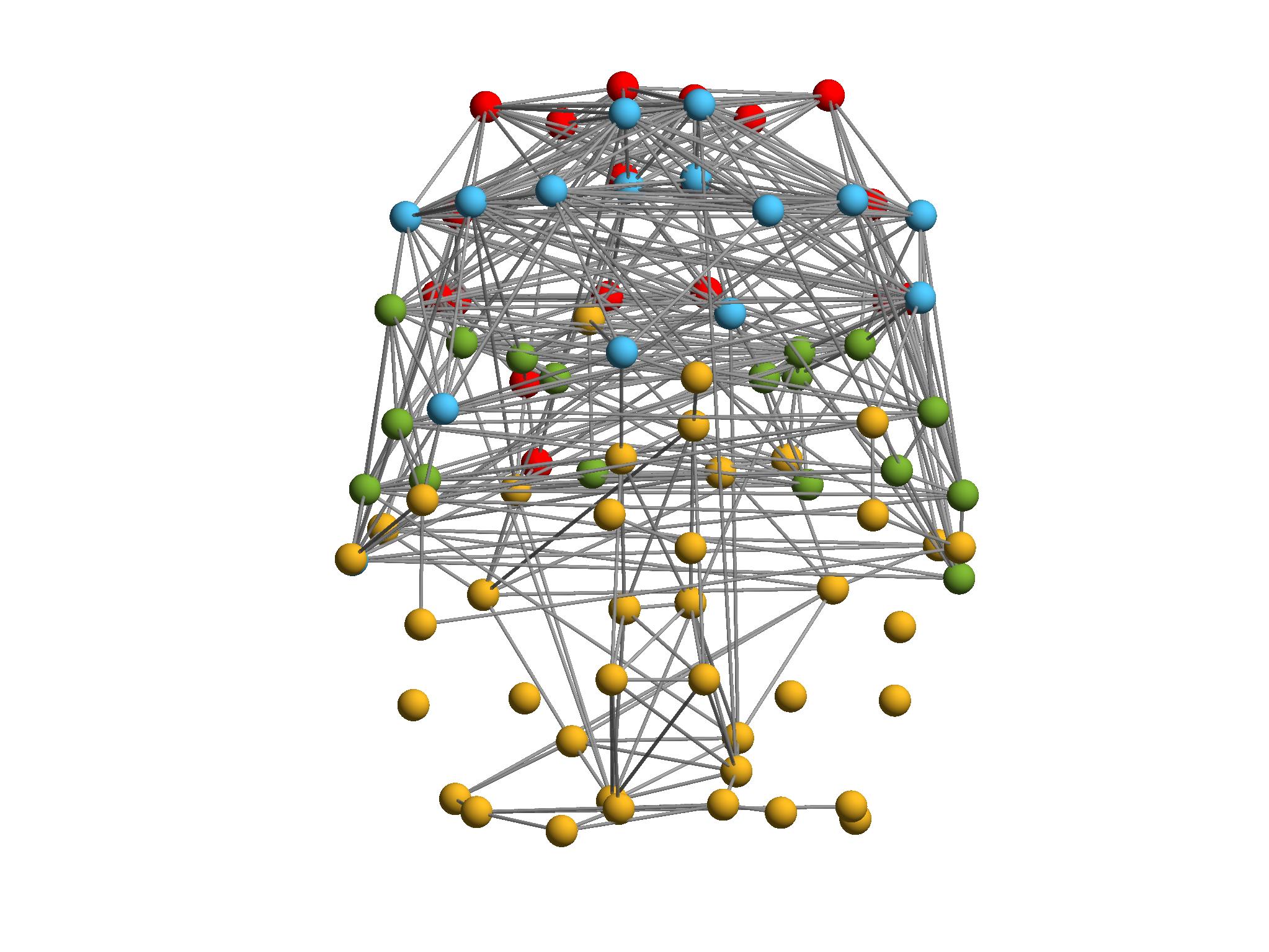}
\end{subfigure}
\begin{subfigure}{0.24 \textwidth}
\includegraphics[width=\linewidth]{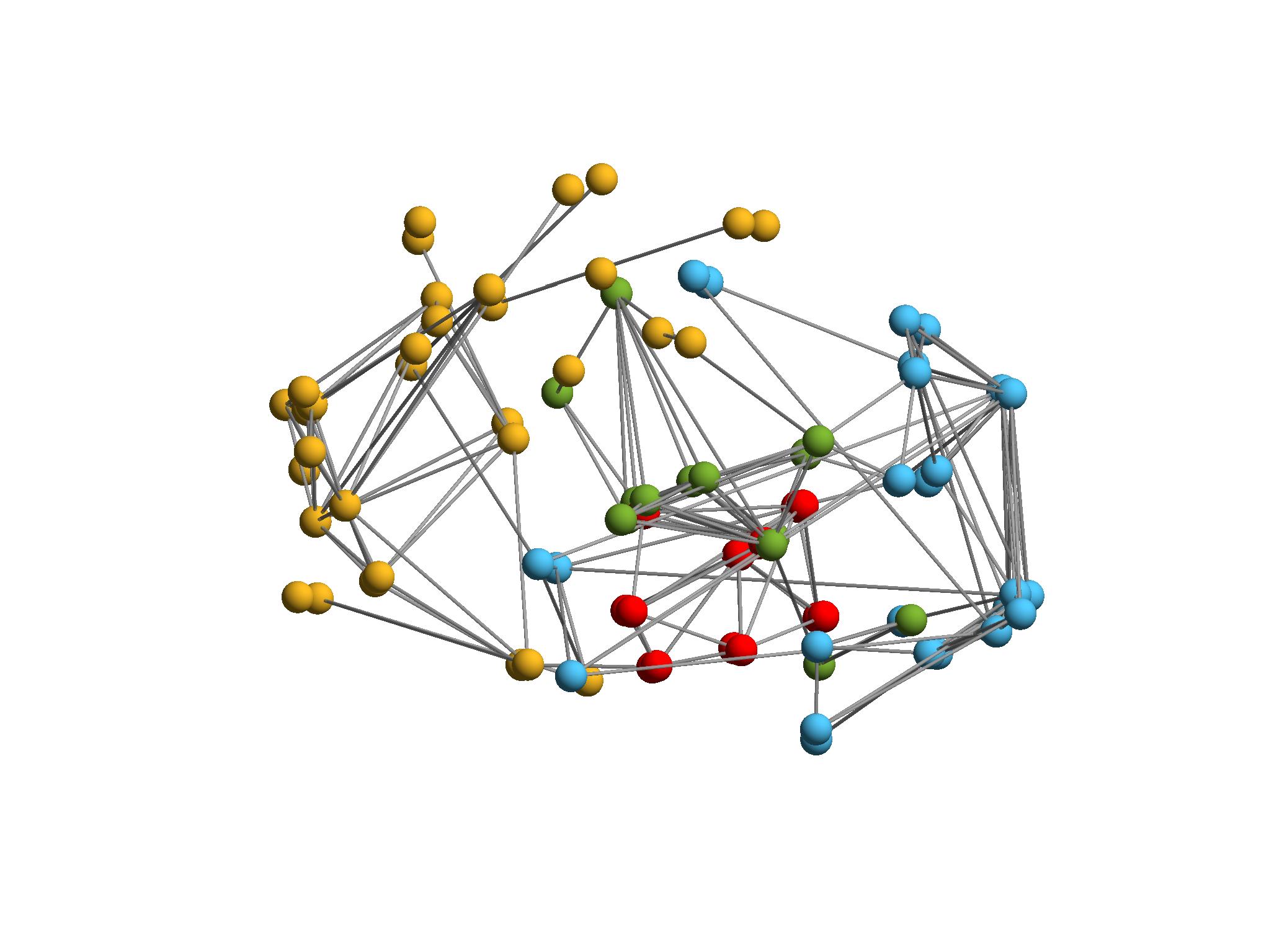}
\end{subfigure}%
\begin{subfigure}{0.24 \textwidth}
\includegraphics[width=\linewidth]{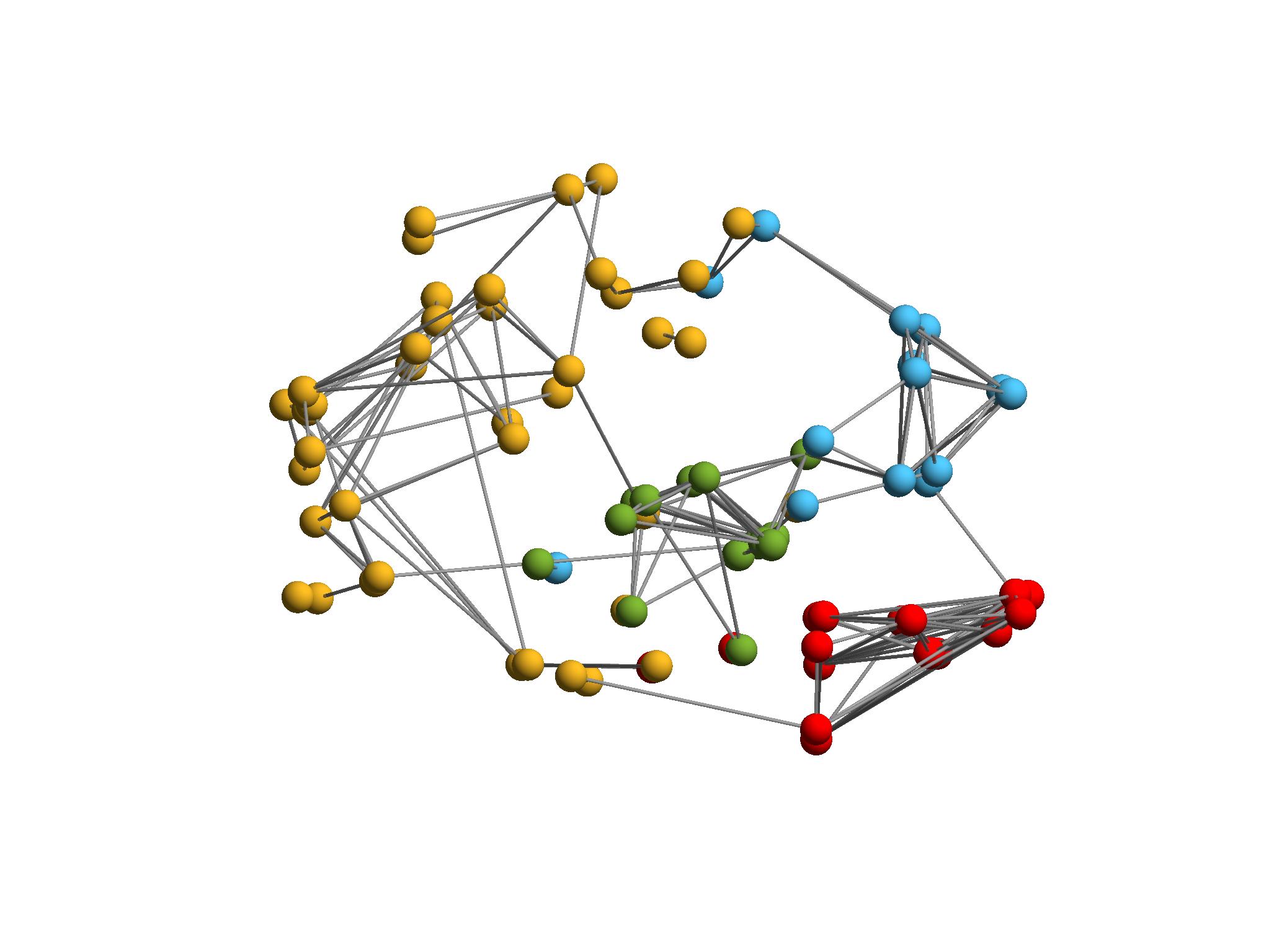}
\end{subfigure}\hfill%
\begin{subfigure}{0.24 \textwidth}
\includegraphics[width=\linewidth]{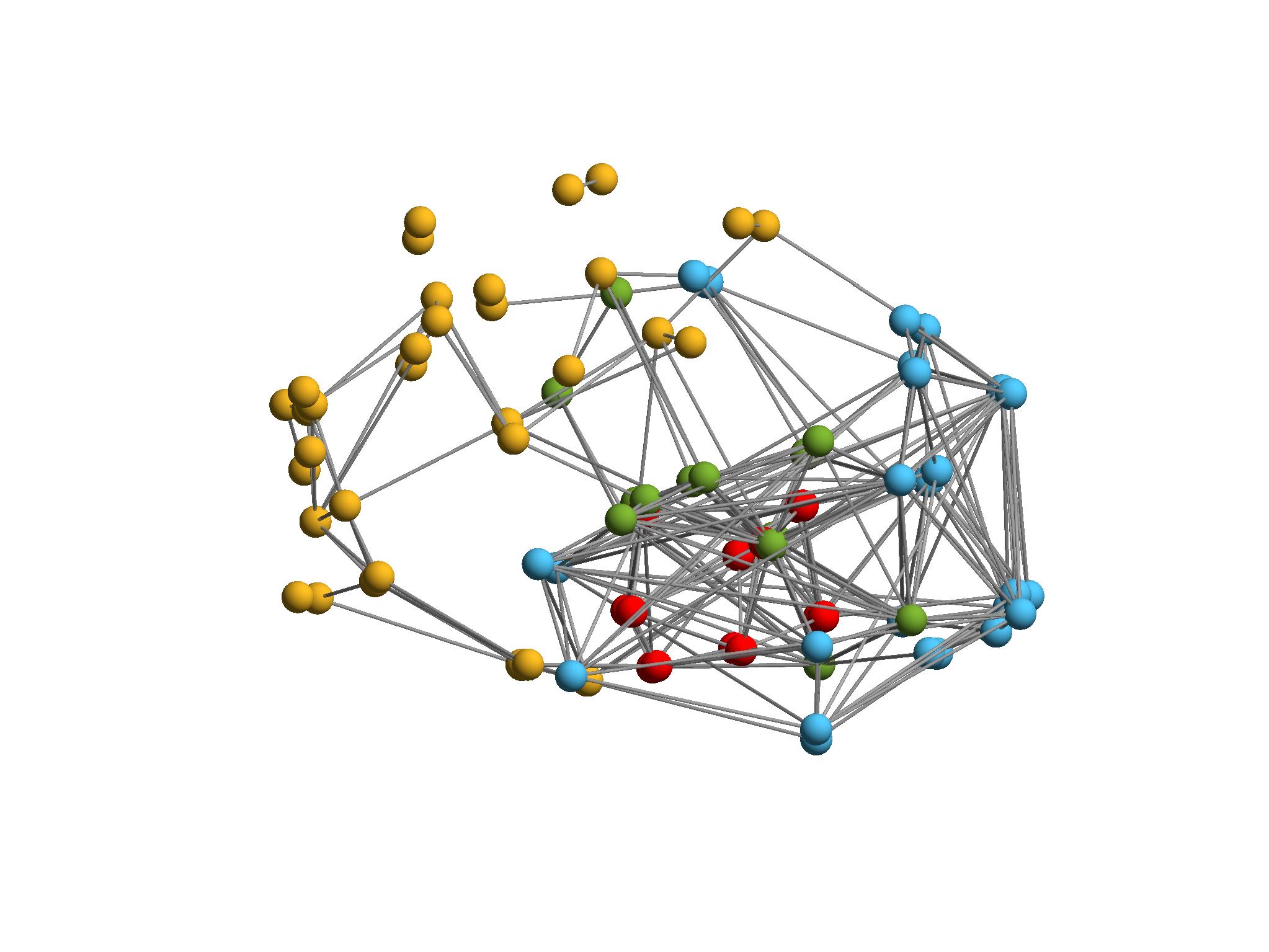}
\end{subfigure}%
\begin{subfigure}{0.24 \textwidth}
\includegraphics[width=\linewidth]{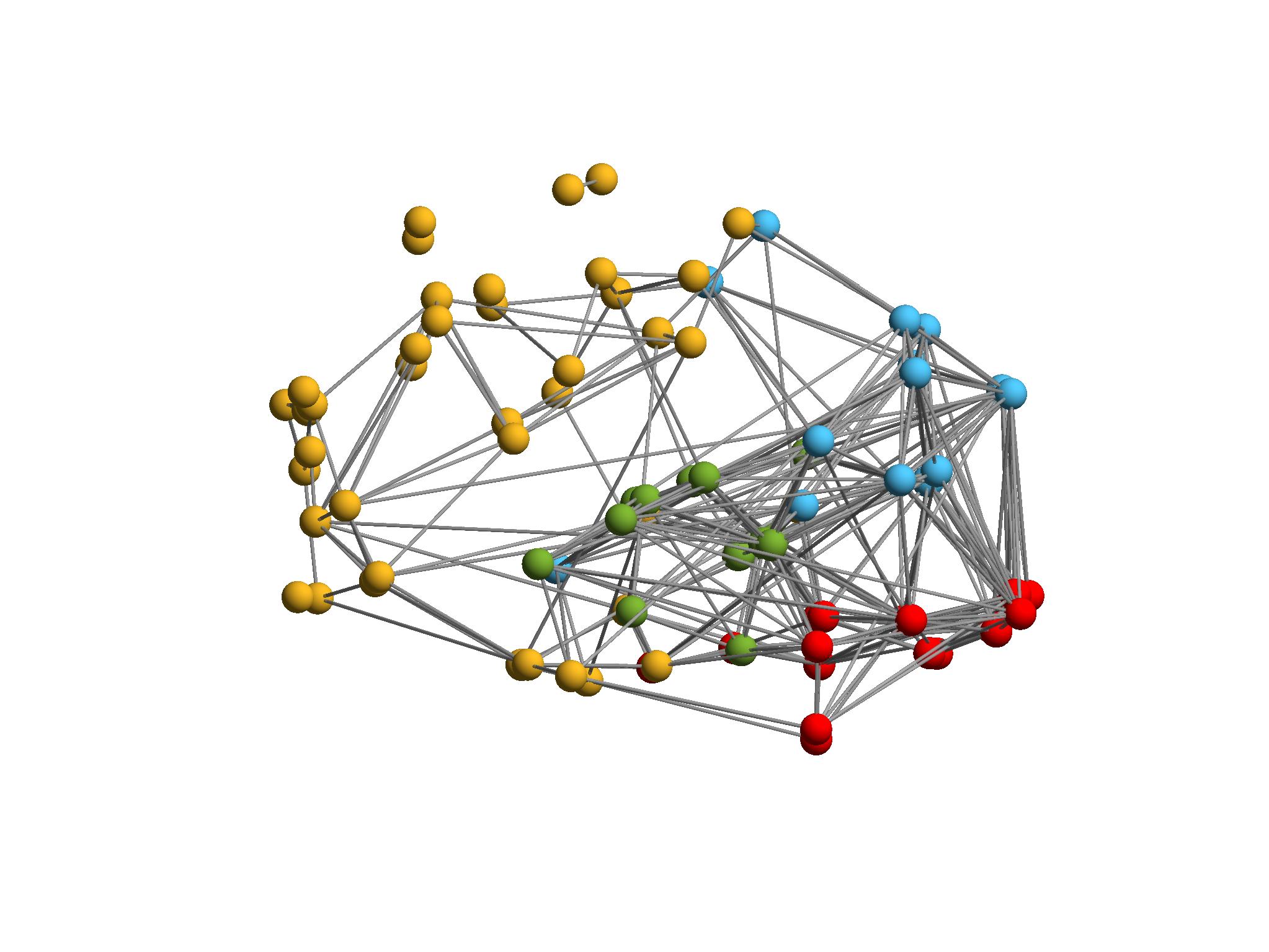}
\end{subfigure}
\vspace{-5pt}
\begin{center} (L) Controls \hspace{30pt} (R) Patients \hspace{60pt} (L) Controls \hspace{30pt} (R) Patients\end{center}
\vspace{-5pt}
\begin{center} (a) Module consistency \hspace{110pt} (b) Mean connectivity\end{center}
\vspace{-5pt}
\caption{Group putative community structure of resting state network from Co-OSNTF visualized through axial (top row) and sagittal (bottom row) brain views. In (a) the edges represent fraction of subjects for which two nodes are in the same community with a threshold of 0.40, while in (b) the edges represent the average connectivity (correlation) between two nodes across all subjects with a threshold of 0.50. In each case nodes are colored according to their group putative community in (L) Controls and (R) Patients.}
\label{SM}
\end{figure}

To better understand the community structure for the entire group, we visualize the structure obtained from Co-OSNTF algorithm in Figure \ref{SM} with the help of two measures: module consistency and mean connectivity. While the community structure is the same (the group putative community structure for controls and patients), the edges in Figure \ref{SM}(a) represent the fraction of subjects for which two nodes are in the same community. The edges in Figure \ref{SM}(b) represent the average connectivity between two nodes across all subjects. In both cases the edges are thresholded at a certain level (i.e., the edge appears if the quantity it represents exceeds a certain value) and are weighted, with thicker edges representing larger values.  Similar to the observation in \citet{moussa12}, we also have the visual module (yellow colored), containing both the primary and secondary visual cortices, as the most consistent module. We also observe that this module is relatively consistent in patients as well, confirming the observation from previous studies \citep{yu2012brain}. The red module in the control group, which roughly corresponds to the default mode network in \citet{moussa12}, is split into two parts with some nodes being part of another module. From Figure \ref{SM}(a) it is clear that the blue group in controls is split in two groups in patients (blue and red) which are almost disjoint in terms of module consistency thresholded at 0.40. The nodes in the red group in controls have lost the tendency to be grouped together in the patients and instead they are more consistently grouped with different modules. The nodes that belonged to the yellow and green groups in controls appear to be unchanged in their co-module relation with the rest of the network.

\begin{figure}[!h]
\centering{}
\begin{subfigure}{0.22 \textwidth}
\includegraphics[width=\linewidth]{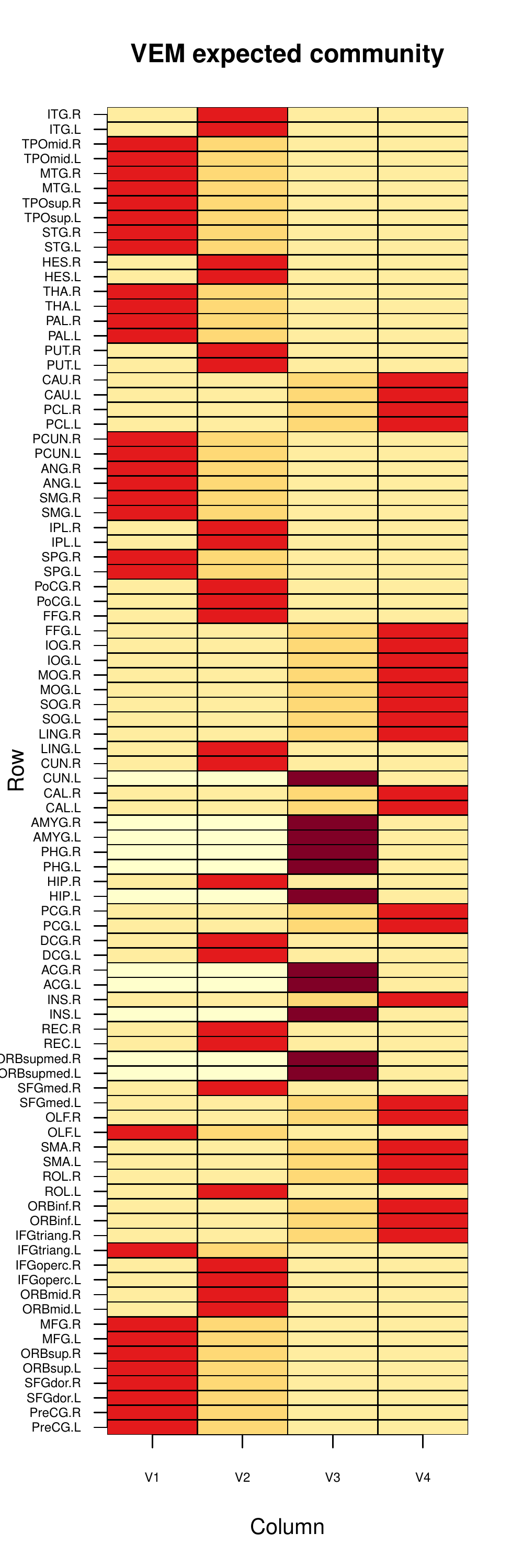}
\end{subfigure}%
\begin{subfigure}{0.22 \textwidth}
\includegraphics[width=\linewidth]{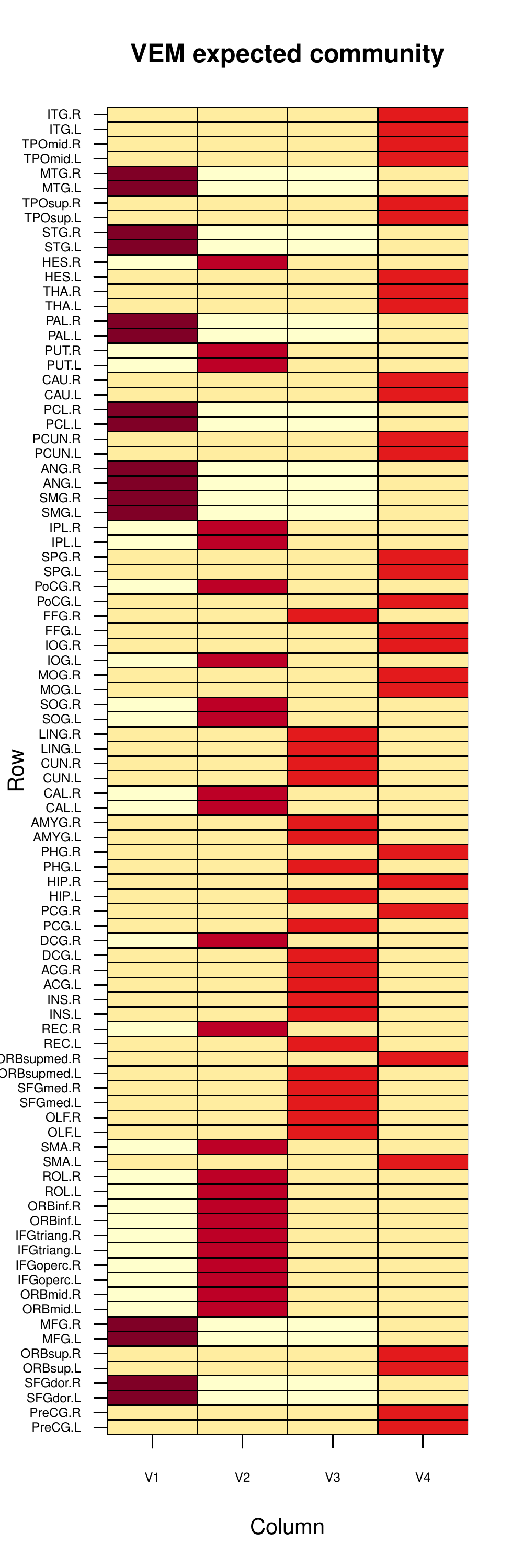}
\end{subfigure}%
\hspace{5pt}
\begin{subfigure}{0.22 \textwidth}
\includegraphics[width=\linewidth]{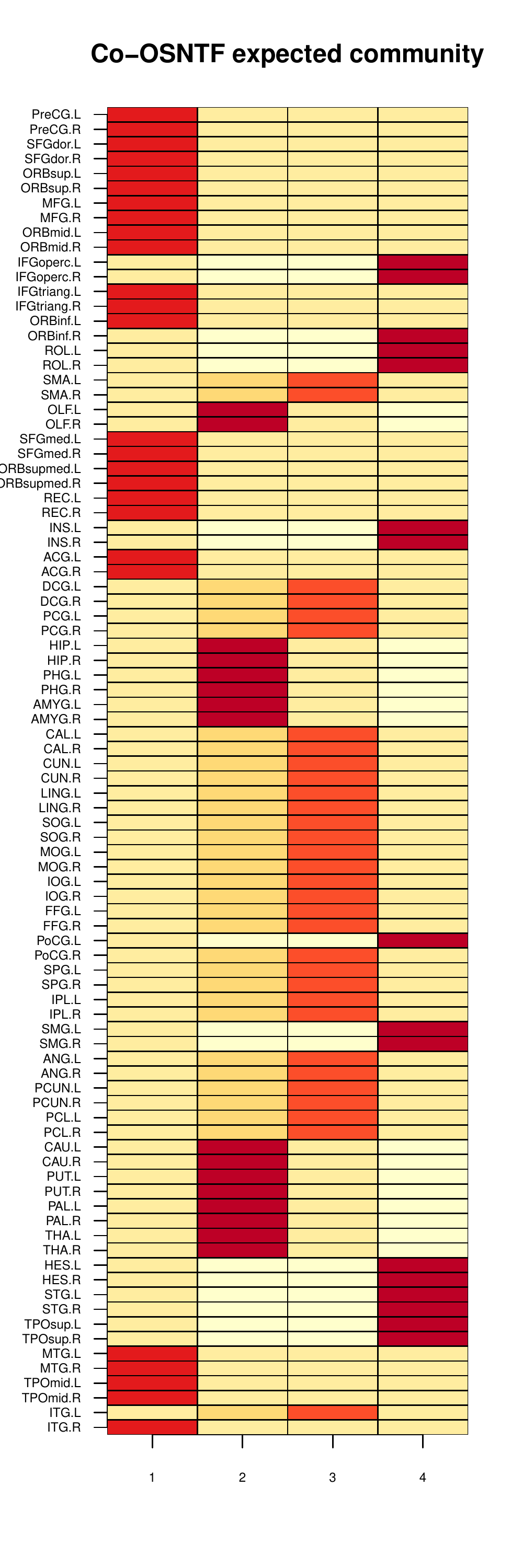}
\end{subfigure}%
\hspace{5pt}
\begin{subfigure}{0.22 \textwidth}
\includegraphics[width=\linewidth]{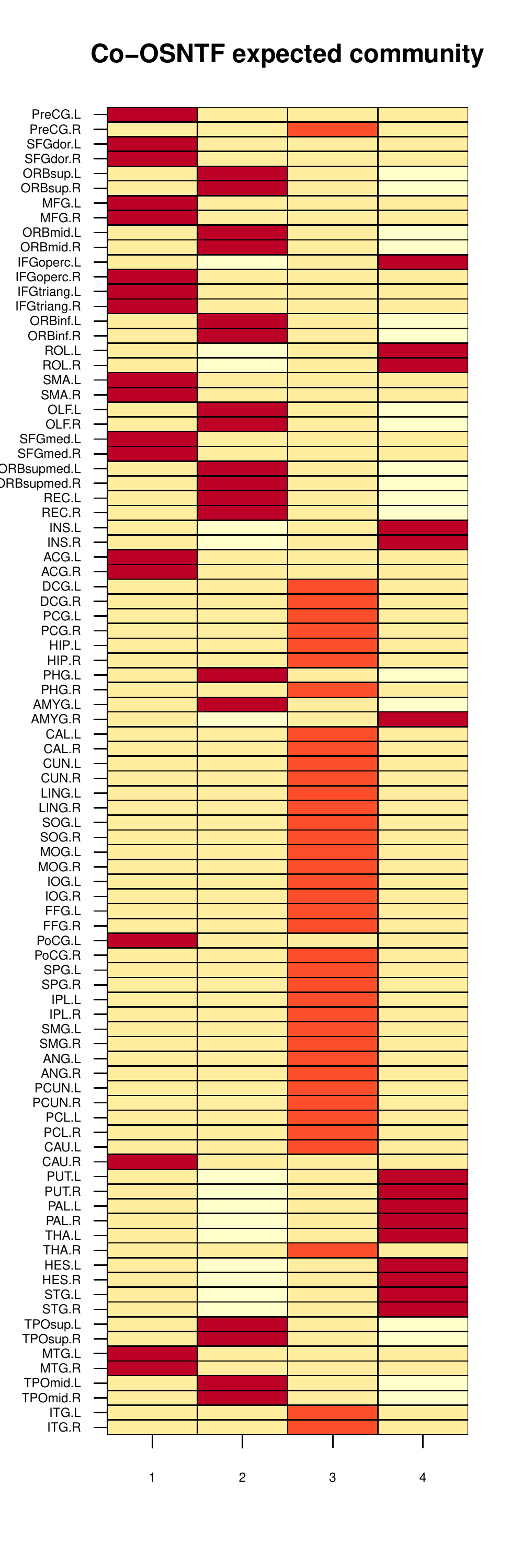}
\end{subfigure}
\begin{center}
\vspace{-5pt}
(a) Controls \hspace{50pt} (b)  Patients \hspace{50pt} (a) Controls \hspace{50pt} (b)  Patients
\end{center}
\vspace{-5pt}
\caption{Community estimate for AAL ROIs in controls and patients: (a) and (b) using the VarEM method, (c) and (d) using the Co-OSNTF method. The colors are proportional to the estimated probabilities that a ROI belongs to one of the four communities. The color scale goes from light yellow (lowest) to dark red (highest)}
\label{nodecomms}
\end{figure}

\subsection{Estimates of expected communities of the ROIs}

We plot the estimated mean community assignments (mixed membership or soft assignments), i.e., estimates  for $\bar{Z}_iT$ for each ROI $i$ for the control and patient groups in Figure \ref{nodecomms}. These mean estimates give us estimated probabilities that a given ROI will belong to one of the 4 communities for a randomly selected subject from the control group or the patient group. We note that for both VarEM and Co-OSNTF, the community labels are separately aligned between controls and patients, and the module numbers are the same as the previous section. Therefore, each row of the four matrices in Figure \ref{nodecomms} represents the estimated community ``profile" for a ROI, with the colors being proportional to the estimated probabilities that the ROI belongs to a community. This enables us to visually detect the ROIs for which there are  differences between the estimated community probabilities. For example, among the ROIs shown in Table \ref{tab:scztest}(B), which have significantly altered expected community assignments using the node level MUV test, we note HIP.R has a very high probability of being in module 2 (red module) in controls and low probability of being in other modules, while it has a medium high probability of being in module 3 in patients and low probability of being in other modules. Similarly, CAU.R and PAL.L have high probability of being in module 2 (red module), low probability of being in module 4 and low to medium probability of being in modules 1 and 3  in controls.  However in patients, their probability profiles are different. CAU.R has high probability of being in module 1, low-medium probability of being in the other three modules, while PAL.L has high probability of being in module 4,  low probability of being in module 2 and low-medium probability of being in modules 1 and 3. The remaining ROI with a significant difference according to our node-level test is TPOsup.L, which gets a very high probability of being in module 4, low-medium probability of being in module 1 and low probability of being in modules 2 and 3 in the control group. In patients it restructures to get high probability of being in module 2, low-medium probability of being in modules 1 and 3 and low probability of being in module 4. We can also see how the probability profiles of ROIs with high probability in module 2 (red module in previous section) in the control group has got disrupted and changed in patient group. While the node level test did not detect them to be significant at the prescribed FDR level, we note the probability profiles of several ROIs, including HIP.L, PHG.R, AMYG.R, CAU.L, PUT.L, PUT.R, THA.L, THA.R, have changed between control and patient groups. We note that some of those ROIs have appeared in Table \ref{tab:scztest}(A) as part of the ROIs detected by the Co-Spectral method. This is an evidence that Co-OSNTF and Co-Spectral detect somewhat consistent set of ROIs as regions where the module structure is disrupted in schizophrenia, even though the ROIs that are found to be statistically significant are not exactly the same. Taken together, Figure \ref{nodecomms} gives a detailed picture and compelling evidence of the disruption and reorganization of module structure in schizophrenia.

\subsection{Prediction  of community structure for new subjects and model fit}

Next we study how the model fits the data, and the predictive power of the model for new subjects. Assessing and comparing the methods in terms of their predictive ability (and model fit) is very important, especially since the methods lead to somewhat different results. In addition, good predictive ability of the model and the methods for new subjects give us confidence about the generalization of the results we have obtained in this study to the greater population.

\begin{figure}[!h]
\centering{}
\begin{subfigure}{0.25 \textwidth}
\includegraphics[width=\linewidth]{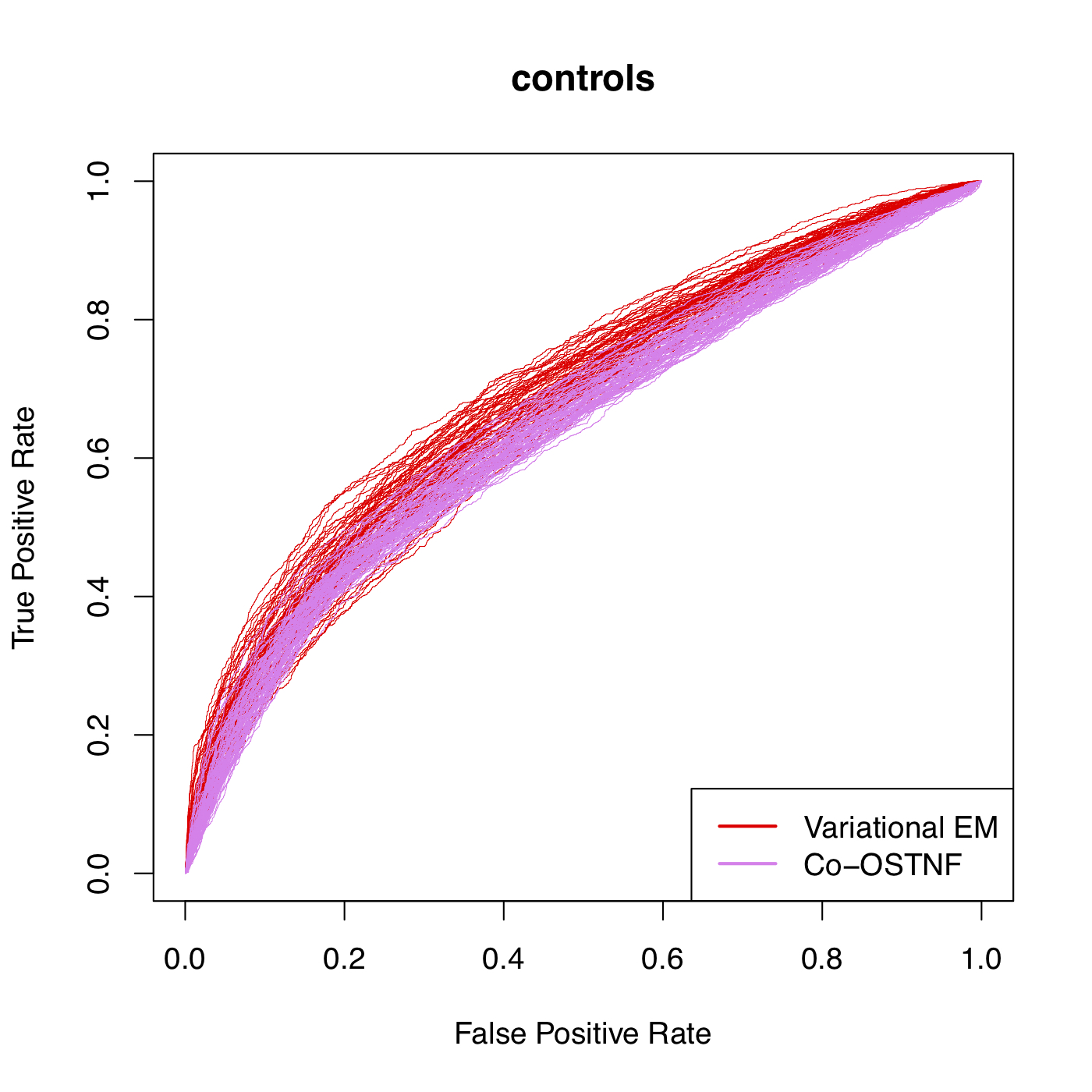}
\end{subfigure}%
\begin{subfigure}{0.25 \textwidth}
\includegraphics[width=\linewidth]{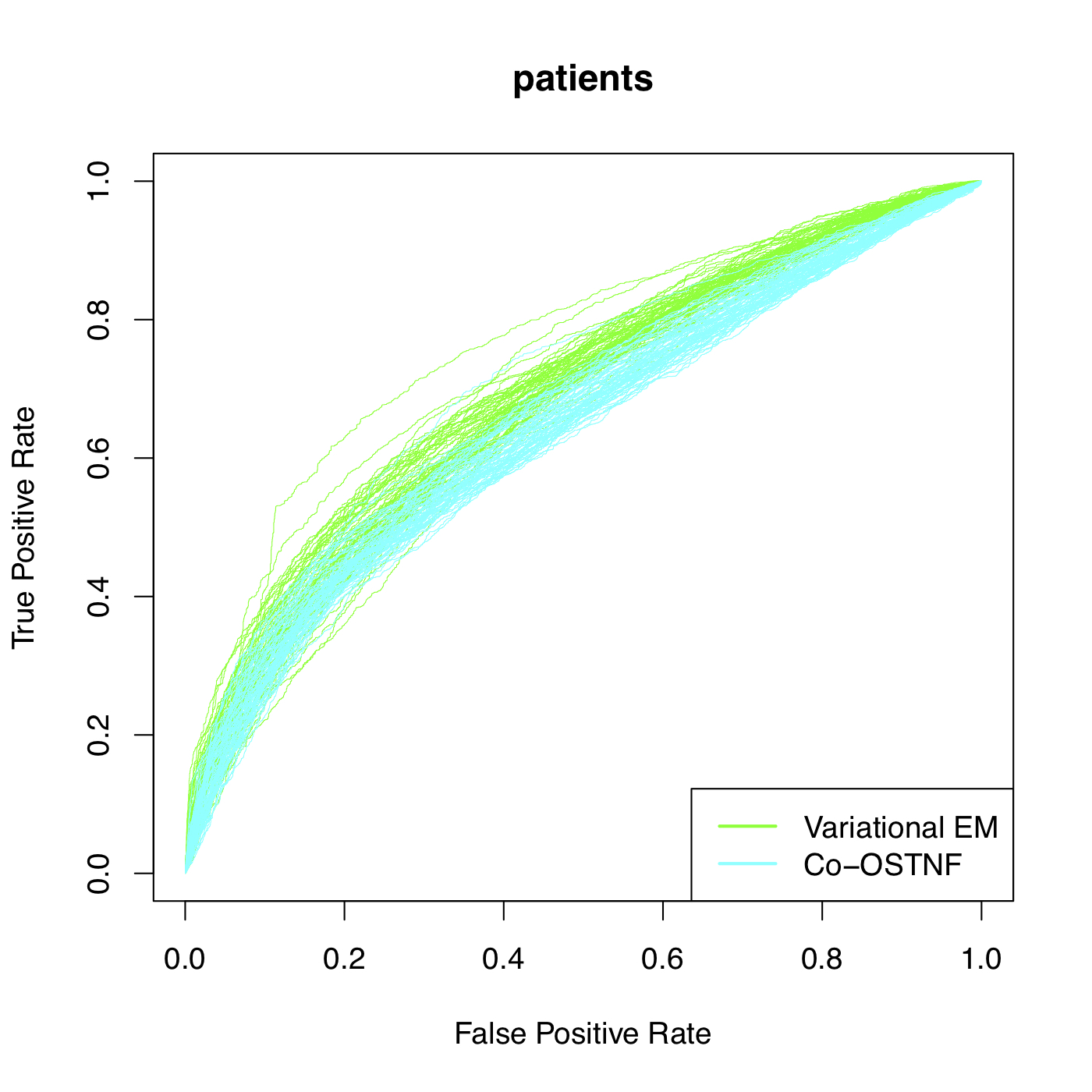}
\end{subfigure}%
\begin{subfigure}{0.25 \textwidth}
\includegraphics[width=\linewidth]{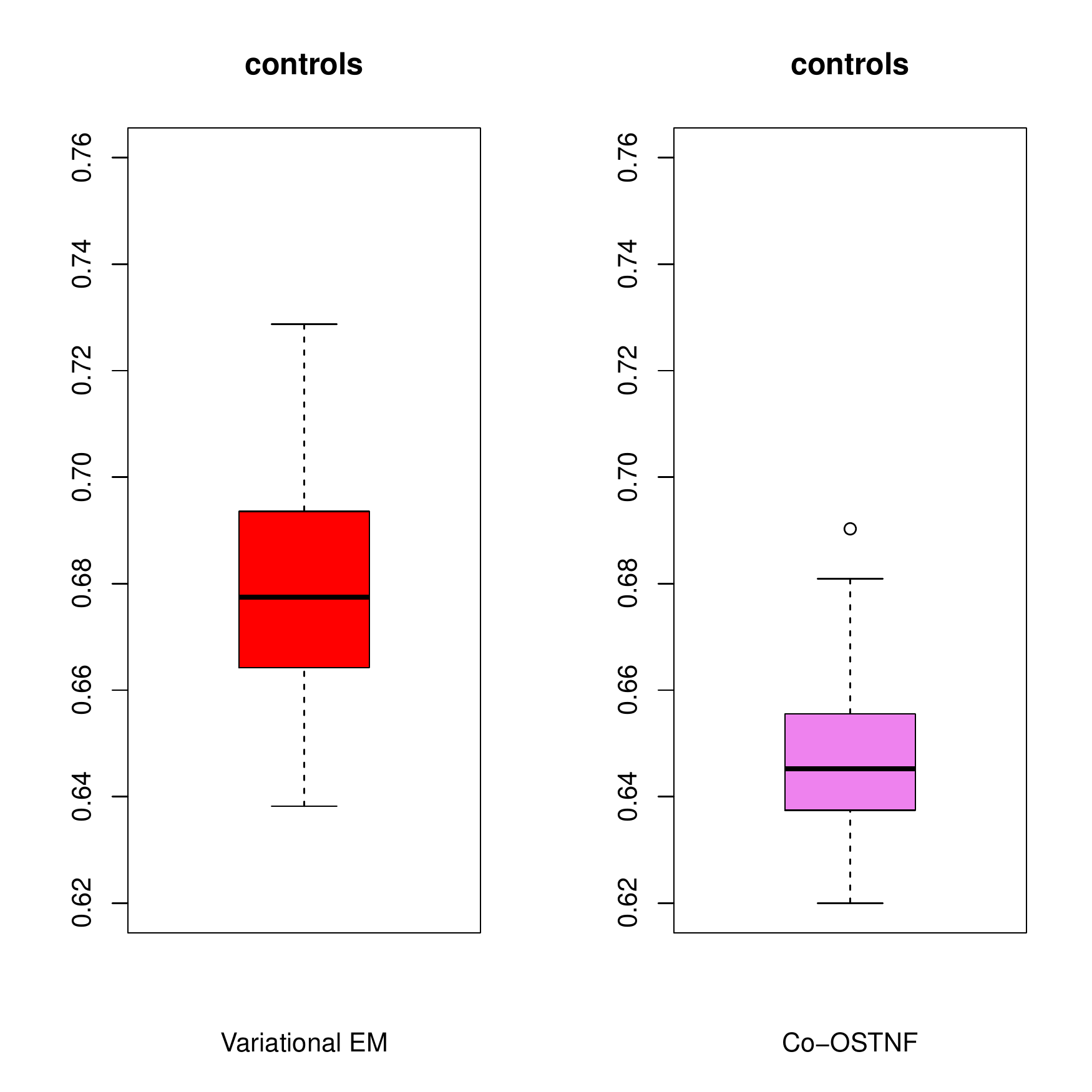}
\end{subfigure}%
\begin{subfigure}{0.25 \textwidth}
\includegraphics[width=\linewidth]{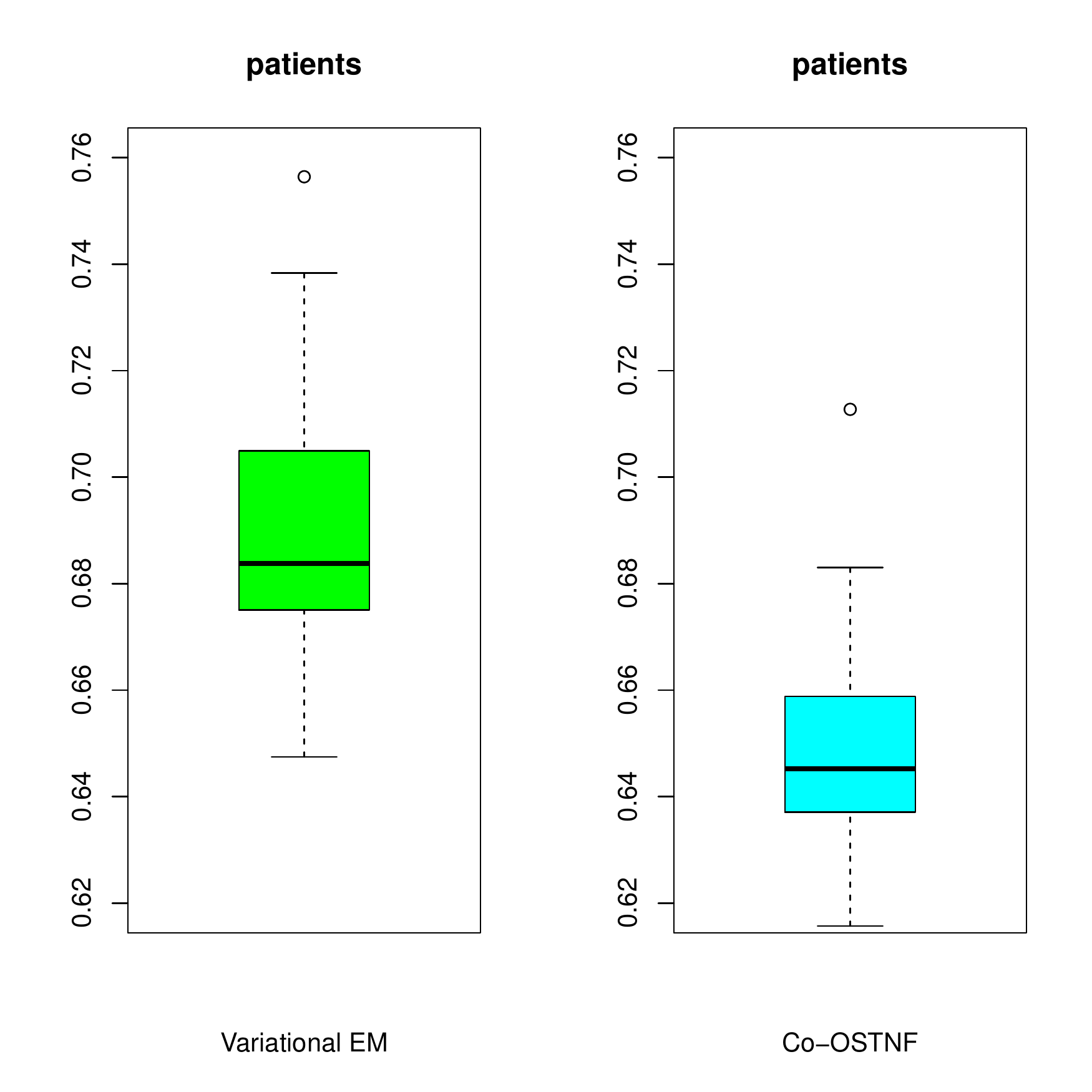}
\end{subfigure}
\begin{center}
(a) ROC curves \hspace{150pt} (b) AUC values
\end{center}
\vspace{-5pt}
\caption{ (a) A plot of ROC curves for all subjects in control and patient groups and (b) a boxplot of distribution of AUC values across subjects.}
\label{ROC}
\end{figure}

We first assess the in-sample model fit of the methods by plotting the Receiver Operating Characteristics (ROC) curves for each of the subject networks in the control and patient groups for VarEM and Co-OSNTF methods in Figure \ref{ROC}(a). The ROC is a plot between the false positive rate and true positive rate for fitting a binary data, where a high area under the curve (AUC) indicates a good model fit. A random guess will have a ROC curve close to the diagonal line and an AUC of 0.5. From the ROC curves it appears that there are a number of ROC curves of the VarEM method which are above the ROC curves of the Co-OSNTF method in both controls and patients. However, we also note that the ROC curves of the VarEM method have a greater variation. This can be seen more clearly from the boxplot of the AUC values across the subject networks in Figure \ref{ROC}(b). For both controls and patients the AUC values are generally higher using the VarEM method compared to the Co-OSNTF method, however the spread in the AUC values is larger for VarEM method. We note the AUC values are in the range of 0.64 to 0.70, indicating a reasonably good model fit for both the methods. However, the VarEM method fits the data better in both control and patient networks.

\begin{figure}[!h]
\begin{subfigure}{0.22 \textwidth}
\includegraphics[width=\linewidth]{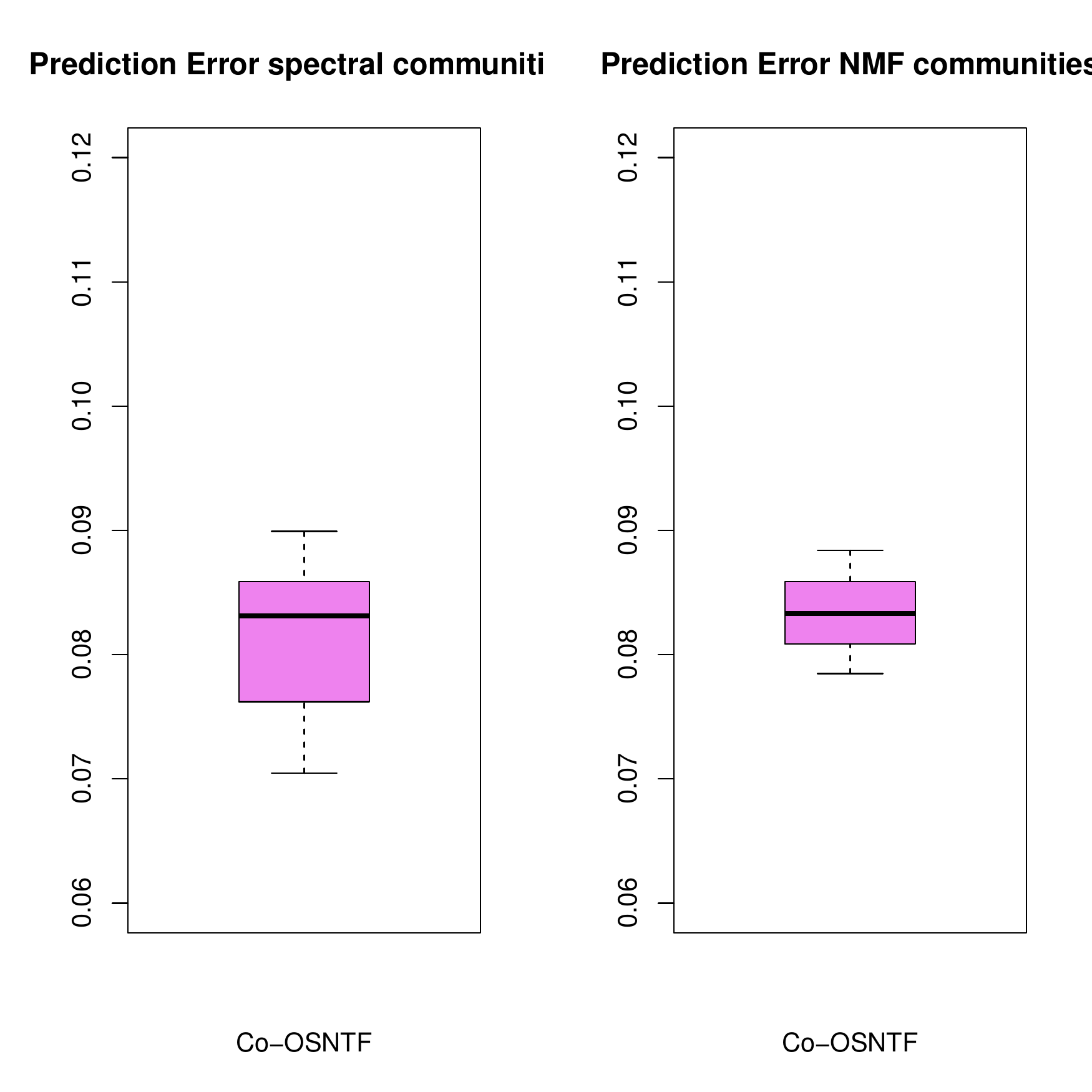}
\end{subfigure}%
\begin{subfigure}{0.22 \textwidth}
\includegraphics[width=\linewidth]{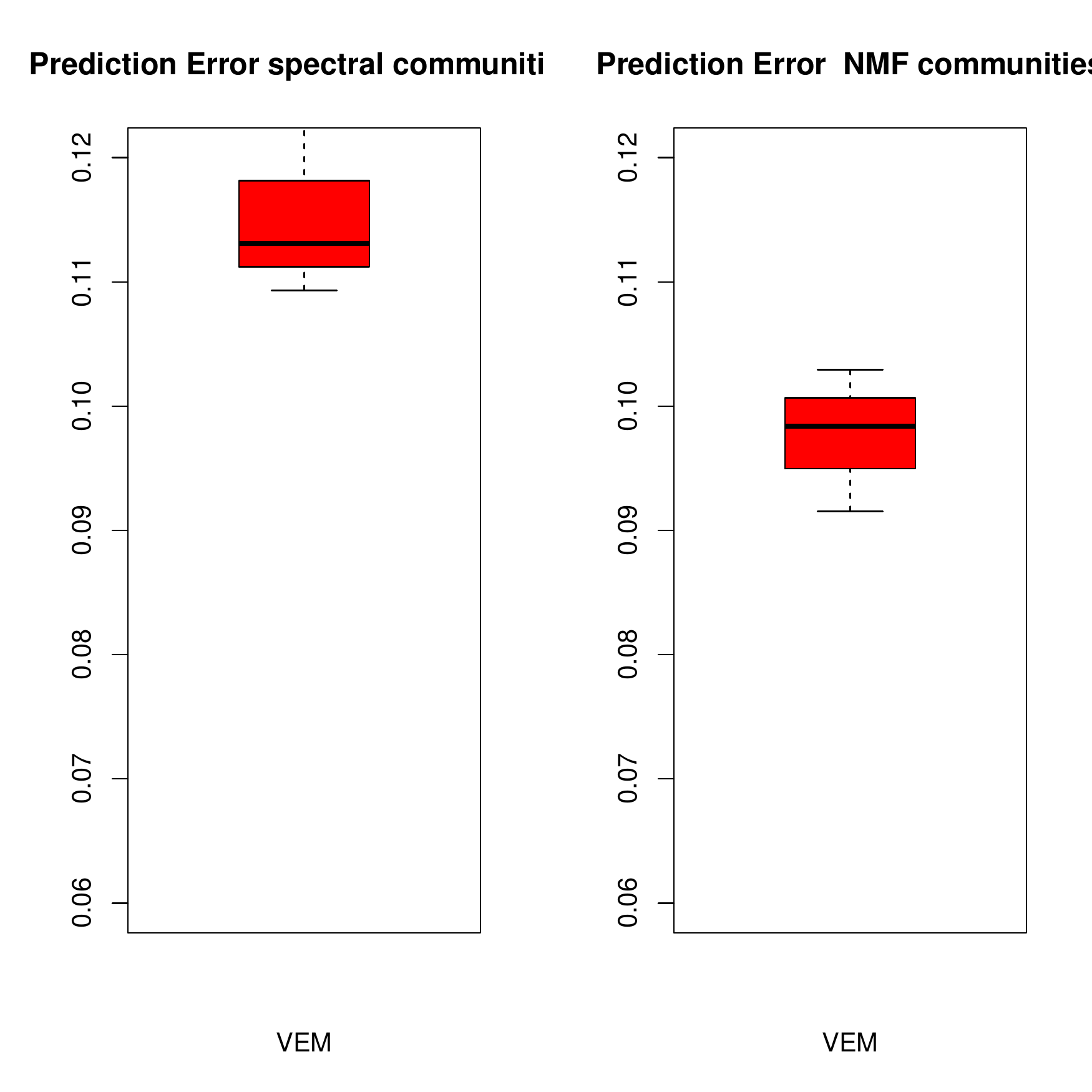}
\end{subfigure}%
\hspace{30pt}
\begin{subfigure}{0.22 \textwidth}
\includegraphics[width=\linewidth]{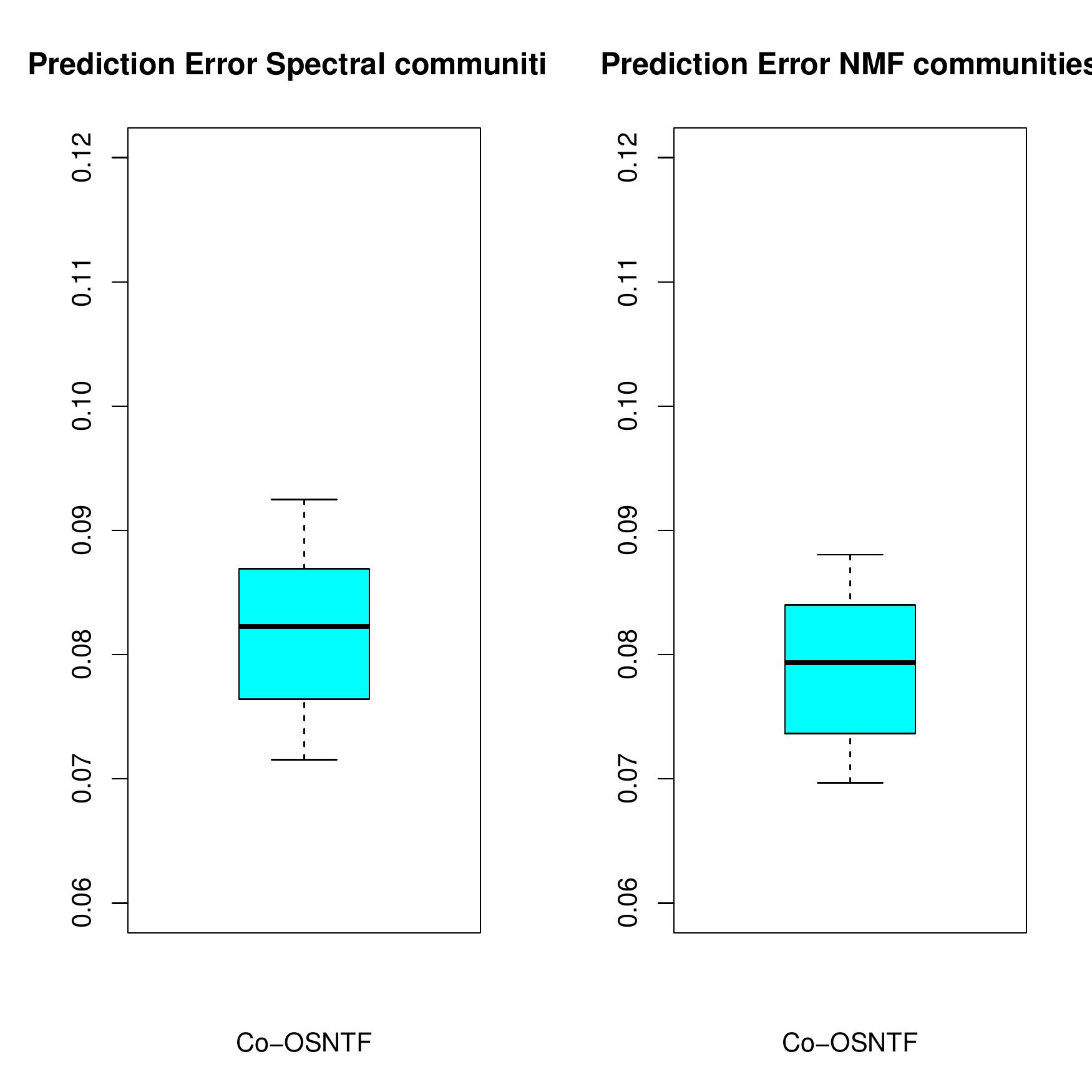}
\end{subfigure}%
\begin{subfigure}{0.22 \textwidth}
\includegraphics[width=\linewidth]{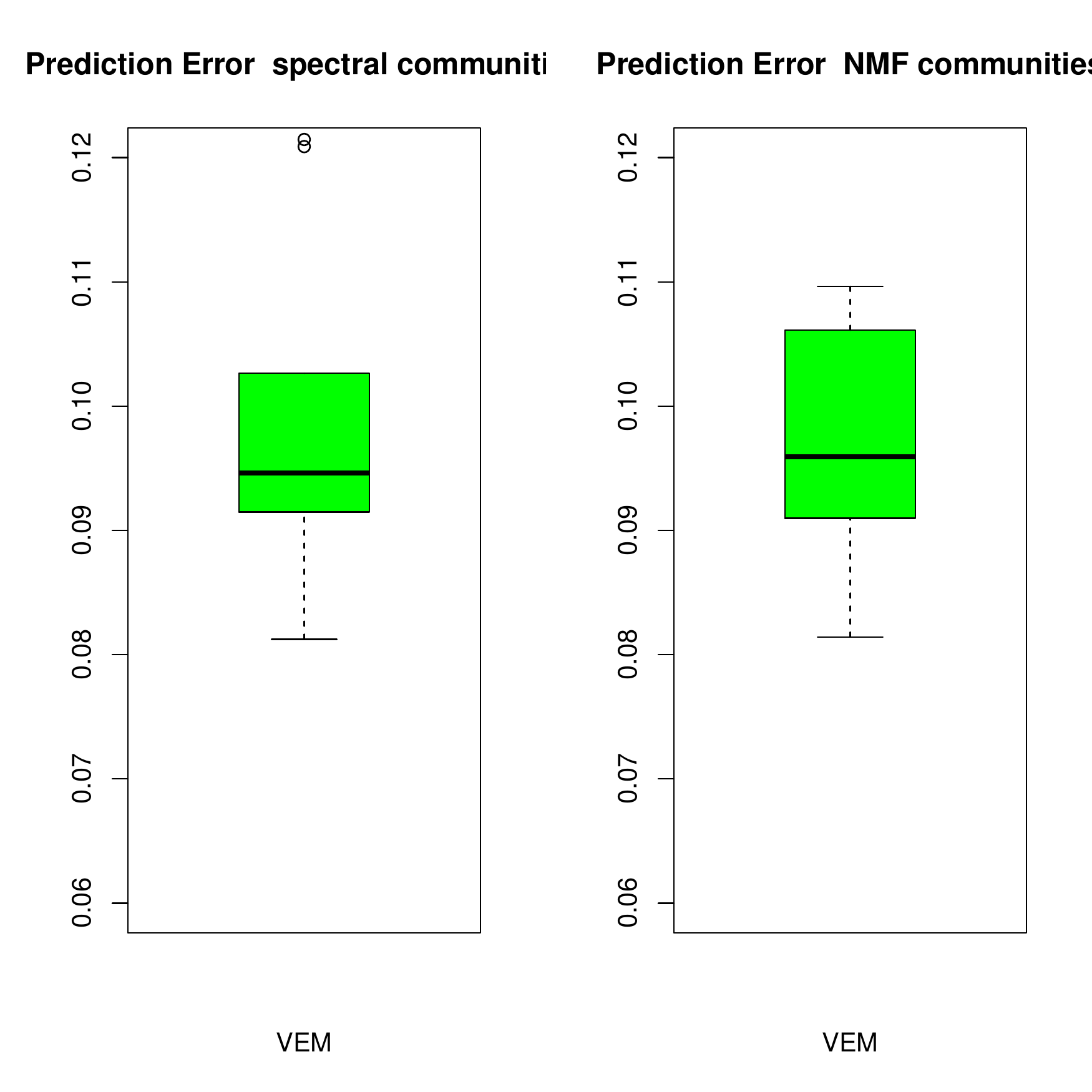}
\end{subfigure}
\begin{center}
Co-OSNTF controls \hspace{20pt} VarEM controls \hspace{30pt}  Co-OSNTF patients \hspace{20pt}  VarEM patients
\end{center}
\vspace{-5pt}
\caption{Predictive performance:  Error of community prediction over 10 fold cross validation using Co-OSNTF and VarEM methods in control and patient groups. }
\label{predictive1}
\end{figure}

\begin{figure}[!h]
\begin{subfigure}{0.22\textwidth}
\includegraphics[width=\linewidth]{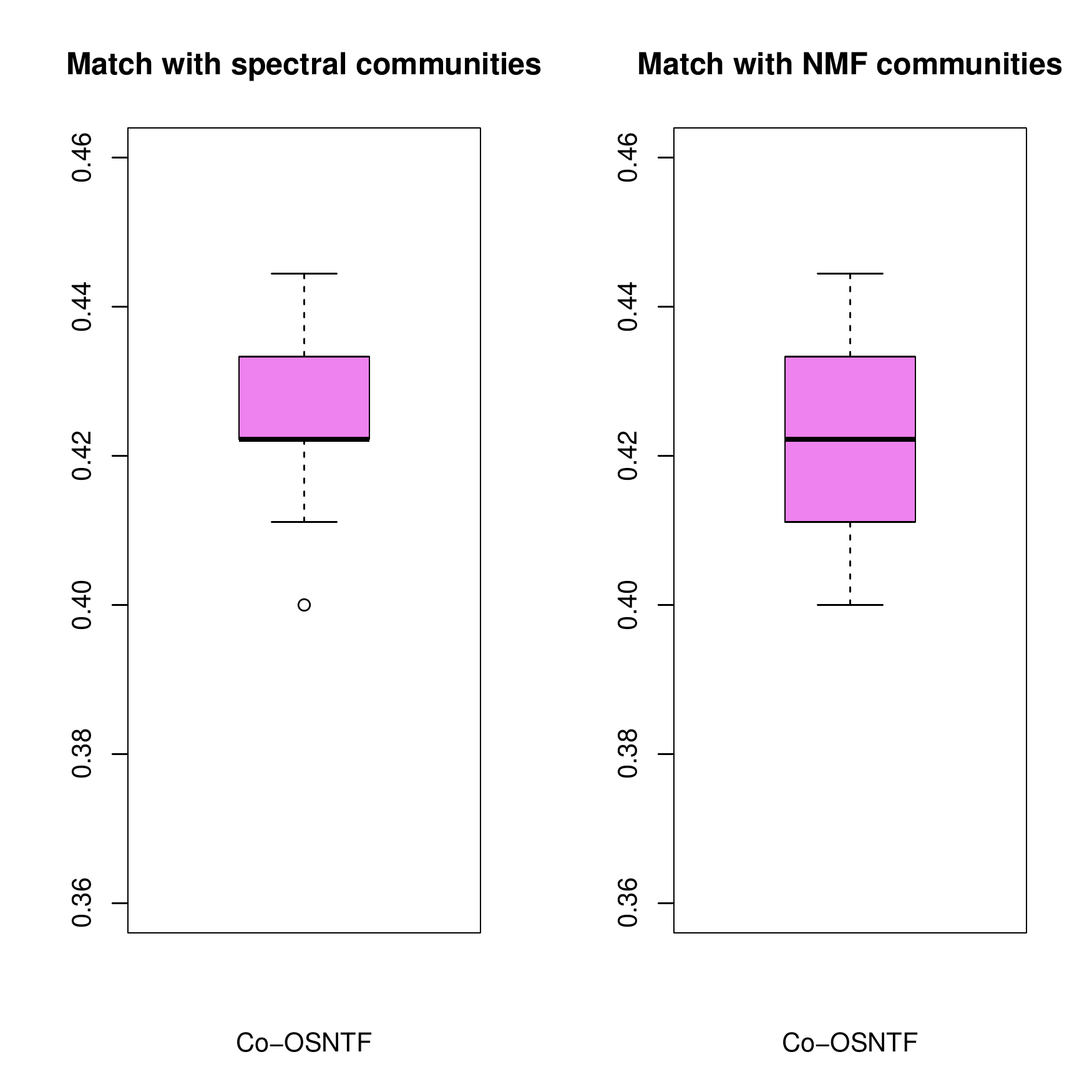}
\end{subfigure}%
\begin{subfigure}{0.22 \textwidth}
\includegraphics[width=\linewidth]{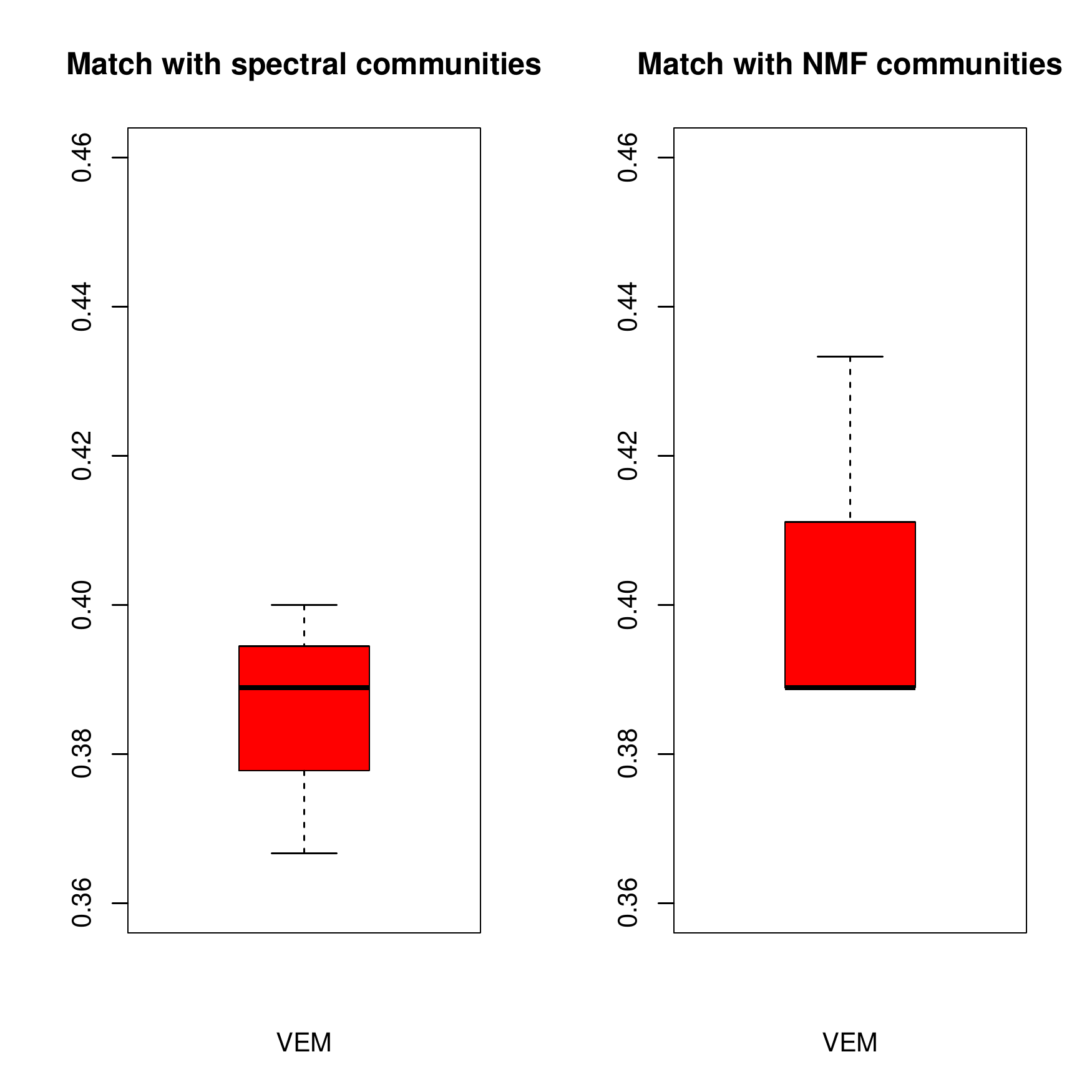}
\end{subfigure}%
\hspace{30pt}
\begin{subfigure}{0.22 \textwidth}
\includegraphics[width=\linewidth]{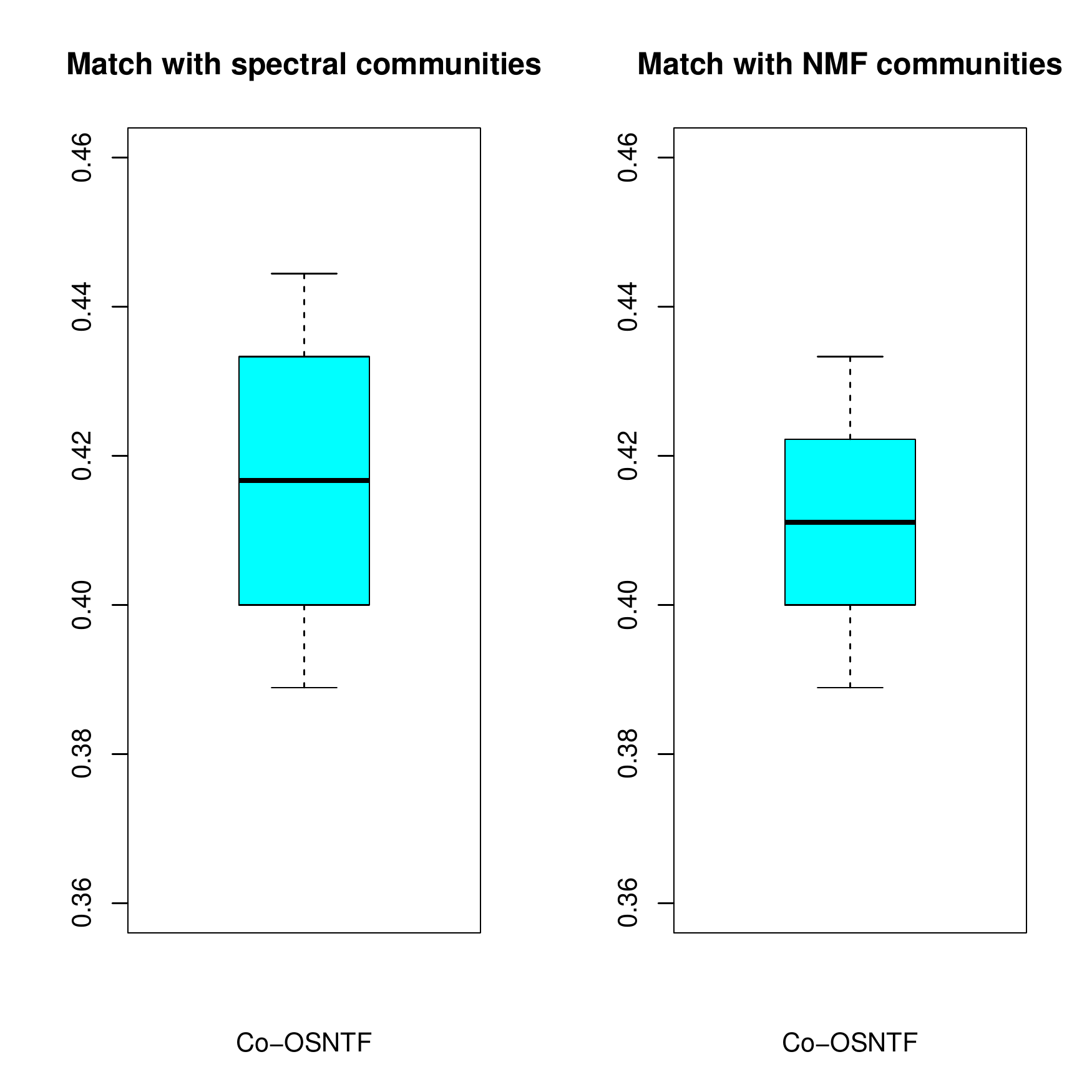}
\end{subfigure}%
\begin{subfigure}{0.22 \textwidth}
\includegraphics[width=\linewidth]{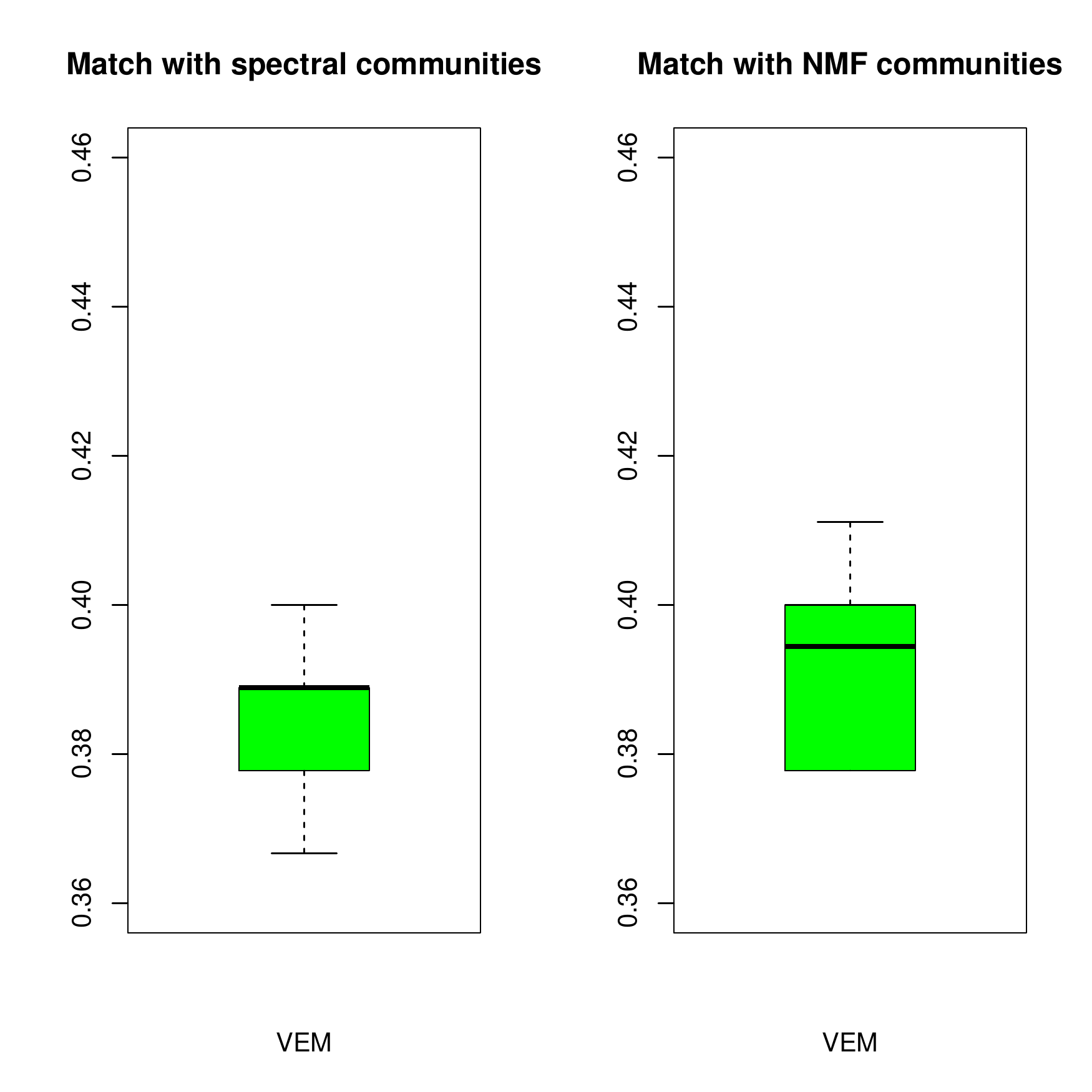}
\end{subfigure}
\begin{center}
Co-OSNTF controls \hspace{20pt}  VarEM controls \hspace{40pt}  Co-OSNTF patients \hspace{20pt} VarEM patients
\end{center}
\vspace{-5pt}
\caption{Predictive performance: overlap (correct classification rate) with mean group putative community structure over 10 fold cross validation using Co-OSNTF and VarEM methods in control and patient groups.}
\label{predictive2}
\end{figure}

While the ROC curves above give us indication of in-sample model fit, our primary tool of assessing the accuracy of the model and the methods is out of sample predictive ability of community structure. We perform a cross validation by splitting both the control and the patient samples into two parts: a test data that holds out 15 samples from each of the groups and a training data consisting of the remaining subjects to which the model is fit. For both control and patient groups, we predict the community structure in the test data using our estimated community structure $\bar{Z}T$ from the training data. Then in each of the subjects of the test data we estimate the community structure separately (independently) using two methods for single network community detection - (a) Spectral clustering with normalized Laplacian matrix and row normalization \cite{lei14,qr13} and (b) the OSNTF method \cite{pc16}. The estimated community labels in each of the subjects are aligned with labels in the predicted community structure, by solving the LSAP problem as before. We assess the predictive performance using two metrics. Let the $k$ dimensional community assignment vector for ROI $i$ in test subject $j$ be $u_{i}^{(j)}$. Then the average community assignment for ROI $i$ across the $J$ test subjects is $\bar{u}_i = \frac{1}{J}\sum_{j}u_{i}^{(j)}$. Then we compute a predictive accuracy by comparing this average community assignment with the predicted community assignment $(\bar{Z}T)_i$:  $\quad \text{median}_{ik}(|\bar{u}_{ik} - (\bar{Z}T)_{ik} |).$
This is our first metric for assessing predictive accuracy.
In Figure \ref{predictive1} we display the box plot of this predictive accuracy over 10 such train-test random sample splits for VarEM and Co-OSNTF methods for control and patient groups. We note that for control group, Co-OSNTF method is more accurate with about 7-9\% median absolute difference between the predicted and observed community assignments compared to VarEM with 9-12\% median absolute difference. We make a similar observation in the patient group as well.

\begin{figure}[!h]
\centering{}
\begin{subfigure}{0.15 \textwidth}
\includegraphics[width=\linewidth]{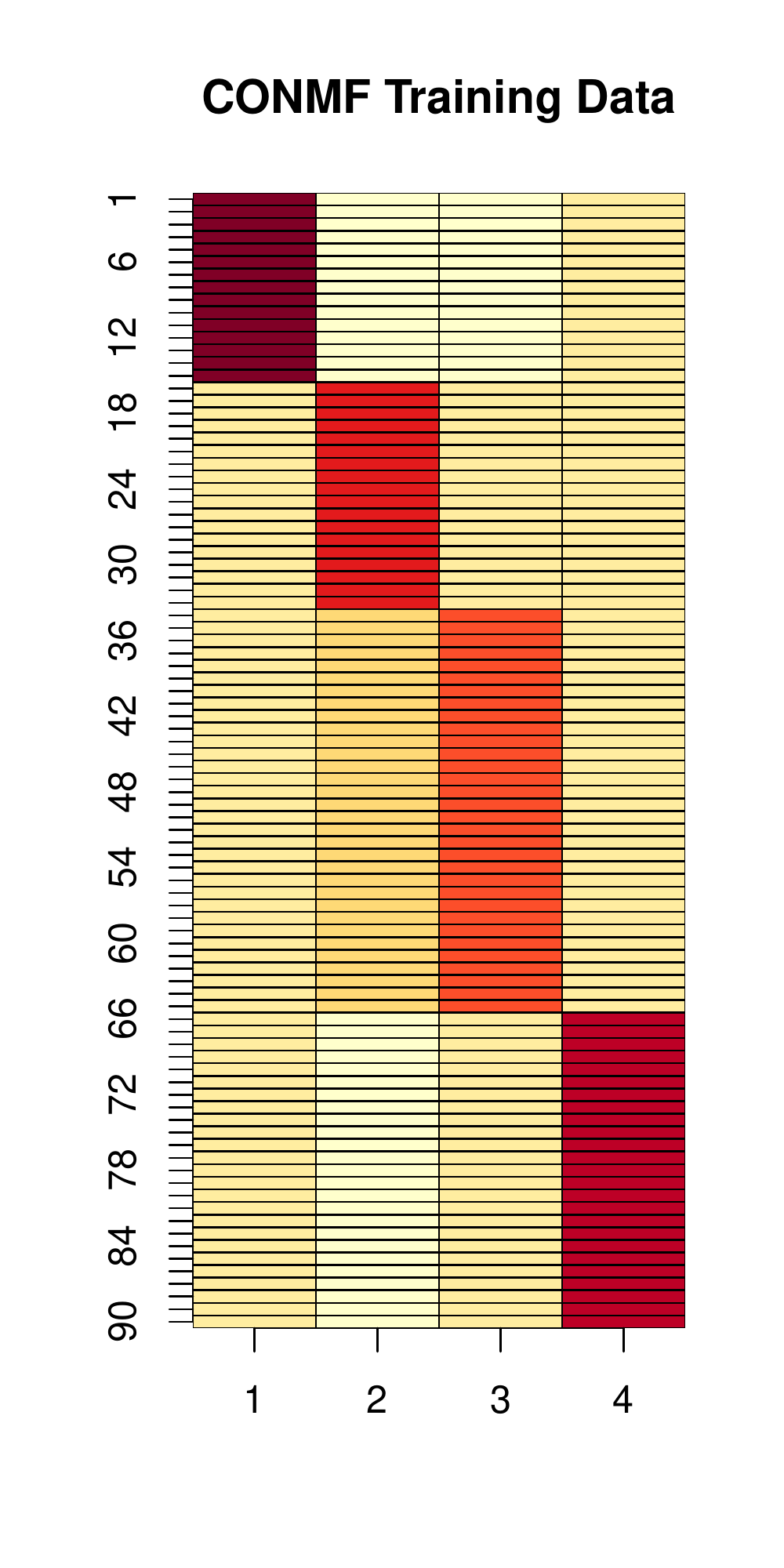}
\end{subfigure}%
\begin{subfigure}{0.15 \textwidth}
\includegraphics[width=\linewidth]{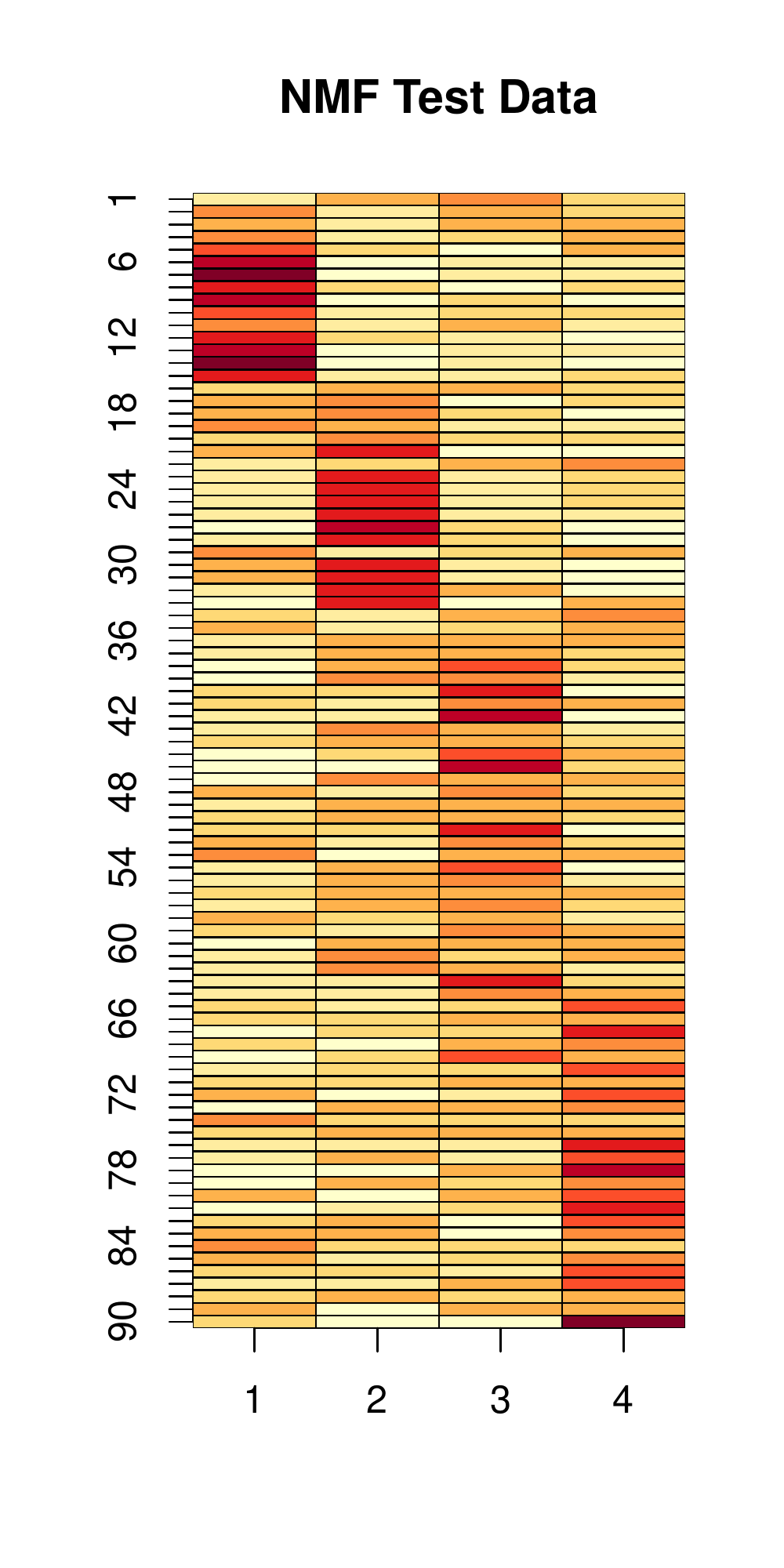}
\end{subfigure}%
\begin{subfigure}{0.15 \textwidth}
\includegraphics[width=\linewidth]{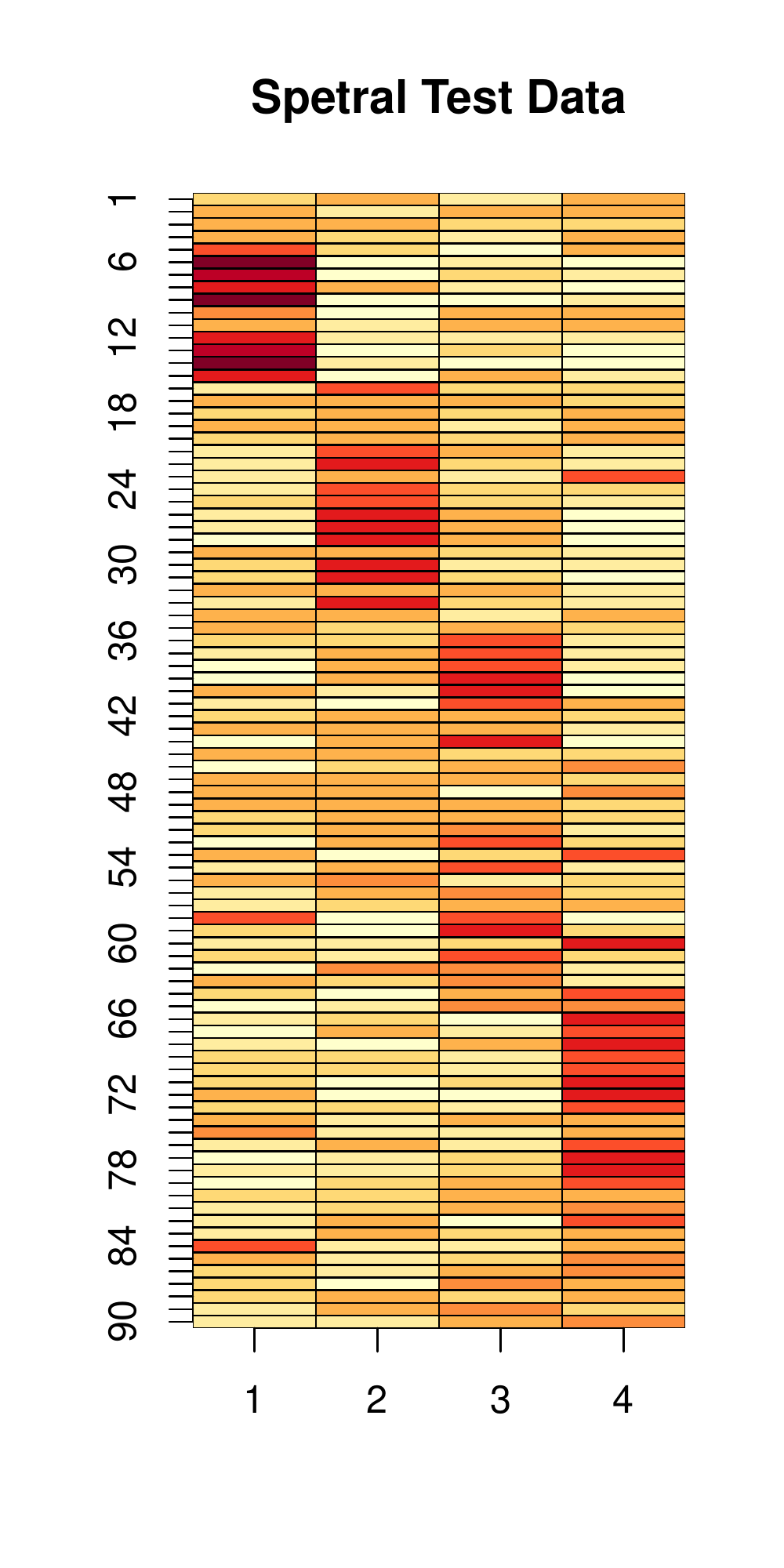}
\end{subfigure}%
\hspace{20pt}
\begin{subfigure}{0.15 \textwidth}
\includegraphics[width=\linewidth]{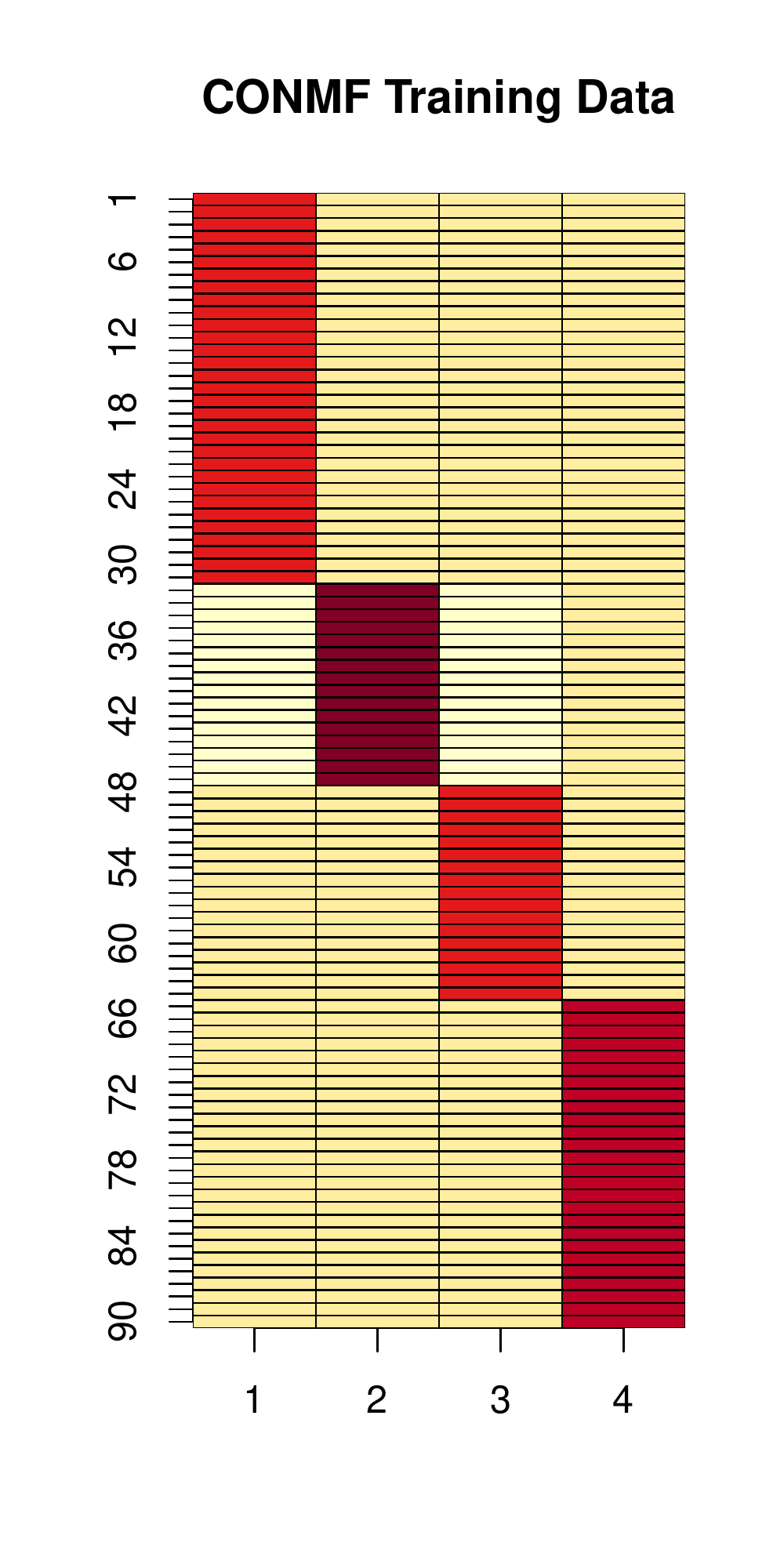}
\end{subfigure}%
\begin{subfigure}{0.15 \textwidth}
\includegraphics[width=\linewidth]{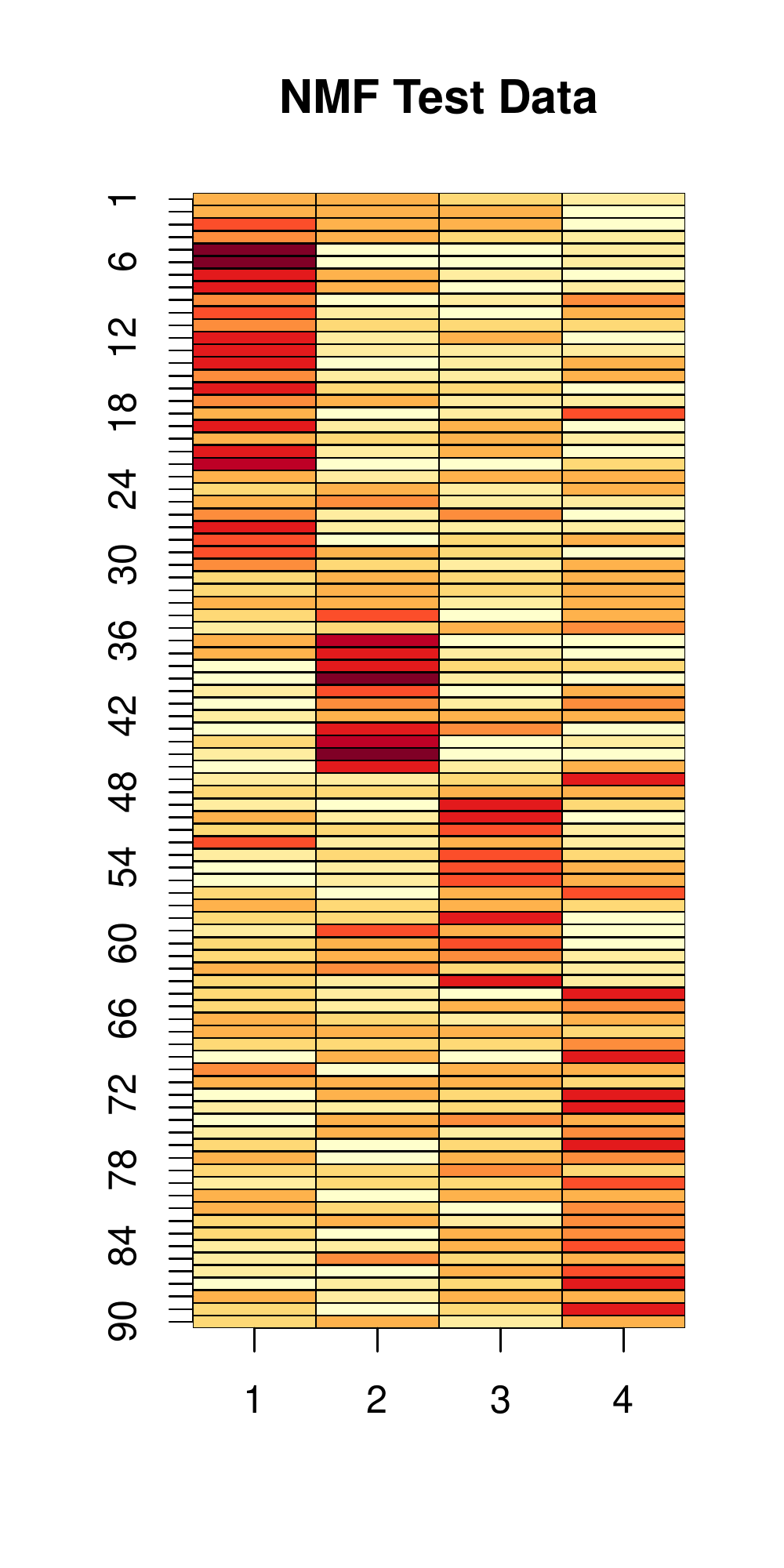}
\end{subfigure}%
\begin{subfigure}{0.15 \textwidth}
\includegraphics[width=\linewidth]{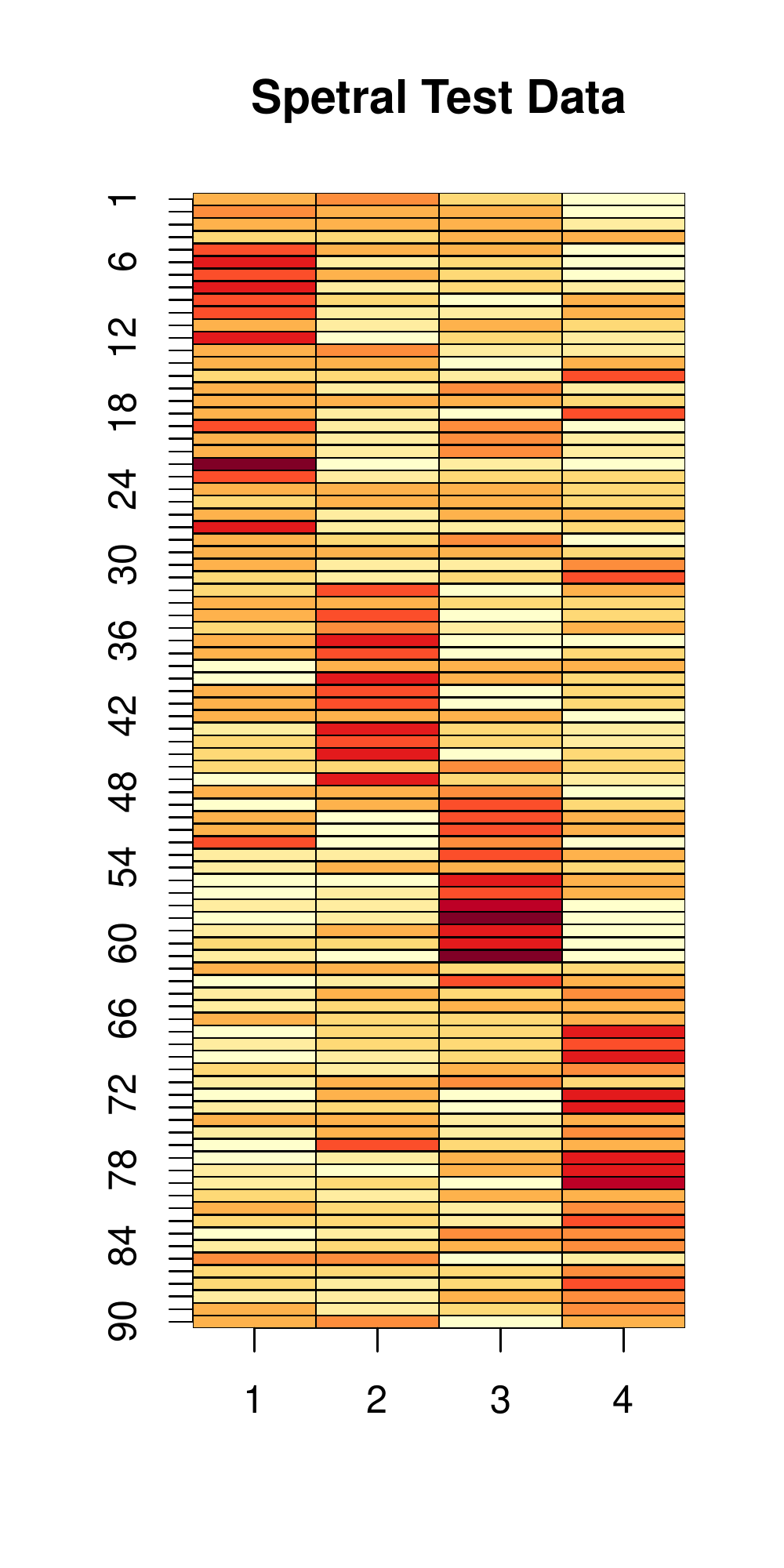}
\end{subfigure}
\begin{center}
\vspace{-10pt}
(a) Control group example 1\hspace{60pt} (b) Control group example 2
\end{center}
\begin{subfigure}{0.15 \textwidth}
\includegraphics[width=\linewidth]{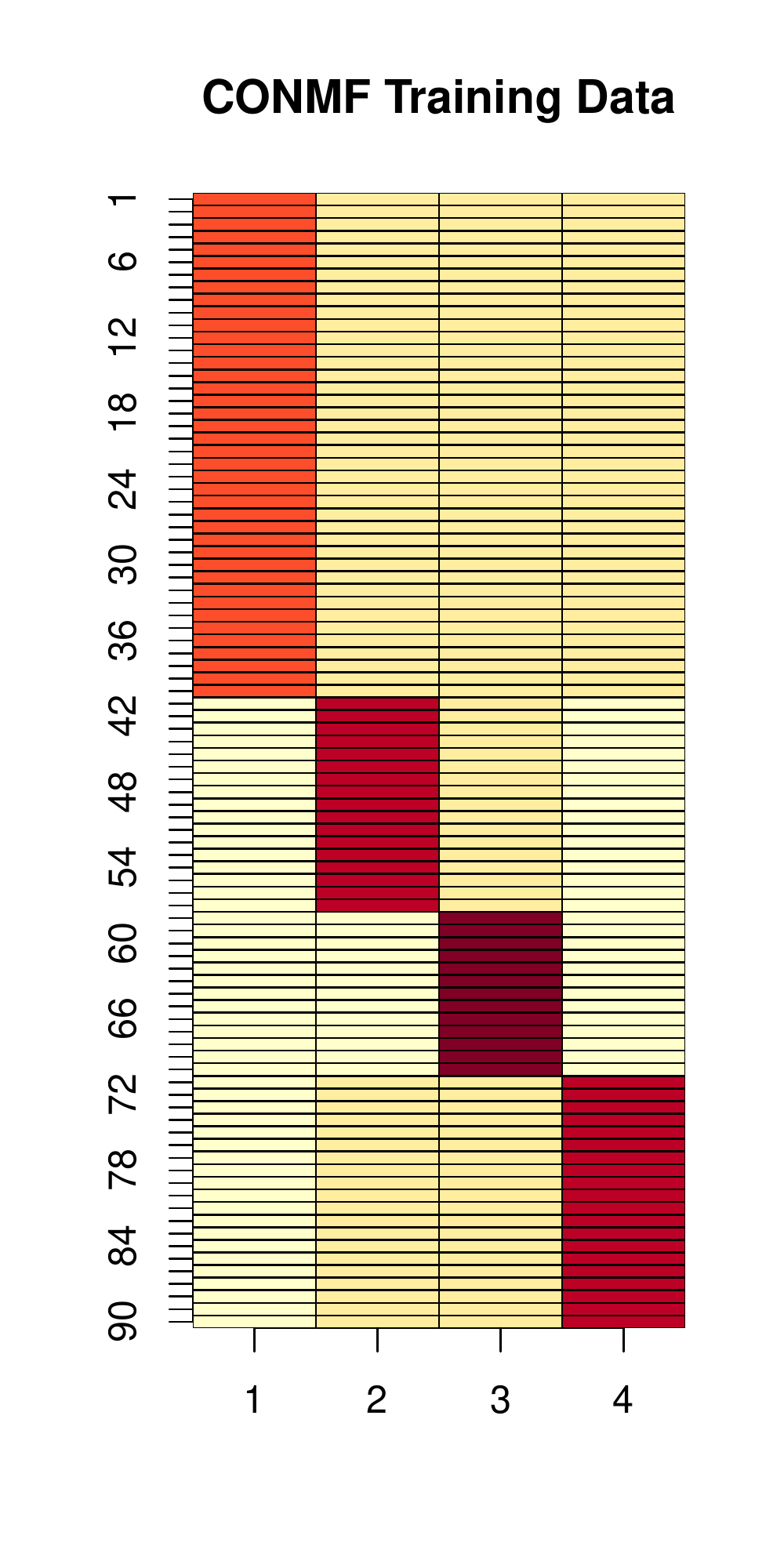}
\end{subfigure}%
\begin{subfigure}{0.15 \textwidth}
\includegraphics[width=\linewidth]{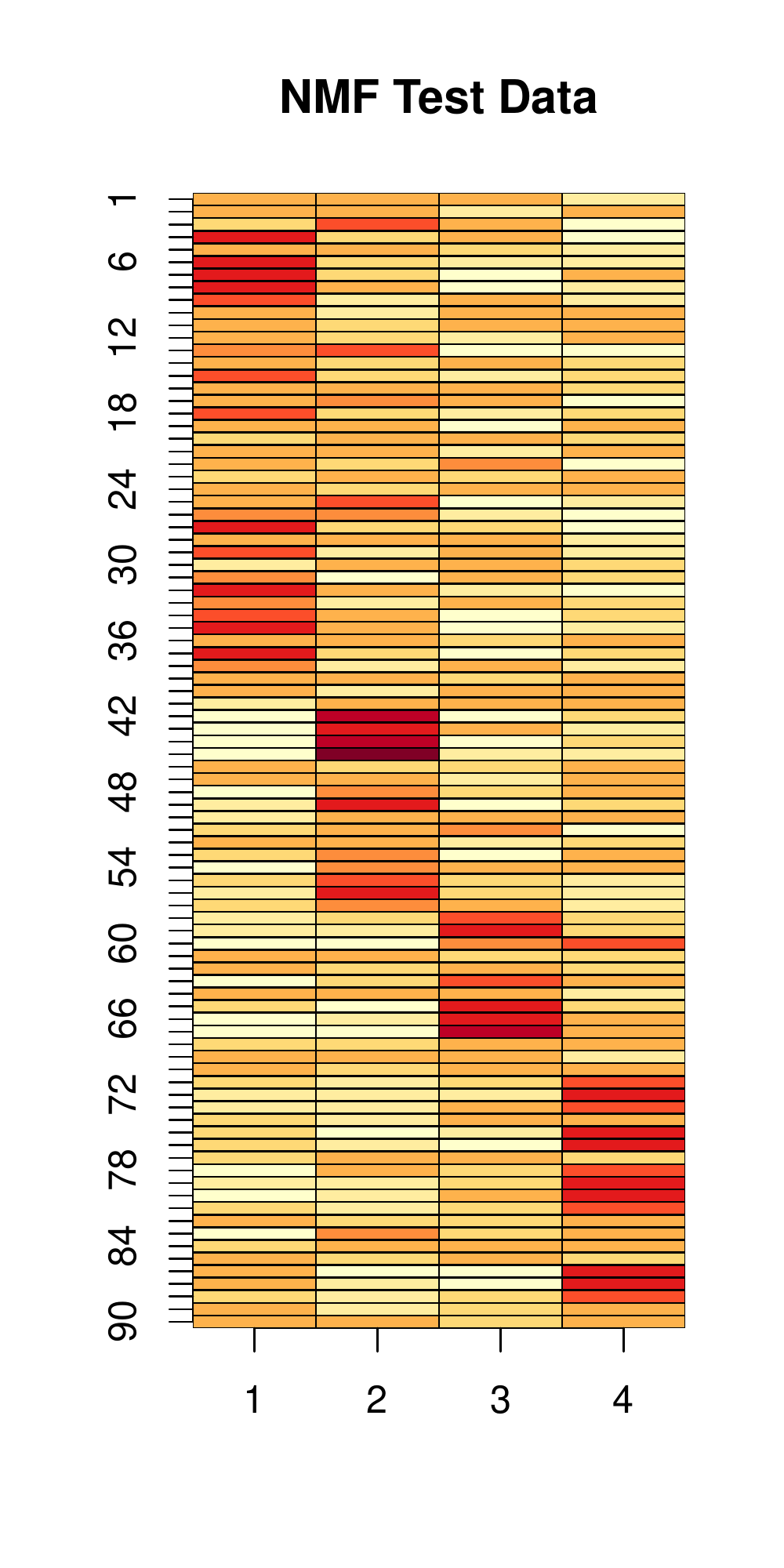}
\end{subfigure}%
\begin{subfigure}{0.15 \textwidth}
\includegraphics[width=\linewidth]{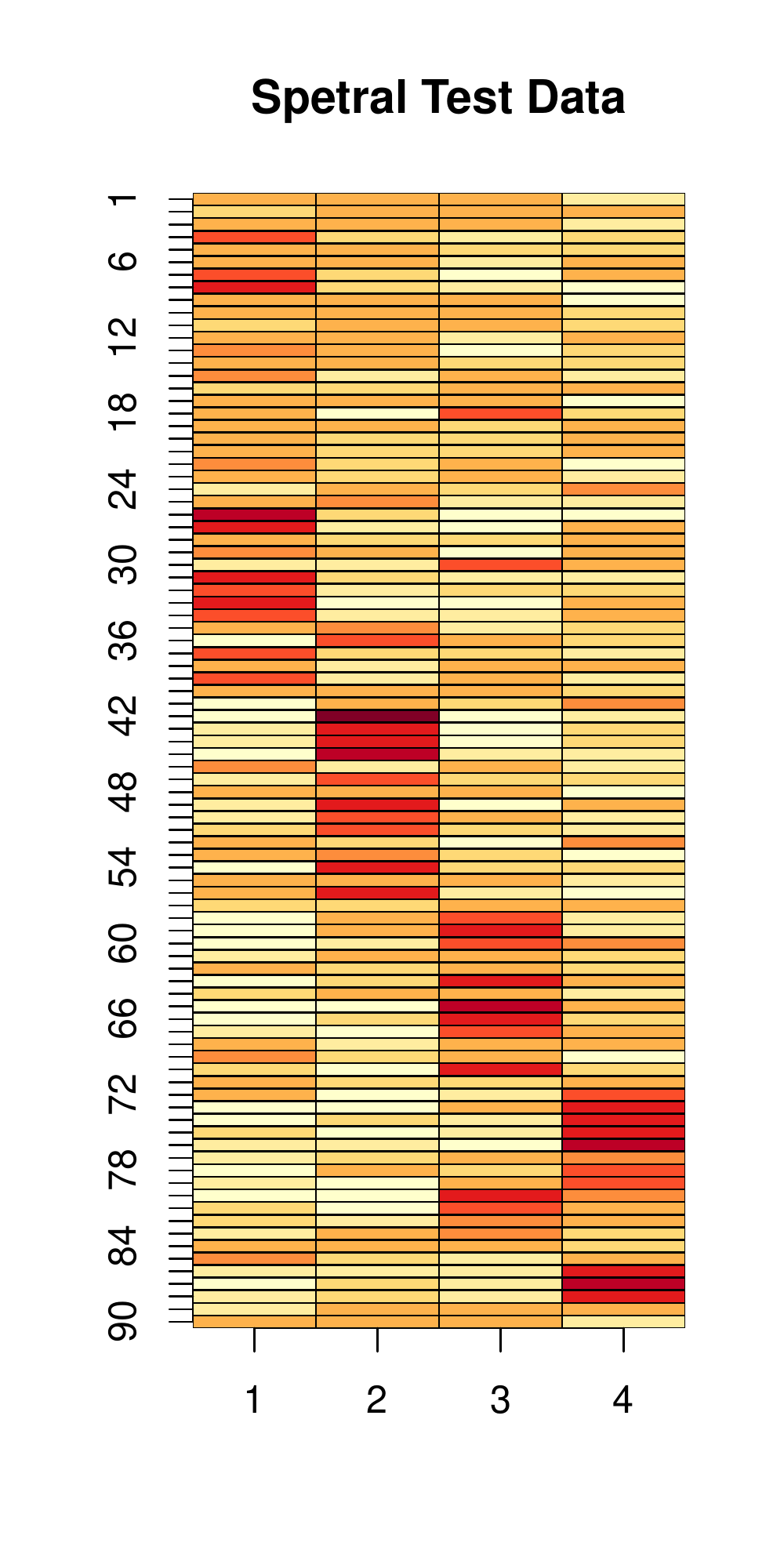}
\end{subfigure}%
\hspace{20pt}
\begin{subfigure}{0.15 \textwidth}
\includegraphics[width=\linewidth]{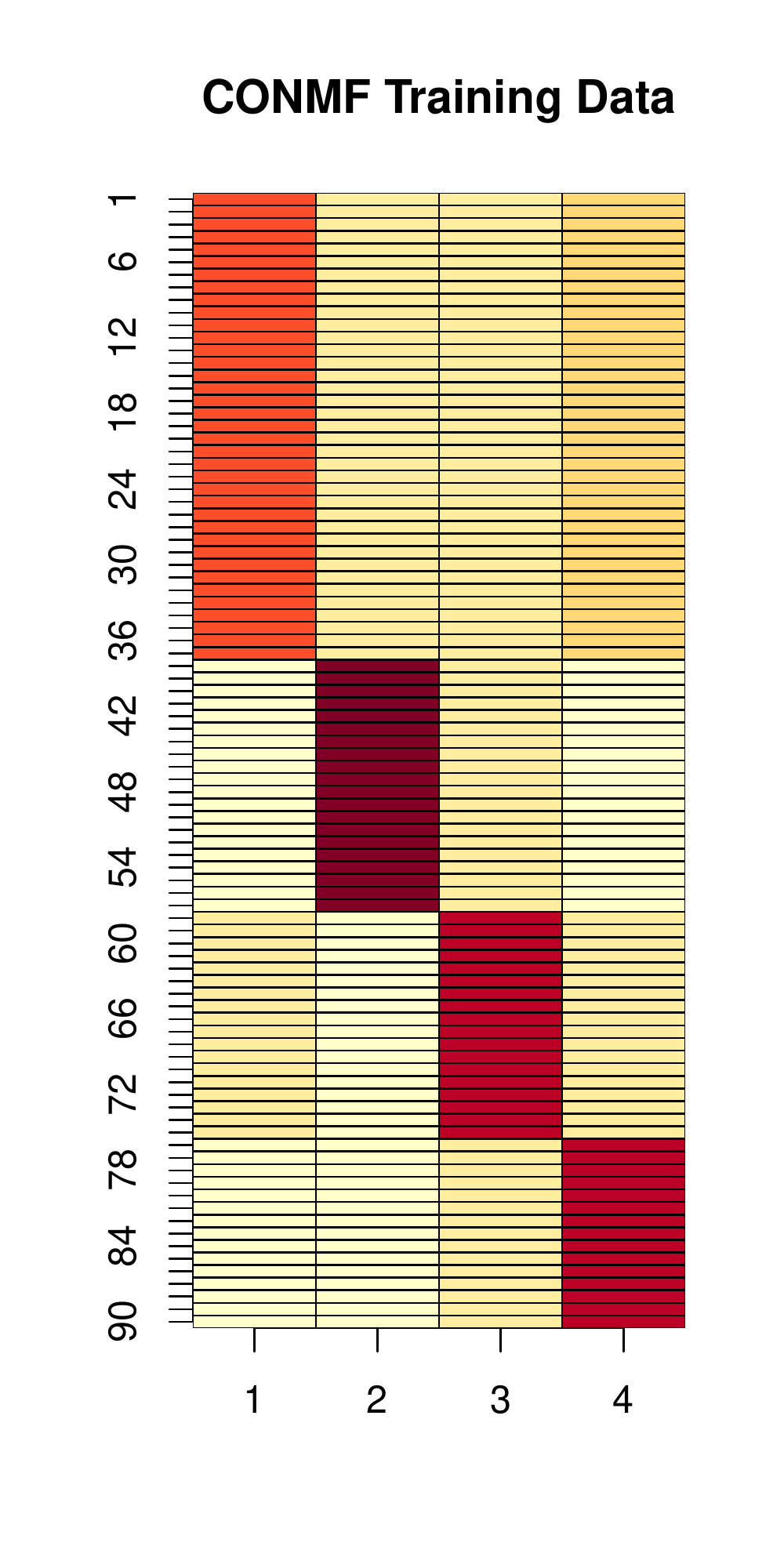}
\end{subfigure}%
\begin{subfigure}{0.15 \textwidth}
\includegraphics[width=\linewidth]{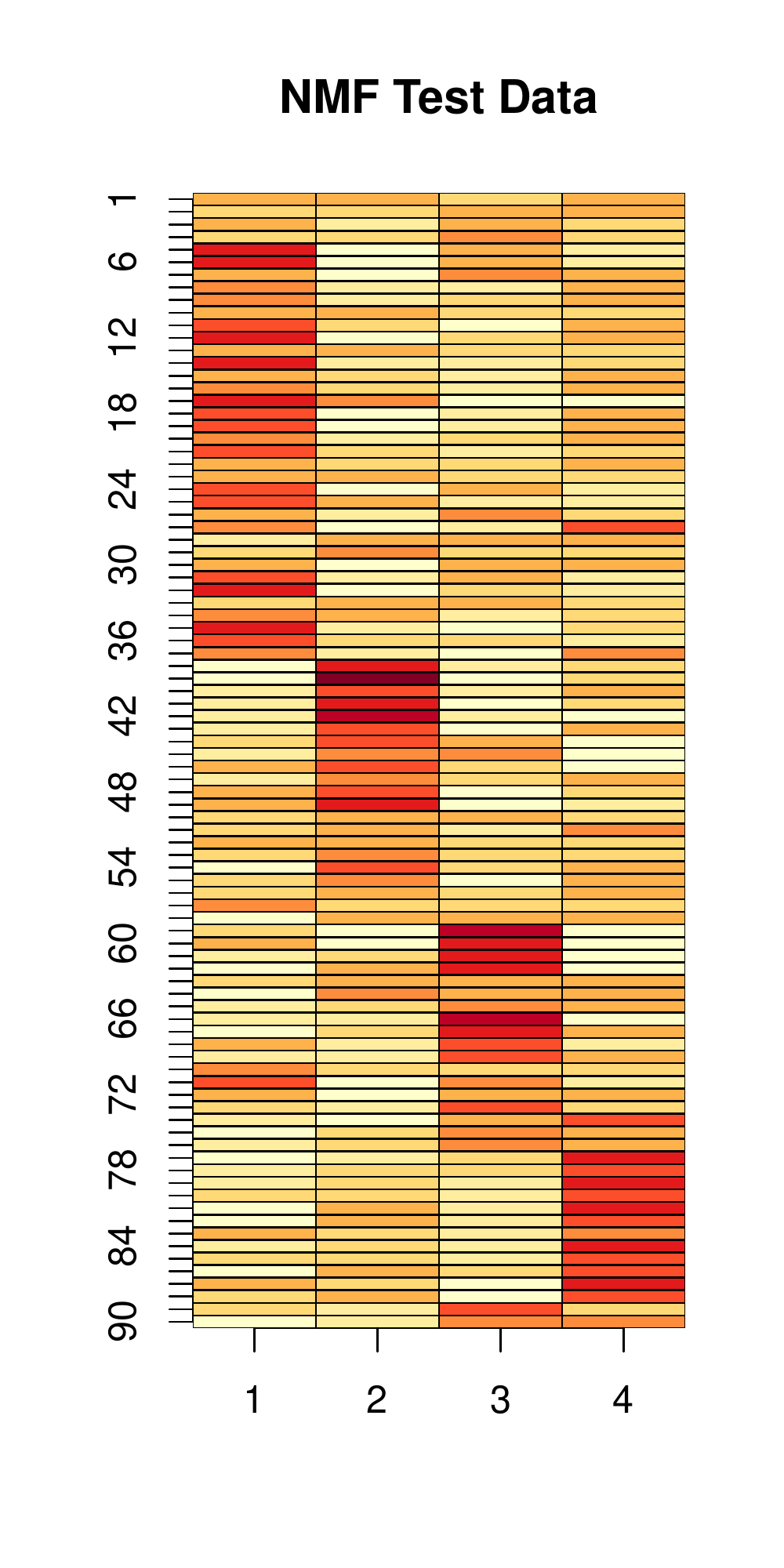}
\end{subfigure}%
\begin{subfigure}{0.15 \textwidth}
\includegraphics[width=\linewidth]{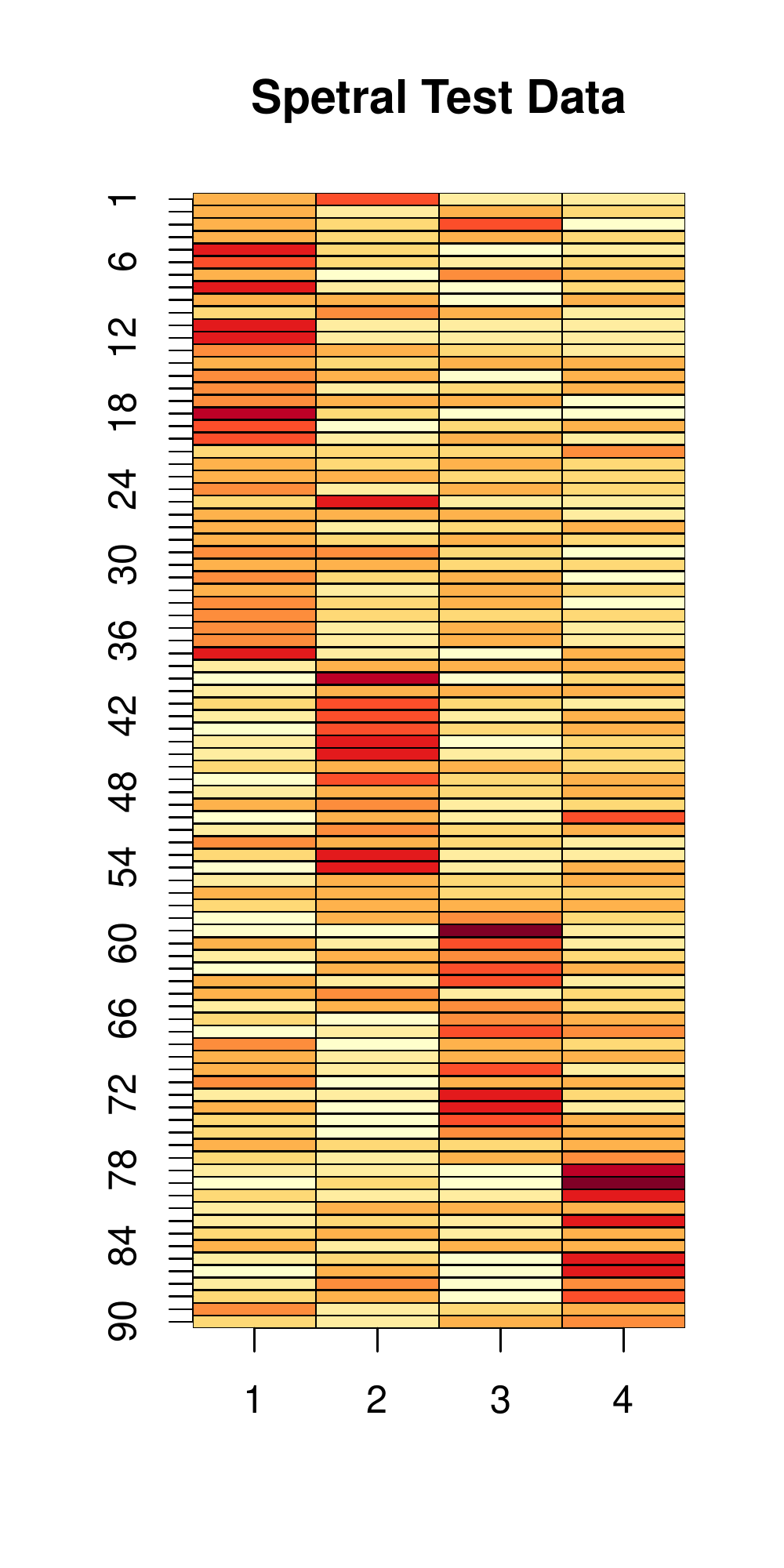}
\end{subfigure}
\begin{center}
\vspace{-10pt}
(c) Partient group example 1\hspace{60pt} (d) Patient group example 2
\end{center}
\vspace{-5pt}
\caption{Examples of predictive performance of Co-OSNTF method. In each case the data is split into a test set comprising of 15 subjects and a training set with the remaining subjects. The Co-OSNTF method is run on the training set to obtain estimated community assignments $\bar{Z}T$ for the ROIs (plotted in left panel in each case). The average of the community assignments for the ROIs obtained separately for each subjects using an NMF and a spectral method are plotted side by side. The plots (a) and (b) are for control group, while (c) and (d) are for patient group.}
\label{prednmf}
\end{figure}

The second metric considers the overlap of the mean putative community structure $\bar{Z}$ with the observed community structures in the test subjects again obtained using the spectral and OSNTF approaches. In Figure \ref{predictive2} we display the box plot of the overlaps (correct classification rate) for VarEM and Co-OSNTF methods for control and patient groups. The correct classification rate for the Co-OSNTF method is better than that of the VarEM method. We note that in both control and patient group, the correct classification rate is around 0.40-0.44 for Co-OSNTF method, which is substantially better than the random guessing rate with 4 communities of $0.25$.

Finally, in Figure \ref{prednmf} we present two examples from each of the control and patient groups, where the sample has been divided into a test set comprising of 15 subjects and a training set comprised of the remaining subjects. For each of the examples (a)-(d), the leftmost figure displays the estimated community membership $(\bar{Z}T)_i$ for each ROI $i$ using the Co-OSNTF method. This estimated community membership is also our prediction for a new subject. The middle and the rightmost figures then display $\bar{u}_i$, the average community membership for each ROI  over the 15 test subjects obtained using the NMF and  spectral methods for single networks. We note that in each case there is strong indication that $\bar{Z}_iT$ from the training sample helps us predict the average community memberships $\bar{u}_i$ from the test samples.

Taken together, the low absolute error in predicting the ROI community assignments and overlap of the actual community assignments with the putative mean assignment, are testaments to the predictive value of the model and the methods. In particular the Co-OSNTF method displays a superior predictive performance, and therefore we primarily present the results from Co-OSNTF method as our findings from this study. This is slightly surprising given the superior performance of the VarEM method in the simulations. However, since the data in the simulations were generated from the RESBM, it is not too surprising that the method which fits the model to the data performed better over a model-free approach. The model is based on SBM and as such inherits the limitations of the SBM, namely a failure to model degree heterogeneity. On the other hand, it has been shown in \cite{pc16} that the OSNTF method is consistent for community detection under degree heterogeneity (degree corrected stochastic block model) as well. The Co-OSNTF method likely inherits this property from OSNTF and is therefore robust against some of the limitations that SBM poses.

\begin{figure}[!h]
\centering{}
\begin{subfigure}{0.23 \textwidth}
\includegraphics[width=\linewidth]{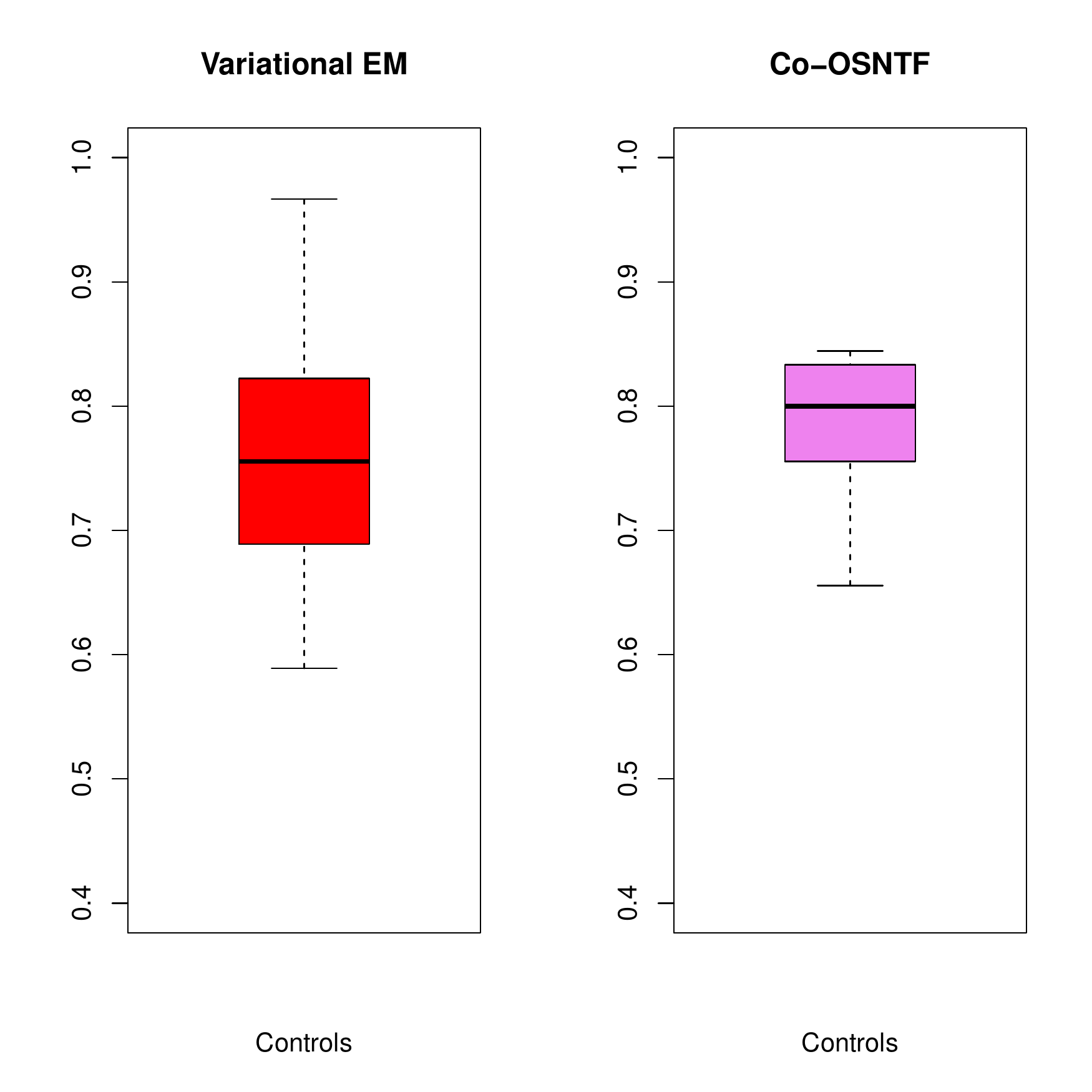}
\end{subfigure}%
\begin{subfigure}{0.23 \textwidth}
\includegraphics[width=\linewidth]{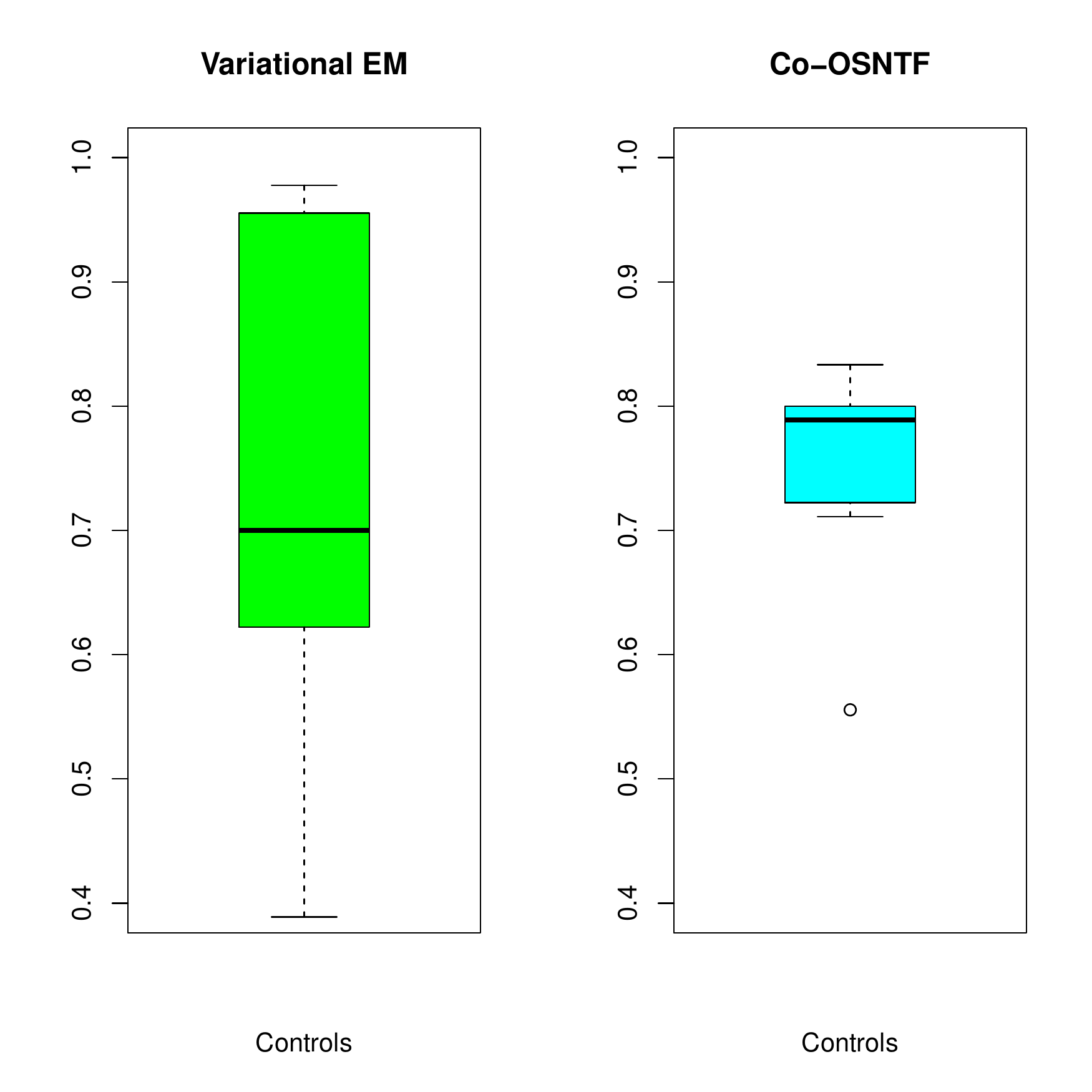}
\end{subfigure}%
\hspace{15pt}
\begin{subfigure}{0.23 \textwidth}
\includegraphics[width=\linewidth]{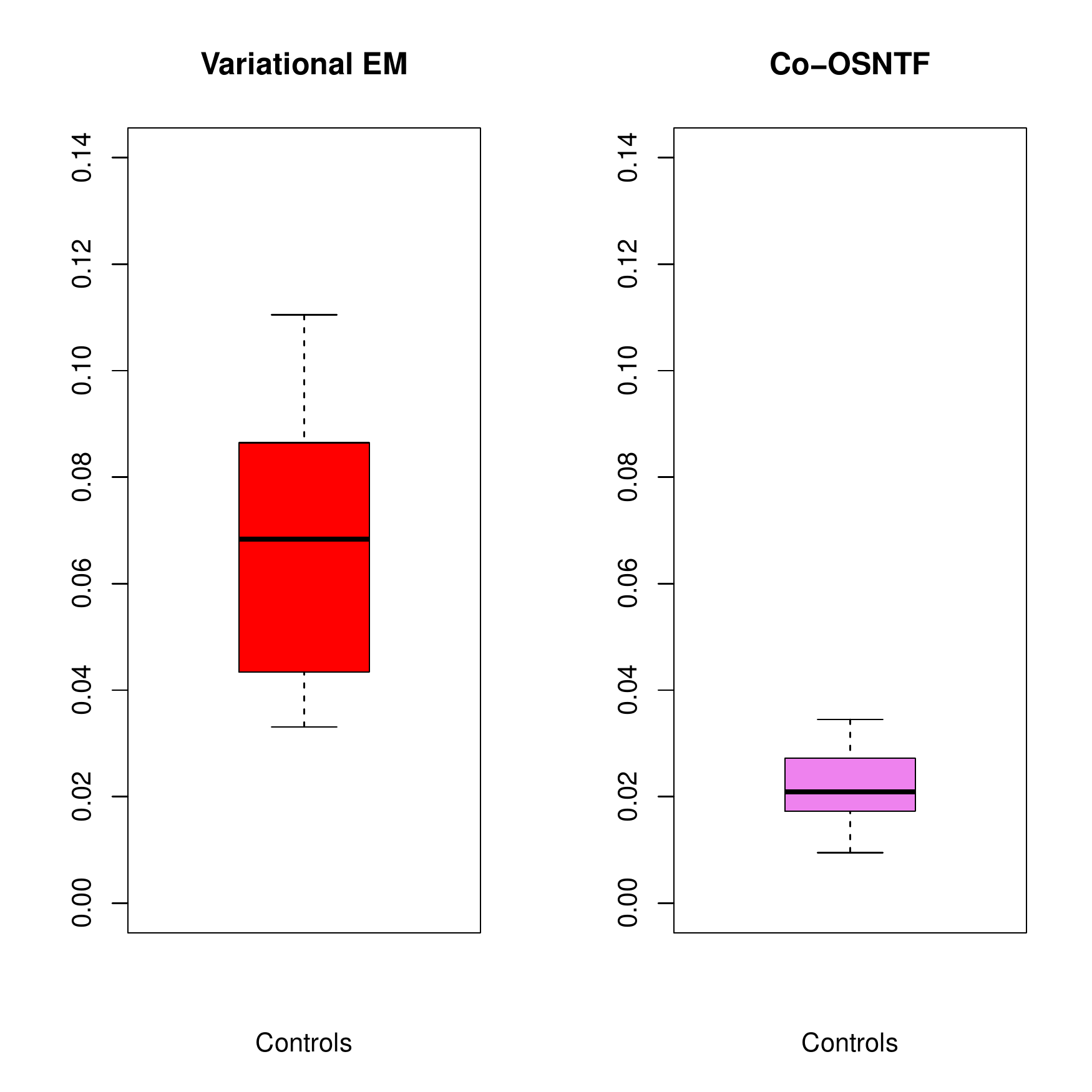}
\end{subfigure}%
\begin{subfigure}{0.23 \textwidth}
\includegraphics[width=\linewidth]{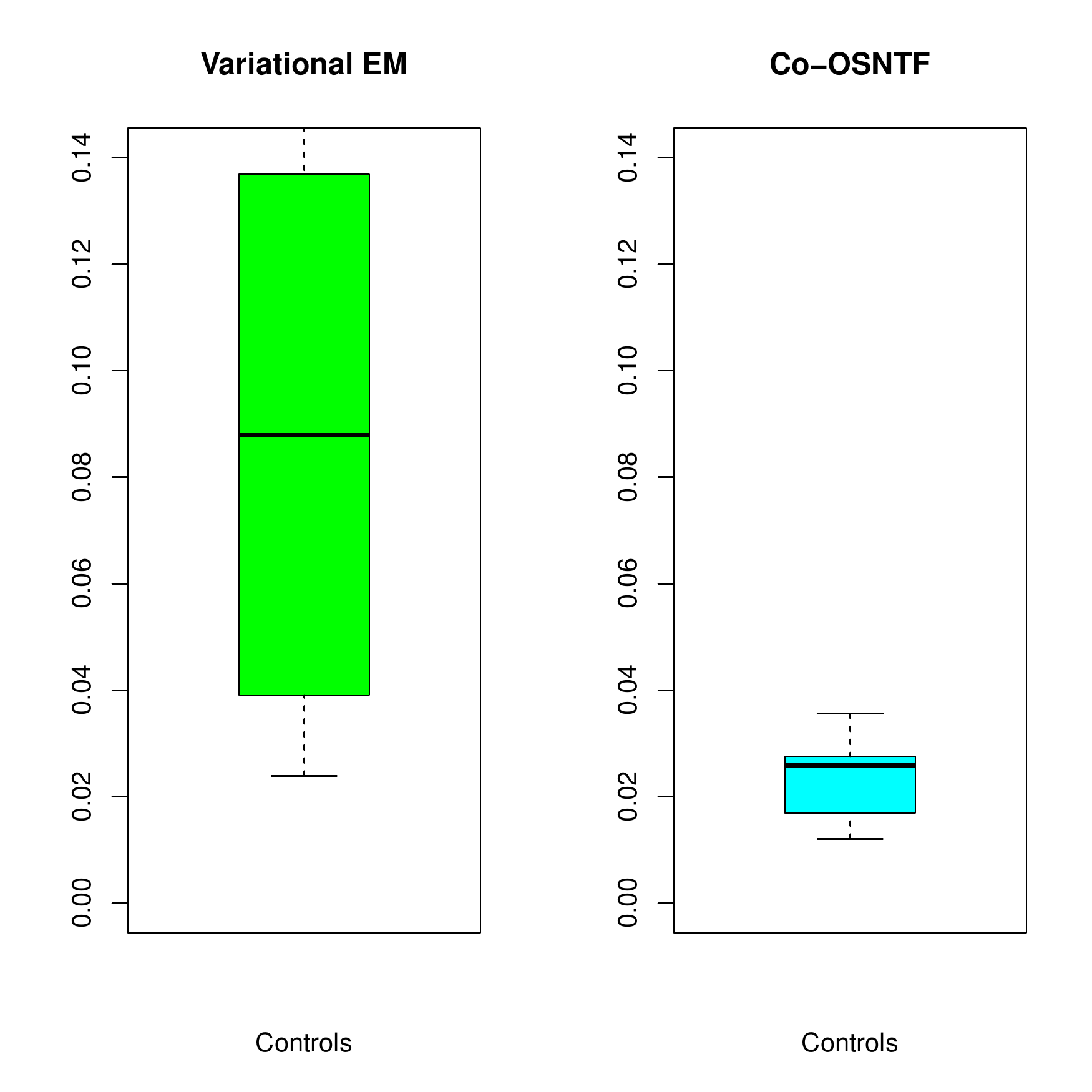}
\end{subfigure}
\begin{center}
(a) comparing $\bar{Z}$ \hspace{150pt} (b) comparing $T$
\end{center}
\vspace{-5pt}
\caption{Split sample validation: The samples are divided into two parts randomly and Co-OSNTF and VarEM are fitted to each part. The (a) $\bar{Z}$ and (b) $T$ estimates from the two splits are compared for control and patient groups.}
\label{valid}
\end{figure}

\subsection{Validation and robustness checks}

The predictive performance validates the use and value of our model and methods. Next we further perform a split sample validation of the results where we split the sample into two roughly equal parts. Then we estimate the parameters $(\bar{Z},T)$ on the two parts separately using VarEM and Co-OSNTF methods. We compare the two sets of estimates. The $\bar{Z}$ estimates are compared using the correct classification rate, while the $T$ estimates are compared using the median absolute difference between the corresponding elements. As before the community labels in the two splits are aligned by solving a LSAP problem. We repeat this procedure 10 times, every time randomly splitting the sample into two parts. Box plots of the correct classification rate for comparing $\bar{Z}$ estimates and the median absolute difference for comparing $T$ estimates are presented in Figure \ref{valid}. We find that there is an extremely large overlap with correct classification rate reaching 80-90\% using Co-OSNTF (Figure \ref{valid}(a)) and the median absolute difference between elements of $T$ matrix being around 0.02-0.03 (Figure \ref{valid}(b)). This validates that our estimates are consistent across splits of the sample.

In our first robustness check we fit RESBM to networks obtained by thresholding the residualized correlation matrices obtained by regressing out the effects of a number of covariates as outline in Section 2.1.2. We obtain $(\bar{Z},T)$ estimates using Co-OSNTF method and compare those with the estimates obtained without residualizing.  We find the estimate for $\bar{Z}$ matches for about 95\% of the ROIs (a mismatch of 4 out of 90 ROIs) for both controls and patients. The median absolute difference for the estimate of $T$ is 0.0063 and 0.0049 for controls and patients respectively. This suggests the estimates obtained from residualized correlation matrices is not much different from the one obtained without residualizing.

\begin{figure}[!h]
\centering{}
\begin{subfigure}{0.23 \textwidth}
\includegraphics[width=\linewidth]{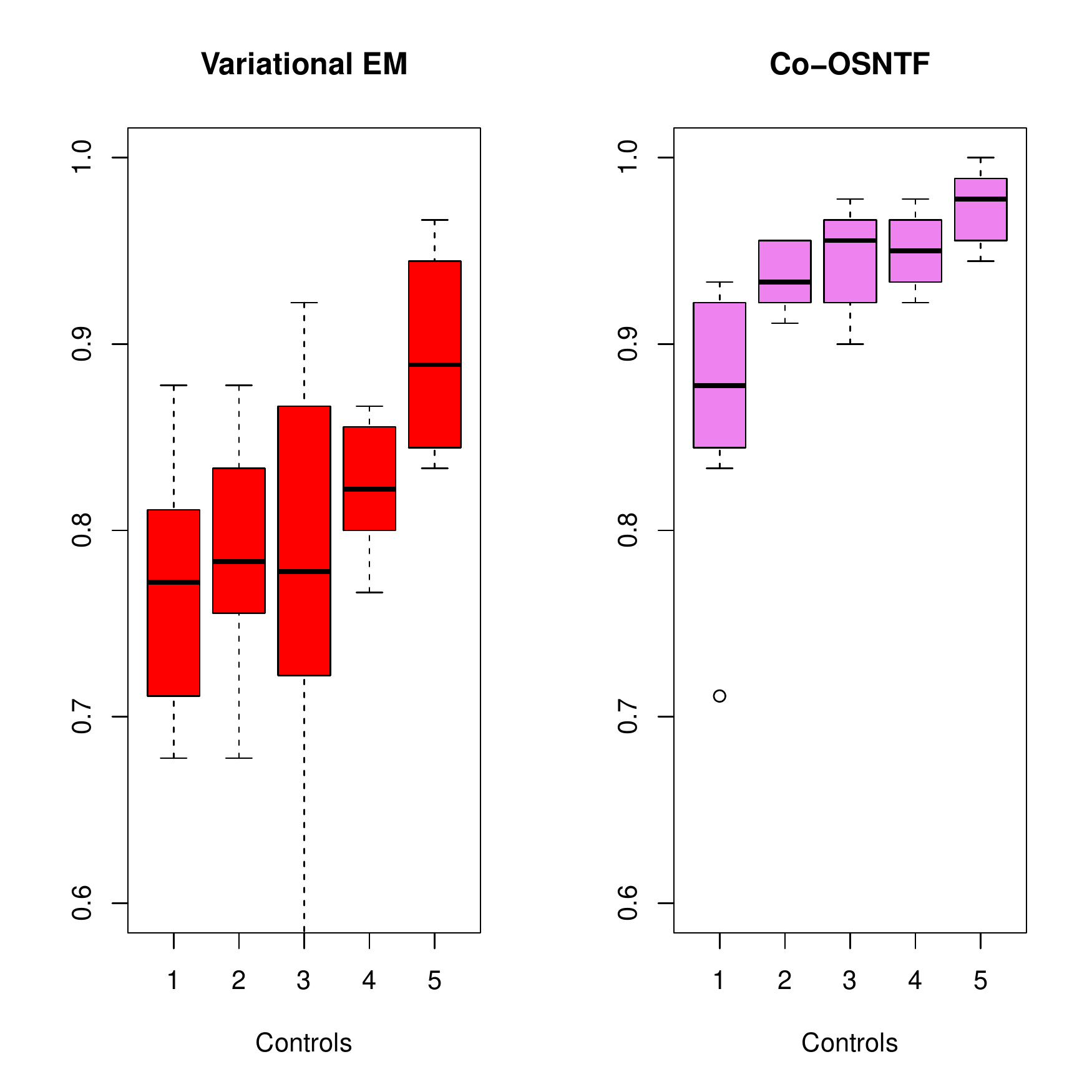}
\end{subfigure}%
\begin{subfigure}{0.23 \textwidth}
\includegraphics[width=\linewidth]{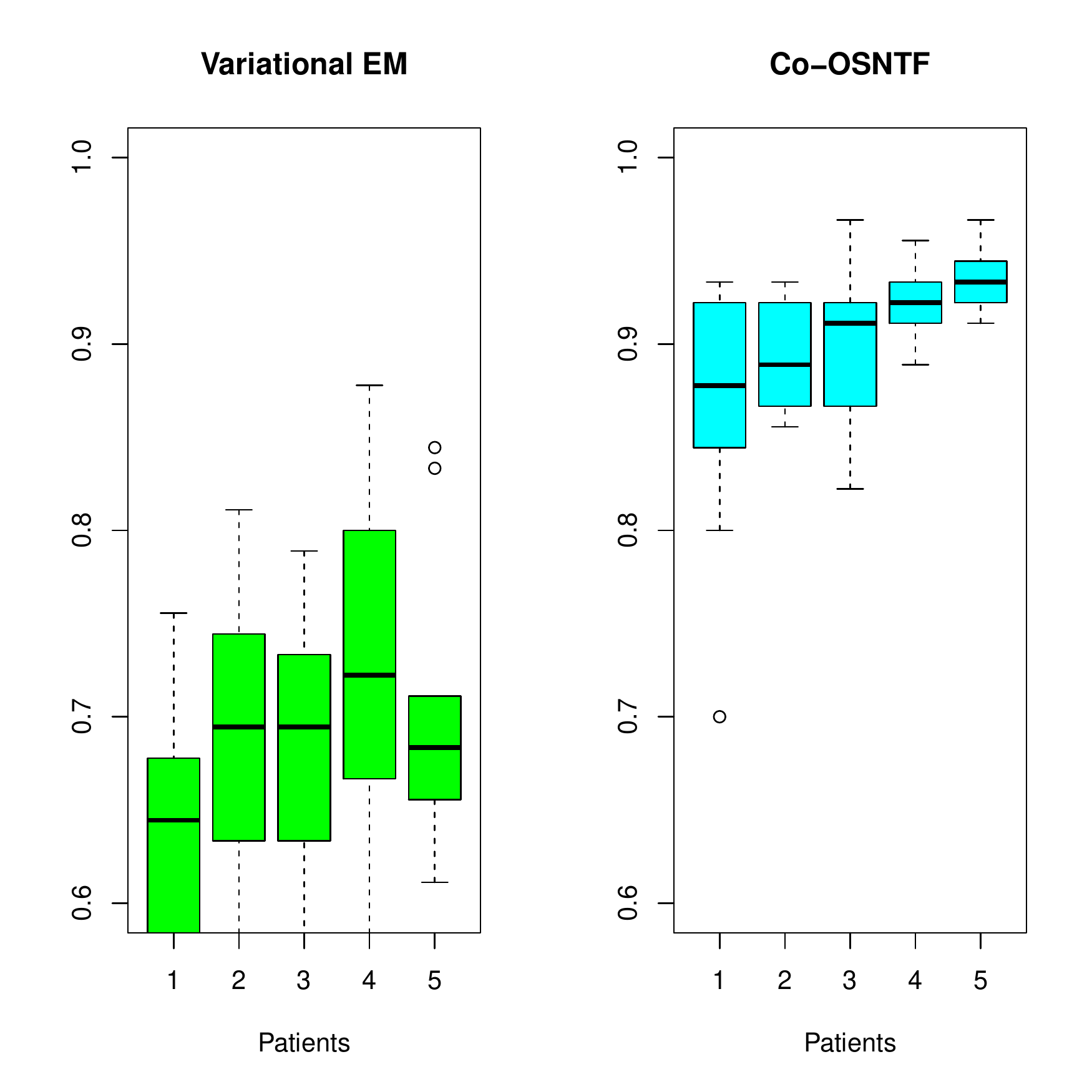}
\end{subfigure}%
\hspace{15pt}
\begin{subfigure}{0.23 \textwidth}
\includegraphics[width=\linewidth]{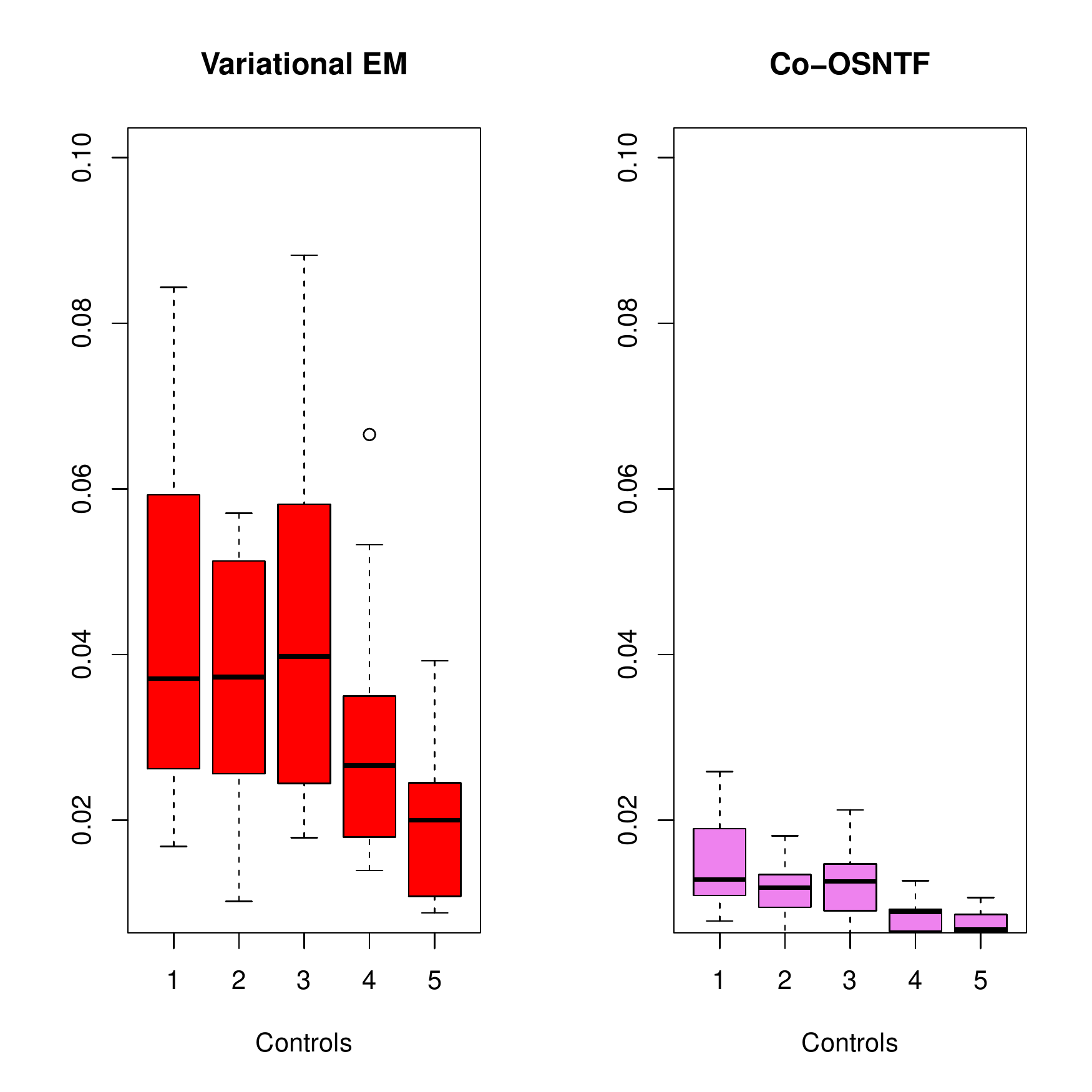}
\end{subfigure}%
\begin{subfigure}{0.23 \textwidth}
\includegraphics[width=\linewidth]{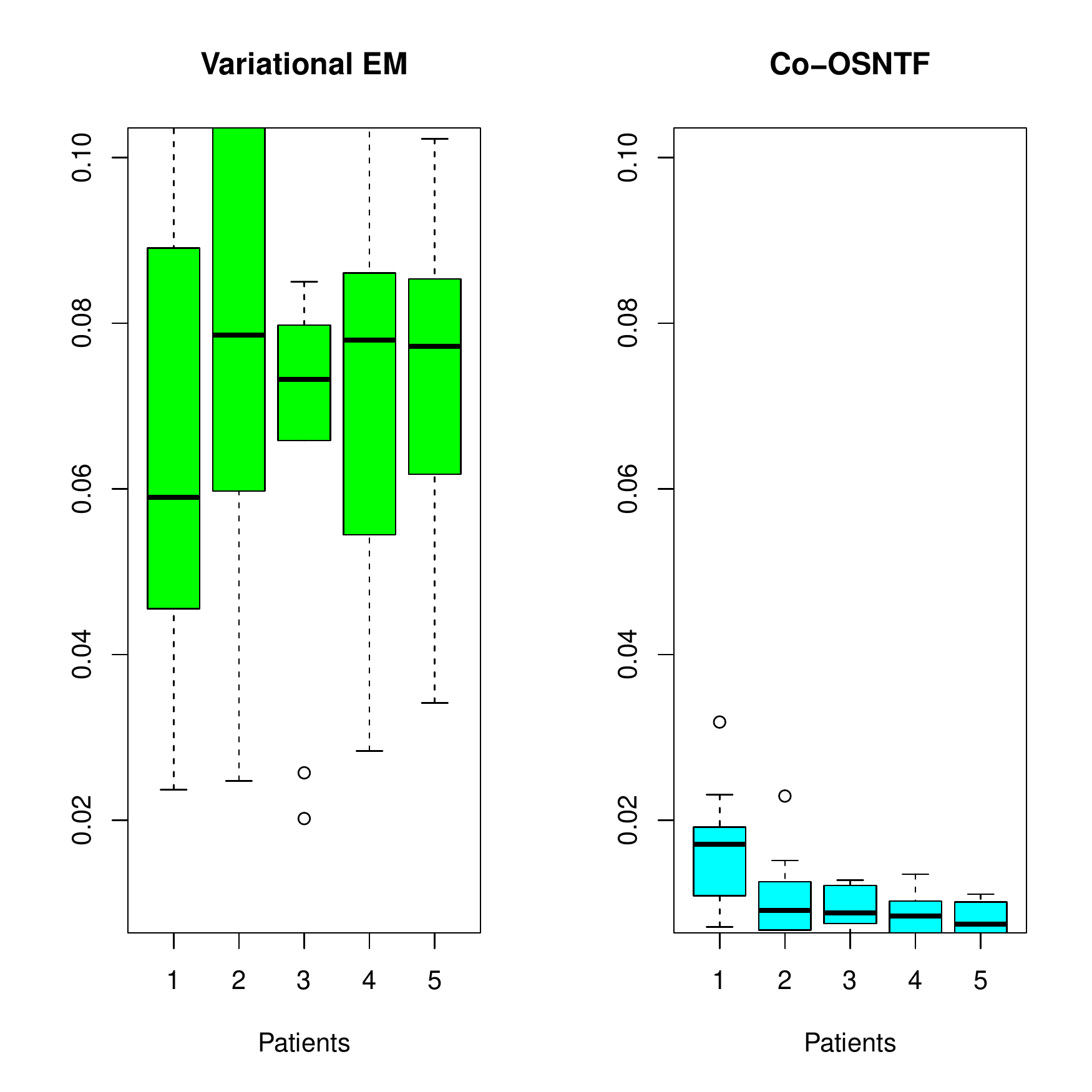}
\end{subfigure}
\begin{center}
(a) accuracy of estimating $\bar{Z}$  \hspace{150pt} (b) accuracy of estimating $T$
\end{center}
\vspace{-5pt}
\caption{Robustness check: Overlap of results on full sample with results from a smaller subsample (a) accuracy of estimating $\bar{Z}$ using correct classification rate and (b) accuracy of estimating $T$ using median absolute error.}
\label{robust}
\end{figure}

Our second robustness check involves sampling a subset of subjects in each group and fitting the model on the subset as if the remaining data did not exist. We then compare the $(\bar{Z},T)$ estimates  from this reduced dataset with those obtained from the full dataset. In Figure \ref{robust} we progressively sample (50\%, 60\%, 70\%, 80\%, 90\%) of the data and compare the correct classification rate of $\bar{Z}$ estimate and median absolute difference of $T$ estimate with the full data. In each case, the boxplot represents the distribution of these values over 10 random samples. We note that the estimates for both $\bar{Z}$ and $T$ from the reduced samples become closer to those obtained from the full sample as the size of the reduced sample increases for both controls and patients. Across different scenarios, Co-OSNTF method gives much better proximity to full sample results compared to  VarEM. For the Co-OSNTF method, the  $\bar{Z}$ estimate from the reduced sample has an overlap of 90\% with 50\% data and it increases to almost 95\% with increasing sample. On the other hand the estimate for $T$ has a median absolute error of 0.01 with 50\% data and decreases even further to almost no error with 90\% of the data. These results can be thought of as another evidence towards validation of the results and an important check for the robustness of our findings in the study.

\begin{figure}[!h]
\begin{subfigure}{0.45 \textwidth}
\includegraphics[width=\linewidth]{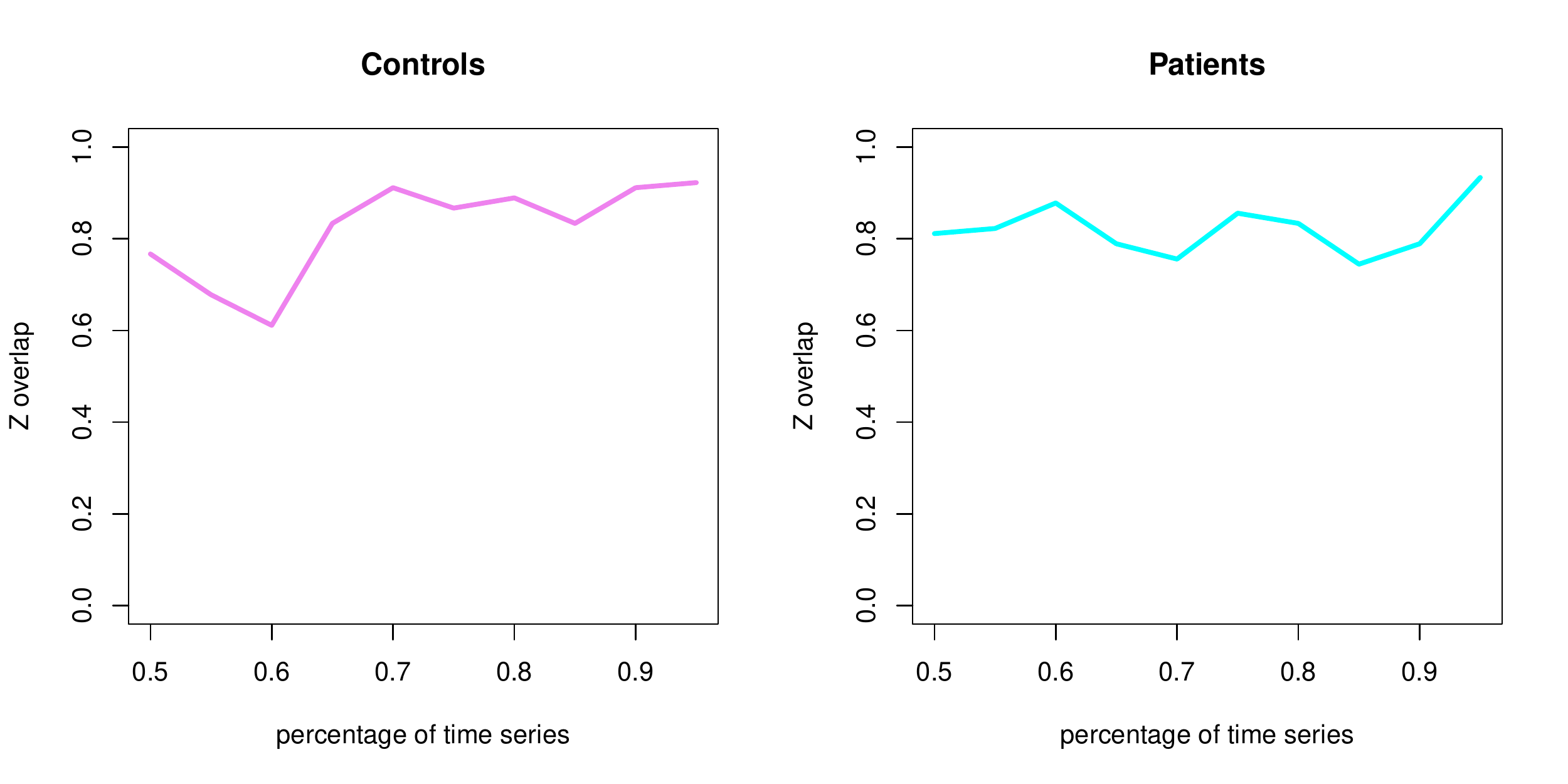}
\end{subfigure}%
\hspace{20pt}
\begin{subfigure}{0.45 \textwidth}
\includegraphics[width=\linewidth]{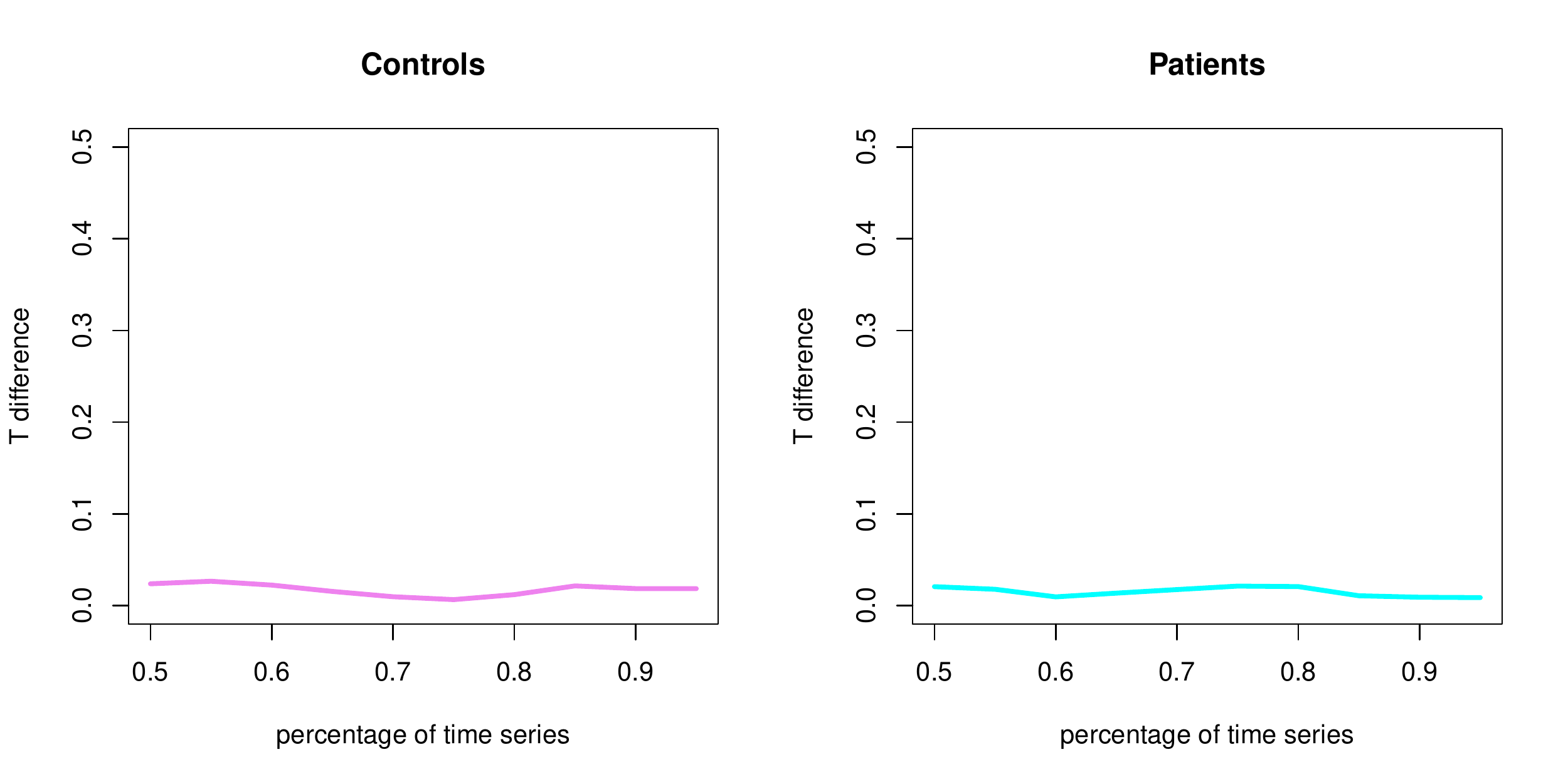}
\end{subfigure}%
\begin{center}
(a) $\bar{Z}$ overlap  \hspace{150pt} (b) Difference in $T$
\end{center}
\vspace{-5pt}
\caption{Robustness check: Overlap between results from shorter time series with those obtained from the full time series.}
\label{robust_ts}
\end{figure}

The final robustness check involves availability of smaller time series data for estimating the correlation matrices. Clearly the accuracy with which the correlation matrices are measured is a function of the sample size of the time series and hence is expected to have an impact in our analysis. In Figure \ref{robust_ts} we start with roughly half the length of the time series and increase the length progressively. We assess the similarity of the estimates for $\bar{Z}$ and $T$ obtained from the smaller time series with the full series in Figures \ref{robust_ts} (a) and (b) respectively. It appears that the length of the time series, despite having some impact, does not drastically change our estimates. The $\bar{Z}$ estimates have strong overlap with the estimate from full sample, while the median absolute difference of the $T$ estimates from that of the full sample is small.

\begin{figure}[h]
\begin{subfigure}{0.35 \textwidth}
\includegraphics[width=\linewidth]{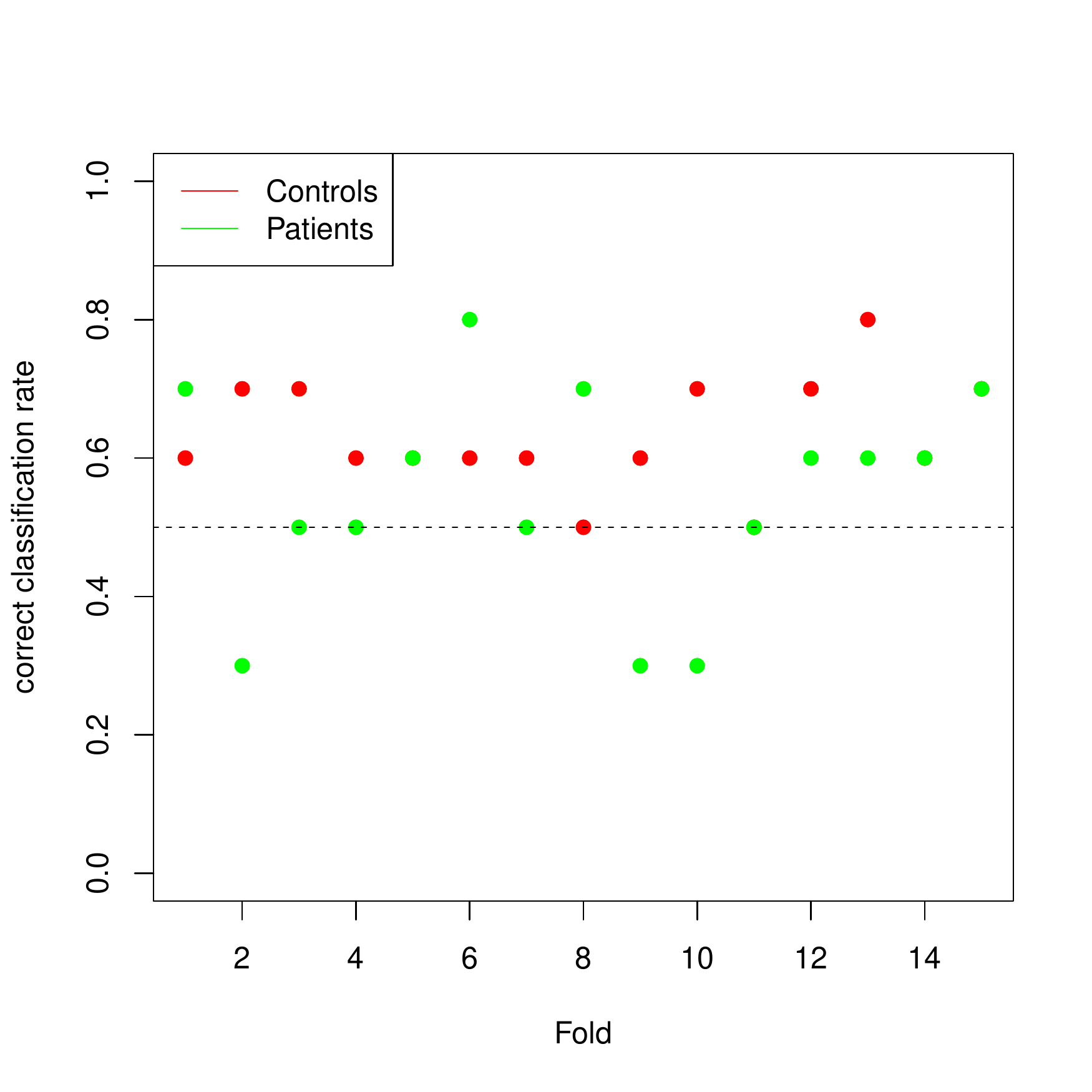}
\end{subfigure}%
\hspace{30pt}
\begin{subfigure}{0.35 \textwidth}
\includegraphics[width=\linewidth]{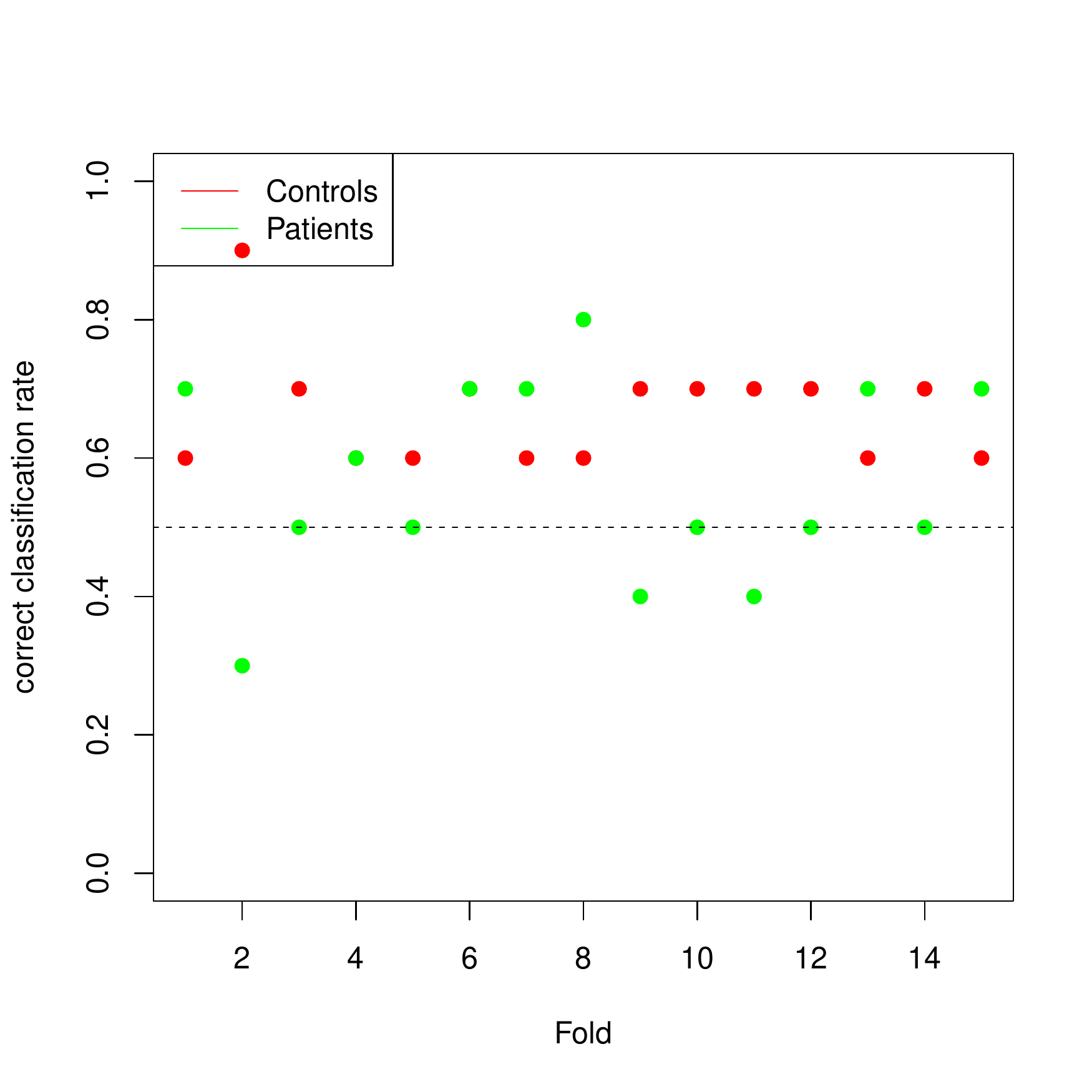}
\end{subfigure}%
\begin{center}
(a) Discrimination with Log-likelihood statistic  \hspace{80pt} (b) Discrimination with MUV statistic
\end{center}
\vspace{-5pt}
\caption{Discrimination power of Co-OSNTF method. We use the (a) log-likelihood and the (b) MUV statistics to classify a new subject to one of the two groups.}
\label{discrimination}
\end{figure}

\subsection{Discrimination analysis with new subjects}

Next we assess the utility of our model and Co-OSNTF method in classifying an unlabeled subject to one of the two groups: controls or patients. Classification of subjects in one of the groups is a fundamental interest in fMRI due to obvious diagnostic and clinical implications. Recently there has been an enhanced interest in classifying subjects in schizophrenia, with many studies reporting high accuracy. We show that the proposed methods can uncover information that are useful in discrimination.

We use two discrimination functions, one based on the MUV statistic described earlier and the other one based on ratio of log-likelihoods. Suppose as before $u_i$ denotes a $k$ dimensional community assignment vector for ROI $i$ in a test subject obtained using either the spectral method or the OSNTF method. Note from Equation (\ref{VCMLSBM}) that the likelihood for  ROI $i$ in a test subject to belong to the population A is $u_i\log (\bar{Z}_{A}T_{A}) $. Therefore the log-likelihood that the test subject belongs to population A is $\sum_i u_i \log (\bar{Z}_{A}T_{A}) $ due to the conditional independence assumption. Similarly, the log-likelihood that the subject belongs to population B is $\sum_i u_i \log (\bar{Z}_{B}T_{B}) $. Therefore the log-likelihood based discrimination rule is to classify a subject to either population $A$ or $B$ based on whether $\sum_i u_i \log (\bar{Z}_{A}T_{A}) >\sum_i u_i \log (\bar{Z}_{B}T_{B})  $ or not. The MUV discrimination function is a simple variant of the MUV test statistic defined earlier,
\[
MUV=\|\bar{Z}_{(A)}T_{(A)}T_{(A)}^{T}\bar{Z}^{T}_{(A)} - UU^{T}\|_F^2.
\]

In Figure \ref{discrimination} we present accuracy of classifying subjects from a 15\% holdout test set on the basis of estimates obtained from training data over 15 repetitions. In majority of cases in both controls and patients, the accuracy of classification is above the random guessing line of 0.5. We note that the classification accuracy is low compared to recent reports. However, given this is a discrimination analysis, and we do not train a classifier for the purpose of supervised classification, it is not so surprising.

\begin{figure}[!h]
\centering{}
\begin{subfigure}{0.32 \textwidth}
\includegraphics[width=\linewidth]{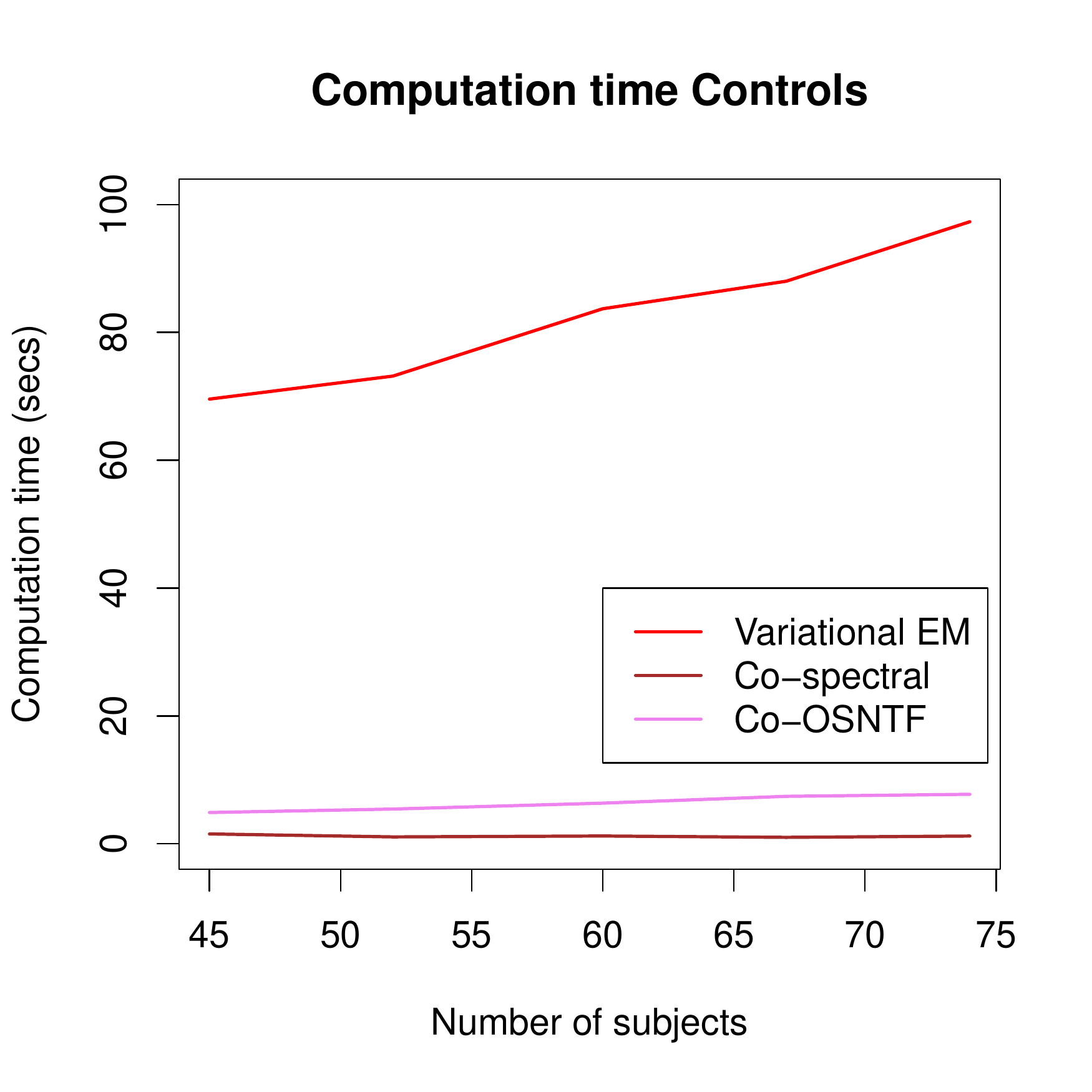}
\end{subfigure}%
\begin{subfigure}{0.32 \textwidth}
\includegraphics[width=\linewidth]{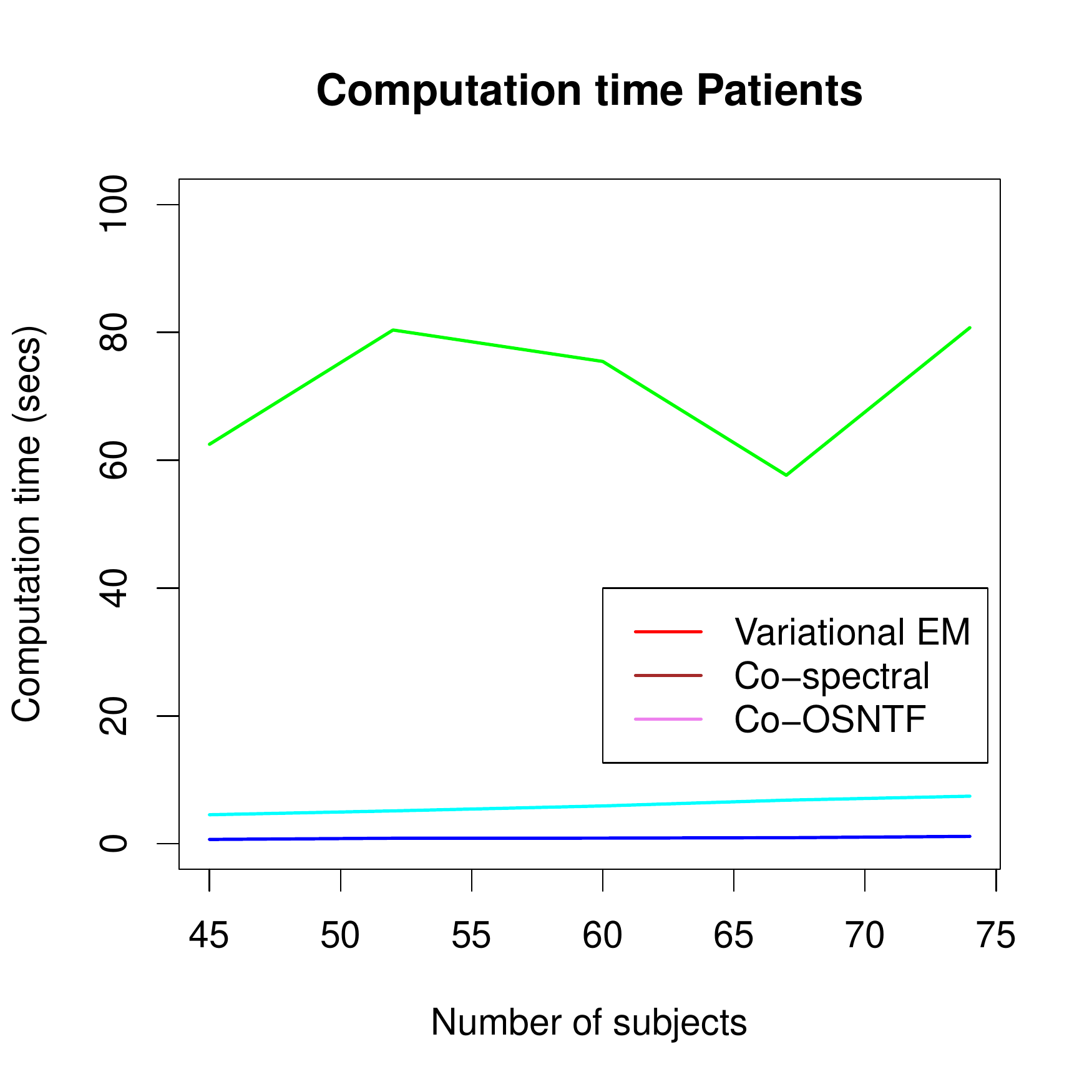}
\end{subfigure}%
\begin{subfigure}{0.32 \textwidth}
\includegraphics[width=\linewidth]{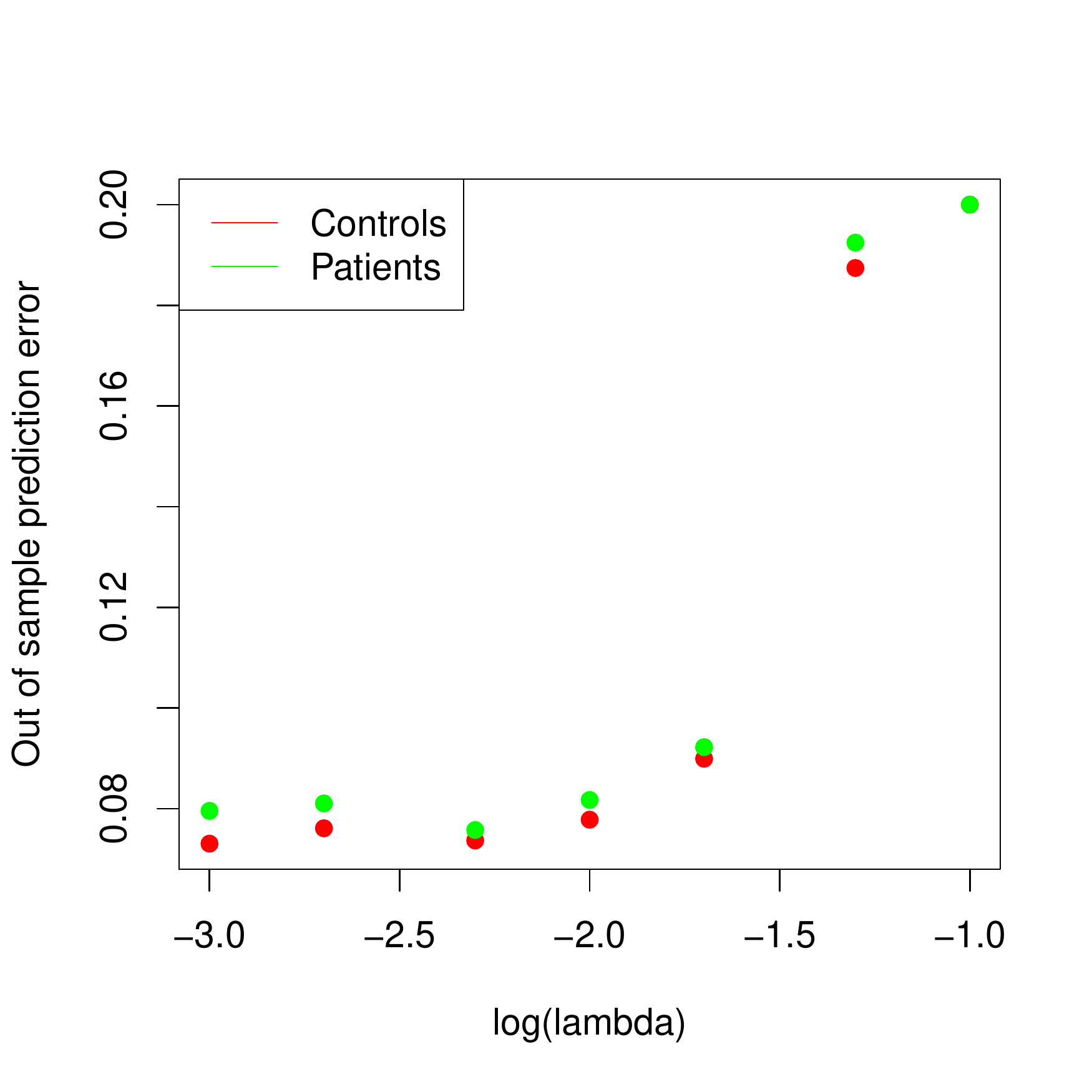}
\end{subfigure}
\begin{center}
\vspace{-5pt}
(a) \hspace{100pt} (b) \hspace{100pt} (c) CV $\lambda$
\end{center}
\vspace{-5pt}
\caption{(a) and (b) Computation time in controls and patients, (c) Cross validation for the choice of $\lambda$.}
\label{comptime}
\end{figure}

\subsection{Computation time and choice of Co-OSNTF tuning parameter $\lambda$ }
Next we comment on the computing time each of our methods take on this data and the choice of the tuning parameter $\lambda$ in the Co-OSNTF method. In Figures \ref{comptime} (a) and (b) we compare the computation time of VarEM, Co-OSNTF and Co-Spectral with increasing number of subjects from each of the control and patient groups. These computations have been carried out using a 4.2 GHz Intel Core i7, 8 core, 32 GB memory system (however, no parallel computing is involved). Generally, the computing time increases as the number of subjects increases in both cases. We also note that both Co-Spectral and Co-OSNTF are faster compare to the VarEM method, with both of them computing their solutions for the whole dataset under 10 seconds, while VarEM requires around 60-80 seconds depending upon the number of subjects. While all the methods for estimation are reasonably fast, the hypothesis testing, however, is computationally intensive because of the requirement of running the methods on 10000 resamples for an accurate estimate of the p-values (this step however is executed in parallel).

While the Co-OSNTF tuning parameter is an user given parameter, it is possible to choose the parameter using cross-validation by treating the prediction accuracy as a loss function. We consider a wide range of lambda values between $[0.001,0.1]$, fit Co-OSNTF with the $\lambda$ values and compute the out of sample prediction error in each case. We display the cross validation prediction error for both controls and patients with log of $\lambda$ (base 10) in Figure \ref{comptime}(c) . It appears that for $\lambda$ between $10^{-3}$ to $10^{-2}$, there is not much difference in the prediction accuracy, and as $\lambda$ increases the accuracy drastically decreases. We also note that the prediction error is least at a point somewhere between $-2.5 $ and $-2$. Therefore, our choice of $0.01$ for $\lambda$ is very close to the optimum.

\begin{figure}[h]
\centering{}
\begin{subfigure}{0.26\textwidth}
\includegraphics[width=1.1\linewidth]{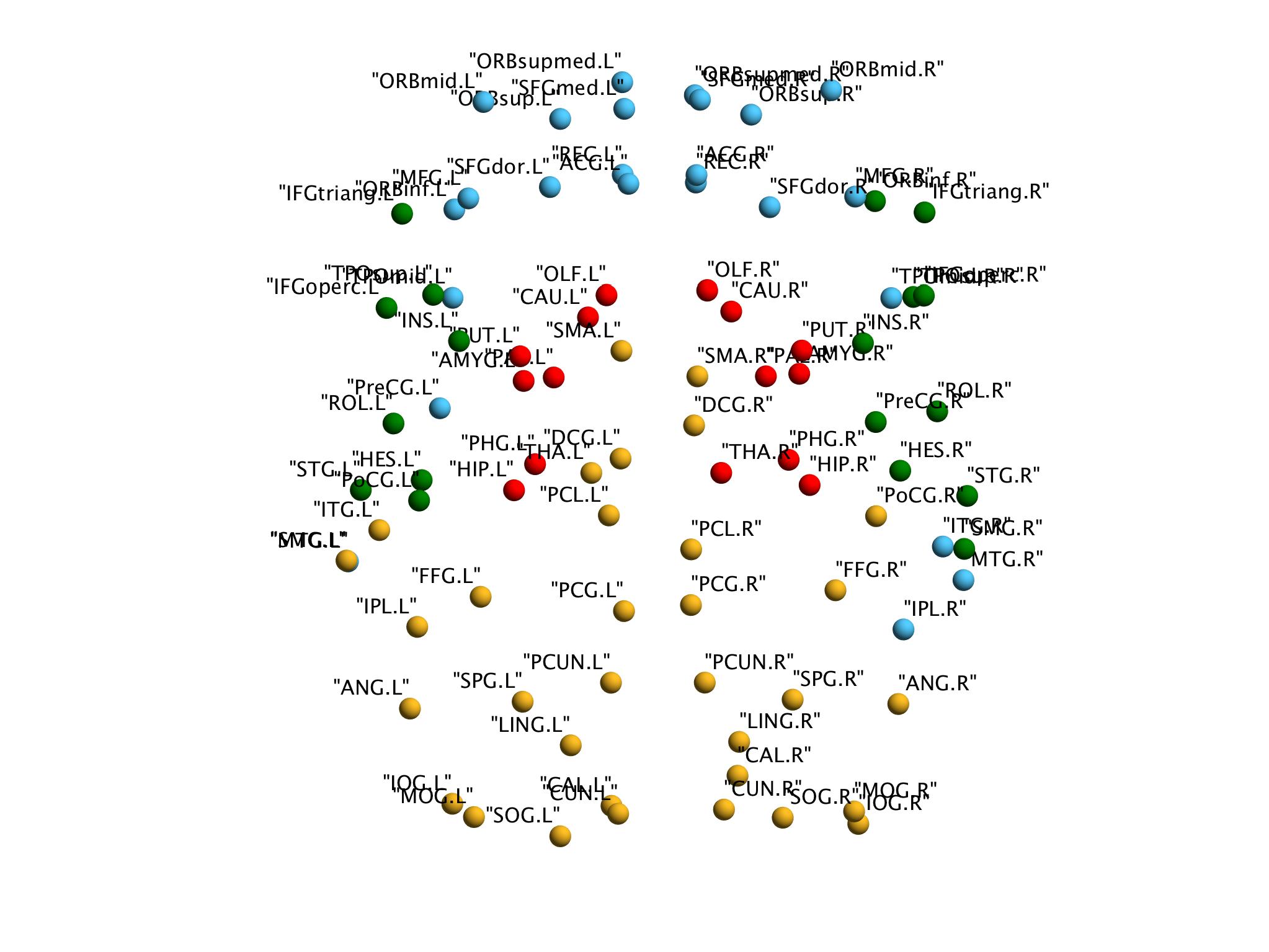}
\end{subfigure}%
\hspace{-20pt}
\begin{subfigure}{0.26\textwidth}
\includegraphics[width=1.1\linewidth]{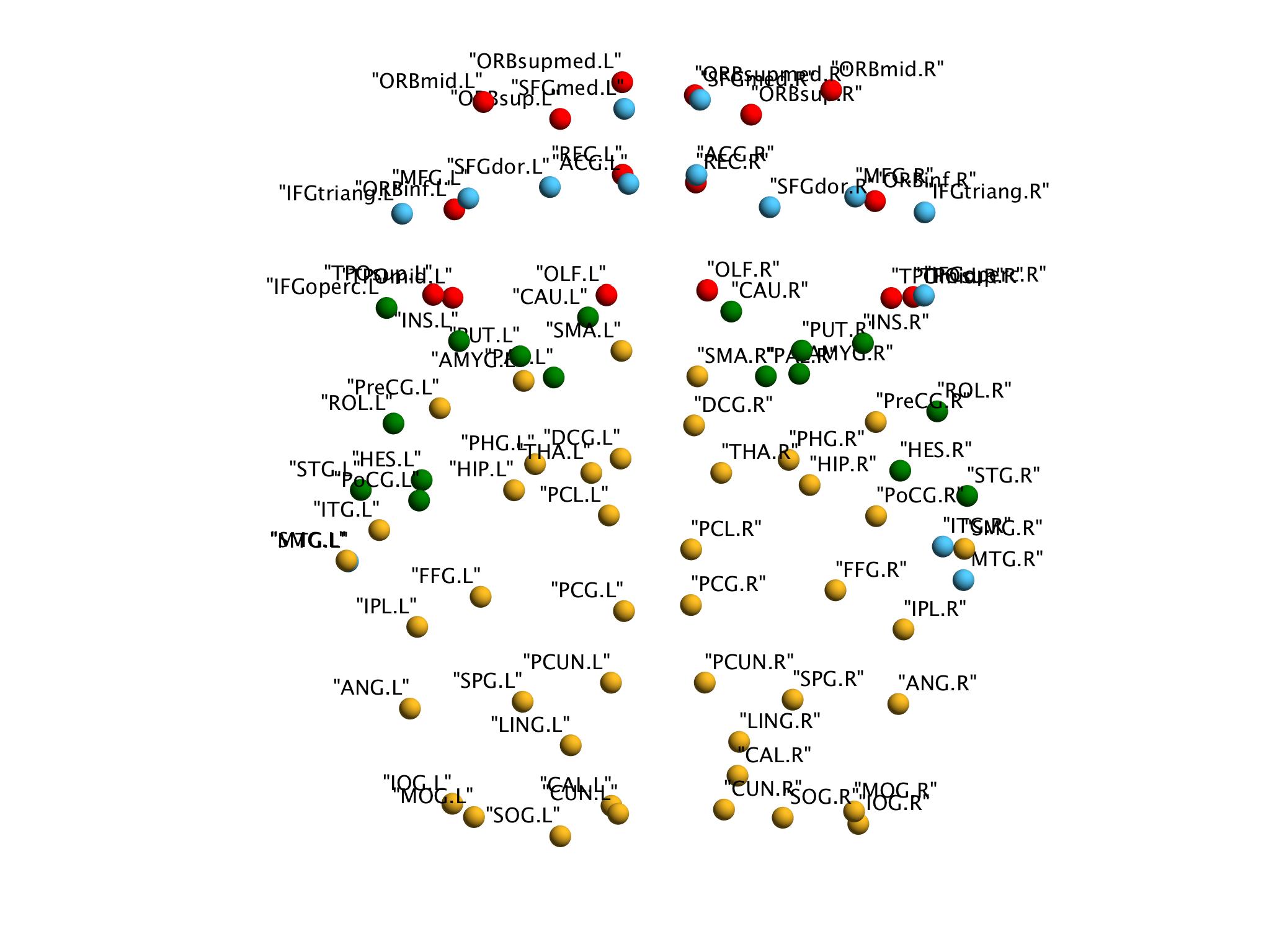}
\end{subfigure}%
\hspace{5pt}
\begin{subfigure}{0.26\textwidth}
\includegraphics[width=1.1\linewidth]{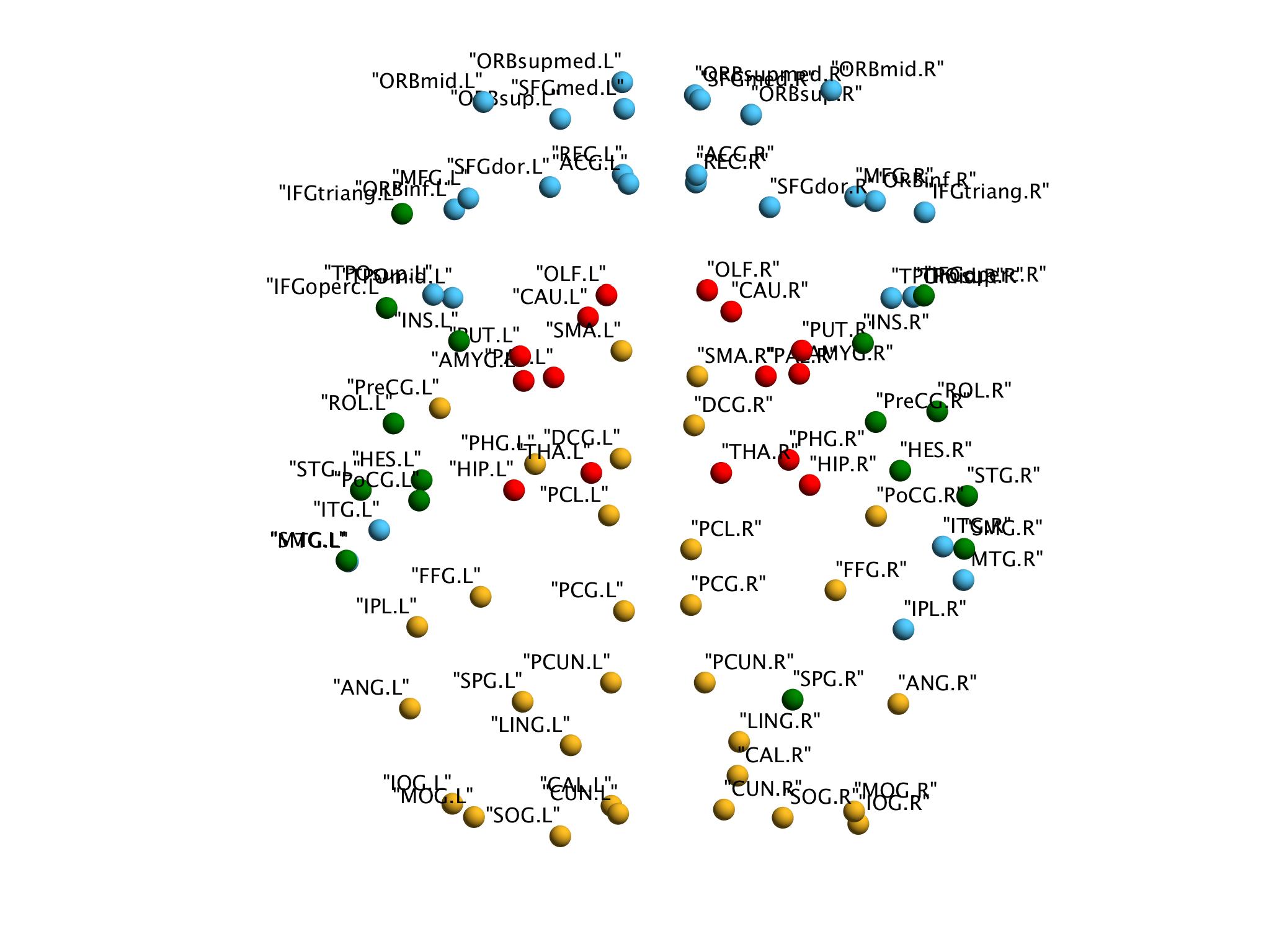}
\end{subfigure}%
\hspace{-20pt}
\begin{subfigure}{0.26\textwidth}
\includegraphics[width=1.1\linewidth]{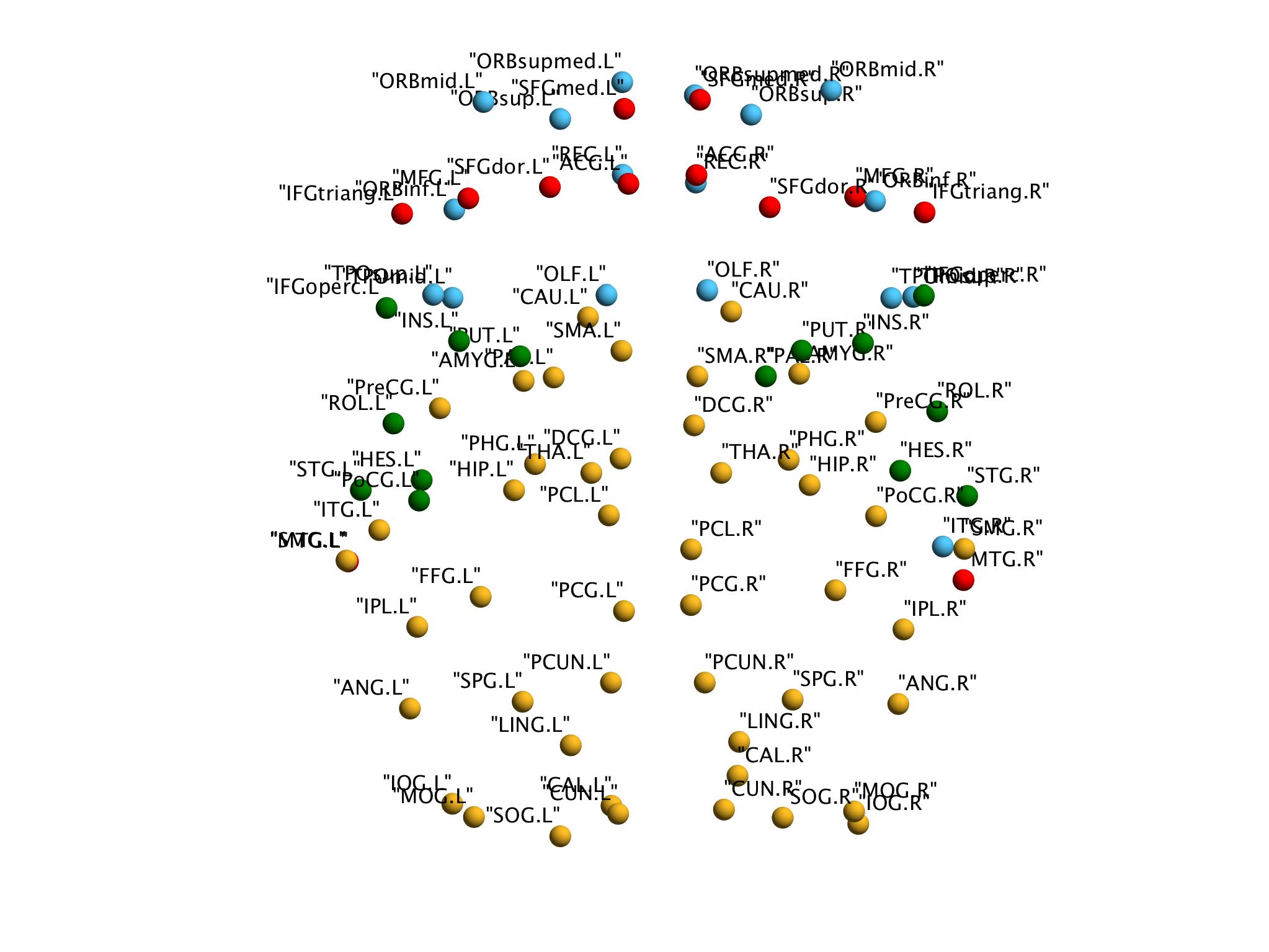}
\end{subfigure}
\begin{center}
\vspace{-10pt}
(a) Control \hspace{40pt} (b) Patient \hspace{70pt} (c) Control \hspace{40pt} (d) Patient
\end{center}
\begin{center}
(A) threshold 0.3 \hspace{100pt} (B) threshold 0.4
\end{center}
\vspace{-5pt}
\caption{Group putative community structure of resting state network based on AAL ROIs for thresholds of (A) 0.3 and (B) 0.4 in (a, c) healthy controls, and (b, d) patients with schizophrenia. Nodes are colored according to their group putative community obtained from Co-OSNTF fitted with 4 communities.}
\label{diffthresh1}
\end{figure}

\begin{figure}[!h]
\centering{}
\begin{subfigure}{0.2\textwidth}
\includegraphics[width=\linewidth]{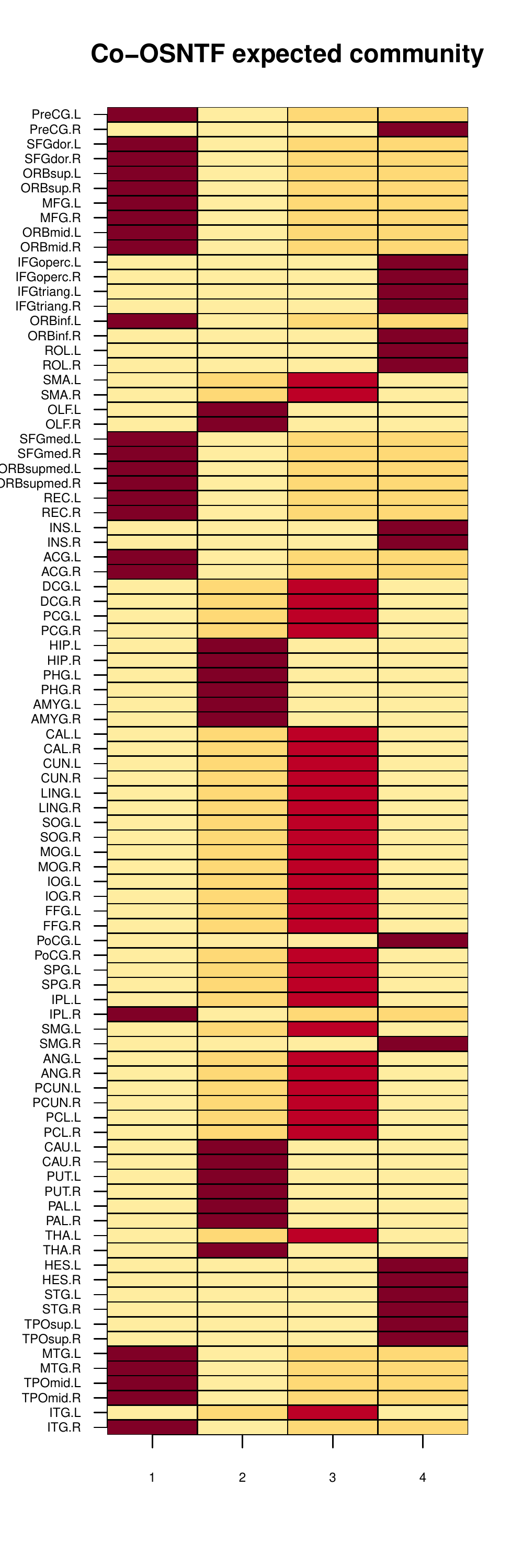}
\end{subfigure}%
\begin{subfigure}{0.2\textwidth}
\includegraphics[width=\linewidth]{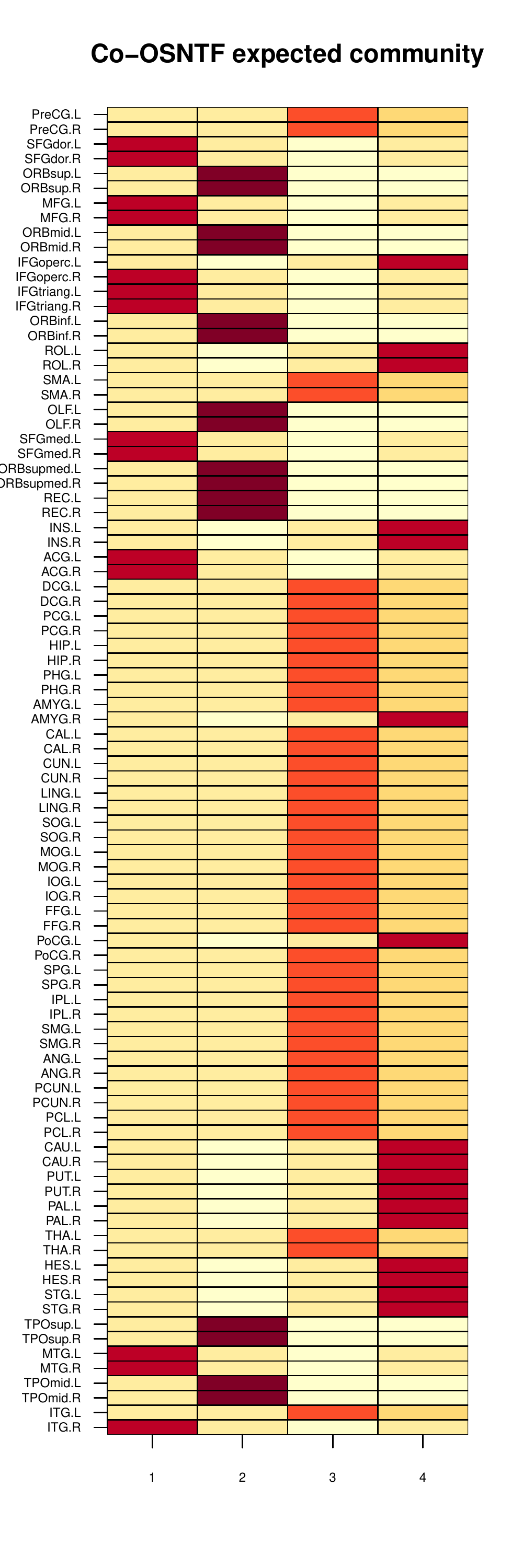}
\end{subfigure}%
\hspace{20pt}
\begin{subfigure}{0.2\textwidth}
\includegraphics[width=\linewidth]{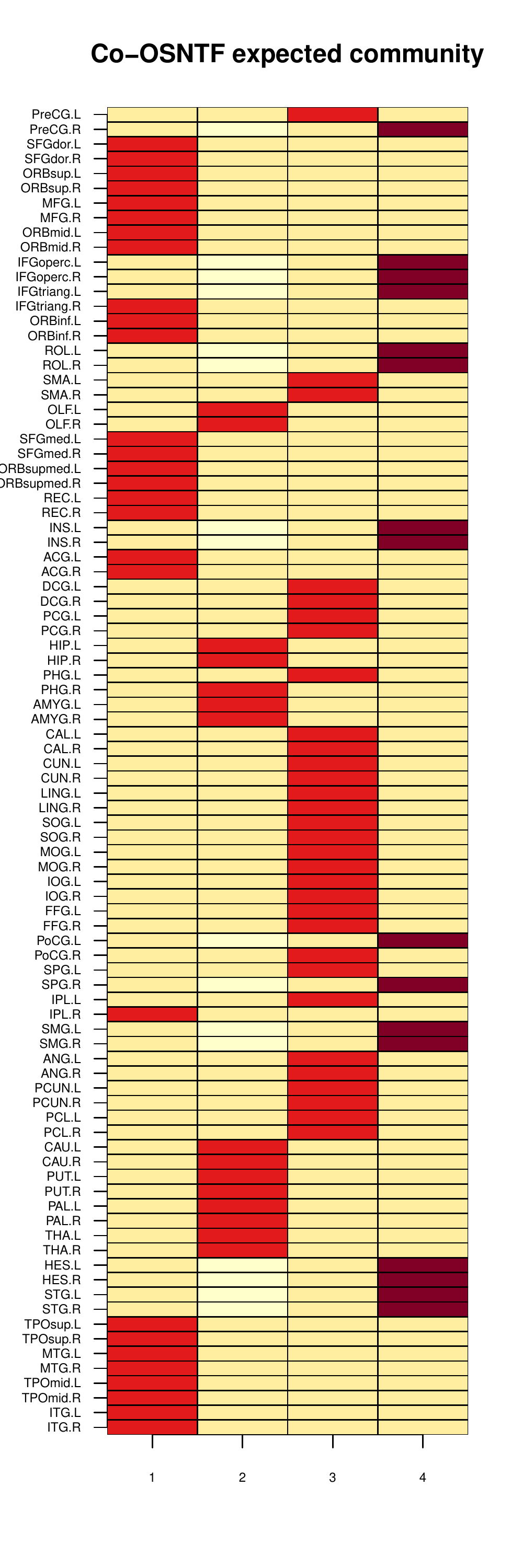}
\end{subfigure}%
\begin{subfigure}{0.2\textwidth}
\includegraphics[width=\linewidth]{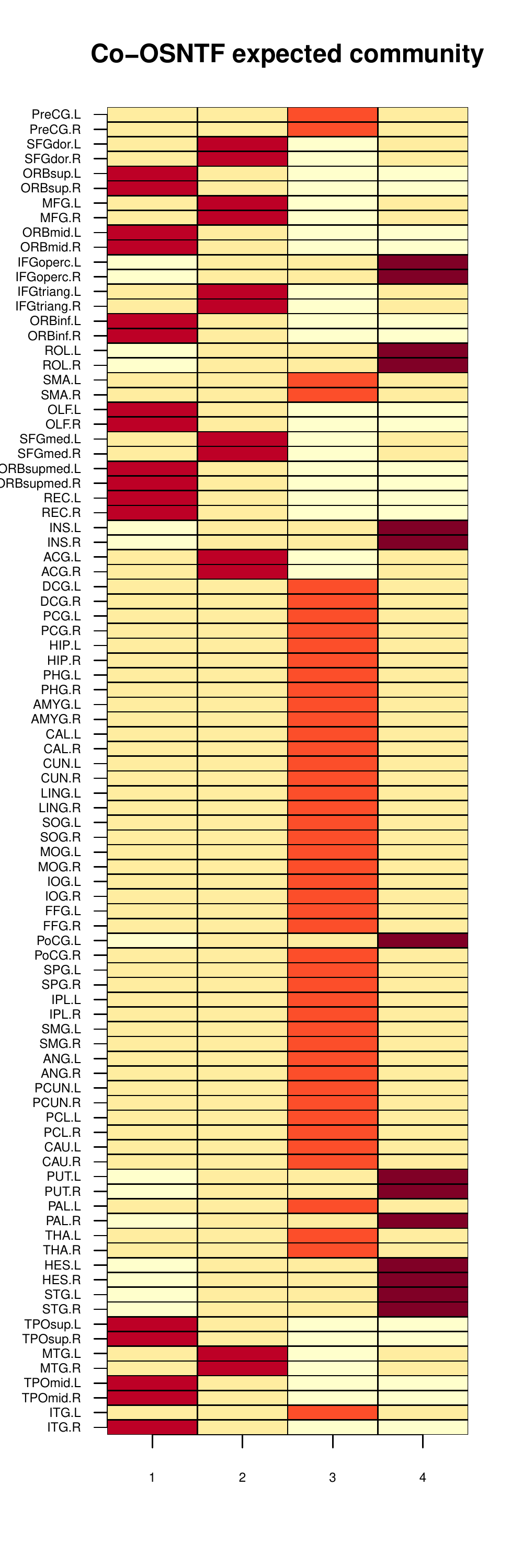}
\end{subfigure}
\begin{center}
\vspace{-20pt}
(a) Controls \hspace{50pt} (b)  Patients \hspace{50pt} (a) Controls \hspace{50pt} (b)  Patients
\end{center}
\begin{center}
(A) threshold 0.3 \hspace{100pt} (B) threshold 0.4
\end{center}
\vspace{-5pt}
\caption{Community estimate for AAL ROIs in controls and patients using the Co-OSNTF method with 4 communities for thresholds (A) 0.3 and (B) 0.4.}
\label{diffthresh2}
\end{figure}

\subsection{Choice of a different threshold}

We close this section by repeating some of the analyses with two different thresholds for the correlation matrices, 0.3 and 0.4. We fit the model using the Co-OSNTF method with $k=4$ communities in each case. The community colors and labels are kept consistent with previous analysis with threshold 0.2, for ease of comparison. The group putative community structure and the estimated community assignment of the ROIs are presented in Figures \ref{diffthresh1} and \ref{diffthresh2} respectively. We note that there is a very high similarity of the results with the ones we have obtained for the threshold of 0.2. We note as before, a disruption in the community structure with nodes from red module in controls being distributed in various modules in patients, the blue module in controls getting divided into two modules in patients, and the yellow module remaining relatively unchanged between controls and patients.

\section{ Discussions and Limitations}
We have proposed a random effects model to jointly model the community structure in a sample from a population of networks. The proposed estimation and hypothesis testing methodology outperforms baseline and competing methods in simulated network samples. Our methods uncover meaningful differences between functional brain networks of healthy controls and schizophrenia patients. We hope the principled approach developed here will be useful in modeling and contrasting network valued samples in terms of their low dimensional latent structures. We conclude by discussing a limitation of our methods and analysis. Another limitation of the methods is discussed in the supplementary materials.

\begin{figure}[!h]
\centering{}
\begin{subfigure}{0.16 \textwidth}
\includegraphics[width=\linewidth]{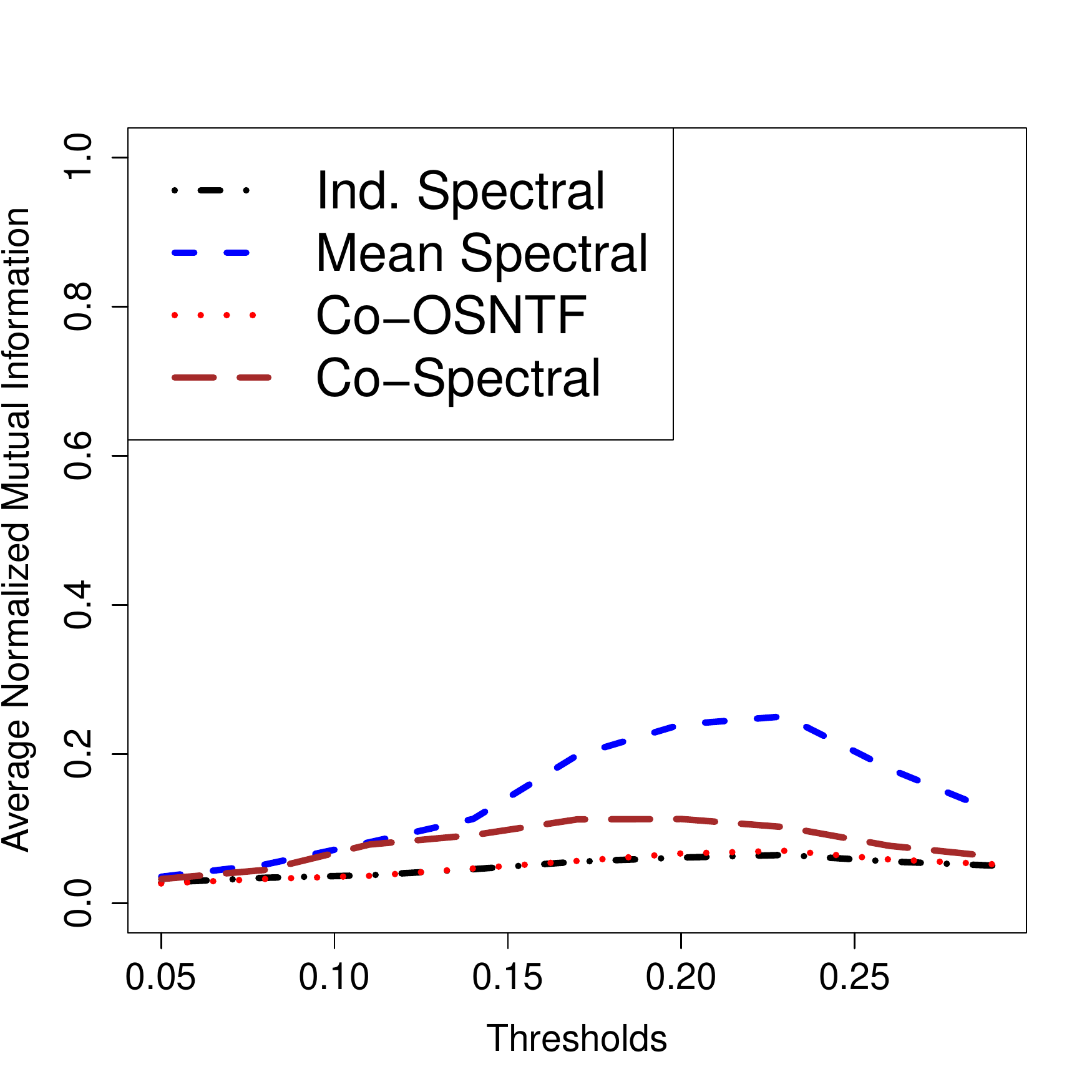}
\end{subfigure}%
\begin{subfigure}{0.16 \textwidth}
\includegraphics[width=\linewidth]{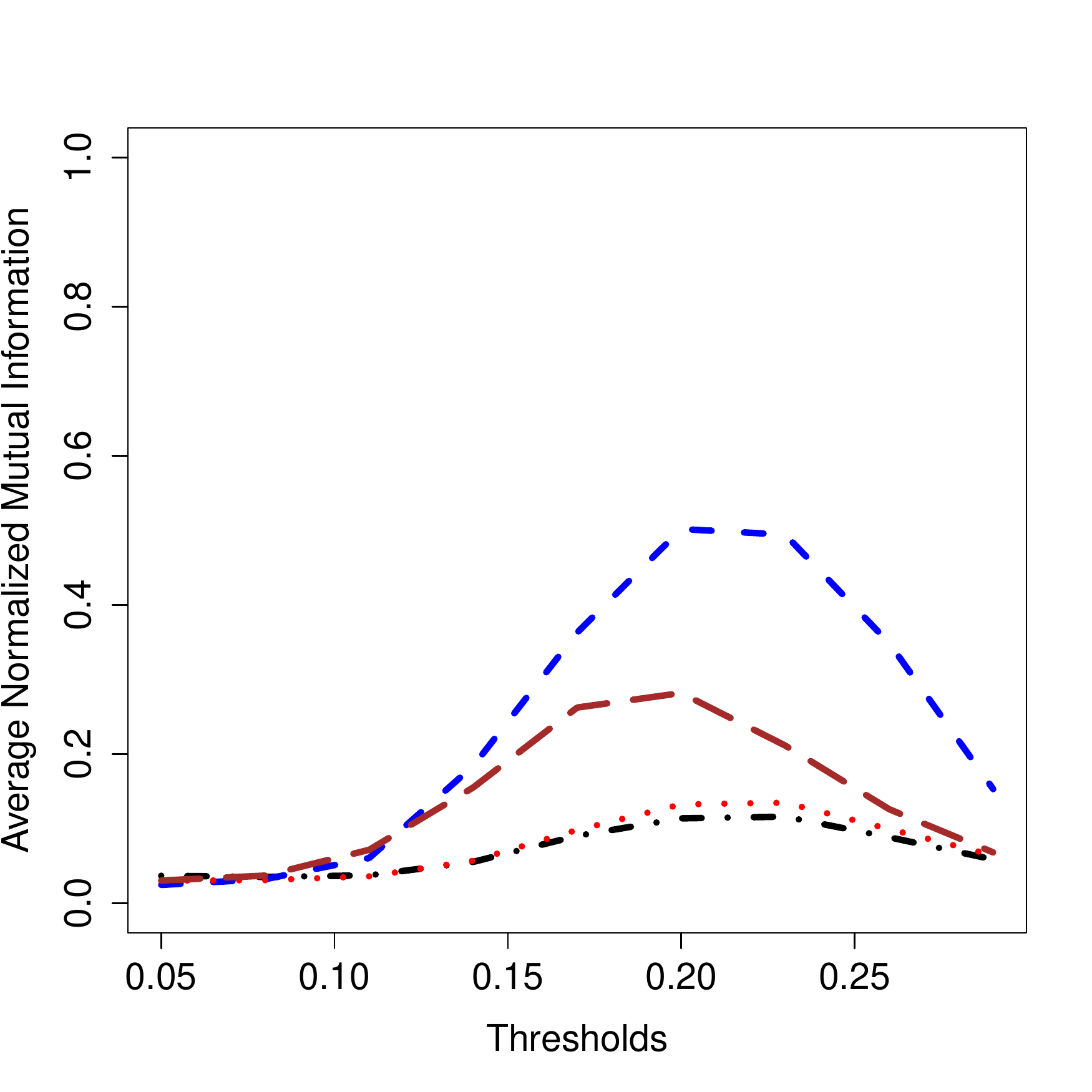}
\end{subfigure}%
\begin{subfigure}{0.16 \textwidth}
\includegraphics[width=\linewidth]{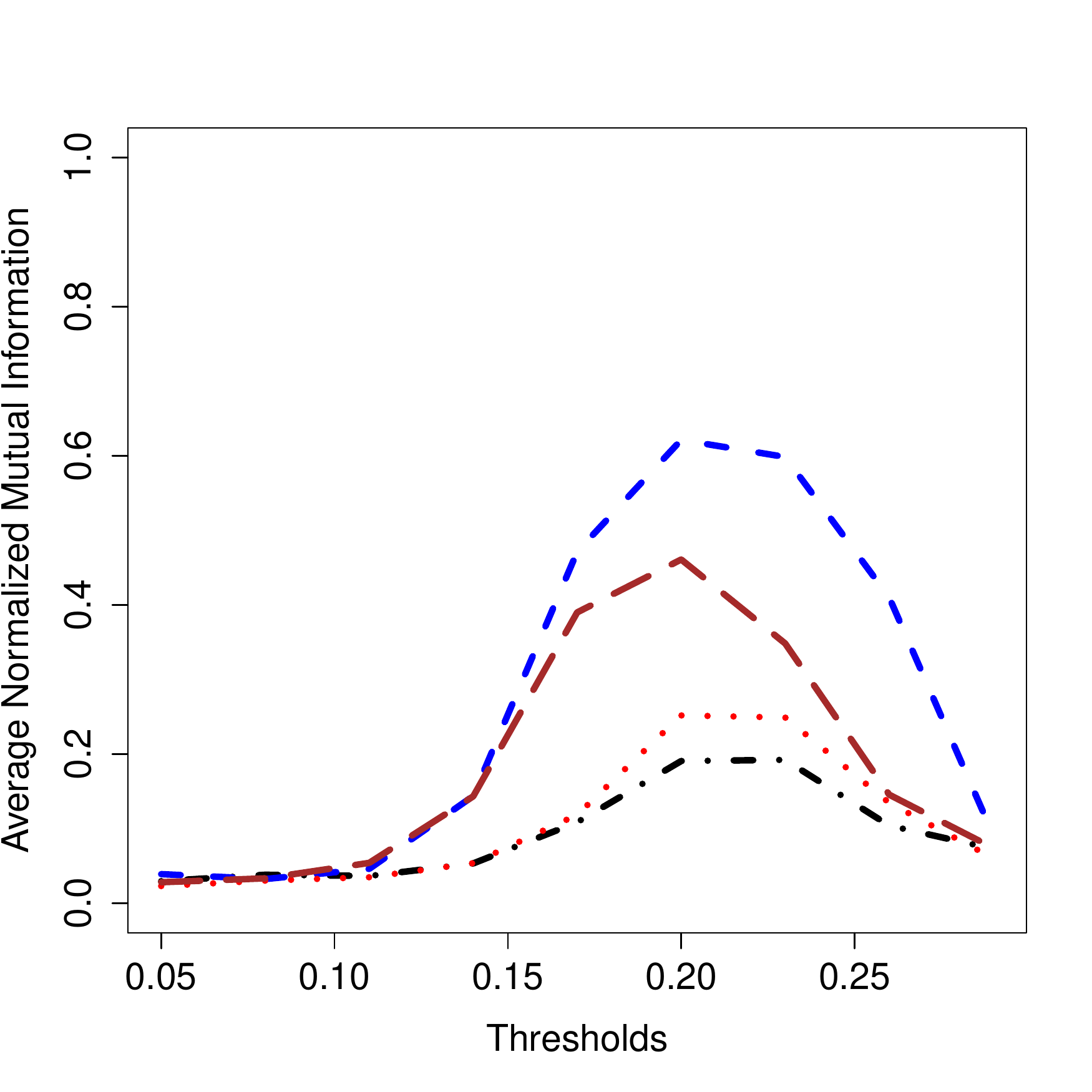}
\end{subfigure}%
\hspace{10pt}
\begin{subfigure}{0.16 \textwidth}
\includegraphics[width=\linewidth]{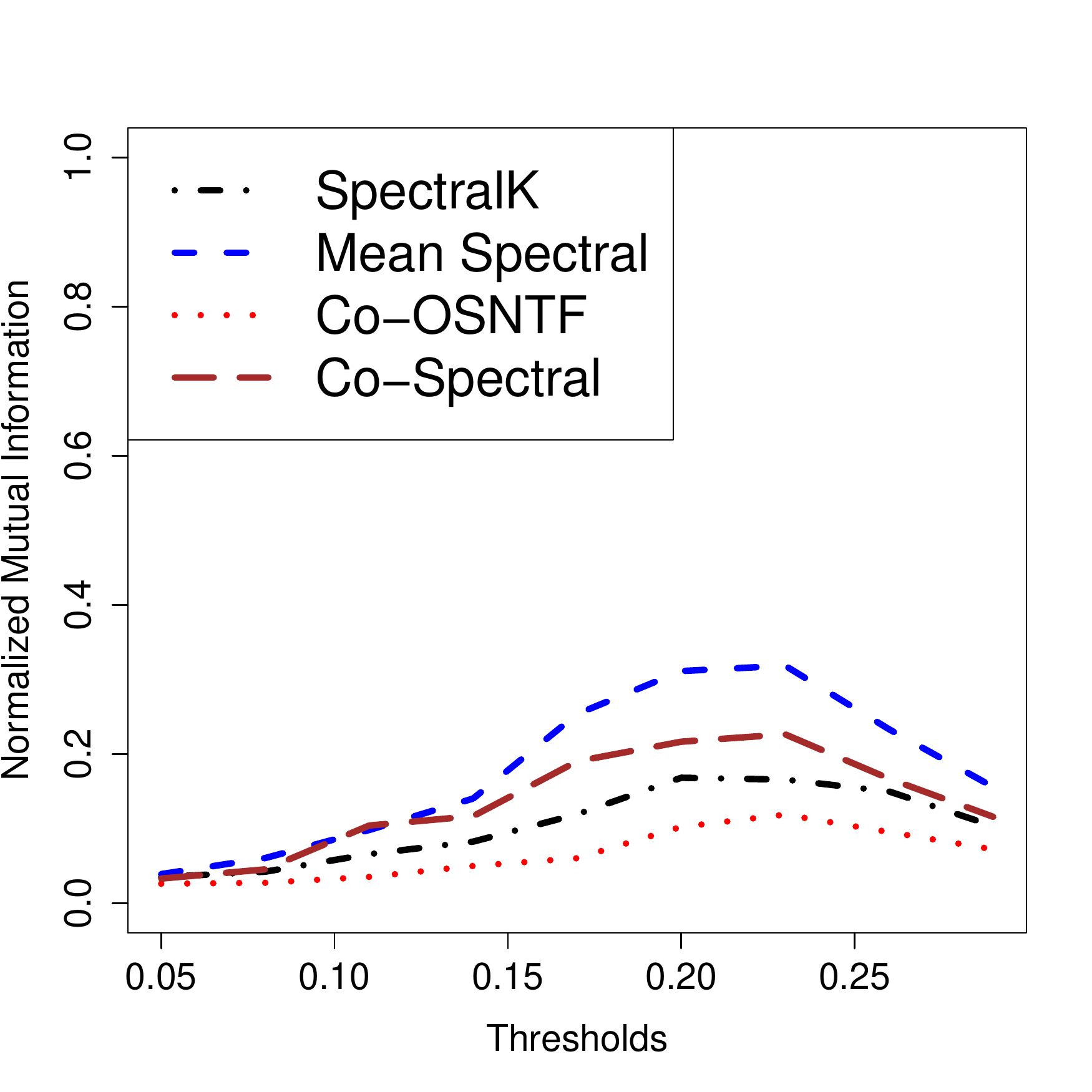}
\end{subfigure}%
\begin{subfigure}{0.16\textwidth}
\includegraphics[width=\linewidth]{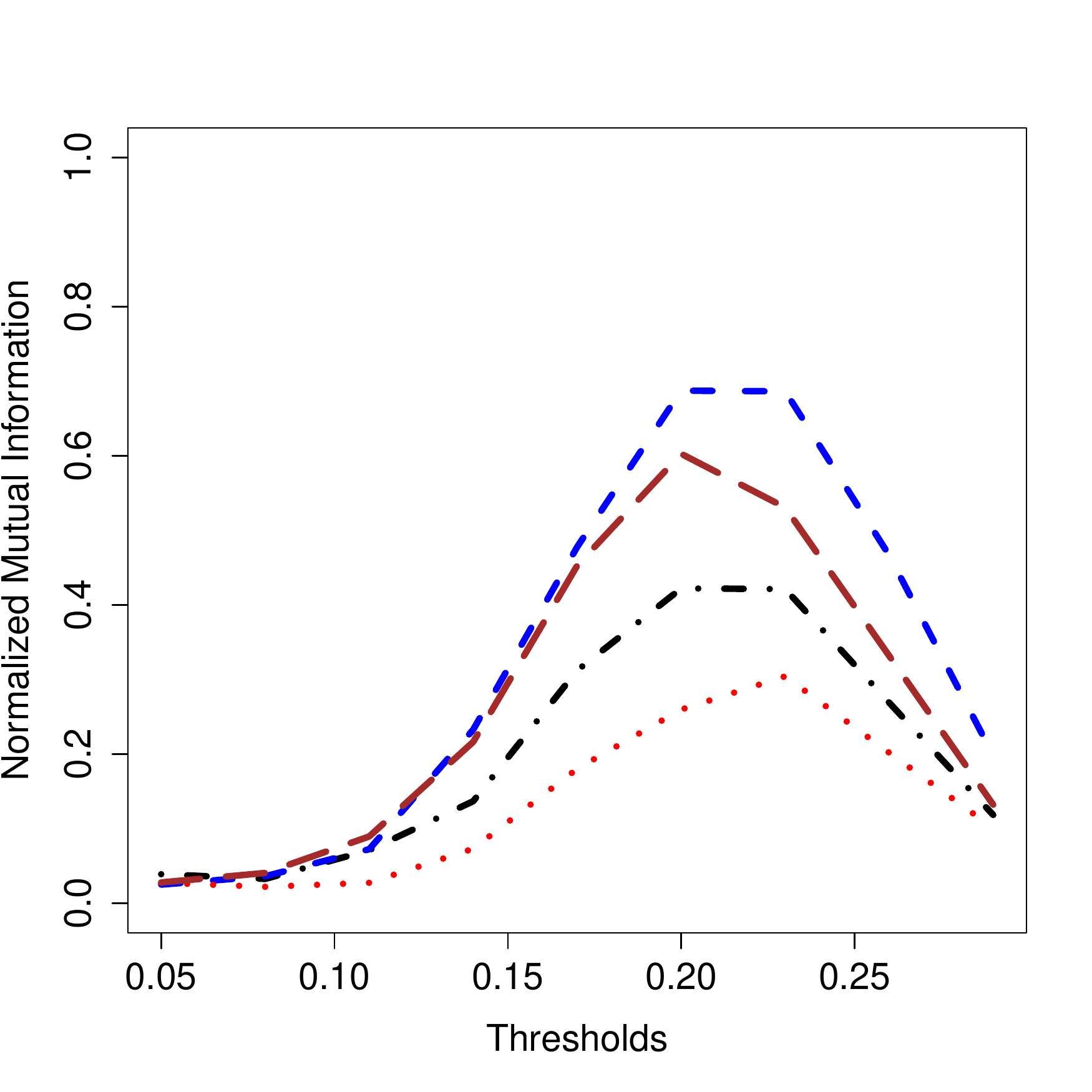}
\end{subfigure}%
\begin{subfigure}{0.16 \textwidth}
\includegraphics[width=\linewidth]{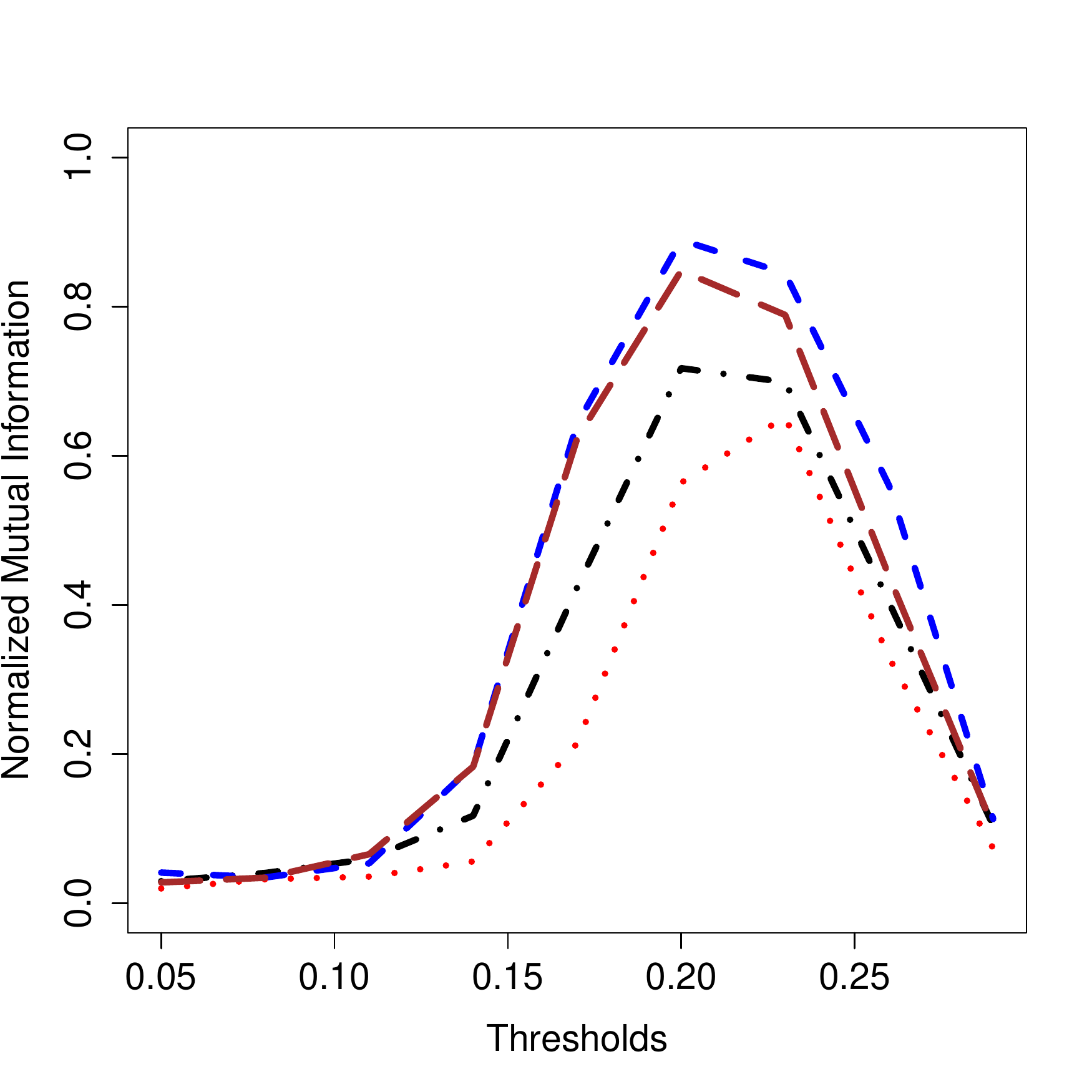}
\end{subfigure}
\begin{subfigure}{0.16 \textwidth}
\includegraphics[width=\linewidth]{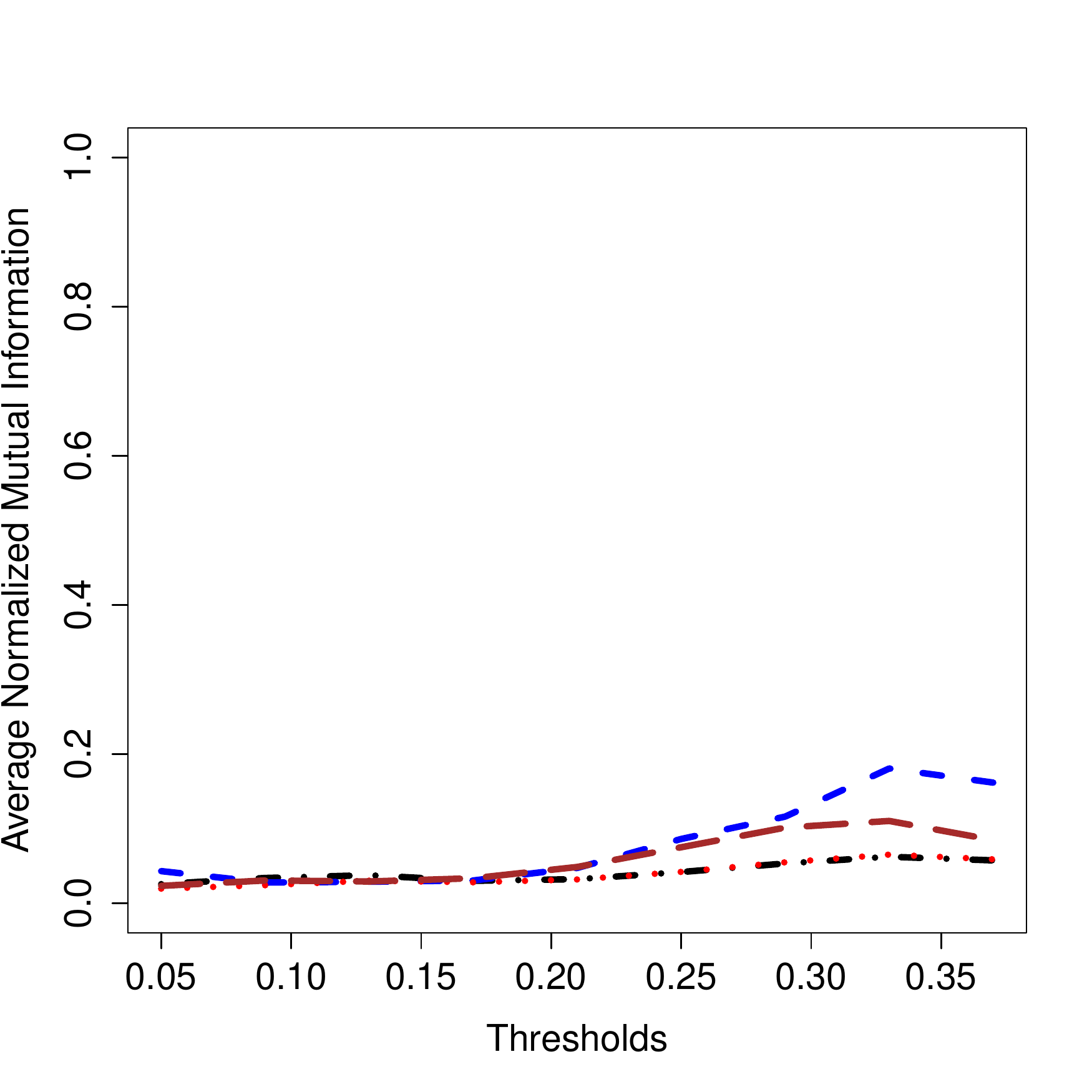}
\end{subfigure}%
\begin{subfigure}{0.16 \textwidth}
\includegraphics[width=\linewidth]{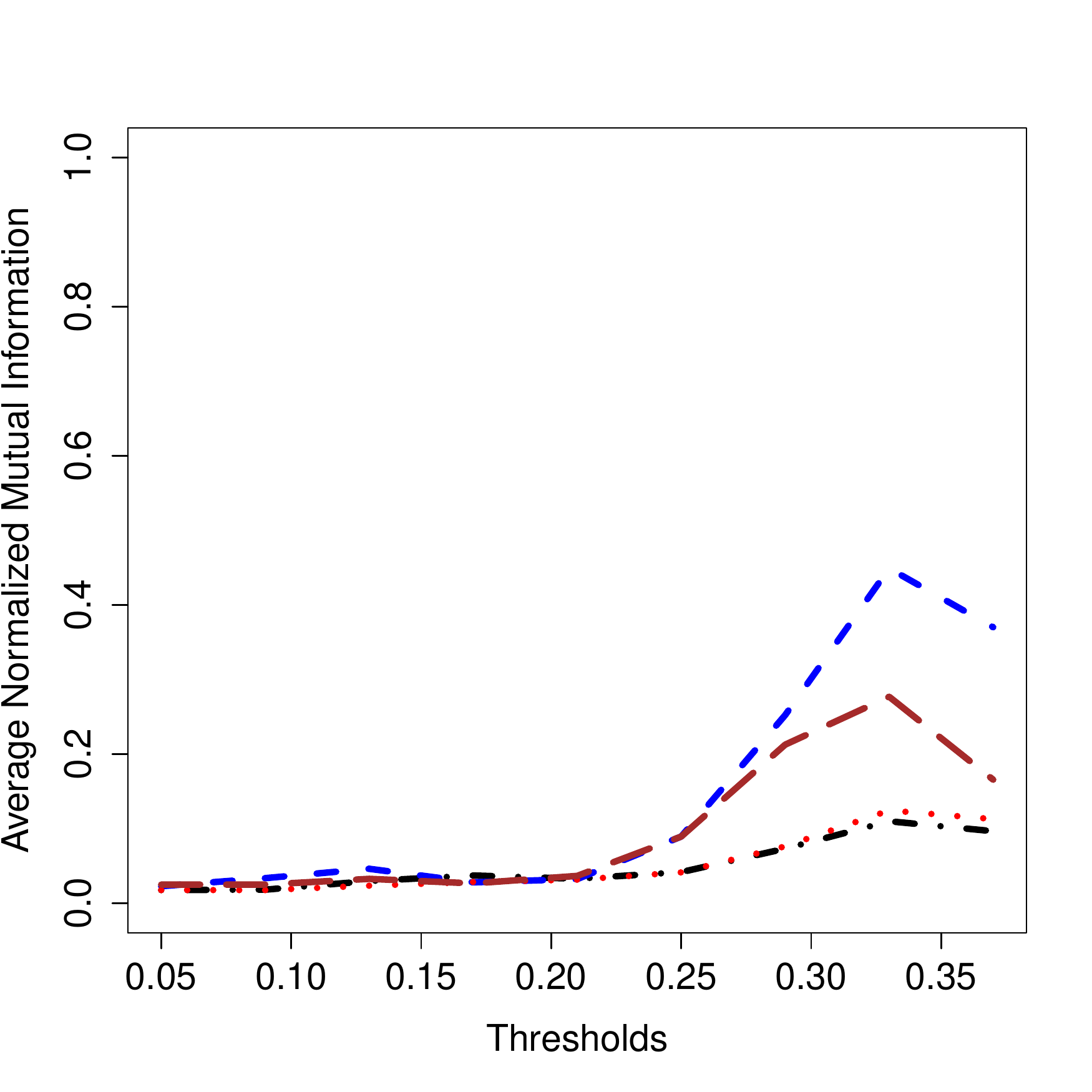}
\end{subfigure}%
\begin{subfigure}{0.16\textwidth}
\includegraphics[width=\linewidth]{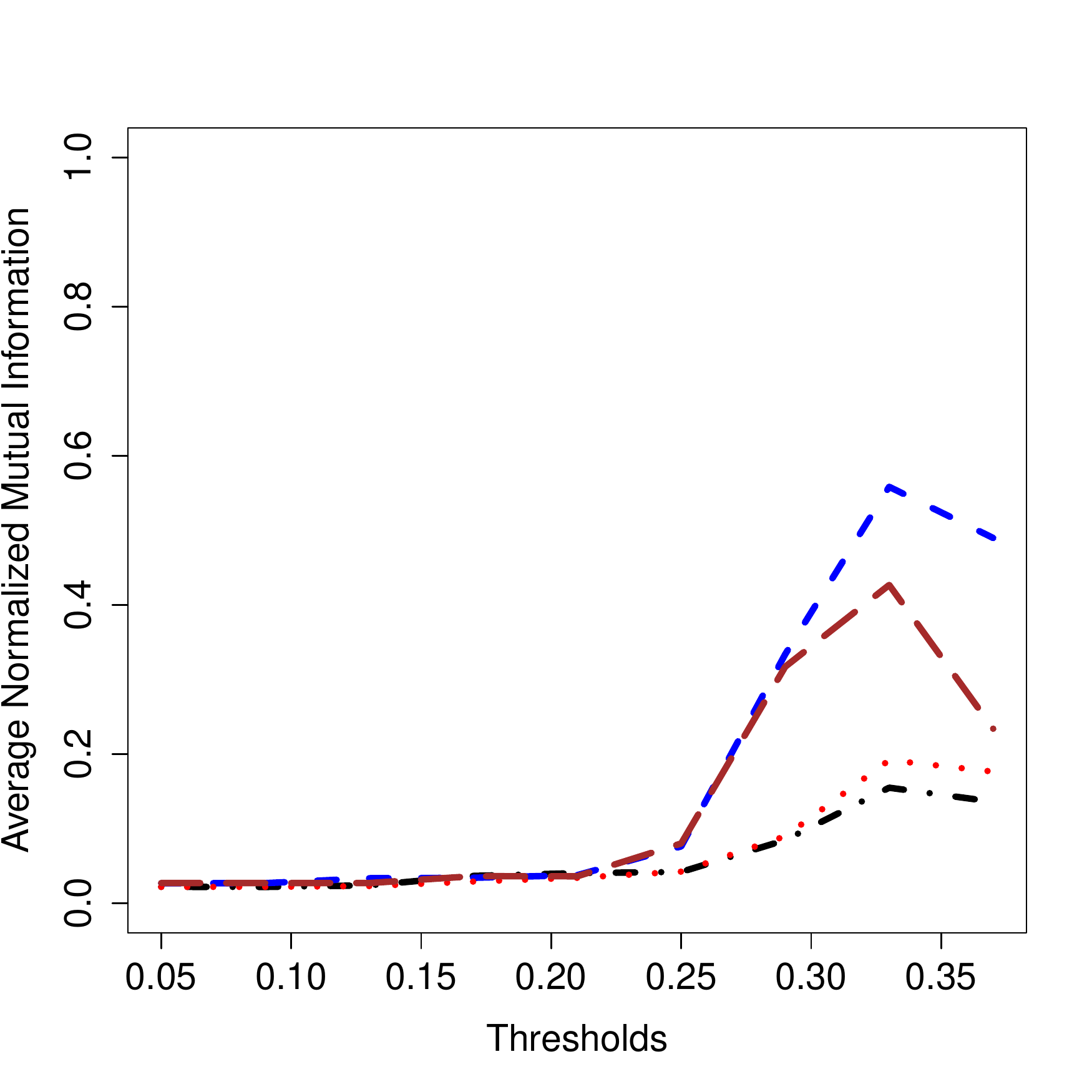}
\end{subfigure}%
\hspace{10pt}
\begin{subfigure}{0.16 \textwidth}
\includegraphics[width=\linewidth]{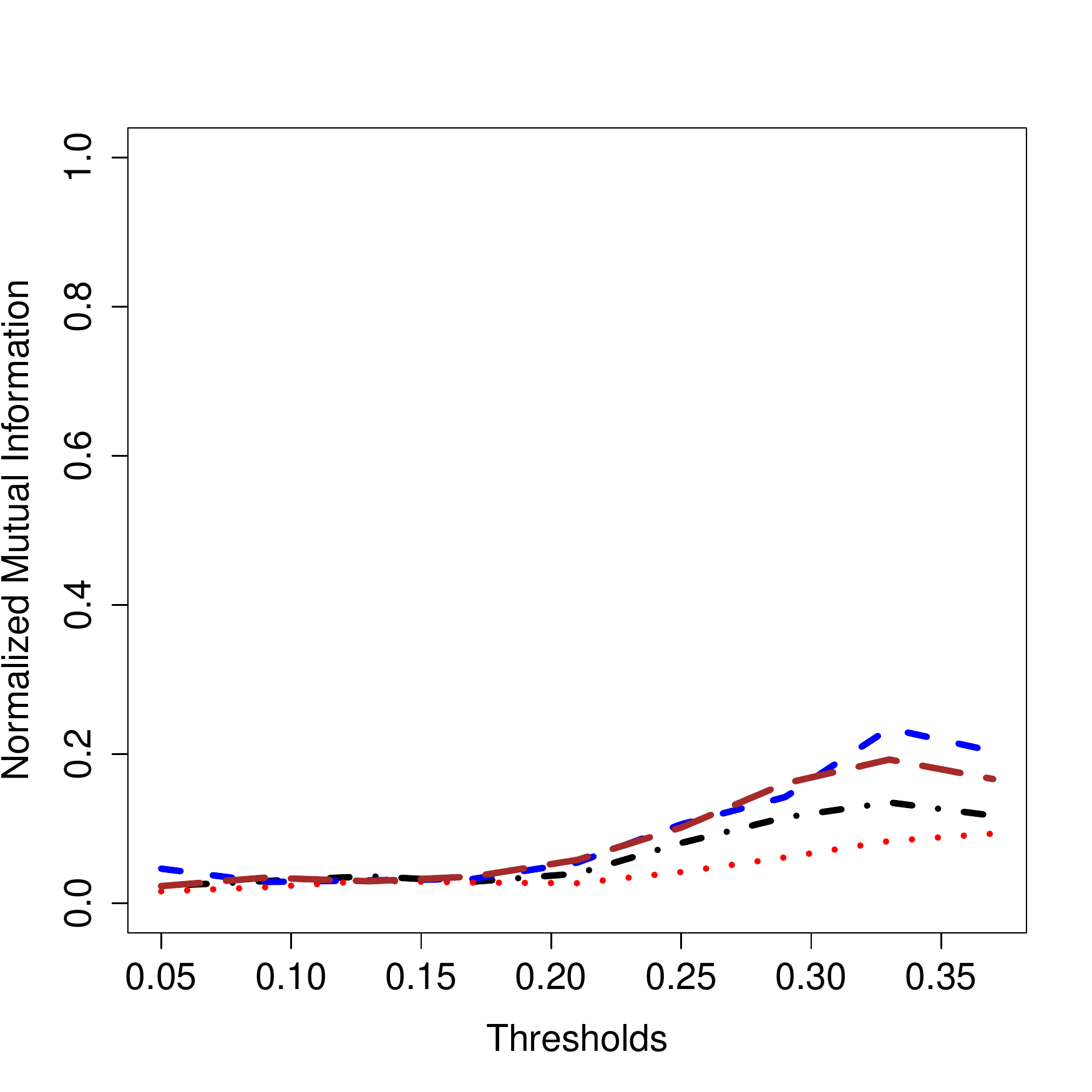}
\end{subfigure}%
\begin{subfigure}{0.16 \textwidth}
\includegraphics[width=\linewidth]{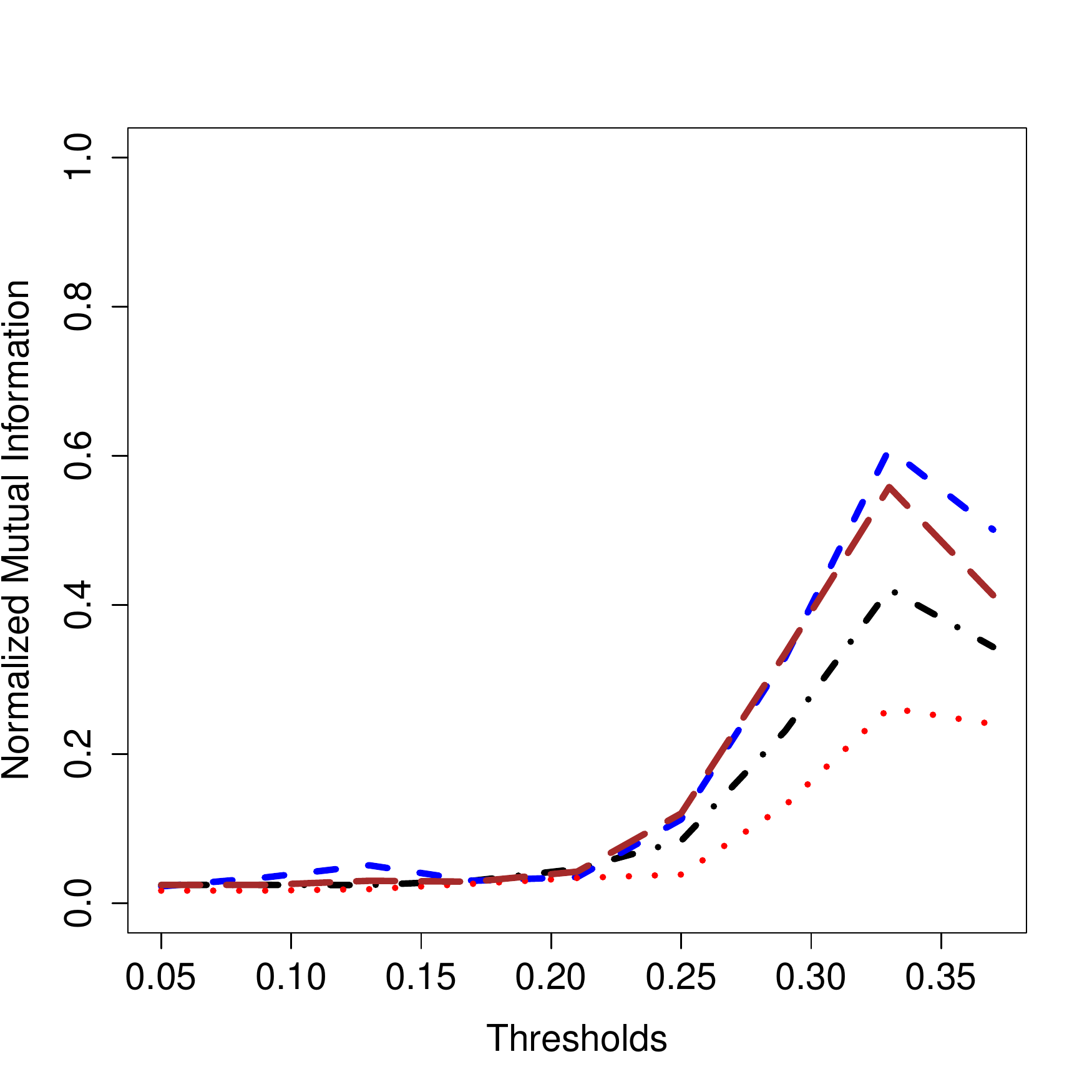}
\end{subfigure}%
\begin{subfigure}{0.16 \textwidth}
\includegraphics[width=\linewidth]{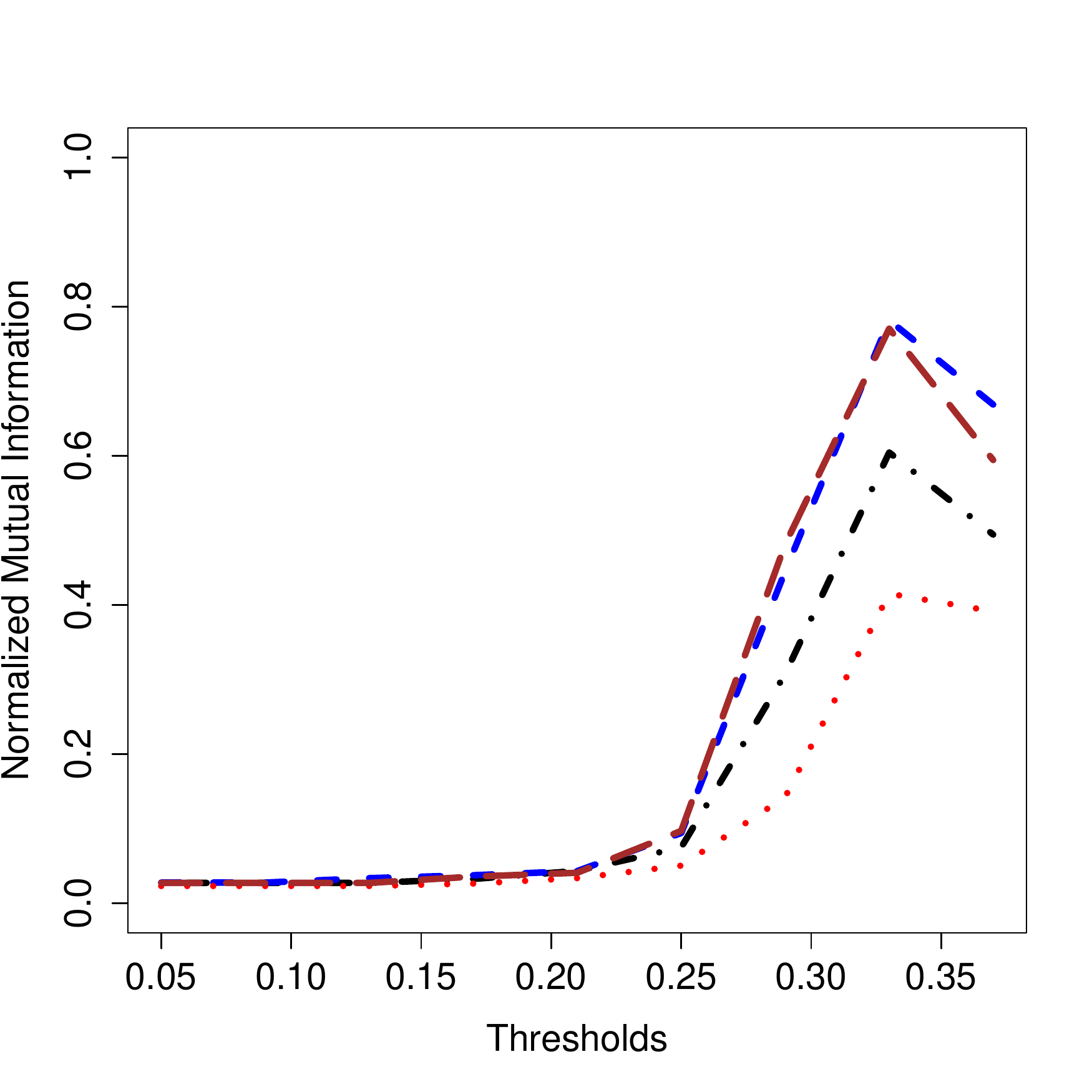}
\end{subfigure}
\begin{subfigure}{0.16 \textwidth}
\includegraphics[width=\linewidth]{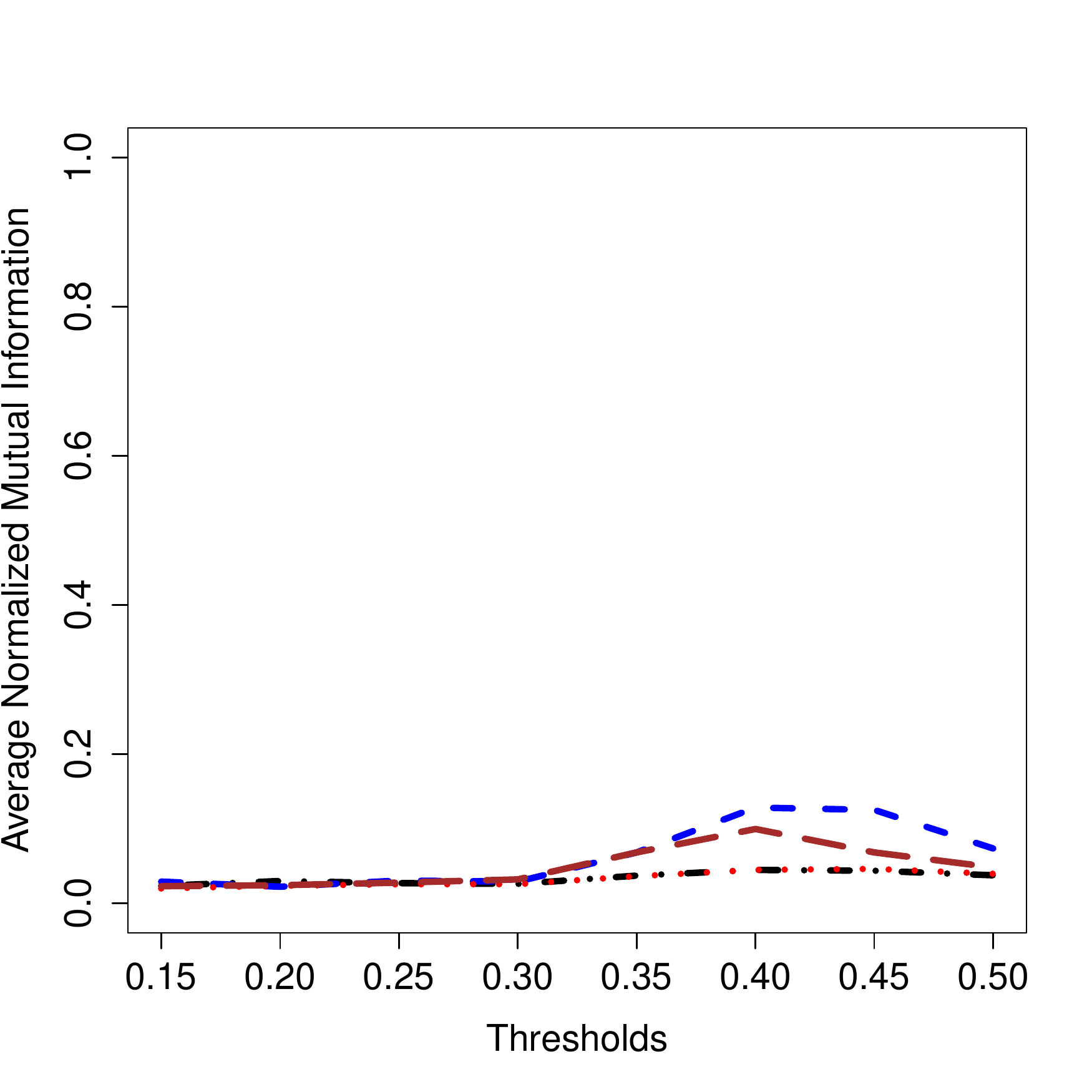}
\end{subfigure}%
\begin{subfigure}{0.16 \textwidth}
\includegraphics[width=\linewidth]{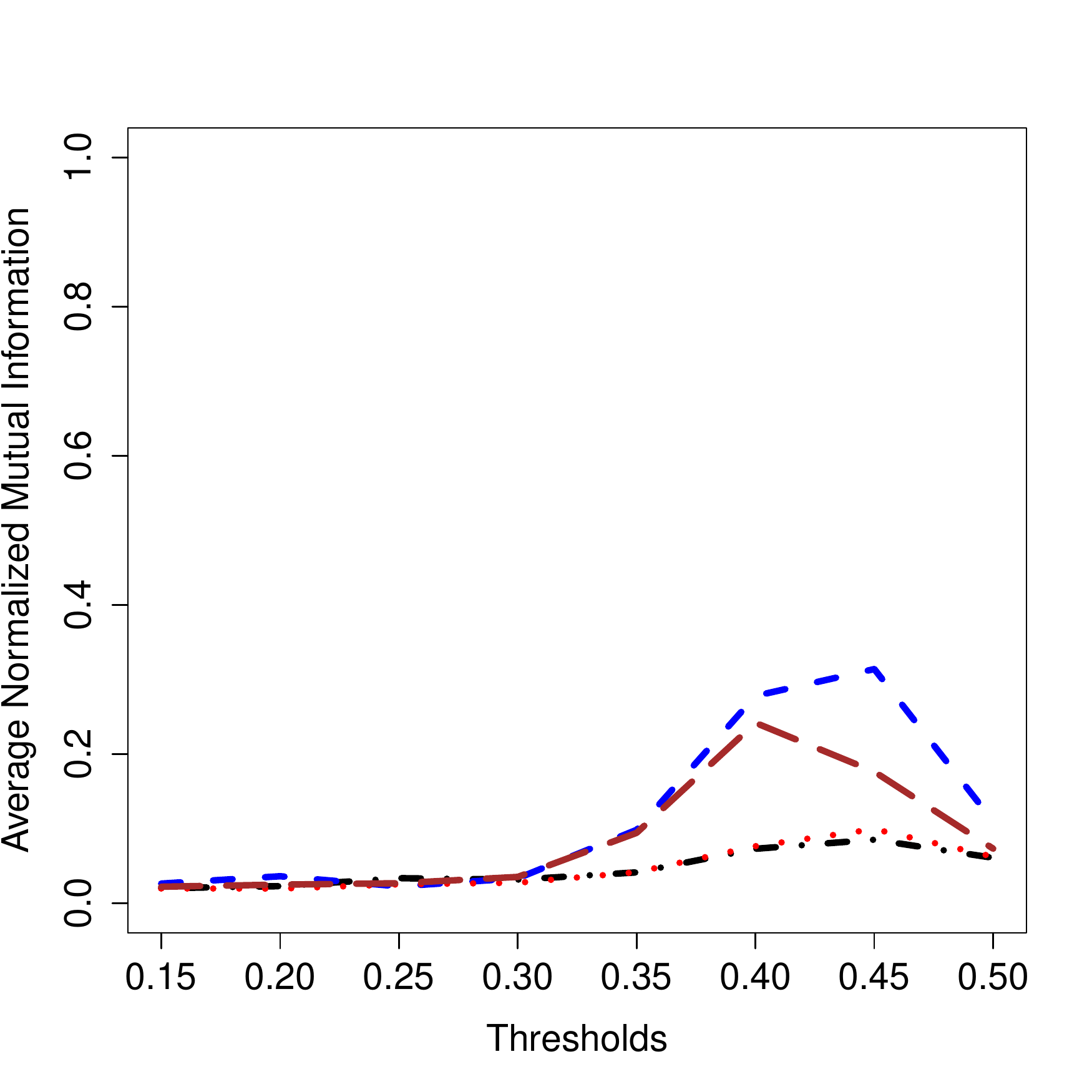}
\end{subfigure}%
\begin{subfigure}{0.16 \textwidth}
\includegraphics[width=\linewidth]{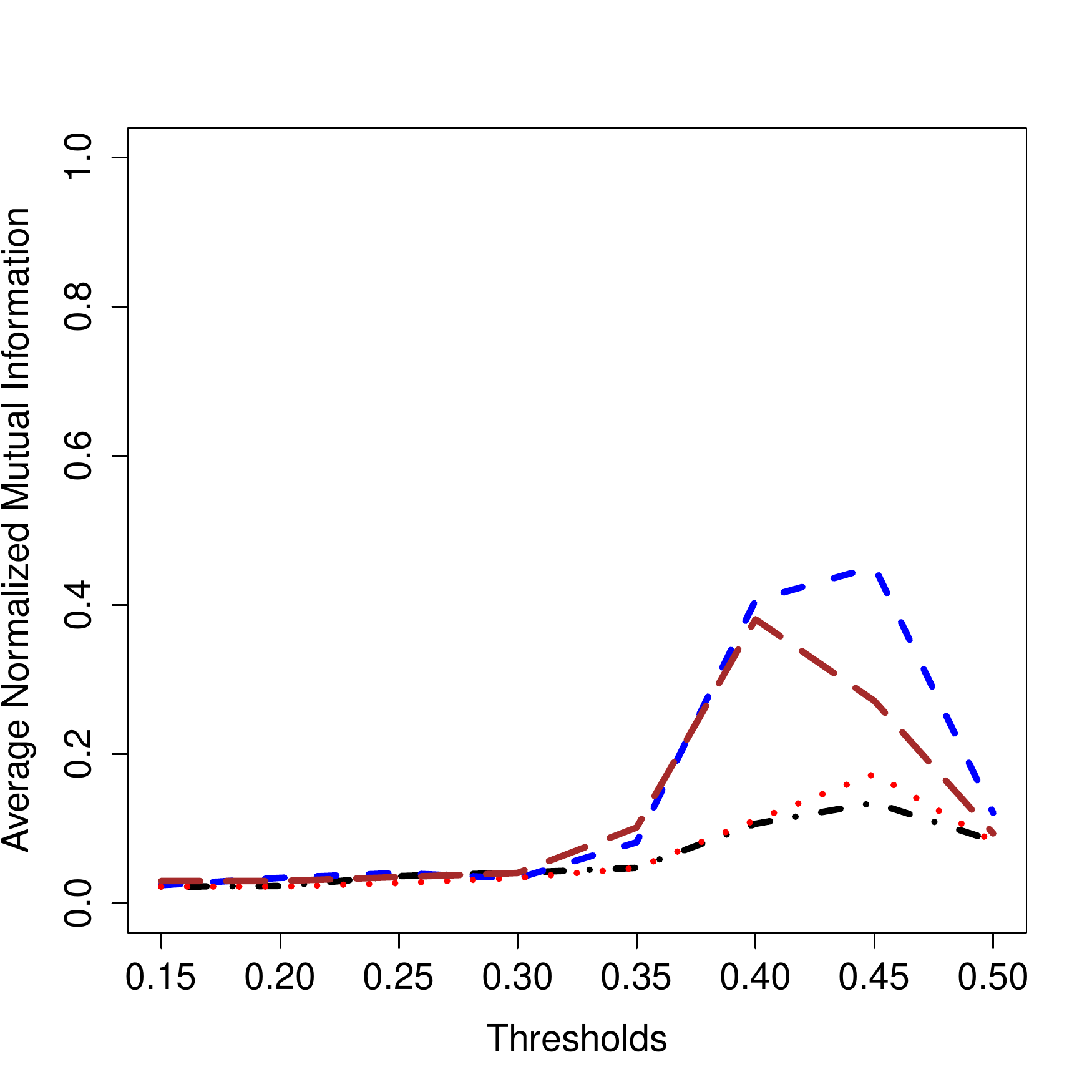}
\end{subfigure}%
\hspace{10pt}
\begin{subfigure}{0.16 \textwidth}
\includegraphics[width=\linewidth]{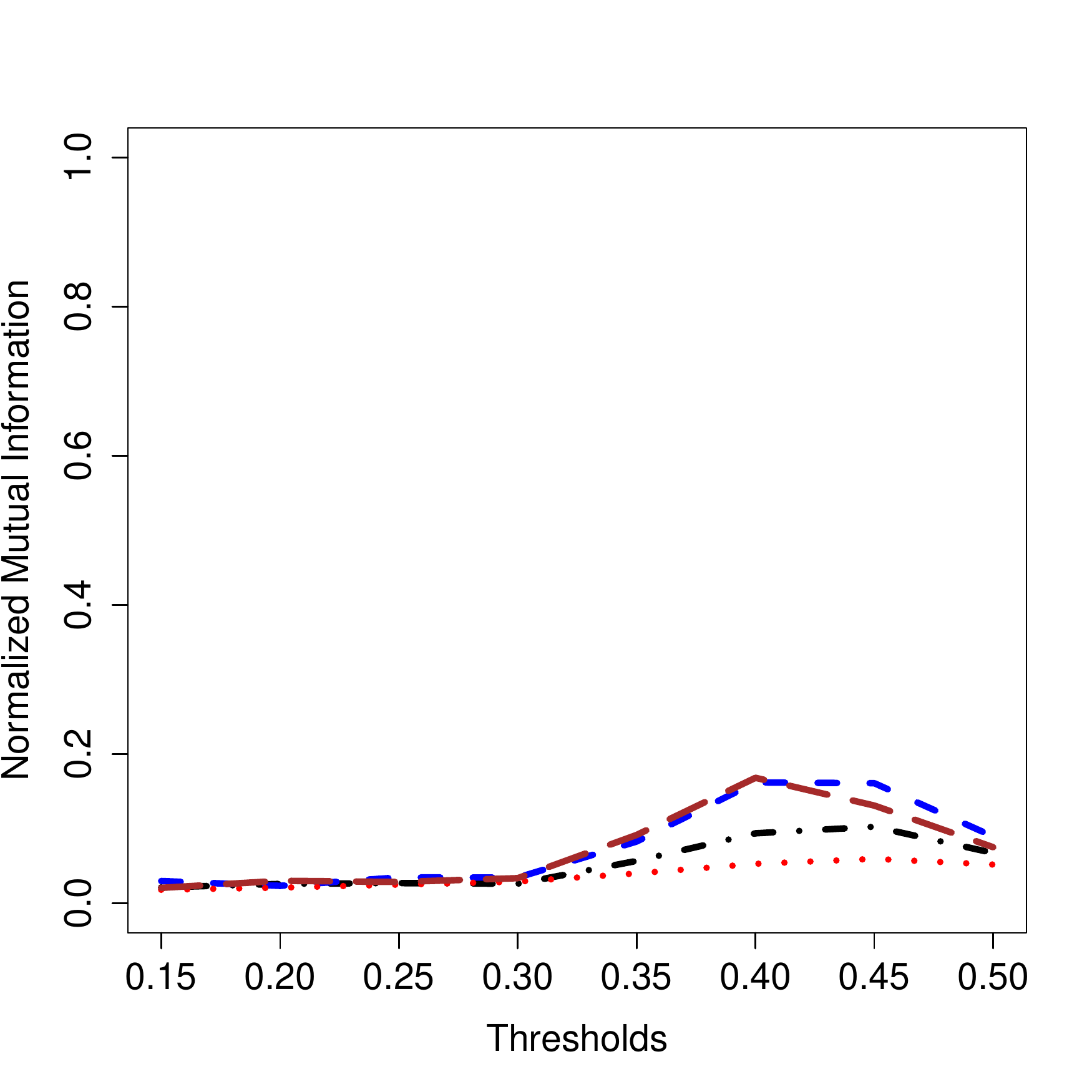}
\end{subfigure}%
\begin{subfigure}{0.16 \textwidth}
\includegraphics[width=\linewidth]{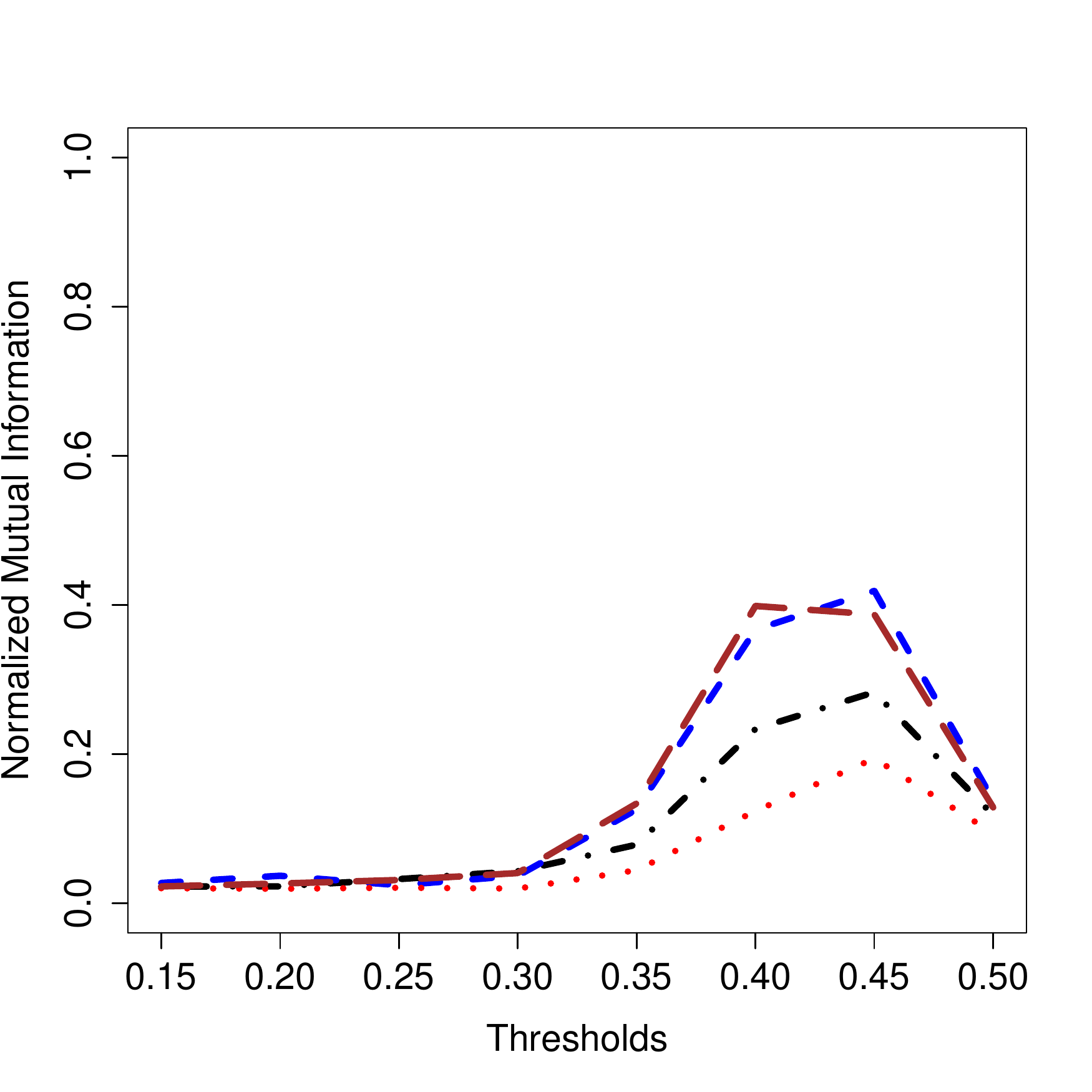}
\end{subfigure}%
\begin{subfigure}{0.16\textwidth}
\includegraphics[width=\linewidth]{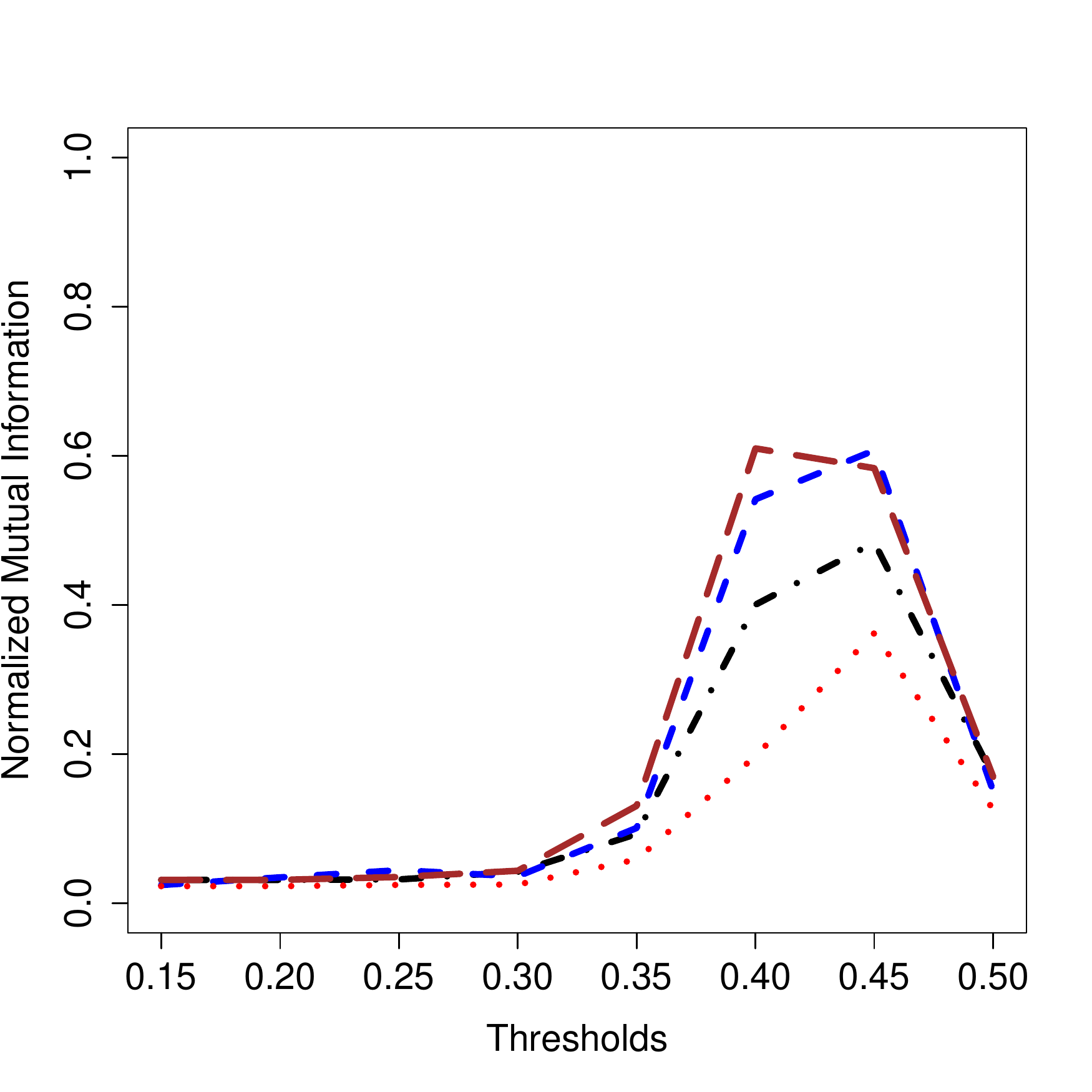}
\end{subfigure}
\begin{center}
(a) $t=200$ \hspace{20pt}  (b) $t=500$  \hspace{20pt}  (c) $t=1000$ \hspace{40pt}  (d) $t=200$  \hspace{20pt} (e) $t=500$ \hspace{20pt}  (f) $t=1000$
\end{center}
\begin{center}
(A) Average accuracy in estimating $Z^{(m)}$s \hspace{100pt}  (B) Accuracy of estimating $\bar{Z}$.
\end{center}
\vspace{-5pt}
\caption{Performance of various methods in terms of (A) average accuracy in estimating the community assignments of the member networks $Z^{(m)}$s and (B) accuracy of estimating the mean putative community assignment $\bar{Z}$ from a sample of multivariate time series data of sizes $200, 500, 1000$ generated using a model which combines RESBM and SBPCM. The three rows of figures correspond to $\phi = 25, 50, 75$ respectively.}
\label{mvts}
\end{figure}

\subsection*{Dependence on selection of thresholds and length of time series}
In Section 4, we have performed comprehensive simulations on the performance of the proposed methods when the networks are directly generated through a RESBM. However, in fMRI application problems the raw data is in the format of a multivariate time series and one needs to estimate the networks from the time series. While thresholding the correlation matrices at an appropriate level is a widely used method, not much attention has been given on how a thresholded estimator performs in estimating the networks. Clearly both the choice of the threshold and the length of the available time series are important considerations for performance of our methods. Therefore we present a limited simulation where we generate multivariate time series data first and then estimate the adjacency matrices. For this purpose we combine the proposed RESBM with a model presented in \cite{brownlees2017community}, called the Stochastic Block Partial Correlation Model (SBPCM). As in Section 4, we generate a sample of $M=10$ networks from the RESBM with $n=100$ vertices and $k=3$ communities. However, instead of directly applying our methods to the sample of networks, we first obtain $n$ dimensional multivariate time series of length $t$ for each of the samples according to the SBPCM. In the terminology of \cite{brownlees2017community} we consider three levels of the parameter $\phi$, which controls the relative weight of the graph Laplacian matrix in creating the inverse covariance (concentration or precision) matrix. The three levels are $25, 50$ and $75$ respectively. We also consider 3 levels of the length of the time series $t$, which are $200, 500$ and $1000$. The accuracy of estimating $\bar{Z}$ and the average accuracy of estimating $Z^{(m)}$s are presented in Figure \ref{mvts}. We note that there is no uniform ``optimum threshold" and the accuracy is highest for different thresholds depending upon the $\phi$ parameter. The accuracies are highly sensitive to the choice of the threshold as well, however there is usually a range of thresholds for which the accuracies are reasonable.  In each case, the accuracies are lower for a very low threshold when the network is dense and full of possibly spurious correlations, and gradually increase with increasing threshold and finally start decreasing as the network becomes too sparse in high thresholds. Therefore the sparse networks obtained at very high thresholds are not necessarily the best ones. We also note from Figure \ref{mvts} that the thresholds at which the accuracies are higher generally increase with increasing $\phi$. This makes intuitive sense, since the parameter $\phi$ effectively controls how much network structure is passed on to the correlation matrix from which the multivariate time series data is generated. Therefore, a higher $\phi$ means the network structure is strongly present and a high cutoff for the correlations produces the most informative networks. Across all values of $\phi$, we note that the performance is poor for the length of the time series $t=200$ and increases as $t$ increases to $1000$. This is of course also not surprising and underscores the dependence of accuracy of our methods on the estimation accuracy of the correlation matrices, which in turn is dependent on the length of the time series.

\begin{supplement}[id=suppA]
 \sname
  {Supplement to ``A random effects stochastic block model for joint community detection in multiple networks with applications to neuroimaging"}
  \slink[doi]{}
  \sdatatype{.pdf}
  \sdescription{The supplementary file contains the details of derivation of the methods, their convergence and implementation details, some additional simulations mentioned in Section \ref{sec:simulation}, and a discussion of a limitation of the algorithms.}
\end{supplement}

\bibliography{vc}

\begin{thebibliography}{85}

\bibitem[\protect\citeauthoryear{Alexander-Bloch
  et~al.}{2010}]{alexander2010disrupted}
\begin{barticle}[author]
\bauthor{\bsnm{Alexander-Bloch},~\bfnm{Aaron~F}\binits{A.~F.}},
  \bauthor{\bsnm{Gogtay},~\bfnm{Nitin}\binits{N.}},
  \bauthor{\bsnm{Meunier},~\bfnm{David}\binits{D.}},
  \bauthor{\bsnm{Birn},~\bfnm{Rasmus}\binits{R.}},
  \bauthor{\bsnm{Clasen},~\bfnm{Liv}\binits{L.}},
  \bauthor{\bsnm{Lalonde},~\bfnm{Francois}\binits{F.}},
  \bauthor{\bsnm{Lenroot},~\bfnm{Rhoshel}\binits{R.}},
  \bauthor{\bsnm{Giedd},~\bfnm{Jay}\binits{J.}} \AND
  \bauthor{\bsnm{Bullmore},~\bfnm{Edward~T}\binits{E.~T.}}
(\byear{2010}).
\btitle{Disrupted modularity and local connectivity of brain functional
  networks in childhood-onset schizophrenia}.
\bjournal{Frontiers in Systems Neuroscience}
\bvolume{4}
\bpages{147}.
\end{barticle}
\endbibitem

\bibitem[\protect\citeauthoryear{Alexander-Bloch
  et~al.}{2012}]{alexander2012discovery}
\begin{barticle}[author]
\bauthor{\bsnm{Alexander-Bloch},~\bfnm{Aaron}\binits{A.}},
  \bauthor{\bsnm{Lambiotte},~\bfnm{Renaud}\binits{R.}},
  \bauthor{\bsnm{Roberts},~\bfnm{Ben}\binits{B.}},
  \bauthor{\bsnm{Giedd},~\bfnm{Jay}\binits{J.}},
  \bauthor{\bsnm{Gogtay},~\bfnm{Nitin}\binits{N.}} \AND
  \bauthor{\bsnm{Bullmore},~\bfnm{Ed}\binits{E.}}
(\byear{2012}).
\btitle{The discovery of population differences in network community structure:
  new methods and applications to brain functional networks in schizophrenia}.
\bjournal{Neuroimage}
\bvolume{59}
\bpages{3889--3900}.
\end{barticle}
\endbibitem

\bibitem[\protect\citeauthoryear{Barbillon
  et~al.}{2017}]{barbillon2017stochastic}
\begin{barticle}[author]
\bauthor{\bsnm{Barbillon},~\bfnm{Pierre}\binits{P.}},
  \bauthor{\bsnm{Donnet},~\bfnm{Sophie}\binits{S.}},
  \bauthor{\bsnm{Lazega},~\bfnm{Emmanuel}\binits{E.}} \AND
  \bauthor{\bsnm{Bar-Hen},~\bfnm{Avner}\binits{A.}}
(\byear{2017}).
\btitle{Stochastic block models for multiplex networks: an application to a
  multilevel network of researchers}.
\bjournal{Journal of the Royal Statistical Society: Series A}
\bvolume{180}
\bpages{295--314}.
\end{barticle}
\endbibitem

\bibitem[\protect\citeauthoryear{Bassett et~al.}{2011}]{bassett11}
\begin{barticle}[author]
\bauthor{\bsnm{Bassett},~\bfnm{Danielle~S}\binits{D.~S.}},
  \bauthor{\bsnm{Wymbs},~\bfnm{Nicholas~F}\binits{N.~F.}},
  \bauthor{\bsnm{Porter},~\bfnm{Mason~A}\binits{M.~A.}},
  \bauthor{\bsnm{Mucha},~\bfnm{Peter~J}\binits{P.~J.}},
  \bauthor{\bsnm{Carlson},~\bfnm{Jean~M}\binits{J.~M.}} \AND
  \bauthor{\bsnm{Grafton},~\bfnm{Scott~T}\binits{S.~T.}}
(\byear{2011}).
\btitle{Dynamic reconfiguration of human brain networks during learning}.
\bjournal{Proceedings of the National Academy of Sciences}
\bvolume{108}
\bpages{7641--7646}.
\end{barticle}
\endbibitem

\bibitem[\protect\citeauthoryear{Bassett et~al.}{2012}]{bassett2012altered}
\begin{barticle}[author]
\bauthor{\bsnm{Bassett},~\bfnm{Danielle~S}\binits{D.~S.}},
  \bauthor{\bsnm{Nelson},~\bfnm{Brent~G}\binits{B.~G.}},
  \bauthor{\bsnm{Mueller},~\bfnm{Bryon~A}\binits{B.~A.}},
  \bauthor{\bsnm{Camchong},~\bfnm{Jazmin}\binits{J.}} \AND
  \bauthor{\bsnm{Lim},~\bfnm{Kelvin~O}\binits{K.~O.}}
(\byear{2012}).
\btitle{Altered resting state complexity in schizophrenia}.
\bjournal{Neuroimage}
\bvolume{59}
\bpages{2196--2207}.
\end{barticle}
\endbibitem

\bibitem[\protect\citeauthoryear{Bazzi et~al.}{2016}]{bazzi2016community}
\begin{barticle}[author]
\bauthor{\bsnm{Bazzi},~\bfnm{Marya}\binits{M.}},
  \bauthor{\bsnm{Porter},~\bfnm{Mason~A}\binits{M.~A.}},
  \bauthor{\bsnm{Williams},~\bfnm{Stacy}\binits{S.}},
  \bauthor{\bsnm{McDonald},~\bfnm{Mark}\binits{M.}},
  \bauthor{\bsnm{Fenn},~\bfnm{Daniel~J}\binits{D.~J.}} \AND
  \bauthor{\bsnm{Howison},~\bfnm{Sam~D}\binits{S.~D.}}
(\byear{2016}).
\btitle{Community detection in temporal multilayer networks, with an
  application to correlation networks}.
\bjournal{Multiscale Modeling \& Simulation}
\bvolume{14}
\bpages{1--41}.
\end{barticle}
\endbibitem

\bibitem[\protect\citeauthoryear{Benjamini and
  Hochberg}{1995}]{benjamini1995controlling}
\begin{barticle}[author]
\bauthor{\bsnm{Benjamini},~\bfnm{Yoav}\binits{Y.}} \AND
  \bauthor{\bsnm{Hochberg},~\bfnm{Yosef}\binits{Y.}}
(\byear{1995}).
\btitle{Controlling the false discovery rate: a practical and powerful approach
  to multiple testing}.
\bjournal{Journal of the Royal Statistical Society. Series B}
\bpages{289--300}.
\end{barticle}
\endbibitem

\bibitem[\protect\citeauthoryear{Betzel et~al.}{2019}]{betzel2019community}
\begin{barticle}[author]
\bauthor{\bsnm{Betzel},~\bfnm{Richard~F}\binits{R.~F.}},
  \bauthor{\bsnm{Bertolero},~\bfnm{Maxwell~A}\binits{M.~A.}},
  \bauthor{\bsnm{Gordon},~\bfnm{Evan~M}\binits{E.~M.}},
  \bauthor{\bsnm{Gratton},~\bfnm{Caterina}\binits{C.}},
  \bauthor{\bsnm{Dosenbach},~\bfnm{Nico~UF}\binits{N.~U.}} \AND
  \bauthor{\bsnm{Bassett},~\bfnm{Danielle~S}\binits{D.~S.}}
(\byear{2019}).
\btitle{The community structure of functional brain networks exhibits
  scale-specific patterns of inter-and intra-subject variability}.
\bjournal{Neuroimage}
\bvolume{202}
\bpages{115990}.
\end{barticle}
\endbibitem

\bibitem[\protect\citeauthoryear{Bickel et~al.}{2013}]{bccz13}
\begin{barticle}[author]
\bauthor{\bsnm{Bickel},~\bfnm{P.~J.}\binits{P.~J.}},
  \bauthor{\bsnm{Choi},~\bfnm{D.}\binits{D.}},
  \bauthor{\bsnm{Chang},~\bfnm{X.}\binits{X.}} \AND
  \bauthor{\bsnm{Zhang},~\bfnm{H.}\binits{H.}}
(\byear{2013}).
\btitle{Asymptotic normality of maximum likelihood and its variational
  approximation for stochastic blockmodels}.
\bjournal{Ann. Statist}
\bvolume{41}
\bpages{1922-1943}.
\end{barticle}
\endbibitem

\bibitem[\protect\citeauthoryear{Blondel et~al.}{2008}]{blondel08}
\begin{barticle}[author]
\bauthor{\bsnm{Blondel},~\bfnm{Vincent~D}\binits{V.~D.}},
  \bauthor{\bsnm{Guillaume},~\bfnm{Jean-Loup}\binits{J.-L.}},
  \bauthor{\bsnm{Lambiotte},~\bfnm{Renaud}\binits{R.}} \AND
  \bauthor{\bsnm{Lefebvre},~\bfnm{Etienne}\binits{E.}}
(\byear{2008}).
\btitle{Fast unfolding of communities in large networks}.
\bjournal{Journal of Statistical Mechanics: Theory and Experiment}
\bvolume{2008}
\bpages{P10008}.
\end{barticle}
\endbibitem

\bibitem[\protect\citeauthoryear{Boccaletti et~al.}{2014}]{boccaletti14}
\begin{barticle}[author]
\bauthor{\bsnm{Boccaletti},~\bfnm{Stefano}\binits{S.}},
  \bauthor{\bsnm{Bianconi},~\bfnm{G}\binits{G.}},
  \bauthor{\bsnm{Criado},~\bfnm{R}\binits{R.}},
  \bauthor{\bsnm{Del~Genio},~\bfnm{Charo~I}\binits{C.~I.}},
  \bauthor{\bsnm{G{\'o}mez-Garde{\~n}es},~\bfnm{J}\binits{J.}},
  \bauthor{\bsnm{Romance},~\bfnm{M}\binits{M.}},
  \bauthor{\bsnm{Sendina-Nadal},~\bfnm{I}\binits{I.}},
  \bauthor{\bsnm{Wang},~\bfnm{Z}\binits{Z.}} \AND
  \bauthor{\bsnm{Zanin},~\bfnm{M}\binits{M.}}
(\byear{2014}).
\btitle{The structure and dynamics of multilayer networks}.
\bjournal{Physics Reports}
\bvolume{544}
\bpages{1--122}.
\end{barticle}
\endbibitem

\bibitem[\protect\citeauthoryear{Braun et~al.}{2015}]{braun15}
\begin{barticle}[author]
\bauthor{\bsnm{Braun},~\bfnm{Urs}\binits{U.}},
  \bauthor{\bsnm{Sch{\"a}fer},~\bfnm{Axel}\binits{A.}},
  \bauthor{\bsnm{Walter},~\bfnm{Henrik}\binits{H.}},
  \bauthor{\bsnm{Erk},~\bfnm{Susanne}\binits{S.}},
  \bauthor{\bsnm{Romanczuk-Seiferth},~\bfnm{Nina}\binits{N.}},
  \bauthor{\bsnm{Haddad},~\bfnm{Leila}\binits{L.}},
  \bauthor{\bsnm{Schweiger},~\bfnm{Janina~I}\binits{J.~I.}},
  \bauthor{\bsnm{Grimm},~\bfnm{Oliver}\binits{O.}},
  \bauthor{\bsnm{Heinz},~\bfnm{Andreas}\binits{A.}} \AND
  \bauthor{\bsnm{Tost},~\bfnm{Heike}\binits{H.}}
(\byear{2015}).
\btitle{Dynamic reconfiguration of frontal brain networks during executive
  cognition in humans}.
\bjournal{Proceedings of the National Academy of Sciences}
\bvolume{112}
\bpages{11678--11683}.
\end{barticle}
\endbibitem

\bibitem[\protect\citeauthoryear{Brownlees, Gudmundsson and
  Lugosi}{2017}]{brownlees2017community}
\begin{barticle}[author]
\bauthor{\bsnm{Brownlees},~\bfnm{Christian~T}\binits{C.~T.}},
  \bauthor{\bsnm{Gudmundsson},~\bfnm{Gudmundur}\binits{G.}} \AND
  \bauthor{\bsnm{Lugosi},~\bfnm{G{\'a}bor}\binits{G.}}
(\byear{2017}).
\btitle{Community detection in partial correlation network models}.
\bjournal{Available at SSRN 2776505}.
\end{barticle}
\endbibitem

\bibitem[\protect\citeauthoryear{Bullmore and Sporns}{2009}]{bullmore09}
\begin{barticle}[author]
\bauthor{\bsnm{Bullmore},~\bfnm{Ed}\binits{E.}} \AND
  \bauthor{\bsnm{Sporns},~\bfnm{Olaf}\binits{O.}}
(\byear{2009}).
\btitle{Complex brain networks: graph theoretical analysis of structural and
  functional systems}.
\bjournal{Nature Reviews Neuroscience}
\bvolume{10}
\bpages{186--198}.
\end{barticle}
\endbibitem

\bibitem[\protect\citeauthoryear{Celisse, Daudin and Pierre}{2012}]{cdp11}
\begin{barticle}[author]
\bauthor{\bsnm{Celisse},~\bfnm{A.}\binits{A.}},
  \bauthor{\bsnm{Daudin},~\bfnm{J.~J.}\binits{J.~J.}} \AND
  \bauthor{\bsnm{Pierre},~\bfnm{L.}\binits{L.}}
(\byear{2012}).
\btitle{Consistency of maximum-likelihood and variational estimators in the
  stochastic block model}.
\bjournal{Electronic Journal of Statistics}
\bvolume{6}
\bpages{1847-1899}.
\end{barticle}
\endbibitem

\bibitem[\protect\citeauthoryear{Chen et~al.}{2015}]{chen2015parsimonious}
\begin{barticle}[author]
\bauthor{\bsnm{Chen},~\bfnm{Shuo}\binits{S.}},
  \bauthor{\bsnm{Kang},~\bfnm{Jian}\binits{J.}},
  \bauthor{\bsnm{Xing},~\bfnm{Yishi}\binits{Y.}} \AND
  \bauthor{\bsnm{Wang},~\bfnm{Guoqing}\binits{G.}}
(\byear{2015}).
\btitle{A parsimonious statistical method to detect groupwise differentially
  expressed functional connectivity networks}.
\bjournal{Human Brain Mapping}
\bvolume{36}
\bpages{5196--5206}.
\end{barticle}
\endbibitem

\bibitem[\protect\citeauthoryear{Daudin, Picard and Robin}{2008}]{dpr08}
\begin{barticle}[author]
\bauthor{\bsnm{Daudin},~\bfnm{J.~J.}\binits{J.~J.}},
  \bauthor{\bsnm{Picard},~\bfnm{F.}\binits{F.}} \AND
  \bauthor{\bsnm{Robin},~\bfnm{S.}\binits{S.}}
(\byear{2008}).
\btitle{A mixture model for random graphs}.
\bjournal{Stat Comput}
\bvolume{18}
\bpages{173-183}.
\end{barticle}
\endbibitem

\bibitem[\protect\citeauthoryear{De~Bacco et~al.}{2017}]{de2017community}
\begin{barticle}[author]
\bauthor{\bsnm{De~Bacco},~\bfnm{Caterina}\binits{C.}},
  \bauthor{\bsnm{Power},~\bfnm{Eleanor~A}\binits{E.~A.}},
  \bauthor{\bsnm{Larremore},~\bfnm{Daniel~B}\binits{D.~B.}} \AND
  \bauthor{\bsnm{Moore},~\bfnm{Cristopher}\binits{C.}}
(\byear{2017}).
\btitle{Community detection, link prediction, and layer interdependence in
  multilayer networks}.
\bjournal{Physical Review E}
\bvolume{95}
\bpages{042317}.
\end{barticle}
\endbibitem

\bibitem[\protect\citeauthoryear{Ding et~al.}{2006}]{ding06}
\begin{binproceedings}[author]
\bauthor{\bsnm{Ding},~\bfnm{Chris}\binits{C.}},
  \bauthor{\bsnm{Li},~\bfnm{Tao}\binits{T.}},
  \bauthor{\bsnm{Peng},~\bfnm{Wei}\binits{W.}} \AND
  \bauthor{\bsnm{Park},~\bfnm{Haesun}\binits{H.}}
(\byear{2006}).
\btitle{Orthogonal nonnegative matrix t-factorizations for clustering}.
In \bbooktitle{Proceedings of the 12th ACM SIGKDD International Conference on
  Knowledge Discovery and Data Mining}
\bpages{126--135}.
\bpublisher{ACM}.
\end{binproceedings}
\endbibitem

\bibitem[\protect\citeauthoryear{Dong et~al.}{2014}]{dong2014clustering}
\begin{barticle}[author]
\bauthor{\bsnm{Dong},~\bfnm{Xiaowen}\binits{X.}},
  \bauthor{\bsnm{Frossard},~\bfnm{Pascal}\binits{P.}},
  \bauthor{\bsnm{Vandergheynst},~\bfnm{Pierre}\binits{P.}} \AND
  \bauthor{\bsnm{Nefedov},~\bfnm{Nikolai}\binits{N.}}
(\byear{2014}).
\btitle{Clustering on multi-layer graphs via subspace analysis on {G}rassmann
  manifolds}.
\bjournal{IEEE Transactions on Signal Processing}
\bvolume{62}
\bpages{905--918}.
\end{barticle}
\endbibitem

\bibitem[\protect\citeauthoryear{Edelman, Arias and
  Smith}{1998}]{edelman1998geometry}
\begin{barticle}[author]
\bauthor{\bsnm{Edelman},~\bfnm{Alan}\binits{A.}},
  \bauthor{\bsnm{Arias},~\bfnm{Tom{\'a}s~A}\binits{T.~A.}} \AND
  \bauthor{\bsnm{Smith},~\bfnm{Steven~T}\binits{S.~T.}}
(\byear{1998}).
\btitle{The geometry of algorithms with orthogonality constraints}.
\bjournal{SIAM Journal on Matrix Analysis and Applications}
\bvolume{20}
\bpages{303--353}.
\end{barticle}
\endbibitem

\bibitem[\protect\citeauthoryear{Fornito, Zalesky and
  Breakspear}{2013}]{fornito2013graph}
\begin{barticle}[author]
\bauthor{\bsnm{Fornito},~\bfnm{Alex}\binits{A.}},
  \bauthor{\bsnm{Zalesky},~\bfnm{Andrew}\binits{A.}} \AND
  \bauthor{\bsnm{Breakspear},~\bfnm{Michael}\binits{M.}}
(\byear{2013}).
\btitle{Graph analysis of the human connectome: promise, progress, and
  pitfalls}.
\bjournal{Neuroimage}
\bvolume{80}
\bpages{426--444}.
\end{barticle}
\endbibitem

\bibitem[\protect\citeauthoryear{Fornito, Zalesky and
  Breakspear}{2015}]{fornito2015connectomics}
\begin{barticle}[author]
\bauthor{\bsnm{Fornito},~\bfnm{Alex}\binits{A.}},
  \bauthor{\bsnm{Zalesky},~\bfnm{Andrew}\binits{A.}} \AND
  \bauthor{\bsnm{Breakspear},~\bfnm{Michael}\binits{M.}}
(\byear{2015}).
\btitle{The connectomics of brain disorders}.
\bjournal{Nature Reviews Neuroscience}
\bvolume{16}
\bpages{159--172}.
\end{barticle}
\endbibitem

\bibitem[\protect\citeauthoryear{Fujita et~al.}{2014}]{fujita2014non}
\begin{barticle}[author]
\bauthor{\bsnm{Fujita},~\bfnm{Andr{\'e}}\binits{A.}},
  \bauthor{\bsnm{Takahashi},~\bfnm{Daniel~Y}\binits{D.~Y.}},
  \bauthor{\bsnm{Patriota},~\bfnm{Alexandre~G}\binits{A.~G.}} \AND
  \bauthor{\bsnm{Sato},~\bfnm{Jo{\~a}o~R}\binits{J.~R.}}
(\byear{2014}).
\btitle{A non-parametric statistical test to compare clusters with applications
  in functional magnetic resonance imaging data}.
\bjournal{Statistics in Medicine}
\bvolume{33}
\bpages{4949--4962}.
\end{barticle}
\endbibitem

\bibitem[\protect\citeauthoryear{GadElkarim
  et~al.}{2012}]{gadelkarim2012framework}
\begin{binproceedings}[author]
\bauthor{\bsnm{GadElkarim},~\bfnm{Johnson~J}\binits{J.~J.}},
  \bauthor{\bsnm{Schonfeld},~\bfnm{Dan}\binits{D.}},
  \bauthor{\bsnm{Ajilore},~\bfnm{Olusola}\binits{O.}},
  \bauthor{\bsnm{Zhan},~\bfnm{Liang}\binits{L.}},
  \bauthor{\bsnm{Zhang},~\bfnm{Aifeng~F}\binits{A.~F.}},
  \bauthor{\bsnm{Feusner},~\bfnm{Jamie~D}\binits{J.~D.}},
  \bauthor{\bsnm{Thompson},~\bfnm{Paul~M}\binits{P.~M.}},
  \bauthor{\bsnm{Simon},~\bfnm{Tony~J}\binits{T.~J.}},
  \bauthor{\bsnm{Kumar},~\bfnm{Anand}\binits{A.}} \AND
  \bauthor{\bsnm{Leow},~\bfnm{Alex~D}\binits{A.~D.}}
(\byear{2012}).
\btitle{A framework for quantifying node-level community structure group
  differences in brain connectivity networks}.
In \bbooktitle{International Conference on Medical Image Computing and
  Computer-Assisted Intervention}
\bpages{196--203}.
\bpublisher{Springer}.
\end{binproceedings}
\endbibitem

\bibitem[\protect\citeauthoryear{Ginestet, Fournel and
  Simmons}{2014}]{ginestet2013statistical}
\begin{barticle}[author]
\bauthor{\bsnm{Ginestet},~\bfnm{Cedric~E}\binits{C.~E.}},
  \bauthor{\bsnm{Fournel},~\bfnm{Arnaud~P}\binits{A.~P.}} \AND
  \bauthor{\bsnm{Simmons},~\bfnm{Andrew}\binits{A.}}
(\byear{2014}).
\btitle{Statistical network analysis for functional {MRI}: summary networks and
  group comparisons}.
\bjournal{Frontiers in Computational Neuroscience}
\bvolume{8}.
\end{barticle}
\endbibitem

\bibitem[\protect\citeauthoryear{Ginestet
  et~al.}{2017}]{ginestet2014hypothesis}
\begin{barticle}[author]
\bauthor{\bsnm{Ginestet},~\bfnm{Cedric~E}\binits{C.~E.}},
  \bauthor{\bsnm{Li},~\bfnm{Jun}\binits{J.}},
  \bauthor{\bsnm{Balachandran},~\bfnm{Prakash}\binits{P.}},
  \bauthor{\bsnm{Rosenberg},~\bfnm{Steven}\binits{S.}},
  \bauthor{\bsnm{Kolaczyk},~\bfnm{Eric~D}\binits{E.~D.}} \betal{et~al.}
(\byear{2017}).
\btitle{Hypothesis testing for network data in functional neuroimaging}.
\bjournal{The Annals of Applied Statistics}
\bvolume{11}
\bpages{725--750}.
\end{barticle}
\endbibitem

\bibitem[\protect\citeauthoryear{Glerean
  et~al.}{2016}]{glerean2016reorganization}
\begin{barticle}[author]
\bauthor{\bsnm{Glerean},~\bfnm{Enrico}\binits{E.}},
  \bauthor{\bsnm{Pan},~\bfnm{Raj~K}\binits{R.~K.}},
  \bauthor{\bsnm{Salmi},~\bfnm{Juha}\binits{J.}},
  \bauthor{\bsnm{Kujala},~\bfnm{Rainer}\binits{R.}},
  \bauthor{\bsnm{Lahnakoski},~\bfnm{Juha~M}\binits{J.~M.}},
  \bauthor{\bsnm{Roine},~\bfnm{Ulrika}\binits{U.}},
  \bauthor{\bsnm{Nummenmaa},~\bfnm{Lauri}\binits{L.}},
  \bauthor{\bsnm{Lepp{\"a}m{\"a}ki},~\bfnm{Sami}\binits{S.}},
  \bauthor{\bparticle{Nieminen-von} \bsnm{Wendt},~\bfnm{Taina}\binits{T.}} \AND
  \bauthor{\bsnm{Tani},~\bfnm{Pekka}\binits{P.}}
(\byear{2016}).
\btitle{Reorganization of functionally connected brain subnetworks in
  high-functioning autism}.
\bjournal{Human Brain Mapping}
\bvolume{37}
\bpages{1066--1079}.
\end{barticle}
\endbibitem

\bibitem[\protect\citeauthoryear{Han, Xu and Airoldi}{2015}]{hxa14}
\begin{binproceedings}[author]
\bauthor{\bsnm{Han},~\bfnm{Qiuyi}\binits{Q.}},
  \bauthor{\bsnm{Xu},~\bfnm{Kevin~S}\binits{K.~S.}} \AND
  \bauthor{\bsnm{Airoldi},~\bfnm{Edoardo~M}\binits{E.~M.}}
(\byear{2015}).
\btitle{Consistent estimation of dynamic and multi-layer block models}.
In \bbooktitle{Proceedings of the 32nd International Conference on Machine
  Learning}
\bpages{1511--1520}.
\end{binproceedings}
\endbibitem

\bibitem[\protect\citeauthoryear{He et~al.}{2009}]{he2009uncovering}
\begin{barticle}[author]
\bauthor{\bsnm{He},~\bfnm{Yong}\binits{Y.}},
  \bauthor{\bsnm{Wang},~\bfnm{Jinhui}\binits{J.}},
  \bauthor{\bsnm{Wang},~\bfnm{Liang}\binits{L.}},
  \bauthor{\bsnm{Chen},~\bfnm{Zhang~J}\binits{Z.~J.}},
  \bauthor{\bsnm{Yan},~\bfnm{Chaogan}\binits{C.}},
  \bauthor{\bsnm{Yang},~\bfnm{Hong}\binits{H.}},
  \bauthor{\bsnm{Tang},~\bfnm{Hehan}\binits{H.}},
  \bauthor{\bsnm{Zhu},~\bfnm{Chaozhe}\binits{C.}},
  \bauthor{\bsnm{Gong},~\bfnm{Qiyong}\binits{Q.}} \AND
  \bauthor{\bsnm{Zang},~\bfnm{Yufeng}\binits{Y.}}
(\byear{2009}).
\btitle{Uncovering intrinsic modular organization of spontaneous brain activity
  in humans}.
\bjournal{PLoS ONE}
\bvolume{4}
\bpages{e5226}.
\end{barticle}
\endbibitem

\bibitem[\protect\citeauthoryear{Hutchison et~al.}{2013}]{hutchison13}
\begin{barticle}[author]
\bauthor{\bsnm{Hutchison},~\bfnm{R~Matthew}\binits{R.~M.}},
  \bauthor{\bsnm{Womelsdorf},~\bfnm{Thilo}\binits{T.}},
  \bauthor{\bsnm{Allen},~\bfnm{Elena~A}\binits{E.~A.}},
  \bauthor{\bsnm{Bandettini},~\bfnm{Peter~A}\binits{P.~A.}},
  \bauthor{\bsnm{Calhoun},~\bfnm{Vince~D}\binits{V.~D.}},
  \bauthor{\bsnm{Corbetta},~\bfnm{Maurizio}\binits{M.}},
  \bauthor{\bsnm{Della~Penna},~\bfnm{Stefania}\binits{S.}},
  \bauthor{\bsnm{Duyn},~\bfnm{Jeff~H}\binits{J.~H.}},
  \bauthor{\bsnm{Glover},~\bfnm{Gary~H}\binits{G.~H.}} \AND
  \bauthor{\bsnm{Gonzalez-Castillo},~\bfnm{Javier}\binits{J.}}
(\byear{2013}).
\btitle{Dynamic functional connectivity: promise, issues, and interpretations}.
\bjournal{Neuroimage}
\bvolume{80}
\bpages{360--378}.
\end{barticle}
\endbibitem

\bibitem[\protect\citeauthoryear{Jones et~al.}{2012}]{jones12}
\begin{barticle}[author]
\bauthor{\bsnm{Jones},~\bfnm{David~T}\binits{D.~T.}},
  \bauthor{\bsnm{Vemuri},~\bfnm{Prashanthi}\binits{P.}},
  \bauthor{\bsnm{Murphy},~\bfnm{Matthew~C}\binits{M.~C.}},
  \bauthor{\bsnm{Gunter},~\bfnm{Jeffrey~L}\binits{J.~L.}},
  \bauthor{\bsnm{Senjem},~\bfnm{Matthew~L}\binits{M.~L.}},
  \bauthor{\bsnm{Machulda},~\bfnm{Mary~M}\binits{M.~M.}},
  \bauthor{\bsnm{Przybelski},~\bfnm{Scott~A}\binits{S.~A.}},
  \bauthor{\bsnm{Gregg},~\bfnm{Brian~E}\binits{B.~E.}},
  \bauthor{\bsnm{Kantarci},~\bfnm{Kejal}\binits{K.}} \AND
  \bauthor{\bsnm{Knopman},~\bfnm{David~S}\binits{D.~S.}}
(\byear{2012}).
\btitle{Non-stationarity in the “resting brain’s” modular architecture}.
\bjournal{PLoS ONE}
\bvolume{7}
\bpages{e39731}.
\end{barticle}
\endbibitem

\bibitem[\protect\citeauthoryear{Kivel{\"a} et~al.}{2014}]{kivela14}
\begin{barticle}[author]
\bauthor{\bsnm{Kivel{\"a}},~\bfnm{Mikko}\binits{M.}},
  \bauthor{\bsnm{Arenas},~\bfnm{Alex}\binits{A.}},
  \bauthor{\bsnm{Barthelemy},~\bfnm{Marc}\binits{M.}},
  \bauthor{\bsnm{Gleeson},~\bfnm{James~P}\binits{J.~P.}},
  \bauthor{\bsnm{Moreno},~\bfnm{Yamir}\binits{Y.}} \AND
  \bauthor{\bsnm{Porter},~\bfnm{Mason~A}\binits{M.~A.}}
(\byear{2014}).
\btitle{Multilayer networks}.
\bjournal{Journal of Complex Networks}
\bvolume{2}
\bpages{203--271}.
\end{barticle}
\endbibitem

\bibitem[\protect\citeauthoryear{Kuhn}{1955}]{kuhn1955hungarian}
\begin{barticle}[author]
\bauthor{\bsnm{Kuhn},~\bfnm{Harold~W}\binits{H.~W.}}
(\byear{1955}).
\btitle{The Hungarian method for the assignment problem}.
\bjournal{Naval Research Logistics Quarterly}
\bvolume{2}
\bpages{83--97}.
\end{barticle}
\endbibitem

\bibitem[\protect\citeauthoryear{Kujala et~al.}{2016}]{kujala2016graph}
\begin{barticle}[author]
\bauthor{\bsnm{Kujala},~\bfnm{Rainer}\binits{R.}},
  \bauthor{\bsnm{Glerean},~\bfnm{Enrico}\binits{E.}},
  \bauthor{\bsnm{Pan},~\bfnm{Raj~Kumar}\binits{R.~K.}},
  \bauthor{\bsnm{J{\"a}{\"a}skel{\"a}inen},~\bfnm{Iiro~P}\binits{I.~P.}},
  \bauthor{\bsnm{Sams},~\bfnm{Mikko}\binits{M.}} \AND
  \bauthor{\bsnm{Saram{\"a}ki},~\bfnm{Jari}\binits{J.}}
(\byear{2016}).
\btitle{Graph coarse-graining reveals differences in the module-level structure
  of functional brain networks}.
\bjournal{European Journal of Neuroscience}
\bvolume{44}
\bpages{2673--2684}.
\end{barticle}
\endbibitem

\bibitem[\protect\citeauthoryear{Kumar, Rai and Daume}{2011}]{kumar2011co}
\begin{binproceedings}[author]
\bauthor{\bsnm{Kumar},~\bfnm{Abhishek}\binits{A.}},
  \bauthor{\bsnm{Rai},~\bfnm{Piyush}\binits{P.}} \AND
  \bauthor{\bsnm{Daume},~\bfnm{Hal}\binits{H.}}
(\byear{2011}).
\btitle{Co-regularized multi-view spectral clustering}.
In \bbooktitle{Advances in Neural Information Processing Systems}
\bpages{1413--1421}.
\end{binproceedings}
\endbibitem

\bibitem[\protect\citeauthoryear{Lee and Seung}{2001}]{lee01}
\begin{binproceedings}[author]
\bauthor{\bsnm{Lee},~\bfnm{Daniel~D}\binits{D.~D.}} \AND
  \bauthor{\bsnm{Seung},~\bfnm{H~Sebastian}\binits{H.~S.}}
(\byear{2001}).
\btitle{Algorithms for non-negative matrix factorization}.
In \bbooktitle{Advances in neural information processing systems}
\bpages{556--562}.
\end{binproceedings}
\endbibitem

\bibitem[\protect\citeauthoryear{Lei and Rinaldo}{2014}]{lei14}
\begin{barticle}[author]
\bauthor{\bsnm{Lei},~\bfnm{Jing}\binits{J.}} \AND
  \bauthor{\bsnm{Rinaldo},~\bfnm{Alessandro}\binits{A.}}
(\byear{2014}).
\btitle{Consistency of spectral clustering in stochastic block models}.
\bjournal{The Annals of Statistics}
\bvolume{43}
\bpages{215--237}.
\end{barticle}
\endbibitem

\bibitem[\protect\citeauthoryear{Liu et~al.}{2008}]{liu2008disrupted}
\begin{barticle}[author]
\bauthor{\bsnm{Liu},~\bfnm{Yong}\binits{Y.}},
  \bauthor{\bsnm{Liang},~\bfnm{Meng}\binits{M.}},
  \bauthor{\bsnm{Zhou},~\bfnm{Yuan}\binits{Y.}},
  \bauthor{\bsnm{He},~\bfnm{Yong}\binits{Y.}},
  \bauthor{\bsnm{Hao},~\bfnm{Yihui}\binits{Y.}},
  \bauthor{\bsnm{Song},~\bfnm{Ming}\binits{M.}},
  \bauthor{\bsnm{Yu},~\bfnm{Chunshui}\binits{C.}},
  \bauthor{\bsnm{Liu},~\bfnm{Haihong}\binits{H.}},
  \bauthor{\bsnm{Liu},~\bfnm{Zhening}\binits{Z.}} \AND
  \bauthor{\bsnm{Jiang},~\bfnm{Tianzi}\binits{T.}}
(\byear{2008}).
\btitle{Disrupted small-world networks in schizophrenia}.
\bjournal{Brain}
\bvolume{131}
\bpages{945--961}.
\end{barticle}
\endbibitem

\bibitem[\protect\citeauthoryear{Liu et~al.}{2013}]{liu13}
\begin{binproceedings}[author]
\bauthor{\bsnm{Liu},~\bfnm{Jialu}\binits{J.}},
  \bauthor{\bsnm{Wang},~\bfnm{Chi}\binits{C.}},
  \bauthor{\bsnm{Gao},~\bfnm{Jing}\binits{J.}} \AND
  \bauthor{\bsnm{Han},~\bfnm{Jiawei}\binits{J.}}
(\byear{2013}).
\btitle{Multi-view clustering via joint nonnegative matrix factorization}.
In \bbooktitle{Proc. of SDM}
\bvolume{13}
\bpages{252--260}.
\bpublisher{SIAM}.
\end{binproceedings}
\endbibitem

\bibitem[\protect\citeauthoryear{Lynall et~al.}{2010}]{lynall2010functional}
\begin{barticle}[author]
\bauthor{\bsnm{Lynall},~\bfnm{Mary-Ellen}\binits{M.-E.}},
  \bauthor{\bsnm{Bassett},~\bfnm{Danielle~S}\binits{D.~S.}},
  \bauthor{\bsnm{Kerwin},~\bfnm{Robert}\binits{R.}},
  \bauthor{\bsnm{McKenna},~\bfnm{Peter~J}\binits{P.~J.}},
  \bauthor{\bsnm{Kitzbichler},~\bfnm{Manfred}\binits{M.}},
  \bauthor{\bsnm{Muller},~\bfnm{Ulrich}\binits{U.}} \AND
  \bauthor{\bsnm{Bullmore},~\bfnm{Ed}\binits{E.}}
(\byear{2010}).
\btitle{Functional connectivity and brain networks in schizophrenia}.
\bjournal{The Journal of Neuroscience}
\bvolume{30}
\bpages{9477--9487}.
\end{barticle}
\endbibitem

\bibitem[\protect\citeauthoryear{Mankad and
  Michailidis}{2013}]{mankad2013structural}
\begin{barticle}[author]
\bauthor{\bsnm{Mankad},~\bfnm{Shawn}\binits{S.}} \AND
  \bauthor{\bsnm{Michailidis},~\bfnm{George}\binits{G.}}
(\byear{2013}).
\btitle{Structural and functional discovery in dynamic networks with
  non-negative matrix factorization}.
\bjournal{Physical Review E}
\bvolume{88}
\bpages{042812}.
\end{barticle}
\endbibitem

\bibitem[\protect\citeauthoryear{Matias and Miele}{2017}]{matias15}
\begin{barticle}[author]
\bauthor{\bsnm{Matias},~\bfnm{Catherine}\binits{C.}} \AND
  \bauthor{\bsnm{Miele},~\bfnm{Vincent}\binits{V.}}
(\byear{2017}).
\btitle{Statistical clustering of temporal networks through a dynamic
  stochastic block model}.
\bjournal{Journal of the Royal Statistical Society: Series B (Statistical
  Methodology)}
\bvolume{79}
\bpages{1119--1141}.
\end{barticle}
\endbibitem

\bibitem[\protect\citeauthoryear{Meunier, Lambiotte and
  Bullmore}{2010}]{meunier2010modular}
\begin{barticle}[author]
\bauthor{\bsnm{Meunier},~\bfnm{David}\binits{D.}},
  \bauthor{\bsnm{Lambiotte},~\bfnm{Renaud}\binits{R.}} \AND
  \bauthor{\bsnm{Bullmore},~\bfnm{Edward~T}\binits{E.~T.}}
(\byear{2010}).
\btitle{Modular and hierarchically modular organization of brain networks}.
\bjournal{Frontiers in Neuroscience}
\bvolume{4}
\bpages{200}.
\end{barticle}
\endbibitem

\bibitem[\protect\citeauthoryear{Mirzal}{2014}]{mirzal2014convergent}
\begin{barticle}[author]
\bauthor{\bsnm{Mirzal},~\bfnm{Andri}\binits{A.}}
(\byear{2014}).
\btitle{A convergent algorithm for orthogonal nonnegative matrix
  factorization}.
\bjournal{Journal of Computational and Applied Mathematics}
\bvolume{260}
\bpages{149--166}.
\end{barticle}
\endbibitem

\bibitem[\protect\citeauthoryear{Moussa et~al.}{2012}]{moussa12}
\begin{barticle}[author]
\bauthor{\bsnm{Moussa},~\bfnm{Malaak~N}\binits{M.~N.}},
  \bauthor{\bsnm{Steen},~\bfnm{Matthew~R}\binits{M.~R.}},
  \bauthor{\bsnm{Laurienti},~\bfnm{Paul~J}\binits{P.~J.}} \AND
  \bauthor{\bsnm{Hayasaka},~\bfnm{Satoru}\binits{S.}}
(\byear{2012}).
\btitle{Consistency of network modules in resting-state f{MRI} connectome
  data}.
\bjournal{PLoS ONE}
\bvolume{7}
\bpages{e44428}.
\end{barticle}
\endbibitem

\bibitem[\protect\citeauthoryear{Mucha et~al.}{2010}]{mucha10}
\begin{barticle}[author]
\bauthor{\bsnm{Mucha},~\bfnm{Peter~J}\binits{P.~J.}},
  \bauthor{\bsnm{Richardson},~\bfnm{Thomas}\binits{T.}},
  \bauthor{\bsnm{Macon},~\bfnm{Kevin}\binits{K.}},
  \bauthor{\bsnm{Porter},~\bfnm{Mason~A}\binits{M.~A.}} \AND
  \bauthor{\bsnm{Onnela},~\bfnm{Jukka~Pekka}\binits{J.~P.}}
(\byear{2010}).
\btitle{Community structure in time-dependent, multiscale, and multiplex
  networks}.
\bjournal{Science}
\bvolume{328}
\bpages{876--878}.
\end{barticle}
\endbibitem

\bibitem[\protect\citeauthoryear{Narayan and Allen}{2016}]{narayan2016mixed}
\begin{barticle}[author]
\bauthor{\bsnm{Narayan},~\bfnm{Manjari}\binits{M.}} \AND
  \bauthor{\bsnm{Allen},~\bfnm{Genevera~I}\binits{G.~I.}}
(\byear{2016}).
\btitle{Mixed effects models for resampled network statistics improves
  statistical power to find differences in multi-subject functional
  connectivity}.
\bjournal{Frontiers in Neuroscience}
\bvolume{10}.
\end{barticle}
\endbibitem

\bibitem[\protect\citeauthoryear{Nicosia and
  Latora}{2015}]{nicosia2015measuring}
\begin{barticle}[author]
\bauthor{\bsnm{Nicosia},~\bfnm{Vincenzo}\binits{V.}} \AND
  \bauthor{\bsnm{Latora},~\bfnm{Vito}\binits{V.}}
(\byear{2015}).
\btitle{Measuring and modeling correlations in multiplex networks}.
\bjournal{Physical Review E}
\bvolume{92}
\bpages{032805}.
\end{barticle}
\endbibitem

\bibitem[\protect\citeauthoryear{Papadimitriou}{2003}]{papadimitriou2003computational}
\begin{bbook}[author]
\bauthor{\bsnm{Papadimitriou},~\bfnm{Christos~H}\binits{C.~H.}}
(\byear{2003}).
\btitle{Computational complexity}.
\bpublisher{John Wiley and Sons Ltd.}
\end{bbook}
\endbibitem

\bibitem[\protect\citeauthoryear{Paul and Chen}{2016a}]{pc15}
\begin{barticle}[author]
\bauthor{\bsnm{Paul},~\bfnm{Subhadeep}\binits{S.}} \AND
  \bauthor{\bsnm{Chen},~\bfnm{Yuguo}\binits{Y.}}
(\byear{2016}a).
\btitle{Consistent community detection in multi-relational data through
  restricted multi-layer stochastic blockmodel}.
\bjournal{Electronic Journal of Statistics}
\bvolume{10}
\bpages{3807--3870}.
\end{barticle}
\endbibitem

\bibitem[\protect\citeauthoryear{Paul and Chen}{2016b}]{paul2016null}
\begin{barticle}[author]
\bauthor{\bsnm{Paul},~\bfnm{Subhadeep}\binits{S.}} \AND
  \bauthor{\bsnm{Chen},~\bfnm{Yuguo}\binits{Y.}}
(\byear{2016}b).
\btitle{Null Models and Modularity Based Community Detection in Multi-Layer
  Networks}.
\bjournal{arXiv preprint arXiv:1608.00623}.
\end{barticle}
\endbibitem

\bibitem[\protect\citeauthoryear{Paul and Chen}{2016c}]{pc16}
\begin{barticle}[author]
\bauthor{\bsnm{Paul},~\bfnm{Subhadeep}\binits{S.}} \AND
  \bauthor{\bsnm{Chen},~\bfnm{Yuguo}\binits{Y.}}
(\byear{2016}c).
\btitle{Orthogonal symmetric non-negative matrix factorization under the
  stochastic block model}.
\bjournal{arXiv preprint arXiv:1605.05349}.
\end{barticle}
\endbibitem

\bibitem[\protect\citeauthoryear{Paul and Chen}{2020}]{paul2017spectral}
\begin{barticle}[author]
\bauthor{\bsnm{Paul},~\bfnm{Subhadeep}\binits{S.}} \AND
  \bauthor{\bsnm{Chen},~\bfnm{Yuguo}\binits{Y.}}
(\byear{2020}).
\btitle{Spectral and matrix factorization methods for consistent community
  detection in multi-layer networks}.
\bjournal{The Annals of Statistics}
\bvolume{48}
\bpages{230--250}.
\end{barticle}
\endbibitem

\bibitem[\protect\citeauthoryear{Peixoto}{2015}]{peixoto15}
\begin{barticle}[author]
\bauthor{\bsnm{Peixoto},~\bfnm{Tiago~P}\binits{T.~P.}}
(\byear{2015}).
\btitle{Inferring the mesoscale structure of layered, edge-valued, and
  time-varying networks}.
\bjournal{Physical Review E}
\bvolume{92}
\bpages{042807}.
\end{barticle}
\endbibitem

\bibitem[\protect\citeauthoryear{Penny et~al.}{2011}]{penny2011statistical}
\begin{bbook}[author]
\bauthor{\bsnm{Penny},~\bfnm{William~D}\binits{W.~D.}},
  \bauthor{\bsnm{Friston},~\bfnm{Karl~J}\binits{K.~J.}},
  \bauthor{\bsnm{Ashburner},~\bfnm{John~T}\binits{J.~T.}},
  \bauthor{\bsnm{Kiebel},~\bfnm{Stefan~J}\binits{S.~J.}} \AND
  \bauthor{\bsnm{Nichols},~\bfnm{Thomas~E}\binits{T.~E.}}
(\byear{2011}).
\btitle{Statistical Parametric Mapping: the Analysis of Functional Brain
  Images}.
\bpublisher{Academic press}.
\end{bbook}
\endbibitem

\bibitem[\protect\citeauthoryear{Percival and
  Walden}{2006}]{percival2006wavelet}
\begin{bbook}[author]
\bauthor{\bsnm{Percival},~\bfnm{Donald~B}\binits{D.~B.}} \AND
  \bauthor{\bsnm{Walden},~\bfnm{Andrew~T}\binits{A.~T.}}
(\byear{2006}).
\btitle{Wavelet Methods for Time Series Analysis}
\bvolume{4}.
\bpublisher{Cambridge University Press}.
\end{bbook}
\endbibitem

\bibitem[\protect\citeauthoryear{Power et~al.}{2011}]{power2011functional}
\begin{barticle}[author]
\bauthor{\bsnm{Power},~\bfnm{Jonathan~D}\binits{J.~D.}},
  \bauthor{\bsnm{Cohen},~\bfnm{Alexander~L}\binits{A.~L.}},
  \bauthor{\bsnm{Nelson},~\bfnm{Steven~M}\binits{S.~M.}},
  \bauthor{\bsnm{Wig},~\bfnm{Gagan~S}\binits{G.~S.}},
  \bauthor{\bsnm{Barnes},~\bfnm{Kelly~Anne}\binits{K.~A.}},
  \bauthor{\bsnm{Church},~\bfnm{Jessica~A}\binits{J.~A.}},
  \bauthor{\bsnm{Vogel},~\bfnm{Alecia~C}\binits{A.~C.}},
  \bauthor{\bsnm{Laumann},~\bfnm{Timothy~O}\binits{T.~O.}},
  \bauthor{\bsnm{Miezin},~\bfnm{Fran~M}\binits{F.~M.}} \AND
  \bauthor{\bsnm{Schlaggar},~\bfnm{Bradley~L}\binits{B.~L.}}
(\byear{2011}).
\btitle{Functional network organization of the human brain}.
\bjournal{Neuron}
\bvolume{72}
\bpages{665--678}.
\end{barticle}
\endbibitem

\bibitem[\protect\citeauthoryear{Qin and Rohe}{2013}]{qr13}
\begin{binproceedings}[author]
\bauthor{\bsnm{Qin},~\bfnm{Tai}\binits{T.}} \AND
  \bauthor{\bsnm{Rohe},~\bfnm{Karl}\binits{K.}}
(\byear{2013}).
\btitle{Regularized spectral clustering under the degree-corrected stochastic
  blockmodel}.
In \bbooktitle{Advances in Neural Information Processing Systems}
\bpages{3120--3128}.
\end{binproceedings}
\endbibitem

\bibitem[\protect\citeauthoryear{Reichardt and
  Bornholdt}{2006}]{reichardt2006statistical}
\begin{barticle}[author]
\bauthor{\bsnm{Reichardt},~\bfnm{J{\"o}rg}\binits{J.}} \AND
  \bauthor{\bsnm{Bornholdt},~\bfnm{Stefan}\binits{S.}}
(\byear{2006}).
\btitle{Statistical mechanics of community detection}.
\bjournal{Physical Review E}
\bvolume{74}
\bpages{016110}.
\end{barticle}
\endbibitem

\bibitem[\protect\citeauthoryear{Reyes and
  Rodriguez}{2016}]{reyes2016stochastic}
\begin{barticle}[author]
\bauthor{\bsnm{Reyes},~\bfnm{Perla}\binits{P.}} \AND
  \bauthor{\bsnm{Rodriguez},~\bfnm{Abel}\binits{A.}}
(\byear{2016}).
\btitle{Stochastic blockmodels for exchangeable collections of networks}.
\bjournal{arXiv preprint arXiv:1606.05277}.
\end{barticle}
\endbibitem

\bibitem[\protect\citeauthoryear{Rohe, Chatterjee and Yu}{2011}]{rcy11}
\begin{barticle}[author]
\bauthor{\bsnm{Rohe},~\bfnm{K.}\binits{K.}},
  \bauthor{\bsnm{Chatterjee},~\bfnm{S.}\binits{S.}} \AND
  \bauthor{\bsnm{Yu},~\bfnm{B.}\binits{B.}}
(\byear{2011}).
\btitle{Spectral clustering and the high-dimensional stochastic blockmodel}.
\bjournal{The Annals of Statistics}
\bvolume{39}
\bpages{1878-1915}.
\end{barticle}
\endbibitem

\bibitem[\protect\citeauthoryear{Rubinov and Sporns}{2010}]{rubinov10}
\begin{barticle}[author]
\bauthor{\bsnm{Rubinov},~\bfnm{Mikail}\binits{M.}} \AND
  \bauthor{\bsnm{Sporns},~\bfnm{Olaf}\binits{O.}}
(\byear{2010}).
\btitle{Complex network measures of brain connectivity: Uses and
  interpretations}.
\bjournal{Neuroimage}
\bvolume{52}
\bpages{1059--1069}.
\end{barticle}
\endbibitem

\bibitem[\protect\citeauthoryear{Simpson, Bowman and
  Laurienti}{2013}]{simpson13}
\begin{barticle}[author]
\bauthor{\bsnm{Simpson},~\bfnm{Sean~L}\binits{S.~L.}},
  \bauthor{\bsnm{Bowman},~\bfnm{F~DuBois}\binits{F.~D.}} \AND
  \bauthor{\bsnm{Laurienti},~\bfnm{Paul~J}\binits{P.~J.}}
(\byear{2013}).
\btitle{Analyzing complex functional brain networks: fusing statistics and
  network science to understand the brain}.
\bjournal{Statistics Surveys}
\bvolume{7}
\bpages{1}.
\end{barticle}
\endbibitem

\bibitem[\protect\citeauthoryear{Simpson et~al.}{2013}]{simpson13perm}
\begin{barticle}[author]
\bauthor{\bsnm{Simpson},~\bfnm{Sean~L}\binits{S.~L.}},
  \bauthor{\bsnm{Lyday},~\bfnm{Robert~G}\binits{R.~G.}},
  \bauthor{\bsnm{Hayasaka},~\bfnm{Satoru}\binits{S.}},
  \bauthor{\bsnm{Marsh},~\bfnm{Anthony~P}\binits{A.~P.}} \AND
  \bauthor{\bsnm{Laurienti},~\bfnm{Paul~J}\binits{P.~J.}}
(\byear{2013}).
\btitle{A permutation testing framework to compare groups of brain networks}.
\bjournal{Frontiers in Computational Neuroscience}
\bvolume{7}
\bpages{171}.
\end{barticle}
\endbibitem

\bibitem[\protect\citeauthoryear{Sporns}{2014}]{sporns2014contributions}
\begin{barticle}[author]
\bauthor{\bsnm{Sporns},~\bfnm{Olaf}\binits{O.}}
(\byear{2014}).
\btitle{Contributions and challenges for network models in cognitive
  neuroscience}.
\bjournal{Nature Neuroscience}
\bvolume{17}
\bpages{652--660}.
\end{barticle}
\endbibitem

\bibitem[\protect\citeauthoryear{Stam}{2014}]{stam2014modern}
\begin{barticle}[author]
\bauthor{\bsnm{Stam},~\bfnm{Cornelis~J}\binits{C.~J.}}
(\byear{2014}).
\btitle{Modern network science of neurological disorders}.
\bjournal{Nature Reviews Neuroscience}
\bvolume{15}
\bpages{683--695}.
\end{barticle}
\endbibitem

\bibitem[\protect\citeauthoryear{Stanley et~al.}{2016}]{stanley15}
\begin{barticle}[author]
\bauthor{\bsnm{Stanley},~\bfnm{Natalie}\binits{N.}},
  \bauthor{\bsnm{Shai},~\bfnm{Saray}\binits{S.}},
  \bauthor{\bsnm{Taylor},~\bfnm{Dane}\binits{D.}} \AND
  \bauthor{\bsnm{Mucha},~\bfnm{Peter~J}\binits{P.~J.}}
(\byear{2016}).
\btitle{Clustering network layers with the strata multilayer stochastic block
  model}.
\bjournal{IEEE Transactions on Network Science and Engineering}
\bvolume{3}
\bpages{95--105}.
\end{barticle}
\endbibitem

\bibitem[\protect\citeauthoryear{Steen et~al.}{2011}]{steen2011assessing}
\begin{barticle}[author]
\bauthor{\bsnm{Steen},~\bfnm{Matthew}\binits{M.}},
  \bauthor{\bsnm{Hayasaka},~\bfnm{Satoru}\binits{S.}},
  \bauthor{\bsnm{Joyce},~\bfnm{Karen}\binits{K.}} \AND
  \bauthor{\bsnm{Laurienti},~\bfnm{Paul}\binits{P.}}
(\byear{2011}).
\btitle{Assessing the consistency of community structure in complex networks}.
\bjournal{Physical Review E}
\bvolume{84}
\bpages{016111}.
\end{barticle}
\endbibitem

\bibitem[\protect\citeauthoryear{Stevens et~al.}{2012}]{stevens12}
\begin{barticle}[author]
\bauthor{\bsnm{Stevens},~\bfnm{Alexander~A}\binits{A.~A.}},
  \bauthor{\bsnm{Tappon},~\bfnm{Sarah~C}\binits{S.~C.}},
  \bauthor{\bsnm{Garg},~\bfnm{Arun}\binits{A.}} \AND
  \bauthor{\bsnm{Fair},~\bfnm{Damien~A}\binits{D.~A.}}
(\byear{2012}).
\btitle{Functional brain network modularity captures inter- and
  intra-individual variation in working memory capacity}.
\bjournal{PLoS ONE}
\bvolume{7}
\bpages{e30468}.
\end{barticle}
\endbibitem

\bibitem[\protect\citeauthoryear{Stewart and Sun}{1990}]{stewart}
\begin{bbook}[author]
\bauthor{\bsnm{Stewart},~\bfnm{Gilbert~W}\binits{G.~W.}} \AND
  \bauthor{\bsnm{Sun},~\bfnm{Ji-Guang}\binits{J.-G.}}
(\byear{1990}).
\btitle{Matrix Perturbation Theory}.
\bpublisher{Academic Press, Boston, MA.}
\end{bbook}
\endbibitem

\bibitem[\protect\citeauthoryear{Sweet, Thomas and
  Junker}{2014}]{sweet2014hierarchical}
\begin{barticle}[author]
\bauthor{\bsnm{Sweet},~\bfnm{Tracy~M}\binits{T.~M.}},
  \bauthor{\bsnm{Thomas},~\bfnm{Andrew~C}\binits{A.~C.}} \AND
  \bauthor{\bsnm{Junker},~\bfnm{Brian~W}\binits{B.~W.}}
(\byear{2014}).
\btitle{Hierarchical mixed membership stochastic blockmodels for multiple
  networks and experimental interventions}.
\bjournal{Handbook on Mixed Membership Models and Their Applications}
\bpages{463--488}.
\end{barticle}
\endbibitem

\bibitem[\protect\citeauthoryear{Tang, Lu and Dhillon}{2009}]{tang09}
\begin{binproceedings}[author]
\bauthor{\bsnm{Tang},~\bfnm{Wei}\binits{W.}},
  \bauthor{\bsnm{Lu},~\bfnm{Zhengdong}\binits{Z.}} \AND
  \bauthor{\bsnm{Dhillon},~\bfnm{Inderjit~S}\binits{I.~S.}}
(\byear{2009}).
\btitle{Clustering with multiple graphs}.
In \bbooktitle{Proceedings of the Ninth IEEE International Conference on Data
  Mining}
\bpages{1016--1021}.
\bpublisher{IEEE}.
\end{binproceedings}
\endbibitem

\bibitem[\protect\citeauthoryear{Tzourio-Mazoyer
  et~al.}{2002}]{tzourio2002automated}
\begin{barticle}[author]
\bauthor{\bsnm{Tzourio-Mazoyer},~\bfnm{Nathalie}\binits{N.}},
  \bauthor{\bsnm{Landeau},~\bfnm{Brigitte}\binits{B.}},
  \bauthor{\bsnm{Papathanassiou},~\bfnm{Dimitri}\binits{D.}},
  \bauthor{\bsnm{Crivello},~\bfnm{Fabrice}\binits{F.}},
  \bauthor{\bsnm{Etard},~\bfnm{Olivier}\binits{O.}},
  \bauthor{\bsnm{Delcroix},~\bfnm{Nicolas}\binits{N.}},
  \bauthor{\bsnm{Mazoyer},~\bfnm{Bernard}\binits{B.}} \AND
  \bauthor{\bsnm{Joliot},~\bfnm{Marc}\binits{M.}}
(\byear{2002}).
\btitle{Automated anatomical labeling of activations in {SPM} using a
  macroscopic anatomical parcellation of the {MNI MRI} single-subject brain}.
\bjournal{Neuroimage}
\bvolume{15}
\bpages{273--289}.
\end{barticle}
\endbibitem

\bibitem[\protect\citeauthoryear{Valles-Catala et~al.}{2016}]{valles14}
\begin{barticle}[author]
\bauthor{\bsnm{Valles-Catala},~\bfnm{Toni}\binits{T.}},
  \bauthor{\bsnm{Massucci},~\bfnm{Francesco~A}\binits{F.~A.}},
  \bauthor{\bsnm{Guimera},~\bfnm{Roger}\binits{R.}} \AND
  \bauthor{\bsnm{Sales-Pardo},~\bfnm{Marta}\binits{M.}}
(\byear{2016}).
\btitle{Multilayer stochastic block models reveal the multilayer structure of
  complex networks}.
\bjournal{Physical Review X}
\bvolume{6}
\bpages{011036}.
\end{barticle}
\endbibitem

\bibitem[\protect\citeauthoryear{Van Den~Heuvel and Pol}{2010}]{van10}
\begin{barticle}[author]
\bauthor{\bsnm{Van Den~Heuvel},~\bfnm{Martijn~P}\binits{M.~P.}} \AND
  \bauthor{\bsnm{Pol},~\bfnm{Hilleke E~Hulshoff}\binits{H.~E.~H.}}
(\byear{2010}).
\btitle{Exploring the brain network: a review on resting-state f{MRI}
  functional connectivity}.
\bjournal{European Neuropsychopharmacology}
\bvolume{20}
\bpages{519--534}.
\end{barticle}
\endbibitem

\bibitem[\protect\citeauthoryear{van~den Heuvel et~al.}{2010}]{van2010aberrant}
\begin{barticle}[author]
\bauthor{\bparticle{van~den} \bsnm{Heuvel},~\bfnm{Martijn~P}\binits{M.~P.}},
  \bauthor{\bsnm{Mandl},~\bfnm{Ren{\'e}~CW}\binits{R.~C.}},
  \bauthor{\bsnm{Stam},~\bfnm{Cornelis~J}\binits{C.~J.}},
  \bauthor{\bsnm{Kahn},~\bfnm{Ren{\'e}~S}\binits{R.~S.}} \AND
  \bauthor{\bsnm{Pol},~\bfnm{Hilleke E~Hulshoff}\binits{H.~E.~H.}}
(\byear{2010}).
\btitle{Aberrant frontal and temporal complex network structure in
  schizophrenia: a graph theoretical analysis}.
\bjournal{The Journal of Neuroscience}
\bvolume{30}
\bpages{15915--15926}.
\end{barticle}
\endbibitem

\bibitem[\protect\citeauthoryear{van~den Heuvel et~al.}{2013}]{van2013abnormal}
\begin{barticle}[author]
\bauthor{\bparticle{van~den} \bsnm{Heuvel},~\bfnm{Martijn~P}\binits{M.~P.}},
  \bauthor{\bsnm{Sporns},~\bfnm{Olaf}\binits{O.}},
  \bauthor{\bsnm{Collin},~\bfnm{Guusje}\binits{G.}},
  \bauthor{\bsnm{Scheewe},~\bfnm{Thomas}\binits{T.}},
  \bauthor{\bsnm{Mandl},~\bfnm{Ren{\'e}~CW}\binits{R.~C.}},
  \bauthor{\bsnm{Cahn},~\bfnm{Wiepke}\binits{W.}},
  \bauthor{\bsnm{Go{\~n}i},~\bfnm{Joaqu{\'\i}n}\binits{J.}},
  \bauthor{\bsnm{Pol},~\bfnm{Hilleke E~Hulshoff}\binits{H.~E.~H.}} \AND
  \bauthor{\bsnm{Kahn},~\bfnm{Ren{\'e}~S}\binits{R.~S.}}
(\byear{2013}).
\btitle{Abnormal rich club organization and functional brain dynamics in
  schizophrenia}.
\bjournal{JAMA Psychiatry}
\bvolume{70}
\bpages{783--792}.
\end{barticle}
\endbibitem

\bibitem[\protect\citeauthoryear{Wang et~al.}{2013}]{wang13}
\begin{barticle}[author]
\bauthor{\bsnm{Wang},~\bfnm{Jinhui}\binits{J.}},
  \bauthor{\bsnm{Zuo},~\bfnm{Xinian}\binits{X.}},
  \bauthor{\bsnm{Dai},~\bfnm{Zhengjia}\binits{Z.}},
  \bauthor{\bsnm{Xia},~\bfnm{Mingrui}\binits{M.}},
  \bauthor{\bsnm{Zhao},~\bfnm{Zhilian}\binits{Z.}},
  \bauthor{\bsnm{Zhao},~\bfnm{Xiaoling}\binits{X.}},
  \bauthor{\bsnm{Jia},~\bfnm{Jianping}\binits{J.}},
  \bauthor{\bsnm{Han},~\bfnm{Ying}\binits{Y.}} \AND
  \bauthor{\bsnm{He},~\bfnm{Yong}\binits{Y.}}
(\byear{2013}).
\btitle{Disrupted functional brain connectome in individuals at risk for
  {A}lzheimer's disease}.
\bjournal{Biological Psychiatry}
\bvolume{73}
\bpages{472--481}.
\end{barticle}
\endbibitem

\bibitem[\protect\citeauthoryear{Weber et~al.}{2013}]{weber13}
\begin{barticle}[author]
\bauthor{\bsnm{Weber},~\bfnm{Matthew~J}\binits{M.~J.}},
  \bauthor{\bsnm{Detre},~\bfnm{John~A}\binits{J.~A.}},
  \bauthor{\bsnm{Thompson-Schill},~\bfnm{Sharon~L}\binits{S.~L.}} \AND
  \bauthor{\bsnm{Avants},~\bfnm{Brian~B}\binits{B.~B.}}
(\byear{2013}).
\btitle{Reproducibility of functional network metrics and network structure: a
  comparison of task-related {BOLD}, resting {ASL} with {BOLD} contrast, and
  resting cerebral blood flow}.
\bjournal{Cognitive, Affective, \& Behavioral Neuroscience}
\bvolume{13}
\bpages{627--640}.
\end{barticle}
\endbibitem

\bibitem[\protect\citeauthoryear{Whitfield-Gabrieli and
  Nieto-Castanon}{2012}]{whitfield2012conn}
\begin{barticle}[author]
\bauthor{\bsnm{Whitfield-Gabrieli},~\bfnm{Susan}\binits{S.}} \AND
  \bauthor{\bsnm{Nieto-Castanon},~\bfnm{Alfonso}\binits{A.}}
(\byear{2012}).
\btitle{{CONN}: a functional connectivity toolbox for correlated and
  anticorrelated brain networks}.
\bjournal{Brain Connectivity}
\bvolume{2}
\bpages{125--141}.
\end{barticle}
\endbibitem

\bibitem[\protect\citeauthoryear{Xia, Wang and He}{2013}]{xia2013brainnet}
\begin{barticle}[author]
\bauthor{\bsnm{Xia},~\bfnm{Mingrui}\binits{M.}},
  \bauthor{\bsnm{Wang},~\bfnm{Jinhui}\binits{J.}} \AND
  \bauthor{\bsnm{He},~\bfnm{Yong}\binits{Y.}}
(\byear{2013}).
\btitle{BrainNet Viewer: a network visualization tool for human brain
  connectomics}.
\bjournal{PLoS ONE}
\bvolume{8}
\bpages{e68910}.
\end{barticle}
\endbibitem

\bibitem[\protect\citeauthoryear{Yu et~al.}{2012a}]{yu2012brain}
\begin{barticle}[author]
\bauthor{\bsnm{Yu},~\bfnm{Qingbao}\binits{Q.}},
  \bauthor{\bsnm{A~Allen},~\bfnm{Elena}\binits{E.}},
  \bauthor{\bsnm{Sui},~\bfnm{Jing}\binits{J.}},
  \bauthor{\bsnm{R~Arbabshirani},~\bfnm{Mohammad}\binits{M.}},
  \bauthor{\bsnm{Pearlson},~\bfnm{Godfrey}\binits{G.}} \AND
  \bauthor{\bsnm{D~Calhoun},~\bfnm{Vince}\binits{V.}}
(\byear{2012}a).
\btitle{Brain connectivity networks in schizophrenia underlying resting state
  functional magnetic resonance imaging}.
\bjournal{Current Topics in Medicinal Chemistry}
\bvolume{12}
\bpages{2415--2425}.
\end{barticle}
\endbibitem

\bibitem[\protect\citeauthoryear{Yu et~al.}{2012b}]{yu2012modular}
\begin{barticle}[author]
\bauthor{\bsnm{Yu},~\bfnm{Qingbao}\binits{Q.}},
  \bauthor{\bsnm{Plis},~\bfnm{Sergey~M}\binits{S.~M.}},
  \bauthor{\bsnm{Erhardt},~\bfnm{Erik~B}\binits{E.~B.}},
  \bauthor{\bsnm{Allen},~\bfnm{Elena~A}\binits{E.~A.}},
  \bauthor{\bsnm{Sui},~\bfnm{Jing}\binits{J.}},
  \bauthor{\bsnm{Kiehl},~\bfnm{Kent~A}\binits{K.~A.}},
  \bauthor{\bsnm{Pearlson},~\bfnm{Godfrey}\binits{G.}} \AND
  \bauthor{\bsnm{Calhoun},~\bfnm{Vince~D}\binits{V.~D.}}
(\byear{2012}b).
\btitle{Modular organization of functional network connectivity in healthy
  controls and patients with schizophrenia during the resting state}.
\bjournal{Frontiers in Systems Neuroscience}
\bvolume{5}
\bpages{103}.
\end{barticle}
\endbibitem

\bibitem[\protect\citeauthoryear{Zalesky, Fornito and
  Bullmore}{2010}]{zalesky2010network}
\begin{barticle}[author]
\bauthor{\bsnm{Zalesky},~\bfnm{Andrew}\binits{A.}},
  \bauthor{\bsnm{Fornito},~\bfnm{Alex}\binits{A.}} \AND
  \bauthor{\bsnm{Bullmore},~\bfnm{Edward~T}\binits{E.~T.}}
(\byear{2010}).
\btitle{Network-based statistic: identifying differences in brain networks}.
\bjournal{Neuroimage}
\bvolume{53}
\bpages{1197--1207}.
\end{barticle}
\endbibitem

\end{thebibliography}

\end{document}